\documentclass[12pt]{article}
 \usepackage{epsfig}
 \usepackage {graphicx}
 \usepackage[latin1]{inputenc}
 \usepackage{amsfonts}
 \usepackage{amsthm}
 \usepackage{amssymb, latexsym, amsmath}
 \newcommand{\inc}{{\it i}}

 \newcommand{\be}{\begin{equation}}
 \newcommand{\ee}{\end{equation}}
 \newcommand{\ba}{\begin{eqnarray}}
 \newcommand{\ea}{\end{eqnarray}}

 \newcommand{\erbold}{\mbox{{\boldmath $\vec r$}}}
 \newcommand{\rbold}{\mbox{{\boldmath $\vec r$}}}

 \newcommand{\omegabold}{\mbox{{\boldmath $
 \vec
 {\omega}$}}}
 \newcommand{\dotomegabold}{\mbox{{\boldmath $\dot{
 \vec
 {\omega}}$}}}

 \newcommand{\xbold}{\mbox{{\boldmath $\vec x$}}}

 \newcommand{\ubold}{\mbox{{\boldmath $\vec u$}}}

 \newcommand{\ddotubold}{\stackrel{\bf{\centerdot\,\centerdot}}{\textbf {\mbox{\boldmath $\vec{\boldmath u}$}}}}
 \newcommand{\dotubold}{\stackrel{\bf{\centerdot}}{\textbf {\mbox{\boldmath $\vec{\boldmath u}$}}}}

 \newcommand{\Fbold}{\mbox{\boldmath $\vec{\boldmath{F}}$}}
 
 \newcommand{\Vbold}{\mbox{\boldmath $\vec {\boldmath{\,V}}$}}

  \newcommand{\eRbold}{\mbox{{\boldmath $\vec{R}$}}}
  \newcommand{\Rbold}{\mbox{{\boldmath $\vec{R}$}}}

 \newcommand{\vbold}{\bf{v}}
  
\begin{document}
 \title{
                 ${{~~~~~~~~~}^{^{^{
 Extended~version~of~a~paper~published~in }}}}$~~~~~~~~~~~~~~~~~~~~~~~~~~~~~~~\\ ${{~~}^{^{^{
                 ``Celestial~Mechanics~\&~Dynamical~Astronomy"\,,~\,Vol.~112\,,~~pp.~283~-~330~~~(March\,~2012)
                  }}}}$
                  ~
 {\Large{\textbf{Bodily tides near spin-orbit resonances
 \\}
            }}}
 \author{
  {\Large{Michael Efroimsky}}\\
 {\small{US Naval Observatory, Washington DC 20392 USA}}\\
 {\small{e-mail: ~michael.efroimsky @ usno.navy.mil~}}
 }
 \date{}

 \maketitle
 \begin{abstract}

 Spin-orbit coupling can be described in two approaches. The first method, known as the {\emph{``MacDonald torque"}},
 is often combined with a convenient assumption that the quality factor $\,Q\,$ is frequency-independent. This makes the method
 inconsistent, for the MacDonald theory tacitly fixes the rheology of the mantle by making $\,Q\,$ scale as the inverse tidal
 frequency.

 Spin-orbit coupling can be treated also in an approach called {\emph{``the Darwin torque"}}. While this theory is general enough to
 accommodate an arbitrary frequency-dependence of $\,Q\,$, this advantage has not yet been fully exploited in the literature, where
 $\,Q\,$ is often assumed constant
 or is set to scale as inverse tidal frequency,
 the latter assertion making the Darwin torque equivalent to a corrected version of the  MacDonald torque.

 However neither a constant nor an inverse-frequency $Q$ reflect the properties of realistic mantles and crusts, because the
 actual frequency-dependence is more complex. Hence it is necessary to enrich the theory of spin-orbit interaction with the right
 frequency-dependence.

 We accomplish this programme for the Darwin-torque-based model near resonances. We derive the frequency-dependence of the tidal torque
 from the first principles of solid-state mechanics, i.e., from the expression for the mantle's compliance in the time domain. We also
 explain that the tidal torque includes not only the customary, secular part, but also an oscillating part.

 We demonstrate that the $\,lmpq\,$ term of the Darwin-Kaula expansion for the tidal torque smoothly passes zero, when the
 secondary traverses the $\,lmpq\,$ resonance (e.g., the principal tidal torque smoothly goes through nil as the secondary crosses the
 synchronous orbit).

 Thus we prepare a foundation for modeling entrapment of a despinning primary into a resonance with its secondary.
 The roles of the primary and secondary may be played, e.g., by Mercury and the Sun, correspondingly, or by an icy moon
 and a Jovian planet.

 We also offer a possible explanation for the unexpected frequency-dependence of the tidal dissipation rate in the Moon,
 discovered by LLR.

 \end{abstract}

  \section{Introduction}

 We continue a critical examination of the tidal-torque techniques, begun in Efroimsky \& Williams (2009), where the empirical
 treatment by 
 and MacDonald (1964) was considered from the viewpoint of a more general and rigorous approach by
 Darwin (1879, 1880) and Kaula (1964).
 Referring the Reader to Efroimsky \& Williams (2009) for proofs and comments, we begin with an inventory of the key
 formulae describing the spin-orbit interaction. While in {\it{Ibid.}} we employed those formulae to explore tidal despinning well
 outside the 1:1 resonance (and in neglect of the intermediate resonances), in the current paper we apply this machinery to the
 case of despinning in the vicinity of a spin-orbit resonance.

 Although the topic has been popular since mid-sixties and has already been addressed in books, the common models are not entirely
 adequate to the actual physics. Just as in the nonresonant case discussed in {\it{Ibid.}}, a
 generic problem with the popular models of libration or of capture into a resonance is that they employ wrong rheologies
 (the work by Rambaux et al. 2010 being the only exception we know of).
 Above that, the model based on the MacDonald torque suffers a defect stemming from a genuine inconsistency inherent in the theory by
 MacDonald (1964). 

 As explained in Efroimsky and Williams (2009) and Williams and Efroimsky (2012), the MacDonald theory, both in its original and
 corrected versions, tacitly fixes an unphysical shape of the functional dependence $\,Q(\chi)\,$, where $\,Q\,$ is the dissipation
 quality factor and $\,\chi\,$ is the
 tidal frequency (Williams \& Efroimsky 2012). So we base our approach on the developments by Darwin (1879,
 1880) and Kaula (1964), combining those with a realistic law of frequency-dependence of the damping rate.

 Since our main purpose is to lay the groundwork for the subsequent study of the process of falling into a resonance, the two
 principal results obtained in this paper are the following:

  (a) Starting with the realistic rheological model (the expression for the compliance in the time domain), we derive the complex Love
 numbers $\,\bar{k}_{\textstyle{_l}}\,$ as functions of the frequency $\,\chi\,$, and write down their negative imaginary parts as functions
 of the frequency: $~\,-\,{\cal{I}}{\it{m}}\left[\,\bar{k}_{\textstyle{_l}}(\chi)\,\right]\,=\,|k_{\textstyle{_l}}(\chi)|~\sin\epsilon_{
 \textstyle{_l}}(\chi)\,$. It is these expressions that appear as factors in the terms of the Darwin-Kaula expansion of tides. These
 factors' frequency-dependencies demonstrate a nontrivial shape, especially near resonances. This shape plays a crucial role in modeling
 of despinning in general, specifically in modeling the process of falling into a spin-orbit resonance.

 (b) We demonstrate that, beside the customary secular part, the Darwin torque contains a usually omitted oscillating part.

 \section{Linear bodily tides
 }

 Linearity of tide means that: (a) under a static load, deformation scales linearly, and (b) under undulatory loading, the same linear
 law applies, separately, to each frequency mode. The latter implies that the deformation magnitude at a certain
 frequency should depend linearly upon the tidal stress at this frequency, and should bear no dependence upon loading at other tidal
 modes. Thence the dissipation rate at that frequency will depend on the stress at that frequency only.

 \subsection{Linearity of the tidal deformation}

 At a point $\Rbold = (R,\lambda,\phi)$, the potential due to a tide-raising secondary
 of mass $M^*_{sec}$, located at $\,{\erbold}^{\;*} = (r^*,\,\lambda^*,\,\phi^*)\,$ with
 $\,r^*\geq R\,$, is expandable over the Legendre polynomials $\,P_{\it l}(\cos\gamma)\;$:
 \ba
 \nonumber
 W(\eRbold\,,\,\erbold^{~*})&=&\sum_{{\it{l}}=2}^{\infty}~W_{\it{l}}(\eRbold\,,~\erbold^{~*})~=~-~\frac{G\;M^*_{sec}}{r^{
 \,*}}~\sum_{{\it{l}}=2}^{\infty}\,\left(\,\frac{R}{r^{~*}}\,\right)^{\textstyle{^{\it{l}}}}\,P_{\it{l}}(\cos \gamma)~~~~\\
 \nonumber\\
 &=&-\,\frac{G~M^*_{sec}}{r^{\,*}}\sum_{{\it{l}}=2}^{\infty}\left(\frac{R}{r^{~*}}\right)^{\textstyle{^{\it{l}}}}\sum_{m=0}^{\it l}
 \frac{({\it l}-m)!}{({\it l}+m)!}(2-\delta_{0m})P_{{\it{l}}m}(\sin\phi)P_{{\it{l}}m}(\sin\phi^*)~\cos m(\lambda-\lambda^*)~~,\quad\,
 \quad
 \label{1}
 \label{L1}
 \ea
 where $G=6.7\times10^{-11}\,\mbox{m}^3\,\mbox{kg}^{-1}\mbox{s}^{-2}\,$ is Newton's gravity constant, and $\gamma\,$ is the angular
 separation between the vectors ${\erbold}^{\;*}$ and $\Rbold$ pointing from the primary's centre. The latitudes $\phi,\,\phi^*$
 are reckoned from the primary's equator, while the longitudes $\lambda,\,\lambda^*$ are reckoned from a fixed meridian.
 ~\\
 $\left.\quad\right.$ Under the assumption of linearity, the $\,{\emph{l}}^{~th}$ term $\,W_{\it{l}}(\eRbold\,,~\erbold^{~*})\,$ in the
 secondary's potential causes a linear deformation of the primary's shape. The subsequent adjustment of the primary's potential being
 linear in the said deformation, the $\,{\emph{l}}^{~th}$ adjustment $\,U_{\it{l}}\,$ of the
 primary's potential is proportional to $\,W_{\it{l}}\,$. The theory of potential requires $\,U_{\it{l}}(\erbold)\,$ to fall
 off, outside the primary, as $\,r^{-(\it{l}+1)}\,$. Thus the overall amendment to the potential of the primary amounts to:
 \ba
 U(\erbold)~=~\sum_{{\it l}=2}^{\infty}~U_{\it{l}}(\erbold)~=~\sum_{{\it l}=2}^{\infty}~k_{\it
 l}\;\left(\,\frac{R}{r}\,\right)^{{\it l}+1}\;W_{\it{l}}(\eRbold\,,\;\erbold^{\;*})~~~,~~~~~~~
 ~~~~~~~~~~~~~~~~
 \label{2}
 \label{L2}
 \ea
 $R\,$ now being the mean equatorial
 radius of the primary, $\,\eRbold\,=\,(R\,,\,\phi\,,\,\lambda)\,$ being a surface point,
 $\,\erbold\,=\,(r\,,\,\phi\,,\,\lambda)\,$ being an exterior point located above
 it at a radius $\,r\,\geq\,R\,$. The coefficients $\,k_{\it l}~$, called Love numbers, are defined by
 the primary's rheology.

 For a homogeneous incompressible spherical primary of density $\,\rho\,$, surface gravity g, and
 rigidity $\,\mu\,$, the {\emph{static}} Love number of degree $\,l\,$ is given by
 \ba
 k_{\it l}\,=\;\frac{3}{2\,({\it l}\,-\,1)}\;\,\frac{1}{1\;+\;A_{\it l}}~~~,~~~~\mbox{where}~~~~A_{\it l}\,\equiv\;\frac{\textstyle{
 (2\,{\it{l}}^{\,2}\,+\,4\,{\it{l}}\,+\,3)\,\mu}}{\textstyle{{\it{l}}\,\mbox{g}\,\rho\,R}}\;=\;\frac{\textstyle{3\;(2\,{\it{l}}^{\,2}
 \,+\,4\,{\it{l}}\,+\,3)\,\mu}}{\textstyle{4\;{\it{l}}\,\pi\,G\,\rho^2\,R^2}}~~~.~~~
 \label{A_def}
 \label{L4}
 \label{love}
 \label{L3}
 \ea
 For $\,R\,\ll\,r\,,\,r^*\,$, consideration of the $\,l=2\,$ input in (\ref{L2}) turns out to be sufficient.\footnote{
 Special is the case of Phobos, for whose orbital evolution the $k_3$ and perhaps even the
 $k_4$ terms may be relevant (Bills et al. 2005). Another class of exceptions is constituted by
 close binary asteroids. The topic is addressed by Taylor \& Margot (2010), who took into account the Love numbers up to $\,k_6\,$.}

 These formulae apply to {\emph{static}} deformations. However an actual tide is never static, except in the case
 of synchronous orbiting with a zero eccentricity and inclination.\footnote{~The case of a permanently deformed moon in a 1:1 spin-orbit
 resonance falls under this description too. Recall that in the tidal context the distorted body is taken to be the primary. So from the
 viewpoint of the satellite its host planet is orbiting the satellite synchronously, thus creating a static tide.} Hence a realistic
 perturbing potential produced by the secondary
 carries a spectrum of modes $\,\omega_{\textstyle{_{{\it{l}}mpq}}}\,$ (positive or negative) numbered with four integers
 $\,{\it{l}}mpq\,$ as in formula (\ref{9}) below. The perturbation causes a spectrum of stresses in the primary, at frequencies
 $\,\chi_{\textstyle{_{{\it{l}}mpq}}}\,=\,|\omega_{\textstyle{_{{\it{l}}mpq}}}|\,$.
 Although in a linear medium strains are generated exactly at the frequencies of the stresses, friction makes each Fourier component of the strain fall
 behind the corresponding component of the stress. Friction also reduces the magnitude of the shape response -- hence
 the deviation of a dynamical Love number $\,k_{\it l}(\chi)\,$ from its static counterpart $\,k_{\it l}\,=\,k_{\it l}(0)
 \,$. Below we shall explain that formulae (\ref{L2} - \ref{L3}) can be easily adjusted to the case of undulatory tidal
 loads in a homogeneous planet or in tidally-despinning homogeneous satellite (treated now as the primary, with its
 planet playing the role of the tide-raising secondary). However generalisation of formulae (\ref{L2} - \ref{L3}) to the
 case of a librating moon (treated as a primary) turns out to be highly nontrivial. As we shall see, the
 standard derivation by Love (1909, 1911) falls apart in the presence of the non-potential inertial
 force containing the time-derivative of the primary's angular velocity.

 The frequency-dependence of a dynamical Love numbers takes its origins in the ``inertia" of strain and,
 therefore, of the shape of the body. Hence the analogy to linear circuits: the
 $\,{\emph{l}}^{\,th}$ components of $\,W\,$ and $\,U\,$ act as a current and voltage, while the
 $\,{\emph{l}}^{\,th}$ Love number plays, up to a factor, the role of impedance. Therefore, under
 a sinusoidal load of frequency $\,\chi\,$, it is convenient to replace the actual Love number
 with its complex counterpart
 \ba
 \bar{k}_{\emph{l}\,}(\chi)\;=\;|\,\bar{k}_{\emph{l}\,}(\chi)\,|\;\exp \left[\,-\;\inc\,
 \epsilon_{\emph{l}\,}(\chi)\,\right]~~~,
 \label{complex_k}
 \label{L5}
 \ea
 $\epsilon_{\emph{l}\,}\,$ being the frequency-dependent phase delay of the reaction relative to the
 load (Munk \& MacDonald 1960, Zschau 1978).
 The ``minus" sign in (\ref{complex_k}) makes $\,U\,$ lag behind $\,W\,$
 for a positive $\,\epsilon_{\emph{l}\,}\,$. (So the situation resembles a circuit with a capacitor, where
 the current leads voltage.)

 In the limit of zero frequency, i.e., for a steady deformation, the lag should vanish, and so
 should the entire imaginary part:
 \ba
 {\cal{I}}{\it{m}}\left[\,\bar{k}_{\it l}(0)\,\right]\;=\;|\,\bar{k}_{\emph{l}\,}(0)\,|\;\sin
 \epsilon_{\emph{l}\,}(0)\;=\;0\;\;\;,
 \label{c1}
 \label{L6}
 \ea
 leaving the complex Love number real:
 \ba
 \bar{k}_{\emph{l}}(0)~=~{\cal{R}}{\it{e}}\left[\,\bar{k}_{\it l}(0)\,\right]~=~
 |\,\bar{k}_{\emph{l}\,}(0)\,|\;\cos\epsilon_{\emph{l}\,}(0)\;\;\;,
 \label{c2}
 \label{L7}
 \ea
 and equal to the customary static Love number:
 \ba
 \bar{k}_{\emph{l}\,}(0)\;=\,k_{\emph{l}}\;\;\;.~~~~~~~~~~~~~~~~~~~~~~~~~~~~~~~~~~~~~~~~
 \label{c4}
 \label{L8}
 \ea

 Solution of the equation of motion combined with the constitutive (rheological) equation renders the complex $\bar{k}_{\emph{l}}
 (\chi)$, as explained in Appendix \ref{3.4}. Once $\bar{k}_{\emph{l}}(\chi)$ is found, its absolute value
 \ba
 k_{\emph{l}\,}(\chi)~\equiv~|\,\bar{k}_{\emph{l}\,}(\chi)\,|
 \label{gog}
 \label{L9}
 \ea
 and negative argument
 \ba
 \epsilon_{\it l}(\chi)\;=\;-\;\arctan\frac{{\cal{I}}{\it{m}}\left[\,\bar{k}_{\it l}(\chi)\,
 \right]}{{\cal{R}}{\it{e}}\left[\,\bar{k}_{\it l}(\chi)\,\right]}
 \label{gog}
 \label{L10}
 \ea
 should be inserted into the $\,{\emph{l}}^{\,th}\,$ term of the Fourier expansion for the tidal potential. Things get simplified
 when we study how the tide, caused on the primary by a secondary, is acting on that same secondary. In this case, the $\,{\emph{l}}^{\,th}\,$ term in the Fourier expansion
 contains $\,|k_{\emph{l}\,}(\chi)|\,$ and $\,\epsilon_{\emph{l}\,}(\chi)\,$ in the
 convenient combination $\,k_{\emph{l}\,}(\chi)\,\sin\epsilon_{\emph{l}\,}(\chi)\,$, which is
 exactly $\;\,-\;{\cal{I}}{\it{m}}\left[\,\bar{k}_{\it l}(\chi)\,\right]\;$.

 Rigorously speaking, we should say not ``the ${\emph{l}}^{\,th}$ term", but ``the ${\emph{l}}^{\,th}$ term{\bf{s}}", as each
 ${\it l}$ corresponds to an infinite set of positive and negative Fourier modes $\,\omega_{{\it l}mpq}\,$, the physical forcing
 frequencies being $\,\chi=\chi_{{\it l}mpq}\equiv|\omega_{{\it l}mpq}|\,$. Thus, while the functional forms of both $|k_{\it l}(\chi)
 |$ and $\sin\epsilon_{\emph{l}\,}(\chi)
 $ depend only on ${\it l}\,$, both functions take values that are different for different sets
 of numbers $mpq$. This happens because $\chi$ assumes different values $\chi_{{\it l}mpq}$
 on these sets. Mind though that for triaxial bodies the functional forms of $|k_{\it l}(\chi)|$
 and $\sin\epsilon_{\emph{l}}(\chi)$ may depend also on $m,\,p,\,q$.

 \subsection{Damping of a linear tide}\label{damp}

 Beside the standard assumption $\;U_{\it l}(\erbold)\propto W_{\it l}(\Rbold,\,\erbold^{\,*})\;$,
 the linearity condition includes the requirement that the functions $\,k_{\it l}(\chi)\,$ and $\,
 \epsilon_{\textstyle{_{{\it{l}}}}}(\chi)\,$ be well defined. This implies that they
 depend solely upon the frequency $\chi$, and not upon the other frequencies involved. Nor shall the Love
 numbers or lags be influenced by the stress or strain magnitudes at this or other frequencies.

 Then, at frequency $\chi$, the mean (over a period) damping rate $\langle\stackrel{
 \centerdot}{E}(\chi)\rangle$ depends on the value of $\chi$ and on the loading at that frequency, and is not
 influenced by the other frequencies:
 \ba
 \langle\,\dot{E}(\chi)\,\rangle \;=\;-\;\frac{\textstyle\chi E_{peak}(\chi)}{\textstyle Q(\chi)}\;
 \label{dissipation_rate}
 \label{L11}
 \ea
 or, equivalently:
 \ba
 \Delta E_{cycle}(\chi)\;=\;-\;\frac{2\;\pi\;E_{peak}(\chi)}{Q(\chi)}\;\;\;,
 \label{dega}
 \label{L12}
 \ea
 $\Delta E_{cycle}(\chi)\,$ being the one-cycle energy loss, and $\,Q(\chi)\,$ being the so-called quality factor.

 If $\,E_{peak}(\chi)\,$ in (\ref{L11} - \ref{L12}) is agreed to denote the peak {\it{energy}} stored at frequency $\,\chi\,$,
 the appropriate $Q$ factor is connected to the phase lag $\,\epsilon(\chi)\,$ through
 \ba
 Q^{-1}_{\textstyle{_{energy}}}&=&\sin|\epsilon|~~~.
 \label{DI6_1}
 \label{DI6}
 \ea
 and {\it{not}} through $~Q^{-1}_{\textstyle{_{energy}}}\,=\,\tan|\epsilon|~$ as often presumed (see Appendix \ref{interconnection} for
 explanation).

 If $E_{peak}(\chi)$ is defined as the peak {\it{work}},
  the corresponding $Q$ factor is related to the lag via
 \ba
 Q^{-1}_{\textstyle{_{work}}}\;=\;\frac{\tan |\epsilon|}{1\;-\;\left(\;\frac{\textstyle \pi}{\textstyle 2}\;-\;
 |\epsilon|\;\right)  \;\tan|\epsilon|}\;\;\;,~~~~~
 \label{A81}
 \label{L14}
 \ea
 as demonstrated in Appendix \ref{interconnection} below.\footnote{Deriving this formula in Appendix to Efroimsky \& Williams (2009),
 we inaccurately termed $\,E_{peak}(\chi)\,$ as peak energy. However our calculation of $Q$ was carried out in understanding that
 $\,E_{peak}(\chi)\,$ is the peak {\it{work}}.} In the limit of a small $\,\epsilon\,$, (\ref{L14}) becomes
 \ba
 Q^{-1}_{\textstyle{_{work}}}&=&\sin|\epsilon|\;+\;O(\epsilon^2)\;=\;|\epsilon|\;+\;O(\epsilon^2)\;\;\;,
 \label{DI6_2}
 \ea
 so definition (\ref{L14}) makes $\,1/Q\,$ a good approximation to $\,\sin\epsilon\,$ for small lags only.

 For the lag approaching $\,\pi/2\,$, the quality factor defined through (\ref{DI6}) attains its minimum, $\,Q_{\textstyle{_{energy}}}=
 1\,$, while definition (\ref{L14}) furnishes $\,Q_{\textstyle{_{work}}}=0\,$. The latter is not surprising, as in the said limit no
 work is carried out on the system.

 Linearity requires the functions $\,\bar{k}_{\emph{l}}(\chi)\,$ and therefore also
 $\,\epsilon_{\emph{l}}(\chi)\,$ to be well-defined, i.e., to be independent from all the other frequencies but $\chi$. We now see, the requirement extends to $\,Q(\chi)\,$.

 The third definition of the quality factor (offered by Golderich 1963) is $\,Q_{\textstyle{_{Goldreich}}}^{-1}=\,\tan|\epsilon|\,$.
 However this definition corresponds neither to the peak work nor to the peak energy. The existing ambiguity in definition of
 $Q$ makes this factor redundant, and we mention it here only as a tribute to the tradition. As we shall
 see, all practical calculations contain the products of the Love numbers by the sines of the phase lags, $\,k_l\,\sin\epsilon_l\,$, where
 $\,l\,$ is the degree of the appropriate spherical harmonic. A possible compromise between this mathematical fact and the historical
 tradition of using $\,Q\,$ would be to define the quality factor through (\ref{DI6}), in which case the quality factor must be
 equipped with the subscript $\,l\,$. (This would reflect the profound difference between the tidal quality factor{\bf{s}} and the
 seismic quality factor -- see Efroimsky 2012.)

 \section{Several basic facts from continuum mechanics}\label{3}

 This section offers a squeezed synopsis of the basic facts from the linear solid-state mechanics. A more detailed
 introduction, including a glossary and examples, is offered in Appendix \ref{3AA3}.

   \subsection{Stationary linear deformation of isotropic incompressible media}\label{evvo}

 Mechanical properties of a medium are furnished by the so-called constitutive equation or constitutive law, which interrelates
 the stress tensor $\,{\mathbb{S}}\,$ with the strain tensor $\,{\mathbb{U}}\,$ defined as
 \ba
 {\mathbb{U}}\,\equiv\,\frac{\textstyle 1}{\textstyle 2}\,\left[\,\left(\nabla\otimes{\bf{u}}\right)\,+\,\left(\nabla
 \otimes{\bf{u}}\right)^{{^{T}}}\,\right]~~~,
 \label{}
 \ea
 where $\,{\bf{u}}\,$ is the vector of displacement.

 As we shall consider only linear deformations, our constitutive laws will be linear, and will be expressed by
 equations which may be algebraic, differential, integral, or integro-differential.

 The elastic stress $\,\stackrel{{{(e)}}}{\mathbb{S}}\,$ is related to $\,{\mathbb{U}}\,$ through the simplest constitutive
 equation
 \ba
 \stackrel{{{(e)}}}{\mathbb{S}}\,=\,{\mathbb{B}}~{\mathbb{U}}~~~,
 \label{}
 \ea
 ${\mathbb{B}}\,$ being a four-dimensional matrix of real numbers called {\it{elasticity moduli}}.

 A hereditary stress $\,\stackrel{{{(h)}}}{\mathbb{S}}\,$ is connected to $\,{\mathbb{U}}\,$ as
 \ba
 \stackrel{{{(h)}}}{\mathbb{S}}\,=\,\tilde{\,\mathbb{B}}~{\mathbb{U}}~~~,
 \label{}
 \ea
 $\tilde{\,\mathbb{B}}\,$ being a four-dimensional integral-operator-valued matrix. Its component $\,\tilde{B}_{ijkl}\,$ acts on an
 element $\,u_{kl}\,$ of the strain not as a mere multiplier but as an integral operator, with integration going from $\,t\,'=-\infty\,$
 through $\,t\,'=t\,$. To furnish the value of $\,\sigma_{ij}=\sum_{kl}\tilde{B}_{ijkl}\,u_{kl}\,$ at time $\,t\,$, the operator
 ``consumes" as arguments all the values of $\,u_{kl}(t\,')\,$ over the interval $~t\,'\,\in\,\left(\right.\,-\,\infty,\,t\left.\right]~$.


 The viscous stress is related to the strain through a differential operator $\,{\mathbb{A}}\,\frac{\textstyle
 \partial~}{\textstyle\partial t}~$:
 \ba
 \stackrel{{{(v)}}}{\mathbb{S}}\,=\,{\mathbb{A}}~\frac{\partial~}{\partial t}~{\mathbb{U}}~~~,
 \label{vi}
 \ea
 $\,{\mathbb{A}}\,$ being a four-dimensional matrix consisting of empirical constants called viscosities.

 In an isotropic medium, each of the three matrices, $\,{\mathbb{B}}\,$, $\,\tilde{\,\mathbb{B}}\,$, and $\,\tilde{\,\mathbb{A}}\,$,
 includes two terms only. The elastic stress becomes:
 \ba
 \stackrel{(e)}{\mathbb{S}}\;=\;\stackrel{(e)}{\mathbb{S}}_{\textstyle{_{volumetric}}}\,+\,\stackrel{(e)}{\mathbb{S}}_{\textstyle{_{deviatoric}}}
 \,=\;3\,K\,
 \left(\frac{\textstyle 1}{\textstyle 3}~{\mathbb{I}}~\mbox{Sp}\,{\mathbb{U}}\,\right)~+~
 2\,\mu\,\left(\,{\mathbb{U}}\;-\;\frac{\textstyle 1}{\textstyle 3}~
 {\mathbb{I}}~\mbox{Sp}\,{\mathbb{U}}\,\right)\quad,\quad
 \label{LL203}
 \ea
 with $\,K\,$ and $\,\mu\,$ being the {\it{bulk elastic modulus}} and the {\it{shear elastic modulus}},
 correspondingly, $\;{\mathbb{I}}\,$ standing for the unity matrix, and $\,\mbox{Sp}\,$ denoting the trace of a matrix:
 $~\mbox{Sp}\,{\mathbb{U}}\,\equiv\,\sum_{i}U_{ii}\;$.

 The hereditary stress becomes:
 \ba
 \stackrel{(h)}{\mathbb{S}}~=~\stackrel{(h)}{\mathbb{S}}_{\textstyle{_{volumetric}}}\,+~\stackrel{(h)}
 {\mathbb{S}}_{\textstyle{_{deviatoric}}}~=\;3\,\tilde{K}~\left(\,\frac{\textstyle{1}}{\textstyle{3}}\,{\mathbb{I}}
 ~\mbox{Sp}\,{\mathbb{U}}\,\right)~+~2\,\tilde{\mu}\,\left(\,{\mathbb{U}}\,-\,\frac{\textstyle{1}}{\textstyle{3}}\,
 {\mathbb{I}}~\mbox{Sp}\,{\mathbb{U}}\,\right)~~~,~~
 \label{LL19}
 \ea
 where $\,\tilde{K}\,$ and $\,\tilde{\mu}\,$ are the {\it{bulk-modulus operator}} and the {\it{shear-modulus operator}},
 accordingly.

 The viscous stress acquires the form:
 \ba
 \stackrel{(v)}{\mathbb{S}}\;=\;\stackrel{(v)}{\mathbb{S}}_{\textstyle{_{volumetric}}}\,+\,\stackrel{(v)}{\mathbb{S}}_{\textstyle{
 _{deviatoric}}}\,=\;3~\zeta\,\frac{\textstyle\partial ~}{\textstyle\partial t}\,
 \left(\frac{\textstyle 1}{\textstyle 3}~{\mathbb{I}}~\mbox{Sp}\,{\mathbb{U}}\,\right)~+~
 2\,\eta\,\frac{\textstyle\partial ~}{\textstyle\partial t}\,\left(\,{\mathbb{U}}\;-\;\frac{\textstyle 1}{\textstyle 3}~
 {\mathbb{I}}~\mbox{Sp}\,{\mathbb{U}}\,\right)\quad,\quad
 \label{LL20}
 \ea
 the quantities $\zeta$ and $\eta$ being termed as the {\it{bulk viscosity}} and the {\it{shear viscosity}},
 correspondingly

 The term $~\frac{\textstyle 1}{\textstyle 3}~{\mathbb{I}}~\mbox{Sp}\,{\mathbb{U}}~$ is called the {\it{volumetric}}
 part of the strain, while $~{\mathbb{U}}\,-\,\frac{\textstyle 1}{\textstyle 3}~{\mathbb{I}}~\mbox{Sp}\,{\mathbb{U}}~$
 is called the {\it{deviatoric}} part. Accordingly, in expressions (\ref{LL203} - \ref{LL20}) for the stresses, the pure-trace
 terms are called {\it{volumetric}}, the other term being named {\it{deviatoric}}.

 If an isotropic medium is also incompressible, the relative change of the volume vanishes: $\,\mbox{Sp}\,{\mathbb{U}}=0\,$, and so does the expansion rate: $\,\nabla\cdot\vbold\,=\,\frac{\textstyle\partial ~}{\textstyle\partial t}\,\mbox{Sp}
 \,{\mathbb{U}}=0\,$. Then the volumetric part of the strain becomes zero, and so do the
 volumetric parts of the elastic, hereditary, and viscous stresses. The incompressibility assumption may be applicable both to
 crusty objects and to large icy moons of low porosity. At least for Iapetus, the low-porosity assumption is likely to be
 correct (Castillo-Rogez et al. 2011).

 \subsection{Approaches to modeling viscoelastic deformations.\\
 Problems with terminology}

 One approach to linear deformations is to assume that the elastic, hereditary and viscous deviatoric stresses simply sum up,
 each of them being linked to the same overall deviatoric strain:
 \ba
 \stackrel{(total)}{\mathbb{S}}\,=~
 \stackrel{{{(e)}}}{\mathbb{S}}\,+\,\stackrel{{{(h)}}}{\mathbb{S}}\,+\,\stackrel{{{{(v)}}}}{\mathbb{S}}~=~
 {\mathbb{B}}~{\mathbb{U}}~+~\tilde{\,\mathbb{B}}~{\mathbb{U}}~+~{\mathbb{A}}\,\frac{\partial~}{\partial t}~{\mathbb{U}}
 ~=~\left(\,{\mathbb{B}}~+~\tilde{\mathbb{B}}~+~{\mathbb{A}}\,\frac{\partial~}{\partial t}\,\right)\;{\mathbb{U}}~~~.
 \label{LL18}
 \ea

 An alternative option, to be used in section \ref{Andrade_section} below, is to start with an overall
 deviatoric stress, and to expand the deviatoric strain into elastic, viscous, and hereditary parts:
 \ba
 {\mathbb{U}}\,=\,\stackrel{(e)}{\mathbb{U}}\,+\,\stackrel{(h)}{\mathbb{U}}\,+\,\stackrel{(v)}{\mathbb{U}}~~~,\quad\quad
 \stackrel{(e)}{\mathbb{U}}\,=\,\frac{1}{\mu}\,{\mathbb{S}}~~~,\quad\quad
 \stackrel{(v)}{\mathbb{U}}\,=\,\frac{1}{\eta}\,\int^t\,{\mathbb{S}}(t\,')\,dt\,'~~~,\quad\quad
 \stackrel{(h)}{\mathbb{U}}\,=\,\tilde{J}\,{\mathbb{S}}~~~,\quad
 \ea
 $\tilde{J}\,$ being an integral operator with a time-dependent kernel.

 An even more general option would be to assume that both the strain and stress are comprised by components of
 different nature -- elastic, hereditary, viscous, or more complicated (plastic).
 Which option to choose -- depends upon the medium studied. The rich variety of materials offered to us by nature
 leaves one no chance to develop a unified theory of deformation.

 As different segments of the continuum-mechanics community use different conventions on the meaning of some terms, we offer a
 glossary of terms in Appendix \ref{3AA3}. Here we would only mention that in our paper the term {\it{viscoelastic}} will be
 applied to a model containing not only viscous and elastic terms, but also an extra term responsible for an anelastic hereditary
 reaction. (A more appropriate term {\it{viscoelastohereditary}} would be way too cumbersome.)

 \subsection{Evolving stresses and strains. Basic notations}\label{evo}

 In the general case, loading varies in time, so one has to deal with the stress and strain tensors as functions of
 time. However, treatment of viscoelasticity turns out to be simpler in the frequency domain, i.e., in the language
 of complex rigidity and complex compliance. To this end, the stress $\,\sigma_{\gamma\nu}\,$ and strain $\,u_{\gamma
 \nu}\,$ in a linear medium can be Fourier-expanded as
 \begin{subequations}
 \ba
 \sigma_{\gamma\nu}(t)\,
 \,=\,\sum_{n=0}^{\infty}\,\sigma_{\gamma\nu}(\chi_{\textstyle{_n}})\,\cos\left[\,\chi_{\textstyle{_n}}
 t+\varphi_{\sigma}(\chi_{\textstyle{_n}})\,\right]
 \,=\,\sum_{n=0}^{\infty}\,\,{\cal{R}}{\it{e}}\left[\;\sigma_{\gamma\nu}(\chi_{\textstyle{_n}})\;\,
 \;e^{\textstyle{^{\inc\chi_{\textstyle{_{\textstyle{_n}}}}{\small{t}}\;+\,
 \inc\varphi_{\textstyle{_{\sigma}}}(\chi_{\textstyle{{{_n}}}})\,}}}
 \;\right]~~~~~
 \label{L15a}
 \ea
 \ba
 \left.~\right.~~~~~~~~~~~~~~~~~~~~~~~~~~~~~~~~~~~~~~~~~~~~~~~~~~~~
 =\;\sum_{n=0}^{\infty}\,
 {\cal{R}}{\it{e}}\left[\,{\bar{\sigma}}_{\gamma\nu}(\chi_{\textstyle{_n}})\,\;e^{\textstyle{^{\,\inc
 \chi_{\textstyle{_n}} t}}}\,\;\right]~~~~~,
 ~~~~~~~~~~~
 \label{L15b}
 \ea
 \label{L15}
 \end{subequations}
 ~\\
 \begin{subequations}
 \ba
 u_{\gamma\nu}(t)
 \,=\,\sum_{n=0}^{\infty}\,u_{\gamma\nu}(\chi_{\textstyle{_n}})\,\cos\left[\,\chi_{\textstyle{_n}}t+
 \varphi_{u}(\chi_{\textstyle{_n}})\,\right]
 \,=\,\sum_{n=0}^{\infty}\,\,{\cal{R}}{\it{e}}\left[\;u_{\gamma\nu}(\chi_{\textstyle{_n}})\;\,
 \;e^{\textstyle{^{\inc\chi_{\textstyle{_{\textstyle{_n}}}}{\small{t}}\;+\,
 \inc\varphi_{\textstyle{_{u}}}(\chi_{\textstyle{{{_n}}}})\,}}}\;\right]~~~~~
 \label{L16a}
 \ea
 \ba
 \left.~\right.~~~~~~~~~~~~~~~~~~~~~~~~~~~~~~~~~~~~~~~
 =\;\sum_{n=0}^{\infty}\,
 {\cal{R}}{\it{e}}\left[\,{\bar{u}}_{\gamma\nu}(\chi_{\textstyle{_n}})\;\,e^{\textstyle{^{\,\inc\chi_{\textstyle{_n}} t}}}\,\;\right]~~~,
 \label{L16b}
 \ea
 \label{L16}
 \end{subequations}
 where the complex amplitudes are:
 \ba
 {\bar{{\sigma}}_{\gamma\nu}}(\chi)={{{\sigma}}_{\gamma\nu}}(\chi)\,\;e^{\inc\varphi_\sigma(\chi)}~~~~~,
 ~~~~~~{\bar{{u}}_{\gamma\nu}}(\chi)={{{u}}_{\gamma\nu}}(\chi)\,\;e^{\inc\varphi_u(\chi)}~~~,
 \label{compamp}
 \label{L17}
 \ea
 while the initial phases $\,\varphi_{\sigma}(\chi)\,$ and $\,\varphi_{u}(\chi)\,$ are chosen in a
 manner that sets the real amplitudes $\,\sigma_{\gamma\nu}(\chi_{\textstyle{_n}})\,$ and $\,u_{\gamma\nu}(
 \chi_{\textstyle{_n}})\,$ non-negative.

 We wrote the above expansions as sums over a discrete spectrum, as the spectrum generated by tides is discrete.
 Generally, the sums can, of course, be replaced with integrals over frequency:
 \ba
 \sigma_{\gamma\nu}(t)~=~\int_{0}^{\infty}\,\bar{\sigma}_{\gamma\nu}(\chi)~e^{\textstyle{^{\,\inc\chi t}}}~d\chi\quad\quad\mbox{and}
 ~\quad~\quad
 u_{\gamma\nu}(t)~=~\int_{0}^{\infty}\,\bar{u}_{\gamma\nu}(\chi)~e^{\textstyle{^{\,\inc\chi t}}}~d\chi~~~.
 \label{phys}
 \ea
 Whenever necessary, the frequency is set to approach the real axis from below: $\,{\cal I}{\it{m}} (\chi)\rightarrow 0-$

 \subsection{Should we consider positive frequencies only?}\label{frequencies}

 At first glance, the above question appears pointless, as a negative frequency is a mere abstraction, while physical processes
 go at positive frequencies. Mathematically, a full Fourier decomposition of a {\it{real}}
 field can always be reduced to a decomposition over positive frequencies only.

 For example, the full Fourier integral for the stress
 can be written as
 \ba
 \sigma_{\gamma\nu}(t)\,=~\int_{-\infty}^{\infty}\,\bar{s}_{\gamma\nu}(\omega)~e^{\textstyle{^{\,\inc\omega t}}}d\omega\,=\,
 \int_{0}^{\infty}\left[\,\bar{s}_{\gamma\nu}(\chi)~e^{\textstyle{^{\,\inc\chi t}}}+~
 \bar{s}_{\gamma\nu}(-\chi)~e^{\textstyle{^{\,-\,\inc\chi t}}}\,\right]\,d\chi~~~,
 \label{fof}
 \ea
 where we define $\,\chi\,\equiv\,|\,\omega\,|\,$. Denoting complex conjugation with asterisk, we write:
 \ba
 \sigma^{*}_{\gamma\nu}(t)\,=
 \int_{0}^{\infty}\left[\,\bar{s}^{\,*}_{\gamma\nu}(-\chi)~e^{\textstyle{^{\,\inc\chi t}}}+~
 \bar{s}^{\,*}_{\gamma\nu}(\chi)~e^{\textstyle{^{\,-\,\inc\chi t}}}\,\right]\,d\chi~~~.
 \label{of}
 \ea
 The stress is real: $\,\sigma^{*}_{\gamma\nu}(t)\,=\,\sigma_{\gamma\nu}(t)\,$. Equating the right-hand sides of (\ref{fof}) and
 (\ref{of}), we obtain
 \ba
 \bar{s}_{\gamma\nu}(-\chi)\,=\,\bar{s}^{\,*}_{\gamma\nu}(\chi)~~~,
 \label{whence}
 \ea
 whence
 \ba
 \sigma_{\gamma\nu}(t)\,=
 \int_{0}^{\infty}\left[\,\bar{s}_{\gamma\nu}(\chi)~e^{\textstyle{^{\,\inc\chi t}}}+~
 \bar{s}^{\,*}_{\gamma\nu}(\chi)~e^{\textstyle{^{\,-\,\inc\chi t}}}\,\right]\,d\chi~=~{\cal{R}}{\it{e}}\int_0^\infty
 2~\bar{s}_{\gamma\nu}(\chi)~e^{\textstyle{^{\,\inc\chi t}}}~d\chi~~~.
 \label{unphys}
 \ea
 This leads us to (\ref{phys}), if we set
 \ba
 \bar{\sigma}_{\gamma\nu}(\chi)\,=\,2\,\bar{s}_{\gamma\nu}(\chi)~~~.
 \label{dott}
 \ea
 While the switch from
 $~\sigma_{\gamma\nu}(t)\,=~\int_{-\infty}^{\infty}\,\bar{s}_{\gamma\nu}(\omega)~e^{\textstyle{^{\,\inc\omega t}}}d\omega~$
 to
 the expansion $~\sigma_{\gamma\nu}(t)\,=~\int_{0}^{\infty}\,\bar{\sigma}_{\gamma\nu}(\omega)~e^{\textstyle{^{\,\inc\chi t}}}d\chi~$
 makes things simpler, the simplification comes at a cost, as we shall see in a second.

 Recall that the tide can be expanded over the modes
 \ba
 \omega_{{\it l}mpq}~\equiv~
 ({\it l}-2p)\;\dot{\omega}\,+\,({\it l}-2p+q)\;\dot{\cal{M}}\,+\,m\;(\dot{\Omega}\,-\,\dot{\theta})~\approx~
 ({\it l}-2p+q)\,n~-~m\,\dot{\theta}~~~,~~~
 \label{ref}
 \ea
 each of which can assume positive or negative values, or be zero. Here $l$, $m$, $p$, $q$ are some integers, ${\theta}$
 is the primary's sidereal angle, $\dot{\theta}$ is its spin rate, while $\omega$, $\Omega$, ${\cal M}$ and $n$ are the
 secondary's periapse, node, mean anomaly, and mean motion. The appropriate tidal frequencies, at which the medium gets
 loaded, are given by the absolute values of the tidal modes: $\,\chi_{\textstyle{_{lmpq}}}\equiv\,|\,\omega_{
 \textstyle{_{lmpq}}}\,|\,$.

 The positively-defined forcing frequencies $\,\chi_{\textstyle{_{lmpq}}}\,$ are the
 actual physical frequency at which the $\,lmpq\,$ term in the expansion for the tidal potential (or stress or strain) oscillates.

 The motivation for keeping also the modes $\,\omega_{\textstyle{_{lmpq}}}\,$ is subtle:
 it depends upon the {\it{sign}} of $\,\omega_{\textstyle{_{lmpq}}}\,$
 whether the $\,lmpq\,$ component of the tide
 lags or advances. Specifically, the phase lag between the $\,lmpq\,$ component of the perturbed
 primary's potential $\,U\,$ and the $\,lmpq\,$ component of the tide-raising potential $\,W\,$ generated by the secondary
 is given by
 \ba
 \epsilon_{\textstyle{_{lmpq}}}~=~\omega_{\textstyle{_{lmpq}}}~\Delta t_{\textstyle{_{lmpq}}}~=~
 |\,\omega_{\textstyle{_{lmpq}}}\,|~\Delta t_{\textstyle{_{lmpq}}}~\mbox{sgn}\,\omega_{\textstyle{_{lmpq}}}~=~
 \chi_{\textstyle{_{lmpq}}}~\Delta t_{\textstyle{_{lmpq}}}~\,\mbox{sgn}\,\omega_{\textstyle{_{lmpq}}}~~~,
 \label{}
 \ea
 where the time lag $~\Delta t_{\textstyle{_{lmpq}}}~$ is always positive.

 While the lag between the applied stress and resulting strain in a sample of a medium is always positive, the case of tides is
 more complex: there, the lag can be either positive or negative. This, of course, in no way implies whatever violation of
 causality (the time lag $~\Delta t_{\textstyle{_{lmpq}}}~$ is always positive). Rather, this is about the directional
 difference between the planetocentric positions of the tide-raising body and the resulting bulge. For example, the
 principal component of the tide, $\,lmpq=2200\,$, stays behind (has a positive phase lag $\,\epsilon_{\textstyle{_{2200}}}\,
 $) when the secondary is below the synchronous orbit, and advances (has a negative phase lag $\,\epsilon_{\textstyle{_{2200}
 }}\,$) when the secondary is at a higher orbit. To summarise, decomposition of a tide over both positive and negative modes
 $\,\omega_{\textstyle{_{lmpq}}}\,$ (and not just over the positive frequencies $\,\chi_{\textstyle{_{lmpq}}}\,$) does have
 a physical meaning, as the sign of a mode $\,\omega_{\textstyle{_{lmpq}}}\,$ carries physical information.

 ~\\
 Thus we arrive at the following conclusions:\\

 \begin{itemize}

 \item[1.] ~As the fields emerging in the tidal theory -- the tidal potential, stress, and strain -- are all real, their expansions
 in the frequency domain may, in principle, be written down using the positive frequencies $\,\chi\,$ only.\\

  \item[2.] ~In the tidal theory, the potential (and, consequently, the tidal torque and force) contain components corresponding to
  the tidal modes $\,\omega_{\textstyle{_{lmpq}}}\,$ of both the positive and negative signs. While the $\,lmpq\,$ components of
  the potential, stress, and strain oscillate at the positive frequencies $\,\chi_{\textstyle{_{lmpq}}}\,=\,|\omega_{\textstyle{_{
  lmpq}}}|~$, the sign of each $\omega_{\textstyle{_{lmpq}}}$ does carry physical information: it distinguishes whether the lagging
  of the $\,lmpq\,$ component of the bulge is positive or negative (falling behind or advancing). Accordingly, this sign enters
  explicitly the expression for the appropriate component of the torque or force. Hence a consistent tidal theory should be
  developed through expansions over both positive and negative tidal modes $\,\omega_{\textstyle{_{lmpq}}}\,$ and not just over the
  positive $\,\chi_{\textstyle{_{lmpq}}}\,$.\\

  \item[3.] ~In order to rewrite the tidal theory in terms of the positively-defined frequencies $~\,\chi_{\textstyle{_{lmpq}
  }}\,$ only, one must inserts ``by hand" the extra multipliers
     \ba
     \mbox{sgn}\,\omega_{\textstyle{_{lmpq}}}~=~\mbox{sgn}\left[~({\it l}-2p+q)\,n~-~m\,\dot{\theta}~\right]
     \label{multi}
     \ea
     into the expressions for the $\,lmpq\,$ components of the tidal torque and force.\\

   \item[4.] ~One can employ a rheological law (constitutive equation interconnecting the strain and stress) and a
       Navier-Stokes equation (the second law of Newton for an element of a viscoelastic medium), to calculate the
       phase lag $\,\epsilon_{\textstyle{_{lmpq}}}\,$ of the primary's potential $\,U_{\textstyle{_{lmpq}}}\,$
       relative to the potential $\,W_{\textstyle{_{lmpq}}}\,$ generated by the secondary. If both these equations
       are expanded, in the frequency domain, via positively-defined forcing frequencies
       $\,\chi_{\textstyle{_{lmpq}}}\,$ only, the resulting phase lag, too, will emerge as a function of
       $\,\chi_{\textstyle{_{lmpq}}}\,$: \ba
       \epsilon_{\textstyle{_{lmpq}}}\,=\,\epsilon_{\textstyle{_{l}}}(\chi_{\textstyle{_{lmpq}}})~~~.
       \label{laglag} \ea
      Within this treatment, one has to equip the lag, ``by hand", with the multiplier (\ref{multi}).\\

 \end{itemize}

 \noindent
 As we saw above, the lag (\ref{laglag}) is the argument of the complex Love number $\,\bar{k}_{\textstyle{_{
 l}}}(\chi_{\textstyle{_{lmpq}}})\,$. Solution of the constitutive and Navier-Stokes equations renders the complex Love
 numbers, from which one can calculate the lags. Hence the above item [4] may be rephrased in the following manner:\\
 \begin{itemize}

 \item[4{$'$}.] ~~Under the convention that $\,U_{\textstyle{_{lmpq}}}=U(\chi_{\textstyle{_{lmpq}}})\,$ and
 $\,W_{\textstyle{_{lmpq}}}=W(\chi_{\textstyle{_{lmpq}}})\,$, ~we have:
 \begin{subequations}
 \ba
 \left.~~\right. U_{\textstyle{_{lmpq}}}&=&\bar{k}_{\textstyle{_{l}}}(\chi_{\textstyle{_{lmpq}}})\,W_{\textstyle{_{lmpq}}}~~~~\quad
 \mbox{when}\quad\omega_{\textstyle{_{lmpq}}}>\,0~,~~\mbox{i.e.}\,,\,~\mbox{when}\quad \omega_{\textstyle{_{lmpq}}}~=~\chi_{\textstyle{_{lmpq}}}~~,\quad~\quad~\quad
 \label{}\\
 \nonumber\\
 \left.~~\right. U_{\textstyle{_{lmpq}}}&=&\bar{k}^{\,*}_{\textstyle{_{l}}}(\chi_{\textstyle{_{lmpq}}})\,W_{\textstyle{_{lmpq}}}~~~\quad
 \mbox{when}\quad\omega_{\textstyle{_{lmpq}}}<\,0~,~~\mbox{i.e.}\,,\,~\mbox{when}\quad \omega_{\textstyle{_{lmpq}}} =\,-\,\chi_{\textstyle{_{lmpq}}}~,\quad~\quad~\quad
 \label{}
 \ea
 \end{subequations}
 $\left.~\,\right.$ asterisk denoting the complex conjugation.\\
 \end{itemize}

 \noindent
 This ugly convention, a switch from $\,\bar{k}_{\textstyle{_{l}}}\,$ to $\,\bar{k}^{\,*}_{\textstyle{_{l}}}\,$, is the price we
 pay for employing only the positive frequencies in our expansions, when solving the constitutive and Navier-Stokes equations,
 to find the Love number. In other words, this is a price for our pretending that $\,W_{\textstyle{_{lmpq}}}\,$ and
 $\,U_{\textstyle{_{lmpq}}}\,$ are functions of $\,\chi_{\textstyle{_{lmpq}}}~$ -- whereas in reality they are functions of
 $\,\omega_{\textstyle{_{lmpq}}}\,$.

 Alternative to this would be expanding the stress, strain, and the potentials over the
 positive and negative modes $\,\omega_{\textstyle{_{lmpq}}}\,$, with the negative frequencies showing up in the equations.
 With the convention that $\,U_{\textstyle{_{lmpq}}}=U(\omega_{\textstyle{_{lmpq}}})\,$ and
  $\,W_{\textstyle{_{lmpq}}}=W(\omega_{\textstyle{_{lmpq}}})\,$, we would have
  \ba
 U_{\textstyle{_{lmpq}}}&=&\bar{k}_{\textstyle{_{l}}}(\omega_{\textstyle{_{lmpq}}})\,W_{\textstyle{_{lmpq}}}~~~~,\quad
 \mbox{for all}\quad\omega_{\textstyle{_{lmpq}}}~~~.
 \label{}
 \ea
 \vspace{1mm}

 All these details can be omitted at the despinning stage, if one keeps only the leading term of the torque and ignores
 the other terms. Things change, though, when one takes these other terms into account. On crossing of an $lmpq$ resonance,
 factor (\ref{multi}) will change its sign. Accordingly, the $lmpq$ term of the tidal torque (and of the tidal force) will
 change its sign too.


 \subsection{The complex rigidity and compliance. Stress-strain relaxation}

 The stress cannot be obtained by means of an integral operator that would map
 the past history of the strain, $\,{\mathbb{U}}(t\,')\,$ over $\,t\,'\,\in\,\left(\right.-\infty,\,t\,\left.\right]\,$, ~to the value
 of $\,{\mathbb{S}}\,$ at time $\,t\,$. The insufficiency of such an operator is evident from the presence of a
 time-derivative on the right-hand side of (\ref{vi}). Exceptional are the cases of no viscosity (e.g., a
 purely elastic material).

 On the other hand, we expect, on physical grounds, that the operator $\,\hat{J}\,$ inverse to $\,\hat{\mu}\,$ {\underline{is}} an
 integral operator. In other words, we assume that the current value of the strain depends only on the present and past values taken by
 the stress and not on the current {\it{rate}} of change of the stress. This assumption works for weak deformations, i.e., insofar
 as no plasticity shows up.
 So we assume that the operator $\,\hat{J}\,$ mapping the stress to the strain is just an integral operator.

 Since the forced medium ``remembers" the history of loading, the strain at time $\,t\,$ must be a sum of small installments $\,\frac{
 \textstyle 1}{\textstyle 2}\,J(t-t\,')\,d{\sigma}_{\gamma\nu}(t\,')\,$, each of which stems from a small change $\,d{\sigma}_{\gamma\nu}
 (t-\tau)\,$ of the stress at an earlier time $\,t\,'\,<\,t$. The entire history of the past loading results, at the time $\,t\,$, in a
 total strain $\,u_{\gamma\nu}(t)\,$ rendered by an integral operator $\,\hat{J}(t)\,$ acting on the entire function $\,\sigma_{\gamma\nu}
 (t\,')\,$ and not on its particular value (Karato 2008):
 \ba
 \label{L18}
 \label{subequations_1}
 2\,u_{\gamma\nu}(t)\;=\;\hat{J}(t)\;\sigma_{\gamma\nu}
 \;=\;\int^{\infty}_{0}J(\tau)\;
 \stackrel{\centerdot}{\sigma}_{\gamma\nu}(t-\tau)\;d\tau
 \;=\;\int^{t}_{-\infty}J(t\,-\,t\,')\;\stackrel{\centerdot}{\sigma}_{\gamma\nu}(t\,')\;dt\,'\;\;\;,
 \;\;\;
 \label{subequations_1_a}
 \label{18a}
 \ea
 where $\,t\,'\,$ is some earlier time ($\,t\,' < t\,$), overdot denotes $\,d/dt\,'\,$, while the ``age variable" $\tau = t-t\,'$
 is reckoned from the current moment $\,t\,$ and is aimed back into the past. The so-defined integral operator $\,\hat{J}(t)\,$ is
 called the {\emph{compliance operator}}, while its kernel $\,J(t-t\,')\,$ goes under the name of the {\it{compliance function}} or
 the {\it{creep-response function}}.

 Integrating (\ref{18a}) by parts, we recast the compliance operator into the form of
 \begin{subequations}
 \ba
 2\,u_{\gamma\nu}(t)\;=\;\hat{J}(t)~\sigma_{\gamma\nu}
 ~=~J(0)\;\sigma_{\gamma\nu}(t)\;-\;J(\infty)\;\sigma_{\gamma\nu}(-\infty)
 ~+~\int^{\infty}_{0}\stackrel{~\centerdot}{J}(\tau)~{\sigma}_{\gamma\nu}(t-\tau)
 ~d\tau~~~~~~~~~~~
 \label{kkllj}
 \label{I12_2}
 \ea
 \ba
 \left.~~~~~~~~~~~~~~~~~~~~~~~~~~\right.=~J(0)\;\sigma_{\gamma\nu}(t)\;-\;J(\infty)\;\sigma_{\gamma\nu}(-\infty)
 ~+~\int^{t}_{-\infty}\stackrel{\;\centerdot}{J}(t\,-\,t\,')~{\sigma}_{\gamma\nu}(t\,')~d t\,'~~\,.~~~~\,
 \label{I12_3}
 \ea
 \label{I12}
 \end{subequations}
 The quantity $\,J(\infty)\,$ is the {\it{relaxed compliance}}. Being the asymptotic value of $\,J(t-t\,')\,$ at $\,t-t\,'\rightarrow
 \,\infty\,$, this parameter corresponds to the strain after complete relaxation. The load in the infinite past may
 be assumed zero, and the term $~-\,J(\infty)\;\sigma_{\gamma\nu}(-\infty)\;$ may be dropped

 The second important quantity emerging in (\ref{I12}) is the {\it{unrelaxed compliance}} $\,J(0)\,$, which is the
 value of the compliance function $\,J(t-t\,')\,$ at $\,t-t\,'=\,0\,$. This parameter describes the instantaneous
 reaction to stressing, and thus defines the {\it{elastic}} part of the deformation (the rest of the deformation
 being viscous and hereditary). Thus the term containing the unrelaxed compliance $\,J(0)\,$ should be kept. The term,
 though, can be absorbed into the integral if we agree that the elastic contribution enters the compliance function not
 as \footnote{~Expressing
 the stress through the strain, we encountered three possibilities: the elastic stress was simply proportional to the strain,
 the viscous stress was proportional to the time-derivative of the strain, while the hereditary stress was expressed by an
 integral operator $\,\tilde{\mu}\,$. However, when we express the strain through the stress, we place the viscosity into
 the integral operator, so the purely viscous reaction also looks like hereditary. It is our convention, though, to apply the
 term {\it{hereditary}} to delayed reactions {\it{other than purely viscous}}.
 }
 \ba
 J(t-t\,')\,=\,J(0)\,+\,\mbox{viscous and hereditary terms}~~~,
 \label{conv}
 \ea
 but as
 \ba
 J(t-t\,')\,=\,J(0)\,\Theta(t\,-\,t\,')\,+\,\mbox{viscous and hereditary terms}~~~,
 \label{convention}
 \ea
 the Heaviside step-function $\,\Theta(t\,-\,t\,')\,$ being unity for $\,t-t\,'\,\geq\,0\,$, and zero for
 $\,t-t\,'\,<\,0\,$. As the derivative of the step-function is the delta-function $\,\delta(t\,-\,t\,')\,$,
 we can write (\ref{I12_3}) simply as
 \ba
 2\,u_{\gamma\nu}(t)\,=\,\hat{J}(t)~\sigma_{\gamma\nu}\,=\,\int^{t}_{-\infty}\stackrel{\;\centerdot}{J}(t-t\,')~
 {\sigma}_{\gamma\nu}(t\,')\,d t\,'~~,~~~\mbox{with}~~
 J(t-t\,')~~\mbox{containing}~~J(0)\,\Theta(t-t\,')~~.~~~
 \label{I12_4}
 \ea
 Equations (\ref{18a}),
 (\ref{I12}),
 (\ref{I12_4}) are but different expressions for the compliance
 operator $\,\hat{J}\,$ acting as
 \ba
 2\,u_{\gamma\nu}\;=\;\hat{J}~\sigma_{\gamma\nu}~~~.
 \label{}
 \ea
 Inverse to the compliance operator is the rigidity operator $\,\hat{\mu}\,$ defined through
 \ba
 \sigma_{\gamma\nu}\;=\;2\,\hat{\mu}~u_{\gamma\nu}~~~.
 \label{kera}
 \ea
 Generally, $\,\hat{\mu}\,$ is not just an integral operator, but is an integro-differential operator. So it cannot take the form of
 $~\sigma_{\gamma\nu}(t)\,=\,2\,\int_{-\infty}^{t}\,\dot{\mu}(t\,-\,t\,')\,u_{\gamma\nu}(t\,')\,dt\,'~$. However it can be written as
 \ba
 \sigma_{\gamma\nu}(t)\,=\,2\,\int_{-\infty}^{t}\,{\mu}(t\,-\,t\,')\,\dot{u}_{\gamma\nu}(t\,')\,dt\,'\;\;\;,
 \label{permitted_1}
 \ea
 if we permit the kernel $\,{\mu}(t\,-\,t\,')\,$ to contain a term $\,\eta\,\delta(t-t\,')\,$, where $\,\delta(t-t\,')
 \,$ is the delta-function. After integration, this term will furnish the viscous part of the stress, $~2\,\eta\,\dot{u}_{
 \gamma\nu}\,$.

 The kernel $\,\mu(t\,-\,t\,')\,$ goes under the name of the {\it{stress-relaxation function}}.
 Its time-independent part is $\,\mu(0)\,\Theta
 (t\,-\,t\,')\,$, where the {\emph{unrelaxed rigidity}} $\,\mu(0)\,$ is inverse to the unrelaxed compliance
 $\,J(0)\,$ and describes the elastic part of deformation.
 Each term in $\,\mu(t-t\,')\,$, which neither is a constant nor contains a delta-function, is responsible for hereditary reaction.

 For more details on the stress-strain relaxation formalism see the book by Karato (2008).

  \subsection{Stress-strain relaxation in the frequency domain}

 Let us introduce the complex compliance $\,\bar{J}(\chi)\,$ and the complex rigidity $\,\bar{\mu}(\chi)\,$, which
 are, by definition, the Fourier images {\bf{not}} of the ${J}(\tau)$ and ${\mu}(\tau)$ functions, but of
 their time-derivatives:\footnote{~Recall that it is the time-derivative of $\,{J}(\tau)\,$ that is the kernel of the integral operator
 (\ref{I12_4}). Hence, to arrive at (\ref{LLJJKK}), we have to define $\,\bar{J}(\chi)\,$ as the Fourier image of $\,\stackrel{\;\bf\centerdot}
 {J}(\tau)\,$.}
 \ba
 \label{L20}
 \int_{0}^{\infty}\bar{J}(\chi)\,e^{\inc\chi \tau}d\chi\,=\,\stackrel{\;\centerdot}{J}(\tau)~~,
 ~~~\mbox{where}~~~
 \bar{J}(\chi)~=\,\int_{0}^{\infty}\stackrel{\;\centerdot}{J}(\tau)\,e^{-\inc\chi \tau}\,d\tau~\,.\,~~
 \ea
 and
 \ba
 \label{L21}
 \int_{0}^{\infty}\bar{\mu}(\chi)\,e^{\inc\chi \tau}d\chi\,=\,{\bf{\dot{\mu}}}(\tau)~~\,,\quad\mbox{where}
 \quad\bar{\mu}(\chi)\,=\,\int_{0}^{\infty}{\bf{\dot{\mu}}}(\tau)\,e^{-\inc\chi \tau}\,d\tau~~\,,~~~
 \ea
 the integrations over $\tau$ spanning the interval $\,\left[\right.0,\infty\left.\right)\,$, as both kernels are nil for $\tau
 <0$ anyway.
 In (\ref{L20}) and (\ref{L21}),
 we made use of the fact (explained in subsection \ref{frequencies}) that, when expanding real fields, it is sufficient to use only
 positive frequencies.

 Expression (\ref{L18}), in combination with the Fourier expansions (\ref{phys}) and with (\ref{L20}), furnishes:
 \ba
 2\,\int_{0}^{\infty}\bar{u}_{\gamma\nu}(\chi)~e^{\textstyle{^{\,\inc\chi t}}}~d\chi
 \;=\;\int_{0}^{\infty}\bar{\sigma}_{\mu\nu}(\chi)~\bar{J}(\chi)~e^{\textstyle{^{\,\inc\chi t}}}~d\chi
  ~~~,
 \label{strain}
 \label{11strainn}
 \ea
 which leads us to:
 \ba
 2\;\bar{u}_{\gamma\nu}(\chi)\,=\;\bar{J}(\chi)\;\bar{\sigma}_{\gamma\nu}(\chi)\;\;\;.
 \label{LLJJKK}
 \ea
 Similarly, insertion of (\ref{phys}) into (\ref{permitted_1}) leads to the relation
 \ba
 \bar{\sigma}_{\gamma\nu}(\chi)\,=\;2\;\bar{\mu}(\chi)\;\bar{u}_{\gamma\nu}(\chi)
 \;\;\;,
 \label{L22}
 \ea
 comparison whereof with (\ref{LLJJKK}) immediately entails:
 \ba
 \bar{J}(\chi)\;{\textstyle\bar{\mu}(\chi)}\;=\;{\textstyle 1}\;\;\;.
 \label{xcx}
 \label{L23}
 \ea

 Writing down the complex rigidity and compliance as
 \ba
 \bar{\mu}(\chi)\;=\;|\bar{\mu}(\chi)|\;\exp\left[\,\inc\,\delta(\chi)\,\right]
 \label{L24}
 \ea
 and
 \ba
 \bar{J}(\chi)\;=\;|\bar{J}(\chi)|\;\exp\left[\,-\,\inc\,\delta(\chi)\,\right]\;\;\;,\;
 \label{L25}
 \ea
 we split (\ref{xcx}) into two expressions:
 \ba
 |\bar{J}(\chi)|\;=\;\frac{1}{|\bar{\mu}(\chi)|}~~~
 \label{L26}
 \ea
 and
 \ba
 \varphi_u(\chi)\;=\;\varphi_{\sigma}(\chi)\;-\;\delta(\chi)\;\;\;.
 \label{L27}
 \ea
 From the latter, we see that the angle $\;\delta(\chi)\,\;$ is a measure of lagging of a strain harmonic mode relative to the
 appropriate harmonic mode of the stress. It is evident from (\ref{L24} - \ref{L25}) that
 \ba
 \tan \delta(\chi)\;\equiv\;-\;\frac{\cal{I}\it{m}\left[ \,\bar{J\,}(\chi)\, \right]}{\cal{R}\it{e}\left[ \,\bar{J\,}(\chi)\,\right]}\;=
 \;\frac{\cal{I}\it{m}\left[ \,\bar{\mu}(\chi)\,\right]}{\cal{R}\it{e}\left[ \,\bar{\mu}(\chi)\,\right]}\;\;\;.
 \label{delta_def}
 \label{L28}
 \ea

 \section{Complex Love numbers}\label{compla}

 The developments presented in this section will rest on a very important theorem from solid-state mechanics. The theorem,
 known as the {\it{correspondence principle}}, also goes under the name of {\it{elastic-viscoelastic analogy}}. The theorem applies to
 linear deformations in the absence of nonconservative (inertial) forces. While the literature attributes the authorship of the theorem
 to different scholars, its true pioneer was Sir George Darwin (1879). One of the corollaries ensuing from this theorem is that, in
 the frequency domain, the complex Love numbers are expressed via the complex rigidity or compliance in the same way as the static Love
 numbers are expressed via the relaxed rigidity or compliance.

 As was pointed out much later by Biot (1954, 1958), the theorem is inapplicable to non-potential forces. Hence the said corollary
 fails in the case of librating bodies, because of the presence of the inertial force\footnote{~The centripetal term is
 potential and causes no troubles, except for the necessity to introduce a degree-0 Love number.} $\;-{\dotomegabold
 }\times\erbold\rho$, where $\rho$ is the density and $\omegabold$ is the libration angular velocity. So the standard expression
 (\ref{L3}) for the Love numbers, generally, cannot be employed for librating bodies.

 Subsection \ref{4.1} below explains the transition from the stationary Love numbers to their dynamical counterparts, the so-called Love
 operators. We present this formalism in the frequency domain, in the spirit of Zahn (1966) who pioneered this approach in application to
 a purely viscous medium. Subsection \ref{4.2} addresses the negative tidal modes emerging in the Darwin-Kaula expansion for tides.
 Employing the correspondence principle, in subsection \ref{4.7} we then write down the expressions for the factors $\,|\bar{k}_{
 \textstyle{_{l}}}(\chi)|\,\sin\epsilon_{\textstyle{_{l}}}(\chi)\,=\,-\,{\cal{I}}{\it{m}}[\bar{k}_{\textstyle{_{l}}}(\chi)]\,$ emerging
 in the expansion for tides. Some technical details of this derivation are discussed in subsections \ref{4.8} - \ref{4.9}.

 For~more~on~the~correspondence~principle~and~its~applicability~to~Phobos~see~Appendix~\ref{corpr}.

 \subsection{From the Love numbers to the Love operators}\label{3.3}\label{4.1}

 A homogeneous incompressible primary, when perturbed by a static secondary, yields its form and,
 consequently, has its potential changed. The $\,{\it{l}}^{th}$ spherical harmonic $\,U_l(\erbold)\,$
 of the resulting increment of the primary's exterior potential is related to the $\,{\it{l}}^{th}$
 spherical harmonic $\,W_l(\Rbold,\erbold)\,$ of the perturbing exterior potential through (\ref{L2}).

 As the realistic disturbances are never static (except for synchronous orbiting), the Love numbers
 become operators:
 \ba
 \label{L29}
 U_{\it l}(\erbold,\,t)\;=\;\left(\,\frac{R}{r}\,
 \right)^{{\it l}+1}\hat{k}_{\it l}(t)\;W_{\it{l}}(\eRbold\,,\;\erbold^{\;*},\,t\,')~~~.
 \ea
 A Love operator acts neither on the value of $\,W\,$ at the current time $\,t\,$, nor at its value at an earlier time $\,t\,'\,$, but
 acts on the entire shape of the function $\,W_{\it{l}}(\eRbold
 \,,\;\erbold^{\;*},\,t\,')\,$, with $\,t\,'\,$ belonging to the semi-interval $\,(-\infty,\,t)\,$.
 This is why we prefer to write $\,\hat{k}_{\it l}(t)\,$ and not $\,\hat{k}_{\it l}(t,\,t\,')\,$.

 Being linear for weak forcing, the operators must read:
 \begin{subequations}
 \label{subequations_3}
 \label{L30}
 \ba
 U_{\it l}(\erbold,\,t)
  =\left(\frac{R}{r}\right)^{{\it l}+1}\int^{\tau=\infty}_{\tau=0}k_{\it l}(\tau)\stackrel{\bf\centerdot}{W}_{\it{l}}(\eRbold\,,\,
  \erbold^{\;*},\,t-\tau)\,d\tau
 =\left(\frac{R}{r}
 \right)^{{\it l}+1}\int_{t\,'
 =-\infty}^{t\,'=t} k_{\it l}(t-t\,')\stackrel{\bf \centerdot}{W}_{\it{l}}
 (\eRbold\,,\,\erbold^{\;*},\,t\,')\,dt\,'~~~~
 \label{churchur}
 \label{L30a}
 \ea
 or, after integration by parts:
 \ba
 U_{\it l}(\erbold,\,t)
  \,=\,\left(\frac{R}{r}\right)^{{\it l}+1}\,\left[\,k_l(0)W(t)\,-\,k_l(\infty)W(-\infty)\,\right]\,+\,\left(\,\frac{R}{r}\,\right)^{
  {\it l}+1}\int^{\infty}_{0}{\bf\dot{\it{k}}}_{\textstyle{_l}}(\tau)~\,W_{\it{l}}(\eRbold\,,\,\erbold^{\;*},\,t-\tau)\,d\tau~~~~~~~~
  \label{}
  \ea
  \ba
  \left.~~~~~~~~~~\right.
 =\left(\frac{R}{r}\right)^{{\it l}+1}\,\left[k_l(0)W(t)\,-\,k_l(\infty)W(-\infty)\right]\,+\,\left(\frac{R}{r}
 \right)^{{\it l}+1}\int_{-\infty}^{t}
 {\bf\dot{\it{k}}}_{\textstyle{_l}}(t-t\,')\,~W_{\it{l}}
 (\eRbold\,,\,\erbold^{\;*},\,t\,')\,dt\,'~~~\,~~~~
 \label{}
 \ea
 \ba
 \nonumber
 \left.~~~~~~~~~~\right.=&-&\left(\frac{R}{r}
 \right)^{{\it l}+1}k_l(\infty)W(-\infty)\\
 \nonumber\\
 &+&\left(\frac{R}{r}\right)^{{\it l}+1}\int_{-\infty}^{t} \frac{d}{dt}\left[~{k}_{\it l}(t\,-\,t\,')~-~{k}_{\it l}(0)~+~
 {k}_{\it l}(0)\Theta(t\,-\,t\,')~\right]~W_{\it{l}}(\eRbold,\erbold^{\;*},t\,')\,dt\,'\quad.\quad\quad\quad\quad
 \label{churban}
 \ea
 \label{L30b}
 \end{subequations}
 Just as in the case of the compliance operator (\ref{L18} - \ref{I12}), in expressions (\ref{L30b}) we obtain the terms
 $~k_l(0)W(t)~$ and $~-k_l(\infty)W(-\infty)~$. Of the latter term, we can get rid by setting $\,W(-\infty)\,$ nil, while the
 former term may be incorporated into the kernel in exactly the same way as in (\ref{conv} - \ref{I12_4}). Thus, dropping the
 unphysical term with $\,W(-\infty)~$, and inserting the elastic term into the Love number not as  $\,k_{l}(0)\,$ but
 as $\,{k}_{l}(0)\,\Theta(t-t\,')\,$, we simplify (\ref{churban}) to
 \ba
 U_{\it l}(\erbold,\,t)\;=\;\left(\frac{R}{r}
 \right)^{{\it l}+1}\int_{-\infty}^{t} {\bf\dot{\it{k}}}_{\textstyle{_l}}(t-t\,')~W_{\it{l}}
 (\eRbold\,,\;\erbold^{\;*},\;t\,')\,dt\,'~,
 \label{chuk}
 \ea
 with $\,{k}_{\it l}(t-t\,')\,$ now including, as its part, $\,{k}_{l}(0)\,\Theta(t-t\,')\,$ instead of $\,{k}_{l}(0)\,$.

 Were the body perfectly elastic, $\,{k}_{\it l}(t-t\,')\,$ would consist of the instantaneous-reaction term
 $\,{k}_{\it l}(0)\,\Theta(t-t\,')\,$ {\emph{only}}. Accordingly, the time-derivative of $\,{k}_{\it l}\,$
 would be: $\,{\bf
 \dot{\it{k}}}_{\textstyle{_{\it{l}}}}(t-t\,')\,=\,k_{\it l}\,\delta(t-t\,')\,$ where $\,k_{\it l}\,
 \equiv\,k_{\it l}(0)\,$, so expressions (\ref{subequations_3} - \ref{chuk}) would coincide with (\ref{L2}).

 Similarly to introducing the complex compliance, one can define the complex Love numbers
 as Fourier transforms of
 $~\stackrel{\bf\centerdot}{k}_{\textstyle{_l}}(\tau)~$:
 \ba
 \label{L31}
 \int_{0}^{\infty}\bar{k}_{\textstyle{_l}}(\chi)e^{\inc\chi \tau}d\chi\;=\;
 \stackrel{\bf\centerdot}{
 k}_{\textstyle{_l}}(\tau)
 ~~~,
 \ea
 the overdot standing for $\,d/d\tau\,$. Churkin (1998) suggested to term the time-derivatives $~\stackrel{\bf\centerdot}{k}_{\textstyle{_{\,
 l}}}(t)~$ as the {\emph{Love functions}}.\footnote{~Churkin (1998) used functions which he called $\,k_{\it l}(t)\,$ and which were, due
 to a difference in notations, the same as our $\,\stackrel{\bf\centerdot}{k}_{\textstyle{_l}}(\tau)\,$.} Inversion of (\ref{L31})
 trivially yields:
 \ba
 \bar{k}_{\textstyle{_{l}}}(\chi)~=~\int_{0}^{\infty}
 {\bf\dot{\mbox{\it{k}}}}_{\textstyle{_{l}}}(\tau)\;\,e^{-\inc\chi \tau}\,d\tau~=~k_{\textstyle{_{l}}}(0)\;+\;
 \inc~\chi~\int_{0}^{\infty}\left[\,k_{\textstyle{_{l}}}(\tau)\,-\,k_{\textstyle{_{l}}}(0)\,\Theta(\tau)\,\right]\;
                e^{-\inc\chi \tau}\,d\tau~~~,~~~~
 \label{L32}
 \label{gig}
 \ea
 where we integrated only from $\;{0}\,$ because the future disturbance contributes nothing to the present
 distortion, so $\,k_{\it l}(\tau)\,$ vanishes at $\,\tau<0\,$. Recall that the time $\,\tau\,$ denotes the
 difference $t-t\,'$. So $\tau$ is reckoned from the present moment $t$ and is directed back into the past.

 Defining in the standard manner the Fourier components $\,\bar{U}_{\textstyle{_{l}}}(\chi)\,$ and $\,
 \bar{W}_{\textstyle{_{l}}}(\chi)\,$ of functions $\,{U}_{\textstyle{_{l}}}(t)\,$ and $\,{W}_{\textstyle{_{l}}}(t)\,$, we
 write (\ref{subequations_3}) in the frequency domain:
 \ba
 \bar{U}_{\textstyle{_{l}}}(\chi)\;=\;\left(\frac{R}{r}\right)^{l+1}\bar{k}_{\textstyle{_{l}}}(\chi)\;\,\bar{W}_{\textstyle{_{l}}}(\chi)\;\;\;,
 \label{VR}
 \label{L33}
 \ea
 where we denote the frequency simply by $\chi$ instead of the awkward $\chi_{\textstyle{_{lmpq}}}$. To employ (\ref{VR}) in the
 tidal theory, one has to know the frequency-dependencies $\,\bar{k}_{\textstyle{_{l}}}(\chi)\,$.

 \subsection{The positive forcing frequencies $\,\chi\,\equiv\,|\omega|\,$ vs.\\ the positive and negative tidal modes
 $\,\omega\,$}\label{more}\label{4.2}

 It should be remembered that, by relying on formula (\ref{L33}), we place ourselves on thin ice, because the similarity of this
 formula to (\ref{LLJJKK}) and (\ref{L22}) is deceptive.

 In (\ref{LLJJKK}) and (\ref{L22}), it was legitimate to limit our expansions of the stress and the strain to positive frequencies
 $\,\chi\,$ only. Had we carried out those expansions over both positive and negative frequencies $\,\omega\,$, we would have
 obtained, instead of (\ref{LLJJKK}) and (\ref{L22}), similar expressions
 \ba
 2\;\bar{u}_{\gamma\nu}(\omega)\,=\;\bar{J}(\omega)\;\bar{\sigma}_{\gamma\nu}(\omega)\quad\quad\mbox{and}\quad\quad
 \bar{\sigma}_{\gamma\nu}(\omega)\,=\;2\;\bar{\mu}(\omega)\;\bar{u}_{\gamma\nu}(\omega)\;\;\;.
 \label{L22_1}
 \ea
 For positive $\,\omega\,$, these would simply coincide with (\ref{LLJJKK}) and (\ref{L22}), if we rename $\,\omega\,$ as
 $\,\chi\,$. For negative $\,\omega\,=\,-\,\chi\,$, the resulting expressions would read as
 \ba
 2\;\bar{u}_{\gamma\nu}(\,-\,\chi)\,=\;\bar{J}(\,-\,\chi)\;\bar{\sigma}_{\gamma\nu}(\,-\,\chi)\quad\quad\mbox{and}\quad\quad
 \bar{\sigma}_{\gamma\nu}(\,-\,\chi)\,=\;2\;\bar{\mu}(\,-\,\chi)\;\bar{u}_{\gamma\nu}(\,-\,\chi)\;\;\;,
 \label{L22_2}
 \ea
 where we stick to the agreement that $\,\chi\,$ always stands for a positive quantity.
 In accordance with (\ref{whence}), complex conjugation of (\ref{L22_2}) would then return us to (\ref{L22_1}).

 Physically, the negative-frequency components of the stress or strain are nonexistent. If brought into consideration,
 they are obliged to obey (\ref{whence}) and, thus, should play no role, except for a harmless renormalisation of
 the Fourier components in (\ref{dott}).

 When we say that the physically measurable stress $\,\sigma_{\gamma\nu}(t)\,$ is equal to $~\sum{\cal{R}}{\it{e}}\left[\,
 \bar{\sigma}_{\gamma\nu}(\chi)\,e^{\textstyle{^{\inc\chi t}}}\,\right]~$, it is unimportant to us whether the
 $\,\chi$-contribution in $\,\sigma_{\gamma\nu}(t)\,$ comes from the term $~\bar{\sigma}_{\gamma\nu}(\chi)\,
 e^{\textstyle{^{\inc\chi t}}}~$ only, or also from the term $~\bar{\sigma}_{\gamma\nu}(\,-\,\chi)\,
 e^{\textstyle{^{\inc(\,-\,\chi) t}}}~$. Indeed, the real part of the latter is a clone of the real part of the former
 (and it is only the former term that is physical). However, things remain that simple only for the stress and the strain.

 As we emphasised in subsection \ref{frequencies}, the situation with the potentials is drastically different. While the physically measurable potential $\,U(t)\,$ is still equal to $~\sum{\cal{R}}{\it{e}}\left[\,\bar{U}(\chi)\,e^{\textstyle
 {^{\inc\chi t}}}\,\right]~$, it is now {\it{important}} to distinguish whether the $\,\chi$-contribution in $\,U(t)\,$ comes from
 the term $~\bar{U}_{\gamma\nu}(\chi)\,e^{\textstyle{^{\inc\chi t}}}~$ or from the term $~\bar{U}(\,-\,\chi)\,e^{\textstyle
 {^{\inc(\,-\,\chi) t}}}~$, or perhaps from both. Although the negative mode $\,-\chi\,$ would bring the same input as the positive
 mode $\,\chi\,$, these inputs will contribute differently into the tidal torque.
 As can be seen from (\ref{T31}), the secular part of the tidal torque is proportional to $\,\sin\epsilon_{\textstyle{_l}}\,$,
 where $\,\epsilon_{\textstyle{_l}}\equiv\,\omega_{\textstyle{_{lmpq}}}\,\Delta t_{\textstyle{_{lmpq}}}\,$, with the time lag
 $\,\Delta t_{\textstyle{_{lmpq}}}\,$ being positively defined -- see formula (\ref{hagen}). Thus the secular part of the tidal
 torque explicitly contains the sign of the tidal mode $\,\omega_{\textstyle{_{lmpq}}}\,$.

 For this reason, as explained in subsection \ref{frequencies}, a more accurate form of formula (\ref{L33}) should be:
 \ba
 \bar{U}_{\textstyle{_{l}}}(\omega)\;=\;\bar{k}_{\textstyle{_{l}}}(\omega)\;\bar{W}_{\textstyle{_{l}}}(\omega)\;\;\;,
 \label{773}
 \ea
 where $\,\omega\,$ can be of any sign.

 If however, we pretend that the potentials depend on the physical frequency $\,\chi\,=\,|\omega|\,$ only, i.e., if we always
 write $\,U(\omega)\,$ as $\,U(\chi)\,$, then (\ref{L33}) must be written as:
 \begin{subequations}
 \ba
 \bar{U}_{\textstyle{_{l}}}(\chi)\;=\;\bar{k}_{\textstyle{_{l}}}(\chi)\;\bar{W}_{\textstyle{_{l}}}(\chi)\;\;\;,~~~\mbox{when}
 ~~~\chi\,=\,|\omega|~~~\mbox{for}~~~\omega\,>\,0~~~,
 \label{7733_1}
 \ea
 and
 \ba
 \bar{U}_{\textstyle{_{l}}}(\chi)\;=\;\bar{k}^{\,*}_{\textstyle{_{l}}}(\chi)\;\bar{W}_{\textstyle{_{l}}}(\chi)\;\;\;,~~~\mbox{when}
 ~~~\chi\,=\,|\omega|~~~\mbox{for}~~~\omega\,<\,0~~~.
 \label{7733_2}
 \ea
 \label{7733}
 \end{subequations}
 Unless we keep this detail in mind, we shall get a wrong sign for the $\,lmpq\,$ component of the torque after the
 despinning secondary crosses the appropriate commensurability. (We shall, of course, be able to
 mend this by simply inserting the sign ~sgn$\,\omega_{\textstyle{_{lmpq}}}\,$ by hand.)

 \subsection{The complex Love number as a function of the complex compliance}\label{4.7}

 While the static Love numbers depend on the static rigidity modulus $\,\mu\,$ via (\ref{L3}), it is not readily apparent
 that the same relation interconnects $\,\bar{k}_{\it l}(\chi)\,$ with $\,\bar{\mu}(\chi)\,$, the quantities that are the
 Fourier components of the time-derivatives of $\,k_2(t\,')\,$ and $\,\mu(t\,')\,$. Fortunately, the {\it{correspondence principle}}
 (discussed in Appendix \ref{corpr}) tells us that, in many situations, the viscoelastic operational moduli $\,\bar{\mu}(\chi)\,$ or
 $\,\bar{J}(\chi)\,$ obey the same algebraic relations as the elastic parameters $\,\mu\,$ or $\,J\,$.
 This is why, in these situations, the Fourier or Laplace transform of our viscoelastic equations will mimic
 (\ref{L34} - \ref{L35}), except that all the functions will acquire overbars:
 $~\bar{\sigma}_{\textstyle{_{\gamma\nu}}}~=~2~\bar{\mu}~\bar{u}_{\textstyle{_{
 \gamma\nu}}}\;$, etc. So their solution, too, will be $\,\bar{U}_{\it l}=
 \bar{k}_{\it l}\,\bar{W}_{\it l}\,$, with $\,\bar{k}_{\it l}\,$ retaining the same functional
 dependence on  $\,\rho\,$, $\,R\,$, and $\,\bar{\mu}\,$ as in (\ref{L3}), except that now $\,\mu\,$ will have an overbar:
 \ba
 \nonumber
 \bar{k}_{\it l}(\chi)&=&\frac{3}{2\,({\it l}\,-\,1)}\;\;\frac{\textstyle 1}{\;\textstyle 1\;+\;
 \frac{\textstyle{(2\,{\it{l}}^{\,2}\,+\,4\,{\it{l}}\,+\,3)\,\bar{\mu}(\chi)}}{\textstyle{{\it{l}}\,\mbox{g}\,
 \rho\,R}}\;}~=~\frac{3}{2\,({\it l}\,-\,1)}\;\,\frac{\textstyle 1}{\textstyle 1\;+\;A_{\it l}\;\bar{\mu}(\chi)/\mu}
 \quad\quad\quad\quad\quad\quad\quad\quad~~\\
 \label{k2bar_2}
 \label{k2bar}
 \label{L335}
 ~\\
 \nonumber
 &=&\frac{3}{2\,({\it l}\,-\,1)}\;\,\frac{\textstyle 1}{\textstyle 1\;+\;A_{\it l}\;J/\bar{J}(\chi)}~
 =~\frac{3}{2\,({\it l}\,-\,1)}\;\,\frac{\textstyle \bar{J}(\chi)}{\textstyle \bar{J}(\chi)\;+\;A_{\it l}\;J}~~~
 \ea
 Here the coefficients $\,A_{\it l}\,$ are defined via the unrelaxed quantities $\;\mu=\mu(0)=1/J=1/J(0)\;$ in the same
 manner as the static $\,A_{\it l}\,$ were introduced through the static (relaxed) $\,\mu=1/J\,$ in formulae (\ref{L3}).

 The moral of the story is that, at low frequencies, each $\,\bar{k}_{\it l}\,$ depends upon $\,\bar{\mu}\,$ (or upon $\,\bar{J}
 \,$) in the same way as its static counterpart $\,k_{\it l}\,$ depends upon the static $\,\mu\,$ (or upon the static $\,J\,$).
 This happens, because at low frequencies we neglect the acceleration term in the equation of motion (\ref{gegg}), so this
 equation still looks like (\ref{L35}).

 Representing a complex Love number as
 \ba
 \bar{k}_{\it{l}}(\chi)\;=\;{\cal{R}}{\it{e}}\left[\bar{k}_{\it{l}}(\chi)\right]\;+\;\inc\;
 {\cal{I}}{\it{m}}\left[\bar{k}_{\it{l}}(\chi)\right]\;=\;|\bar{k}_{\it{l}}(\chi)|\;
 e^{\textstyle{^{-\inc\epsilon_{\it l}(\chi)}}}
 \label{L36}
 \ea
 we can write for the phase lag $\,\epsilon_{\it l}(\chi)\,$:
 \ba
 \tan\epsilon_{\it l}(\chi)\;\equiv\;-\;\frac{{\cal{I}}{\it{m}}
 \left[\bar{k}_{\it{l}}(\chi)\right]}{{\cal{R}}{\it{e}}\left[\bar{k}_{\it{l}}(\chi)\right]}\;\;\;
 \label{epsi}
 \ea
 or, equivalently:
 \ba
 |\bar{k}_{\it{l}}(\chi)|\;\sin\epsilon_{\it l}(\chi)\;=\;-\;{\cal{I}}{\it{m}}\left[\bar{k}_{\it{l}}(\chi)
 \right]\;\;\;.
 \label{ggffrr}
 \ea
 The products $\;|\bar{k}_{\it{l}}(\chi)|\;\sin\epsilon_{\it l}(\chi)\;$ standing on the left-hand
 side in (\ref{ggffrr}) emerge also in the Fourier series for the tidal potential. Therefore it is
 these products (and not $\;k_{\it l}/Q\;$) that should enter the expansions for forces, torques, and
 the damping rate. This is the link between the body's rheology and the history of its spin: from
 $\,\bar{J}(\chi)\,$ to $\,\bar{k}_{\it{l}}(\chi)\,$ to $\,|\bar{k}_{\it{l}}(\chi)|\;\sin\epsilon(
 \chi)\,$, the latter being employed in the theory of bodily tides.

 Through simple algebra, expressions  (\ref{L335}) entail:
 \ba
 |\bar{k}_{\it l}(\chi)|\;\sin\epsilon_{\it l}(\chi)\;=\;-\;{\cal{I}}{\it{m}}\left[\bar{k}_{\it l}(\chi)
 \right]\;=\;\frac{3}{2\,({\it l}\,-\,1)}\;\,\frac{-\;A_l\;J\;{\cal{I}}{\it{m}}\left[\bar{J}(\chi)\right]}{\left(\;{\cal{R}}{
 \it{e}}\left[\bar{J}(\chi)\right]\;+\;A_l\;J\;\right)^2\;+\;\left(\;{\cal{I}}{\it{m}}
 \left[\bar{J}(\chi)\right]\;\right)^2} ~~~.~~~~~
 \label{L39}
 \ea

 As we know from subsections \ref{frequencies} and \ref{more}, formulae (\ref{epsi} - \ref{L39}) should be used with care.
 Since in reality the potential $\,\bar{U}\,$ and therefore also $\,\bar{k}_l\,$ are functions not of $\,\chi\,$ but of $\,\omega
 \,$, then formulae (\ref{L39}) should be equipped with multipliers ~sgn$\,\omega_{\textstyle{_{lmpq}}}\,$, when plugged into
 the expression for the $lmpq$ component of the tidal force or torque. This prescription is equivalent to (\ref{7733}).

 \subsection{Should we write $\,\bar{k}_{{\it l}mpq}\;$ and $\,\epsilon_{{\it l}mpq}\,$,
 or would $\,\bar{k}_{{\it l}}\,$ and $\,\epsilon_{{\it l}}\,$ be enough?}\label{3.5}\label{4.8}

 In the preceding subsection, the static relation (\ref{2}) was generalised to evolving settings as
 \ba
 U_{{\it l}mpq}(\erbold,\,t)\;=\;\left(\,\frac{R}{r}\,
 \right)^{{\it l}+1}\hat{k}_{{\it l}}(t)~W_{\it{l}mpq}(\eRbold\,,\;\erbold^{\;*},\,t\,')~~~,
 \label{L40}
 \ea
 where $~{\it l}mpq~$ is a quadruple of integers employed to number a Fourier mode in the Darwin-Kaula expansion (\ref{L51}) of the
 tide, while $\,U_{{\it l}mpq}(\erbold,\,t)\,$ and $\,W_{\it{l}mpq}(\eRbold\,,\;\erbold^{\;*},\,t\,')\,$ are the harmonics containing
 $\,\cos(\chi_{{\it l}mpq}t  - \epsilon_{{\it l}mpq})\,$ and $\,\cos(\chi_{{\it l}mpq}t\,')\,$ correspondingly.

 One might be tempted to generalise (\ref{2}) even further to
 \ba
 \nonumber
 U_{{\it l}mpq}(\erbold,\,t)\;=\;\left(\,\frac{R}{r}\,
 \right)^{{\it l}+1}\hat{k}_{{\it l}mpq}(t)~W_{\it{l}mpq}(\eRbold\,,\;\erbold^{\;*},\,t\,')~~~,
 \label{L41}
 \ea
 with the Love operator (and, consequently, its kernel, the Love function) bearing dependence upon
 $\,m$, $\,p$, $\,$and $\,q\,$. Accordingly, (\ref{VR}) would become
 \ba
 \bar{U}_{\it{l}mpq}(\chi)\;=\;\bar{k}_{{\it l}mpq}(\chi)\;\bar{W}_{\it{l}mpq}(\chi)\;\;\;.
 \label{L42}
 \ea
 Fortunately, insofar as the Correspondence Principle is valid, the functional form of the function
 $\,\bar{k}_{{\it l}mpq}(\chi)\,$ depends upon $\,\it l\,$ only and, thus, can be written down simply as
 $\;\bar{k}_{{\it l}}(\chi_{{\it l}mpq})\;$. We know this from the considerations offered after equations
 (\ref{L34} - \ref{L35}). There we explained that $\,\bar{k}_{{\it l}}\,$
 depends on $\,\chi=\chi_{{\it l}mpq}\,$ only via $\,\bar{J}(\chi)\,$, while {\emph{the functional
 form}} of $\,\bar{k}_{{\it l}}\,$ bears no dependence on $\,\chi=\chi_{{\it l}mpq}\,$ and, therefore,
 no dependence on $\,m,\,p,\,q\,$.

 The phase lag is often denoted as $\,\epsilon_{{\it l}mpq}\,$, a time-honoured tradition established by Kaula (1964). However,
 as the lag is expressed through $\,\bar{k}_{\it l}\,$ via (\ref{epsi}), we see that all said above about $\,\bar{k}_{\it l}\,$
 applies to the lag too: while the functional form of the dependency $\,\epsilon_{{\it l}mpq}(\chi)\,$ may be different for different
 {\it{l}}{\small{$\,$s}}, it is invariant under the other three integers, so the notation $~\epsilon_{{\it l}}(\chi_{{\it l}mpq})
 ~$ would be more adequate.

 It should be mentioned, though, that for bodies of pronounced non-sphericity
 coupling between the spherical harmonics furnishes the Love numbers and lags whose expressions through the frequency, for a fixed
 $\,l\,$, have different functional forms for different $\,m,\,p,\,q\,$. In these cases, the notations $\,\bar{k}_{{\it l}mpq}\,$ and
 $\,\epsilon_{{\it l}mpq}\,$ become necessary (Smith 1974; Wahr 1981a,b,c; Dehant 1987a,b). For a slightly non-spherical body, the
 Love numbers differ from the Love numbers of the spherical reference body by a term of the order of the flattening, so a small
 non-sphericity can usually be neglected.

 \subsection{Rigidity vs self-gravitation}\label{3.6}\label{4.9}

 For small bodies and small terrestrial planets, the values of $\,A_{\textstyle{_l}}\,$ vary from about unity to dozens to hundreds.
 For example, $\,A_2\,$ is about 2 for the Earth (Efroimsky 2012), about 20 for Mars (Efroimsky \& Lainey 2007), about 80 for
 the Moon (Efroimsky 2012), and about 200 for Iapetus (Castillo-Rogez et al. 2011). For superearths, the values will be much smaller
 than unity, though.

 Insofar as
 \ba
 A_l\;\,
 \frac{J}{{\bf{\Large|}}\stackrel{-~~~}{J(\chi)}{\bf{\Large|}}}\;\,\gg\;1~~~,
 \label{inequality}
 \label{I30}
 \ea
 one can approximate (\ref{k2bar}) with
 \ba
 \bar{k}_l(\chi)\,=\;-\;\frac{3}{2(l-1)}\;\frac{{\textstyle\,\stackrel{\mbox{\bf \it \_}}{J}
 (\chi)}}{{\textstyle\,\stackrel{\mbox{\bf \it \_}}{J}
 (\chi)}\;+\;A_l\;{\textstyle J}}\;=\;-\;\frac{3}{2}\;\frac{{\textstyle\,\stackrel{\mbox{\bf \it \_}}{J}
 (\chi)}}{A_l\;{\textstyle J}}~+~O\left(~|{\textstyle\,\stackrel{\mbox{\bf \it \_}}{J}
 }/(A_l\,J)\,|^2~\right)\;\;\;,~~~~~
 \label{pivotal}
 \ea
 except in the closest vicinity of an $\,lmpq\,$ resonance, where the tidal frequency $\,\chi_{\textstyle{_{lmpq}}}\,$ approaches
 nil, and $\,\bar{J}\,$ diverges for some rheologies -- like, for example, for those of Maxwell or Andrade.

 Whenever the approximate formula (\ref{pivotal}) is applicable, we can rewrite (\ref{epsi}) as
 \ba
 \tan\epsilon(\chi)\;\equiv\;
 -\;\frac{{\cal{I}}{\it{m}}\left[\bar{k}_{\it{l}}(\chi)\right]}{{\cal{R}}{\it{e}}
 \left[\bar{k}_{\it{l}}(\chi)\right]}\;\approx\;
 -\;\frac{{\cal{I}}{\it{m}}\left[\bar{J}(\chi)\right]}{{\cal{R}}{\it{e}}\left[\bar{J}(\chi)\right]}
 \;=\;\tan\delta(\chi)\;\;\;,
 \label{epsidelta}
 \ea
 wherefrom we readily deduce that the phase lag $\,\epsilon(\chi)\,$ of the tidal frequency $\,\chi\,$
 coincides with the phase lag of the complex compliance:
 \ba
 \epsilon(\chi)\;\approx\;\delta(\chi)\;\;\;,
 \label{deltaepsi}
 \ea
 provided $\,\chi\,$ is not too close to nil (i.e., provided we are not too close to the commensurability). This way, insofar as the condition (71) is fulfilled, the component $\,\bar{U}_{\it{l}}(\chi)\,$ of the primary's
 potential lags behind the component $\,\bar{W}_{\it{l}}(\chi)\,$ of the perturbed potential by the same phase angle
 as the strain lags behind the stress at frequency $\,\chi\,$ in a sample of the material. Dependent upon the rheology, a
 vanishing tidal frequency may or may not limit the applicability of (71) and thus cause a considerable difference between $\,\epsilon\,$ and $\,\delta\,$.

 In other words, the suggested approximation is valid insofar as changes of shape are determined solely by the local material
 properties, and not by self-gravitation of the object as a whole. Whether this is so or not -- depends upon the rheological model. For
 a Voigt or SAS$\,$\footnote{~The acronym {\it{SAS}} stands for the {\it{Standard Anelastic Solid}}, which is another name for the
 Hohenemser-Prager viscoelastic model. See the Appendix for details.} solid in the limit of $~{\chi\rightarrow 0}~$, we have $~\bar{J}(\chi)\,\rightarrow\,J~$, so the zero-frequency limit of
 $\,\bar{k}_l(\chi)\,$ is the static Love number $\,k_l\,\equiv\,|\bar{k}(0)|\,$. In this case, approximation (\ref{pivotal} - \ref{deltaepsi}) remains
 applicable all the way down to $\,\chi=0\,$. For the Maxwell and Andrade models, however, one obtains, for vanishing frequency: $~
 \bar{J}(\chi)\,\sim\,1/(\eta\chi)~$, whence $\,\bar{\mu}\,\sim\,\eta\chi\,$ and $\,\bar{k}_2(\chi)\,$ approaches the hydrodynamical Love
 number $\,k_2^{(hyd)} =3/2\,$.

 We see that, for the Voigt and SAS models, approximation $\,(\ref{deltaepsi})\,$ can work, for $A_l\gg 1$,  at all frequencies, because the condition $\,A_L\gg 1\,$ can
 be set for all frequencies. For the Maxwell and Andrade solids, this condition holds only at frequencies larger than $~\tau_{_M}^{-1}A_{\textstyle
  {_l}}^{-1}\,=\,\frac{\textstyle\mu}{\textstyle\eta}\,A_{\textstyle{_l}}^{-1}$, and so does the approximation (\ref{deltaepsi}). ~Indeed, at
 frequencies below this threshold, self-gravitation ``beats" the local material properties of the body, and the behaviour of the tidal lag deviates
 from that of the lag in a sample. This deviation will be indicated more clearly by formula
 (\ref{cccc}) in the next section. The fact that, for some models, the tidal lag $\,\epsilon\,$ deviates from the material lag angle
 $\,\delta\,$ at the lowest frequencies should be kept in mind when one wants to explore {\it{crossing}} of a resonance.

 A standard caveat is in order, concerning formulae (\ref{pivotal} - \ref{deltaepsi}). Since in reality the
 potential $\,\bar{U}\,$ is a function of $\,\omega\,$ and not $\,\chi\,$, our illegitimate use of $\,\chi\,$ should be
 compensated by multiplying the function $\,\epsilon_{\textstyle{_l}}(\chi_{\textstyle{_{lmpq}}})\,$ with ~sgn$\,\omega_{\textstyle
 {_{lmpq}}}\,$, when the lag shows up in the expression for the tidal force or torque.

 \subsection{The case of inhomogeneous bodies}\label{3.7}

 Tidal dissipation within a multilayer near-spherical body is studied through expanding the involved fields over the
 spherical harmonics in each layer, setting the boundary conditions on the outer surface, and using the matching conditions
 on boundaries between layers. This formalism was developed by Alterman et al (1959). An updated discussion of the method
 can be found in Sabadini \& Vermeersen (2004). For a brief review, see Legros et al (2006).

 Calculation of tidal dissipation in a Jovian planet is an even more formidable task (see Remus et al. 2012a and
 references therein). However dissipation in a giant planet with a solid core may turn out to be approachable by analytic means
 (Remus et al. 2011, 2012b).

 \section{Dissipation at different frequencies}\label{dissip}

 \subsection{The data collected on the Earth: ~in the lab,\\ over seismological basins, and through geodetic measurements}\label{previous}

 In Efroimsky \& Lainey (2007), we considered the generic rheological model
 \begin{subequations}
 \ba
 Q\;=\;\left(\,{\cal{E}}\,\chi\,\right)^{\textstyle{^{\alpha}}}~~,
 \label{generic_1}
 \ea
 where $\,\chi\,$ is the tidal frequency and $\,{\cal{E}}\,$ is a parameter having the dimensions of time. The physical
 meaning of this parameter is elucidated in {\it{Ibid.}}. Under the special choice of $\,\alpha = - 1\,$
 and for sufficiently large values of $\,Q\,$, this parameter coincides with the time lag $\,\Delta t\,$ which, for this
 special rheology, turns out to be the same at all frequencies.

 Actual experiments register not the inverse quality factor but the phase lag between the reaction and the action.
 So the empirical law should rather be written down as
 \ba
 \frac{1}{\sin\delta}\;=\;\left(\,{\cal{E}}\,\chi\,\right)^{\textstyle{^{\alpha}}}~~,
 \label{generic_2}
 \ea
 \label{generic}
 \end{subequations}
 which is equivalent to (\ref{generic_1}), provided the $Q$ factor is defined there as $\,Q_{\textstyle{_{energy}}}\,$ and not as
 $\,Q_{\textstyle{_{work}}}~$ -- see subsection \ref{damp} for details.

 The applicability realm of the empirical power law (\ref{generic}) is remarkably broad -- in terms of both the physical constituency
 of the bodies and their chemical composition. Most intriguing is the robust universality of the values taken by the index $\,\alpha
 \,$ for very different materials: between $\,0.2\,$ and $\,0.4\,$ for ices and silicates, and between $\,0.14\,$ and $\,0.2\,$ for
 partial melts.  Historically, two communities independently converged on this form of dependence.

 In the material sciences, the rheological model (\ref{112}), wherefrom the power law (\ref{generic_2}) stems,
 traces its lineage to the groundbreaking work by Andrade (1910) who explored creep in metals.
 Through the subsequent century, this law  was found to be applicable to a vast variety of other materials, including minerals (Weertman
 \& Weertman 1975, Tan et al. 1997) and their partial melts (Fontaine et al. 2005). As recently discovered by McCarthy et al. (2007) and
 Castillo-Rogez (2009), the same law, with almost the same values of $\,\alpha\,$, also applies to ices. The result is milestone, taken
 the physical and chemical differences between ices and silicates. It is agreed upon that in crystalline materials the Andrade regime
 can find its microscopic origin both in the dynamics of dislocations (Karato \& Spetzler 1990) and in the grain-boundary diffusional
 creep (Gribb \& Cooper 1998). As the same behaviour is inherent in metals, silicates, ices, and even glass-polyester composites (Nechada
 et al. 2005), it should stem from a single underlying phenomenon determined by some principles more general than specific material
 properties. An attempt to find such a universal mechanism was undertaken by Miguel et al. (2002). See also the theoretical
 considerations offered in Karato \& Spetzler (1990).

 In seismology, the power law (\ref{generic}) became popular in the second part of the XX$^{th}$ century, with the progress of precise
 measurements on large seismological basins (Mitchell 1995, Stachnik et al. 2004, Shito et al. 2004). Further confirmation of this law
 came from geodetic experiments that included: (a) satellite laser ranging (SLR) measurements of tidal variations in the $\,J_2\,$
 component of the gravity field of the Earth; (b) space-based observations of tidal variations in the Earth's rotation rate; and (c)
 space-based measurements of the Chandler Wobble period and damping (Benjamin et al. 2006, Eanes \& Bettadpur 1996, Eanes 1995). Not
 surprisingly, the Andrade law became a key element in the recent attempt to construct a universal rheological model of the Earth's
 mantle (Birger 2007). This law also became a component of the non-hydrostatic-equilibrium model for the
 zonal tides in an inelastic Earth by Defraigne \& Smits (1999), a model that became the basis for the
 IERS Conventions (Petit \& Luzum 2010). While the lab experiments give for $\,\alpha\,$
 values within $\,0.2\,-\,0.4\,$, the geodetic techniques favour the interval $\,0.14\,-\,0.2\,$. This
 minor discrepancy may have emerged due to the presence of partial melt in the mantle and, possibly, due to
 nonlinearity at high bounding pressures in the lower mantle. The universality of the Andrade law compels us to assume that (\ref{generic}) works equally well for other terrestrial
 bodies. Similarly, the applicability of (\ref{generic}) to samples of ices in the lab is likely to
 indicate that this law can be employed for description of an icy moon as a whole.

 Karato \& Spetzler (1990) argue that at frequencies below a certain threshold $\,\chi_0\,$ anelasticity gives way to purely viscoelastic
 behaviour, so the
 parameter $\,\alpha\,$ becomes close to unity.\footnote{~This circumstance was ignored by Defraigne \& Smits (1999). Accordingly, if
 the claims by Karato \& Spetzler (1990) are correct, the table of corrections for the tidal variations in the Earth's rotation in the
 IERS Conventions is likely to contain increasing errors for periods of about a year and longer.

 This detail is missing in the theory of the Chandler wobble of Mars, by Zharkov \& Gudkova (2009).} For the Earth's mantle, the
 threshold corresponds to the time-scale about a year or slightly longer. Although in Karato \& Spetzler (1990) the rheological law is
 written in terms of $\,1/Q\,$, we shall substitute it with a law more appropriate to the studies of tides:
 \ba
 \nonumber
 k_{\textstyle{_l}}\,\sin\epsilon_{\textstyle{_l}}
 \;=\;\left(\,{\cal E}\,\chi\,\right)^{\textstyle{^{\,-\,p}}}
 \;\;\;,
 ~~~~~\mbox{where}&\;&\;\;p\,=\,0.2\;-\;0.4\;\;\;\mbox{for}\;\;\;\chi\,>\,\chi_0\;\;\;\\
  \mbox{and~~~}&\;&\;\;p~\,\sim~\, 1~\quad~\quad~\quad\mbox{for}\;\;\;\chi\,<\,\chi_0\;\;\;,\;
 \label{rheology}
 \ea
 $\chi\,$ being the frequency, and $\,\chi_0\,$ being the frequency threshold below which viscosity takes over anelasticity.

 The reason why we write the power scaling law as (\ref{rheology}) and not as (\ref{generic}) is that at the lowest frequencies the
 geodetic measurements give us actually $\,k_l\,\sin\epsilon_l\,=\,-\,{\cal{I}}{\it{m}}\left[\,\bar{k}_{\textstyle{_l}}(\chi)\,\right]
 \,$ and not the lag angle $\,\delta\,$ in a sample (e.g., Benjamin et al. 2006). For this same reason, we denoted the exponents
 in (\ref{generic}) and (\ref{rheology}) with different letters, $\,\alpha\,$ and $\,p\,$. Below we shall see that these exponents
 do not always coincide. Another reason for giving preference to (\ref{rheology}) is that not only the sine of the lag but also the
 absolute value of the Love number is frequency dependent.

 \subsection{Tidal damping in the Moon, from laser ranging}

 Fitting of the LLR data to the power scaling law (\ref{generic}), which was carried out by Williams et al. (2001), has demonstrated
 that the lunar mantle possesses quite an abnormal value of the exponent: $\,-\,0.19\,$. A later reexamination in Williams et al.
 (2008) rendered a less embarrassing value, $\,-\,0.09\,$, which nevertheless was still negative and thus seemed to contradict our
 knowledge about microphysical damping mechanisms in minerals. Thereupon, Williams \& Boggs (2009) commented:\\
 ~\\
 ``{\it{There is a weak dependence of tidal specific dissipation $\,Q\,$ on period. The $\,Q\,$ increases from $\,\sim 30\,$ at a
 month to $\,\sim 35\,$ at one year. $~Q\,$ for rock is expected to have a weak dependence on tidal period, but it is expected to
 decrease with period rather than increase. The frequency dependence of $\,Q\,$ deserves further attention and should be improved.}}"\\
 ~\\
 While there always remains a possibility of the raw data being insufficient or of the fitting procedure being imperfect, the fact
 is that the negative exponent obtained in {\it{Ibid.}} does {\bf{not}} necessarily contradict the scaling law (\ref{generic})
 proven for minerals and partial melts. Indeed, the exponent obtained by the LLR Team was not the $\,\alpha\,$ from (\ref{generic})
 but was the $\,p\,$ from (\ref{rheology}). The distinction is critical due to the difference in frequency-dependence of the seismic
 and tidal dissipation. It turns out that the near-viscous value $\,p\,\sim\,1\,$ from the second line of (\ref{rheology}),
 appropriate for low frequencies, does not retain its value all the way to the zero frequency. Specifically, in subsection
 \ref{behaviour} we shall see that at the frequency  $~\frac{\textstyle 1}{\textstyle{\tau_{_M}\,{A_{\textstyle{_l}}}}}~$
 (where $\,\tau_{_M}=\eta/\mu\,$ is the Maxwell time, with $\,\eta\,$ and $\,\mu\,$ being the lunar mantle's viscosity and rigidity),
 the exponent $\,p\,$ begins to decrease with the decrease of the frequency. As the frequency becomes lower, $\,p\,$ changes its
 sign and eventually becomes $\,-\,1\,$ in a close vicinity of $\,\chi\,=\,0\,$. This behaviour follows from calculations based on a
 realistic rheology (see formulae (\ref{aaaa} - \ref{cccc}) below), and it goes along well with the evident physical fact that the
 average tidal torque must vanish in a resonance.\footnote{~For example, the principal tidal torque $\,\tau_{\textstyle{_{lmpq}}}=\,\tau_{
 \textstyle{_{2200}}}\,$ acting on a secondary must vanish when the secondary is crossing the synchronous orbit. Naturally, this
 happens because $\,p\,$ becomes $\,-\,1\,$ in the close vicinity of $\,\chi_{\textstyle{_{2200}}}=\,0\,$.} In subsection \ref{Moon},
 comparison of this behaviour with the LLR results will yield us an estimate for the mean lunar viscosity.

 \subsection{The Andrade model as an example of viscoelastic behaviour}\label{Andrade_section}

 The complex compliance of a Maxwell material contains a term $\,J=J(0)\,$ responsible for the elastic part of the deformation and a
 term $\,-\,\frac{\textstyle\inc}{\textstyle\chi\eta}\,$ describing the viscosity. Whatever other terms get incorporated into the
 compliance, these will correspond to other forms of hereditary reaction. The available geophysical data strongly favour a particular
 extension of the Maxwell approach, the {\it{Andrade model}}
 (Cottrell \& Aytekin 1947, Duval 1976).
 In modern notations, the model can be expressed as \footnote{As long as we agree to integrate over $~\,t-t\,'\in\left[\right.0,
 \,\infty\left.\right)~$, the terms $\,\beta (t-t\,')^{\alpha}~$ and $~{\textstyle \eta}^{-1}\left(t-t\,'\right)
 \,$ can do without the Heaviside step-function $\,\Theta(t-t\,')\,$. We remind though that the first term, $\,J\,$, does need this
 multiplier, so that insertion of (\ref{I6412}) into (\ref{I12_4}) renders the desired $\,J\,\delta(t-t\,')\,$ under the integral,
 after the differentiation in (\ref{I12_4}) is performed.}
 \ba
 J(t-t\,')\;=\;\left[\;J\;+\;\beta\,(t-t\,')^{\alpha}\,+\;{\eta}^{-1}\left(t-t\,'\right)\;\right]\,
 \Theta(t-t\,')\;\;\;,
 \label{I64}
 \label{I6412}
 \ea
 $\alpha\,$ being a dimensionless parameter, $\,\beta\,$ being a dimensional parameter, $\eta\,$ denoting the steady-state viscosity, and
 $\,J\,$ standing for the unrelaxed compliance, which is inverse to the unrelaxed rigidity: $\,J\equiv J(0)=1/\mu(0)\,=\,1/\mu\,$. We see
 that (\ref{I6412}) is the Maxwell model amended with an extra term of a hereditary nature.

 A simple example illustrating how the model works is rendered by deformation under constant loading. In this case, the anelastic term
 dominates at short times, the strain thus being a convex function of $\,t\,$ (the so-called primary or transient creep). As time goes
 on and the applied loading is kept constant, the viscous term becomes larger, and the strain becomes almost linear in time -- a
 phenomenon called the secondary creep.

 Remarkably, for all minerals (including ices) the values of $\,\alpha\,$ belong to the interval from $\,0.14\,$ through $\,0.4\,$
 (more often, through $\,0.3\,$) -- see the references in subsection \ref{previous} above. The
 other parameter, $\,\beta\,$, may be rewritten as
 \ba
 \beta\,=\,J~\tau_{_A}^{-\alpha}\,=~ \mu^{-1}\,\tau_{_A}^{-\alpha}~~~,
 \label{beta_0}
 \ea
 the quantity $\,\tau_{_A}\,$ having dimensions of time. This quantity is the timescale associated with the Andrade creep, and it
 may be termed as the ``Andrade time" or the ``anelastic time". It is clear from (\ref{beta_0}) that a short $\,\tau_{_A}\,$ makes
 the anelasticity more pronounced, while a long $\,\tau_{_A}\,$ makes the anelasticity weak.\footnote{~While the Andrade creep is
 likely to be caused by ``unpinning" of jammed dislocations (Karato \& Spetzler 1990, Miguel et al 2002), it is not apparently
 clear if the Andrade time can be identified with the typical time of unpinning of defects.}

 It is known from Castillo-Rogez et al. (2011) and Castillo-Rogez \& Choukroun (2010) that for some minerals, within some frequency bands, the
 Andrade time gets very close to the Maxwell time:
 \ba
 \tau_{_A}\,\approx~\tau_{_M}~\quad~\Longrightarrow~\quad~\beta\,\approx~J~\tau_{_M}^{-\alpha}\,=\,J^{1-\alpha}\,\eta^{-\alpha}\,=\,\mu^{\alpha-1}
 \,\eta^{-\alpha}~~~,
 \label{beta}
 \ea
 where the relaxation Maxwell time is given by:
 \ba
 \tau_{_M}\,\equiv\,\frac{\eta}{\mu}\,=\,\eta\,J~~.
 \label{}
 \ea
 On general grounds, though, one cannot expect the anelastic timescale $\,\tau_{_A}\,$ and the viscoelastic timescale $\,\tau_{_M}\,$ to
 coincide in all situations. This is especially so due to the fact that both these times may possess some degree of frequency-dependence.
 Specifically, there exist indications that in the Earth's mantle the role of anelasticity (compared to viscoelasticity) undergoes a
 decrease when the frequencies become lower than $\,1/$yr -- see the miscrophysical model suggested in subsection 5.2.3 of Karato \&
 Spetzler (1990). It should be remembered, though, that the relation between $\,\tau_{_A}\,$ and $\,\tau_{_M}\,$ may depend also upon the
 intensity of loading, i.e., upon the damping mechanisms involved. The microphysical model considered in {\it{Ibid.}} was applicable to
 strong deformations, with anelastic dissipation being dominated by dislocations unpinning. Accordingly, the dominance of viscosity over
 anelasticity ($\,\tau_{_A}\ll\tau_{_M}\,$) at low frequencies may be regarded proven for strong deformations only. At low stresses, when
 the grain-boundary diffusion mechanism is dominant, the values of $\,\tau_{_A}\,$ and $\,\tau_{_M}\,$ may remain comparable at low
 frequencies. The topic needs further research.

 In terms of the Andrade and Maxwell times, the compliance becomes:
 \ba
 J(t-t\,')\;=\;J~\left[~1~+~\left(\frac{t-t\,'}{\tau_{_A}}\right)^{\alpha}\,+~\frac{t-t\,'}{\tau_{_M}}\;\right]\,
 \Theta(t-t\,')\;\;\;.
 \label{I64}
 \ea

 In the frequency domain, compliance (\ref{I64}) will look:
 \begin{subequations}
 \ba
 {\bar{\mathit{J\,}}}(\chi)&=&J\,+\,\beta\,(i\chi)^{-\alpha}\;\Gamma\,(1+\alpha)\,-\,\frac{i}{\eta\chi}
 \label{112_1}\\
 \nonumber\\
 &=&J\,\left[\,1\,+\,(i\,\chi\,\tau_{_A})^{-\alpha}\;\Gamma\,(1+\alpha)~-~i~(\chi\,\tau_{_M})^{-1}
 \right]\;\;\;,
 \label{112_2}
 \ea
 \label{112}
 \label{LL44}
 \end{subequations}
 $\chi\,$ being the frequency, and $\,\Gamma\,$ denoting the Gamma function. The imaginary and real parts of the complex
 compliance are:
 \begin{subequations}
 \ba
 {\cal I}{\it m} [ \bar{J}(\chi)]&=&-\;\frac{1}{\eta\,\chi}\;-\;\chi^{-\alpha}\,\beta\;\sin\left(
 \,\frac{\alpha\,\pi}{2}\,\right)
 \;\Gamma(\alpha\,+\,1)
 ~~~~~~~~~~~~~~~\quad\quad\quad
 \label{A3a}\\
 \nonumber\\
 &=&-\;J\,(\chi\tau_{_M})^{-1}\;-\;J\,(\chi\tau_{_A})^{-\alpha}\;\sin\left(
 \,\frac{\alpha\,\pi}{2}\,\right)\;\Gamma(\alpha\,+\,1)~~~~~~~~~~~~~~~\quad\quad\quad
 \label{A3b}
 \ea
 \label{A3}
 \label{A3c}
 \end{subequations}
 and
 \begin{subequations}
 \ba
 {\cal R}{\it e} [ \bar{J}(\chi)]&=&J\;+\;\chi^{-\alpha}\,\beta\;\cos\left(\,\frac{\alpha\,\pi}{2}
 \,\right)\;\Gamma(\alpha\,+\,1)~~~~~~~~~~~~~~~~~~~~~~~~~~~~~~
 \label{A4a}\\
 \nonumber\\
 &=&J\;+\;J\,(\chi\tau_{_A})^{-\alpha}\;\cos\left(\,\frac{\alpha\,\pi}{2}\,\right)
 \;\Gamma(\alpha\,+\,1)~~~,\quad\quad\quad\quad\quad~\quad\quad\quad\quad\quad\quad
 \label{A4b}
 \ea
 \label{A4}
 \label{A4c}
 \end{subequations}
 whence we obtain the following dependence of the phase lag upon the frequency:
 \begin{subequations}
 \ba
 \tan\delta(\chi)~=~-~\frac{{{\cal{I}}\textit{m}\left[\bar{J}(\chi)\right]}}{{\cal{R}}{\textit{e}\left[\bar{J}(\chi)\right]}}
 &=&\frac{(\eta\;\chi)^{\textstyle{^{\,-1}}}\,+~\chi^{\textstyle{^{\,-\alpha}}}\;\beta\;\sin\left(\frac{
 \textstyle\alpha~\pi}{\textstyle 2}\right)\Gamma\left(\alpha\,+\,1\right)}{\mu^{\textstyle{^{~-1}}
 }\,+\;\chi^{\textstyle{^{\,-\alpha}}}\;\beta\;\cos\left(\frac{\textstyle\alpha\;\pi}{\textstyle 2}
 \right)\;\Gamma\left(\textstyle\alpha\,+\,1\right)}
 \label{113a}
 \label{LL45a}\\
 \nonumber\\
 \nonumber\\
 &=&\frac{z^{-1}\,\zeta~+~z^{-\alpha}\,\sin\left(\frac{
 \textstyle\alpha~\pi}{\textstyle 2}\right)\Gamma\left(\alpha\,+\,1\right)}{1\,+~z^{-\alpha}
 \,\cos\left(\frac{\textstyle\alpha\;\pi}{\textstyle 2}\right)~\Gamma\left(\textstyle\alpha\,+\,1\right)}~~~.\quad\quad
 \label{113b}
 \label{LL45b}
 \ea
 \label{113}
 \label{LL45}
 \end{subequations}
 Here $\,z\,$ is the dimensionless frequency defined as
 \ba
 z~\equiv~\chi~\tau_{_A}~=~\chi~\tau_{_M}~\zeta~~~,
 \label{sim}
 \ea
 while $\,\zeta\,$ is a dimensionless parameter of the Andrade model:
 \ba
 \zeta~\equiv~\frac{\textstyle\tau_{_A}}{\textstyle\tau_{_M}}~~~.
 \label{ham}
 \ea

 \subsection{Tidal response of viscoelastic
 near-spherical bodies obeying the Andrade and Maxwell models}\label{behaviour}

 An $~lmpq\,$ term in the expansion for the tidal torque is proportional to the factor $~k_{\textstyle{_l}}(\chi)\,\sin\epsilon_{
 \textstyle{_l}}(\chi)\,=\,|\bar{k}_{\textstyle{_l}}(\chi_{\textstyle{_{lmpq}}})|\,\sin\epsilon_{\textstyle{_l}}(\chi_{\textstyle{_{lmpq}
 }})~$. Hence the tidal response of a body is determined by the frequency-dependence of these factors.

 Combining (\ref{L39}) with (\ref{112}), and keeping in mind that $\,A_l\gg 1\,$, it is easy to write down the frequency-dependencies
 of the products $~|\bar{k}_{\textstyle{_l}}(\chi)|~\sin\epsilon_{\textstyle{_l}}(\chi)~$. Referring the reader to Appendix
 \ref{Andarde_Maxwell} for details, we present the results, without the sign multiplier.\\

 $\boldmath \bullet $ ~{\underline{In the high-frequency band:}}
 \ba
 |\bar{k}_{\textstyle{_l}}(\chi)|\,\sin\epsilon_{\textstyle{_l}}(\chi)\,\approx\,\frac{3}{2\,(l-1)}\;\frac{A_{\textstyle{_l}}}{(A_{
 \textstyle{_l}}+\,1)^2}~\sin\left(\frac{\alpha\pi}{2}\right)~\Gamma(\alpha+1)~\,\zeta^{-\alpha}\,\left(
 \,\tau_{_M}\,\chi\,\right)^{-\alpha}\quad,~\quad\mbox{for}~\quad~\chi\,\gg\,\tau_{_M}^{-1}~~.~\quad
 \label{aaaa}
 \ea
 For small bodies and small terrestrial planets (i.e., for $\,A_{\textstyle{_{l}}}\gg 1\,$), the boundary between the high and
 intermediate frequencies turns out to be
 $$\,\chi_{{_{HI}}}\,=\,\tau_{_M}^{-1}\,\zeta^{\textstyle{^{\textstyle\,\frac{\alpha}{1-\alpha}}}}~~~.$$
 For large terrestrial planets (i.e., for $\,A_{\textstyle{_{l}}}\ll 1\,$) the boundary frequency is
 $$\,\chi_{{_{HI}}}\,=\,\tau_{_A}^{-1}\,=\,\tau_{_M}^{-1}\,\zeta^{-1}~~~.$$
 At high frequencies, anelasticity dominates. So, dependent upon the microphysics of the mantle, the parameter $\,\zeta\,$ may be of
 order unity or slightly lower. We say {\it{slightly}}, because we expect both anelasticity and viscosity to be present near the
 transitional zone. (A too low $\,\zeta\,$ would eliminate viscosity from the picture completely.) This said, we may assume that
 the boundary $\,\chi_{{_{HI}}}\,$ is comparable to $\,\tau_{_M}^{-1}\,$ for both small and large solid objects. This is why in
 (\ref{aaaa}) we set the inequality simply as $~\chi\,\gg\,\tau_{_M}^{-1}~$.\\
  ~\\

 $\boldmath \bullet $ ~{\underline{In the intermediate-frequency band:}}
 \ba
 |\bar{k}_{\textstyle{_l}}(\chi)|\,\sin\epsilon_{\textstyle{_l}}(\chi)\,\approx\,\frac{3}{2\,(l-1)}\;\frac{A_{\textstyle{_l}}}{(A_{
 \textstyle{_l}}+1)^2}\,~
 \left(\,\tau_{_M}\,\chi\,\right)^{-1}\quad,\quad~\quad\mbox{for}\quad\quad\tau_{_M}^{-1}\gg\chi\gg
 \tau_{_M}^{-1}\,(A_{\textstyle{_l}}+1)^{-1}~\,~.\quad\quad
  \label{bbbb}
 \ea
 While the consideration in the Appendix \ref{Andarde_Maxwell} renders $\,\tau_{_M}^{-1}\,\zeta^{\textstyle{^{\textstyle\,\frac{\alpha}{
 1-\alpha}}}}\,$ for the upper bound, here we approximate it with $\,\tau_{_M}^{-1}\,$ in understanding that $\,\zeta\,$ does not differ
 from unity too much near the transitional zone. Further advances of rheology may challenge this convenient simplification.\\
 ~\\

 $\boldmath \bullet $ ~{\underline{In the low-frequency band:}}
  \ba
 |\bar{k}_{\textstyle{_l}}(\chi)|~\sin\epsilon_{\textstyle{_l}}(\chi)\,\approx\,\frac{3}{2\,(l-1)\textbf{}}~{A_{\textstyle{_l}}}~\,
 \tau_{_M}~\chi\quad\quad,\quad\quad\quad\quad~\quad\quad\,\mbox{for}~\quad~\quad\,\tau_{_M}^{-1}\,(A_{\textstyle{_l}}+1)^{
 -1}\,\gg\,\chi~~~.\quad\quad~\quad
 \label{cccc}
 \ea
 ~\\

 Scaling laws (\ref{aaaa}) and (\ref{bbbb}) mimic, up to constant factors, the frequency-dependencies of $\,|\bar{J}(\chi)|\,\sin\delta(
 \chi)\,=\,-\,{\cal I}{\it m} [ \bar{J}(\chi)]~$ at high and low frequencies, correspondingly, -- this can be seen from (\ref{A3}).

 Expression (\ref{cccc}) however shows a remarkable phenomenon inherent only in the {\it{tidal}} lagging, and not in the lagging in a
 sample of material: at frequencies below $~\tau_{_M}^{-1}(A_{\textstyle{_l}}+1)^{-1}\,=\,\frac{\textstyle\mu}{\textstyle\eta}\,(A_{\textstyle{_l}}+1)^{-1}$, the product $\;|\bar{k}_{\it{l}}(\chi)|\;\sin
 \epsilon_{\it{l}}(\chi)\;$ changes its behaviour and becomes linear in $\,\chi\,$.

 While elsewhere the $~|\bar{k}_{\textstyle{_l}}(\chi)|~\sin\epsilon_{\textstyle{_l}}(\chi)~$ factor increases with decreasing $\,\chi\,$,
 it changes its behaviour drastically on close approach to the zero frequency. Having reached a finite maximum at about $~\chi\,=\,\tau_{
 _M}^{-1}(A_{\textstyle{_l}}+1)^{-1}$, the said factor begins to scale linearly in $\,\chi\,$ as $\,\chi\,$ approaches zero. This way, the
 factor $~|\bar{k}_{\textstyle{_l}}(\chi)|~\sin\epsilon_{\textstyle{_l}}(\chi)~$ decreases continuously on close approach to a resonance,
 becomes nil together with the frequency at the point of resonance. So neither the tidal torque nor the tidal force explodes in resonances. In a somewhat
 heuristic manner, this change in the frequency-dependence was pointed out, for $\,{\it l} =2\,$, in Section 9 of Efroimsky \& Williams
 (2009).

 \subsection{Example}

\begin{figure}
 \begin{center}
 \includegraphics[width=16.2cm]{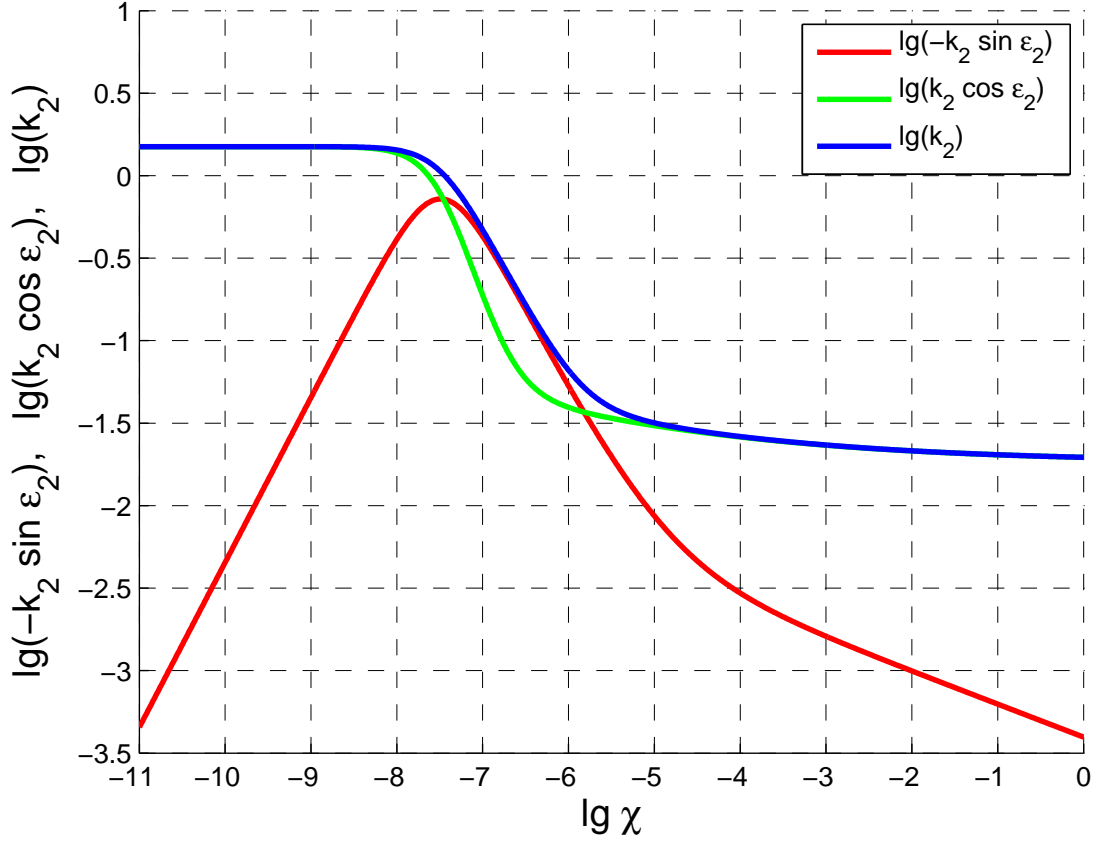}
 \caption{Tidal response of a homogeneous spherical Andrade body, set against the decadic logarithm of the forcing frequency $\chi$
 (in Hz).
 ~The blue curve renders the decadic logarithm of the absolute value of the quadrupole complex Love number, $\,\lg k_2 = \lg |\bar{k}_2
 (\chi)|\,$. The green and red curves depict the logarithms of the real  and the negative imaginary parts of the Love number:
 $\,\lg{\cal{R}}{\it{e}}\left[\bar{k}_2(\chi)\right]=\lg \left(k_2\,\cos\epsilon_2\right)\,$ and $\,\lg\left\{
 -{\cal{I}}{\it{m}}\left[\bar{k}_2(\chi)\right]\right\}=\lg\left(-k_2\,\sin\epsilon_2\right)\,$, accordingly.
 The change in the slope of the red curve (the ``elbow"), which takes place to the right of the maximum, corresponds to
 the switch from viscosity dominance at lower frequencies to anelasticity dominance at higher frequencies. The parameters $\,A_2\,$ and $\,\tau_{_M}\,$ were given values
 appropriate to a homogeneous Moon with a low viscosity, as described in subsection \ref{Moon}. The plots
 were generated for an Andrade body with $\,\zeta=1\,$ at all frequencies. Setting the body Maxwell at lower frequencies will only slightly change the shape of
 the ``elbow" and will have virtually no effect on the maximum.
 \label{Figure}}
 \end{center}
 \end{figure}

 Figure \ref{Figure} shows the absolute value, $~k_2 \equiv |\bar{k}_2(\chi)|~$, as well as the real part, $~{\cal{R}}{\it{e}}\left[
 \bar{k}_2(\chi)\right]=k_2\,\cos\epsilon_2~$, and the negative imaginary part, $~\,-{\cal{I}}{\it{m}}\left[\bar{k}_2(\chi)\right]=k_2
 \,\sin\epsilon_2~$, of the complex quadrupole Love number. Each of the three quantities is represented by its decadic logarithm as a
 function of the decadic logarithm of the forcing frequency $\,\chi\,$ (given in Hz). The curves were obtained by insertion of formulae
 (\ref{A3} - \ref{A4}) into (\ref{k2bar_2}). As an example, the case of $~\,-{\cal{I}}{\it{m}}\left[\bar{k}_2(\chi)\right]~$ is worked
 out in Appendix \ref{Andarde_Maxwell}, see formulae (\ref{DRR} - \ref{A4c}).

 Both in the high- and low-frequency limits, the negative imaginary part of $\,\bar{k}_2(\chi)\,$, given on Figure \ref{Figure} by the
 red curve, approaches zero. Accordingly, over the low- and high-frequency bands the real part (the green line) virtually coincides with
 the absolute value (the blue line).

 While on the left and on the close right of the peak, dissipation is mainly due to viscosity, friction at higher frequencies is mainly
 due to anelasticity. This switch corresponds to the change of the slope of the red curve at high frequencies (for our choice of
 parameters, at around $\,10^{-5}\,$ Hz). This change of the slope is often called {\it{the elbow}}.

 Figure \ref{Figure} was generated for $\,A_2=80.5\,$ and $\,\tau_{_M} = 3.75\times 10^{5}\,$s. The value of $\,A_2\,$ corresponds to
 the Moon modeled by a homogeneous sphere of rigidity $\,\mu=0.8\times 10^{11}\,$Pa. Our choice of the value of $\,\tau_{_M}\equiv\eta/
 \mu\,$ corresponds to a homogeneous Moon with the said value of rigidity and with viscosity set to be $\,\eta=3\times 10^{16}\,$ Pa s.
 The reason why we consider an example with such a low value of $\,\eta\,$ will be explained in subsection \ref{Moon}. Finally, it was
 assumed for simplicity that $\,\zeta=1\,$, i.e., that $\,\tau_{_A}=\tau_{_M}\,$. Although unphysical at low frequencies, this
 simplification only slightly changes the shape of the ``elbow" and exerts virtually no influence upon the maximum of the
 red curve, provided the maximum is located well into the viscosity zone.

 \subsection{Crossing a resonance -- with a chance for entrapment}\label{crossing}

 As ever, we recall that in the expansion for the tidal torque factors (\ref{aaaa}
 - \ref{cccc}) should appear in the company of multipliers ~sgn$\,\omega\,$. For example, the factor (\ref{cccc}) describing
 dissipation near an $lmpq$ resonance will enter the expansions as
 \ba
 |\bar{k}_{\textstyle{_l}}(\chi_{\textstyle{_{lmpq}}})|~\sin\epsilon_{\textstyle{_l}}(\chi_{\textstyle{_{lmpq}}})~\mbox{sgn}\,
 \omega_{\textstyle{_{lmpq}}}\,\approx\,
 \frac{3}{2\,(l-1)}~{A_{\textstyle{_l}}}~\,\tau_{_M}~\chi_{\textstyle{_{lmpq}}}~\mbox{sgn}\,\omega_{\textstyle{_{lmpq}}}
 ~=~\frac{3}{2\,(l-1)}~{A_{\textstyle{_l}}}~\,\tau_{_M}~\omega_{\textstyle{_{lmpq}}}~~.\quad
 \label{lmpq}
 \ea
 Naturally, the $lmpq$ term of the torque depends on $\,\omega_{\textstyle{_{lmpq}}}\,$ and not just on $\,\chi_{\textstyle{_{lmpq}}}
 =|\omega_{\textstyle{_{lmpq}}}|\,$. This term then goes continuously through zero, and changes its sign as the $lmpq$
 resonance is crossed (i.e., as $\,\omega_{\textstyle{_{lmpq}}}\,$ goes through nil and changes its sign).

 Formula (\ref{lmpq}) tells us that an {\emph{$~$lmpq$\,$ component of the tidal torque goes continuously through zero as the
 satellite is traversing the commensurability which corresponds to vanishing of a tidal frequency $\,\chi_{{\it l}mpq}\,$}}. This gets
 along with the physically evident fact that the principal (i.e., $\,2200\,$) term of the tidal torque should vanish as the secondary
 approaches the synchronous orbit.

 It is important that a $lmpq$ term of the torque changes its sign and thus creates a chance for entrapment. As the value of an $lmpq$
 term of the torque is much lower than that of the principal, $2200$ term, we see that a perfectly spherical body will never get stuck
 in a resonance other than $2200$. (The latter is, of course, the $\,1:1\,$ resonance, i.e., the one in which the principal term of the
 torque vanishes.) However, the presence of the triaxiality-generated torque is known to contribute to the probabilities of entrapment
 into other resonances (provided the eccentricity is not zero). Typically, in the literature they consider a superposition of the
 triaxiality-generated torque with the principal tidal term. We would point out that the ``trap" shape of the $lmpq$ term (\ref{lmpq})
 makes this term relevant for the study of entrapment in the $lmpq$ resonance. In some situations, one has to take into
 account also the non-principal terms of the tidal torque.

 \subsection{Comparison with the LLR results}\label{Moon}

 As we mentioned above, fitting of the lunar laser ranging (LLR) data to the power law has resulted in a very small {\it{negative}}
 exponent $\,p=\,-\,0.19$ (Williams et al. 2001). Since the measurements of the lunar damping described in {\it{Ibid.}} rendered
 information on the {\it{tidal}} and not seismic dissipation, those results can and should be compared to the scaling law (\ref{aaaa}
 - \ref{cccc}). As the small negative exponent was devised from observations over periods of a month to a year, it is natural to
 presume that the appropriate frequencies were close to or slightly below the frequency $~\frac{\textstyle 1}{\textstyle{\tau_{_M}\,
 {A_{\textstyle{_2}}}}}~$ at which the factor $\,k_2\,\sin\epsilon_2\,$ has its peak:
 \ba
 3\times 10^6~\mbox{s}~\approx~0.1~\mbox{yr}~=~\tau_{_M}\,{A_{\textstyle{_2}}}\,=~\frac{\eta}{\mu}~{A_{\textstyle{_2}}}\,=~\frac{57~
 \eta}{8\,\pi\,G\,(\rho\,R)^2}~~~,
 \label{}
 \ea
 as on Figure \ref{Figure}. Hence, were the Moon a uniform viscoelastic body, its viscosity would be only
 \ba
 \eta~=~3\,\times\,10^{16}~\mbox{Pa~}\,\mbox{s}~~~.
 \label{visco}
 \ea
 For the actual Moon, the estimate means that the lower mantle contains a high percentage of partial melt, a fact which goes along
 well with the model suggested in Weber et al. (2011), and which was anticipated yet in Williams et al. (2001) and Williams \& Boggs
 (2009), following an earlier work by Nakamura et al. (1974).

 \section{The polar component of the tidal torque acting on the primary}

 Let vector $\,\erbold\,=\,(r,\,\lambda,\,\phi)\,$ point from the centre of the primary toward a point-like secondary of mass $\,M_{sec
 }\,$. Associating the coordinate system with the primary, we reckon the latitude $\,\phi\,$ from the equator. Setting the coordinate
 system to corotate, we determine the longitude $\,\lambda\,$ from a fixed meridian. The tidally induced component
 of the primary's potential, $\,U\,$, can be generated either by this secondary itself or by some other secondary of mass $\,M_{sec}^*
 \,$ located at $\,\erbold^{\,*}\,=\,(r^*,\,\lambda^*,\,\phi^*)\,$. In either situation, the tidally induced potential $\,U\,$
 generates a tidal force and a tidal torque  wherewith the secondary of mass $\,M_{sec}\,$ acts on the primary.


 The scope of this paper is limited to low values of $\inc$. When the role of the primary is played by a planet, the secondary being its
 satellite, $\inc$ is the satellite's inclination. When the role of the primary is played by the satellite, the planet acting as its
 secondary, $\inc$ acquires the meaning of the satellite's obliquity. Similarly, when the planet is regarded as a primary and its host
 star is treated as its secondary, $\inc$ is the obliquity of the planet. In all these cases, the smallness of $\,\inc\,$ indicates that
 the tidal torque acting on the primary can be identified with its polar component, the one orthogonal to the equator of the primary. The
 other components of the torque will be neglected in this approximation.

 The polar component of the torque acting on the primary is the negative of the partial derivative of the tidal potential, with respect
 to the primary's sidereal angle:
 \ba
 {\cal T}(\erbold)\;=\;-\;M_{sec}\;\frac{\partial U(\erbold)}{\partial \theta}\;\;\;,
 \label{hh}
 \ea
 $\theta\,$ standing for the primary's sidereal angle. This formula is convenient when the tidal potential $\,U\,$ is expressed through
 the secondary's orbital elements and the primary's sidereal angle.\footnote{~Were the potential written down in the spherical
 coordinates associated with the primary's equator and corotating with the primary, the polar component of the tidal torque could be
 calculated with aid of the expression
  \ba
  \nonumber
  {\cal T}(\erbold)\;=\;\,M_{sec}\;\frac{\partial U(\erbold)}{\partial\lambda}~~~
  \label{15}
  \ea
 derived, for example, in Williams \& Efroimsky (2012). That the expression agrees with (\ref{hh}) can be seen from the
 formula
  \ba
  \nonumber
  \lambda\;=\;-\;\theta\;+\;\Omega\;+\;\omega\;+\;\nu\;+\;O(\inc^2)\;=\;-\;\theta\;+\;\Omega\;+\;\omega\;+\;
  {\cal{M}}\;+\;2\;e\;\sin{\cal{M}}\;+\;O(e^2)\;+\;O(\inc^2)\;\;\;,\;\;\;\;
  \label{38}
  \ea
  $e\,,\,i\,,\,\omega\,,\,\Omega\,,\,\nu\,$ and $\,{\cal M}\,$ being the eccentricity, inclination, argument of the pericentre, longitude of the node,
  true anomaly, and mean anomaly of the tide-raising secondary.}

 Here and hereafter we are deliberately referring to {\emph{a primary and a secondary}} in lieu of {\emph{a planet and a satellite}}.
 The preference stems from our intention to extend the formalism to setting where a moon is playing the role of a tidally-perturbed primary, the planet being its tide-producing secondary. Similarly, when addressing the rotation of Mercury, we interpret the Sun as a secondary that is causing a tide on the effectively primary body, Mercury.

 \section{The tidal potential}

  \subsection{Darwin (1879) and Kaula (1964)}

 The potential produced at point $\eRbold=(R\,,\,\lambda\,,\,\phi)\,$ by a secondary body of mass $\,M^*$, located~at~$\,\erbold^{\;*}
 =(r^*,\,\lambda^*\,,\,\phi^*)\,$ with $\,r^*\geq R\,$, is given by (\ref{1}). When a tide-raising secondary located at $\,\erbold^{\;*}
 \,$ distorts the shape of the primary, the potential generated by the primary at some exterior point $\,\erbold\,$ gets changed. In the
 linear approximation, its variation is given by (\ref{2}). Insertion of (\ref{1}) into (\ref{2}) entails
 \ba
 U(\erbold)\;=\;\,-\,{G\;M^*_{sec}}
 \sum_{{\it{l}}=2}^{\infty}k_{\it l}\;
 \frac{R^{
 \textstyle{^{2\it{l}+1}}}}{r^{
 \textstyle{^{\it{l}+1}}}{r^{\;*}}^{
 \textstyle{^{\it{l}+1}}}}\sum_{m=0}^{\it l}\frac{({\it l} - m)!
 }{({\it l} + m)!}(2-\delta_{0m})P_{{\it{l}}m}(\sin\phi)P_{{
 \it{l}}m}(\sin\phi^*)\;\cos m(\lambda-\lambda^*)~~~.~~~~~
 \label{4}
 \ea
 A different expression for the tidal potential was offered by Kaula (1961, 1964), who developed a powerful technique that enabled him to switch from
 the spherical coordinates to the Kepler elements $\,(\,a^*,\,e^*,\,\inc^*,\,\Omega^*,\,\omega^*,\,{\cal M}^*\,)\,$ and $\,(\,a,\,e,\,\inc,\,\Omega,\,
 \omega,\,{\cal M}\,)\,$ of the secondaries located at $\,\erbold^{\;*}\,$ and $\,\erbold\,$.
 Application of this technique to (\ref{4}) results in
 \ba
 \nonumber
 U(\erbold)\;=\;-\;\sum_{{\it
  l}=2}^{\infty}\;k_{\it l}\;\left(\,\frac{R}{a}\,\right)^{\textstyle{^{{\it
  l}+1}}}\frac{G\,M^*_{sec}}{a^*}\;\left(\,\frac{R}{a^*}\,\right)^{\textstyle{^{\it
  l}}}\sum_{m=0}^{\it l}\;\frac{({\it l} - m)!}{({\it l} + m)!}\;
  \left(\,2\;\right. ~~~~~~~~~~~~~~~~~~~~~~~~~~~~~~~~~~~~~~~~~~~~~~~~~~~~~\\
                                   \label{3}
                                   \label{L51}\\
                                   \nonumber\\
                                    \nonumber
 ~~~~~\left.-\,\delta_{0m}\,\right)\,\sum_{p=0}^{\it
  l}F_{{\it l}mp}(\inc^*)\sum_{q=-\infty}^{\infty}G_{{\it l}pq}
  (e^*) \sum_{h=0}^{\it l}F_{{\it
  l}mh}(\inc)\sum_{j=-\infty}^{\infty}
  G_{{\it l}hj}(e)\;\cos\left[
  \left(v_{{\it l}mpq}^*-m\theta^*\right)-
  \left(v_{{\it l}mhj}-m\theta\right) \right]
 ~~_{\textstyle{_{\textstyle ,}}}
 \ea
 where
 \ba
 v_{{\it l}mpq}^*\;\equiv\;({\it l}-2p)\omega^*\,+\,
 ({\it l}-2p+q){\cal M}^*\,+\,m\,\Omega^*~~~,
 \label{6}
 \ea
  \ba
 v_{{\it l}mhj}\;\equiv\;({\it l}-2h)\omega\,+\,
 ({\it l}-2h+j){\cal M}\,+\,m\,\Omega~~~,
 \label{7}
 \ea
 $\theta\,=\,\theta^*\,$ being the sidereal angle, $\,G_{lpq}(e)\,$ signifying the eccentricity functions,\footnote{~Functions
 $\,G_{lhj}(e)\,$ coincide with the Hansen polynomials $\,X^{\textstyle{^{(-l-1),\,(l-2p)}}}_{\textstyle{_{(l-2p+q)}}}(e)~$. In
 Appendix \ref{functions}, we provide a table of the $\,G_{lhj}(e)\,$ required for expansion of tides up to $\,e^6\,$, inclusively.}
 and $\,F_{lmp}(\inc)\,$ denoting the inclination functions (Gooding \& Wagner 2008).

 Being equivalent for a planet with an instant response of the shape, (\ref{4}) and (\ref{L51}) disagree when dissipation-caused delays
 come into play. Kaula's expression (\ref{L51}), as well as its truncated, Darwin's
 version,\footnote{~While the treatment by Kaula (1964) entails the infinite Fourier series (\ref{L51}), the development by Darwin
 (1879, 1880) renders its partial sum with $\,|{\it{l}}|,\,|q|,\,|j|\,\leq\,2\,.$ For a simple introduction into Darwin's method see
 Ferraz-Mello et al. (2008). Be mindful that in {\it{Ibid}.} the convention on the notations $\erbold$
 and $\erbold^{\,*}$ is opposite to ours.} is capable of accommodating separate phase lags for each mode:
 \ba
 \nonumber
 U(\erbold)\;=\;-\;\sum_{{\it
 l}=2}^{\infty}\;k_{\it l}\;\left(\,\frac{R}{a}\,\right)^{\textstyle{^{{\it
 l}+1}}}\frac{G\,M^*_{sec}}{a^*}\;\left(\,\frac{R}{a^*}\,\right)^{\textstyle{^{\it
 l}}}\sum_{m=0}^{\it l}\;\frac{({\it l} - m)!}{({\it l} + m)!}\;
 \left(\,2\;-\right. ~~~~~~~~~~~~~~~~~~~~~~~~~~~~~~~~~~~~~~~~~~~~~~~~~\\
                                   \label{5}
                                   \label{LLLL54}
                                   \label{L54}\\
                                   \nonumber\\
                                    \nonumber
 \left.\delta_{0m}\,\right)\,\sum_{p=0}^{\it
 l}F_{{\it l}mp}(\inc^*)\sum_{q=-\infty}^{\infty}G_{{\it l}pq}
 (e^*)
 \sum_{h=0}^{\it l}F_{{\it
 l}mh}(\inc)\sum_{j=-\infty}^{\infty}
 G_{{\it l}hj}(e)\;\cos\left[
 \left(v_{{\it l}mpq}^*-m\theta^*\right)-
 \left(v_{{\it l}mhj}-m\theta\right)-
 \epsilon_{{\it l}mpq} \right]
 ~~_{\textstyle{_{\textstyle ,}}}
  \ea
 where
 \ba
 \epsilon_{{\it l}mpq}=\left[\,({\it l}-2p)\,\dot{\omega}^*\,+\,
 ({\it l}-2p+q)\,\dot{\cal{M}}^*\,+\,m\,(\dot{\Omega}^*\,-\,\dot{\theta}^*)\,
 \right]\,\Delta t_{\it{l}mpq}=\,\omega^*_{\it{l}mpq}\,\Delta t_{\it{l}mpq}
 =\,\pm\,\chi^*_{\it{l}mpq}\,\Delta t_{\it{l}mpq}~~~~~~
 \label{8}
 \label{L55}
 \ea
 is the phase lag. The tidal mode $\omega^*_{\it{l}mpq}$ introduced in
 (\ref{8}) is
 \ba
 \omega^*_{{\it l}mpq}\;\equiv\;({\it l}-2p)\;\dot{\omega}^*\,+\,({\it l}-
 2p+q)\;\dot{\cal{M}}^*\,+\,m\;(\dot{\Omega}^*\,-\,\dot{\theta}^*)\;~~,~~~
 \label{9}
 \ea
 while the positively-defined quantity
 \ba
 \chi^*_{{\it l}mpq}\,\equiv\,|\,\omega^*_{{\it l}mpq}\,|\,=\,|\,({\it
 l}-2p)\,\dot{\omega}^*\,+\,({\it l}-2p+q)\,\dot{\cal{M}}^*\,+\,m\,(\dot{\Omega}^*
 \,-\,\dot{\theta}^*)\;|~~~~~
 \label{10}
 \ea
 is the actual physical $\,{{\it l}mpq}\,$ frequency excited by the tide in the primary.
 The corresponding positively-defined time delay $\,\Delta t_{\it{l}mpq}=\,\Delta t_l(\chi_{{\it l}mpq})\,$ depends on
 this physical frequency, the functional forms of this dependence being different for
 different materials.

 In neglect of the apsidal and nodal precessions, and also of $\,\dot{\cal{M}}_0\,$, the above formulae become:
 \ba
 \omega_{{\it l}mpq}\,=\,({\it l}-2p+q)\,n\,-\,m\,\dot{\theta}~~~,~~
 \label{}
 \ea
 \ba
 \chi_{{\it l}mpq}\,\equiv\,|\,\omega_{{\it l}mpq}\,|\,=\,|\,({\it l}-2p+q)\,n\,-\,m\,\dot{\theta}\;|~~~,~~
 \label{}
 \ea
 and
 \begin{subequations}
 \ba
 \epsilon_{{\it l}mpq}\,\equiv\,\omega_{{\it l}mpq}\,\Delta t_{{\it l}mpq}&=&\left[\,({\it l}-2p+q)\,n\,-\,m\,\dot{\theta}\,\right]
 \,\Delta t_{{\it l}mpq}
 \label{}\\
 \nonumber\\
 &=&\chi_{{\it l}mpq}~\,\Delta t_l(\chi_{{\it l}mpq})~\,\mbox{sgn}\,\left[\,({\it l}-2p+q)\,n\,-\,m\,\dot{\theta}\,\right]~~~,~~
 \label{1000000}
 \ea
 \label{hagen}
 \end{subequations}

 Formulae (\ref{L51}) and (\ref{5}) constitute the pivotal result of Kaula's theory of
 tides. Importantly, Kaula's theory imposes no {\emph{a priori}} constraint on
 the form of frequency-dependence of the lags.

 \section{The Darwin torque}

 As explained in Williams \& Efroimsky (2012), the empirical model by 
 MacDonald (1964), called {\it{MacDonald
 torque}}, tacitly sets an unphysical rheology of the satellite's material. The rheology is given by (\ref{generic}) with $\,\alpha\,=\,
 -\,1\,$. More realistic is the dissipation law (\ref{rheology}). An even more accurate and practical formulation of the damping law,
 stemming from the Andrade formula for the compliance, is rendered by (\ref{aaaa} - \ref{cccc}). These formulae should be inserted into
 the Darwin-Kaula theory of tides.

 \subsection{The secular and oscillating parts of the Darwin torque}\label{kerr}

 \subsubsection{The general formula}

 Direct differentiation of (\ref{LLLL54}) with respect to $\;-\,\theta\,$ will
 result in the expression\footnote{~For justification of this operation,
 see Section 6 in Efroimsky \& Williams (2009).}
 \ba
 \nonumber
 {\cal T}= -\,\sum_{{\it
 l}=2}^{\infty}\left(\frac{R}{a}\right)^{\textstyle{^{{\it
 l}+1}}}\frac{G\,M^*_{sec}\,M_{sec}}{a^*}\left(\frac{R}{a^*}\right)^{\textstyle{^{\it
 l}}}\sum_{m=0}^{\it l}\frac{({\it l} - m)!}{({\it l} + m)!}
 \;2\,m\,\sum_{p=0}^{\it
 l}F_{{\it l}mp}(\inc^*)\sum_{q=-\infty}^{\infty}G_{{\it l}pq}
 (e^*)~~~~~~~~~~~~~~~\\
                                   \nonumber\\
 \sum_{h=0}^{\it l}F_{{\it l}mh}(\inc)\sum_{j=-\infty}^{\infty} G_{{\it l}hj}(e)
 \;k_{\it l}\;\sin\left[\,
 v^*_{{\it l}mpq}\,-\;v_{{\it l}mhj}\,-\;\epsilon_{{\it l}mpq}\,\right]
 ~~_{\textstyle{_{\textstyle .}}}~~~~~~~~~~~~~~~~~~~~~~~~~~~~~~~~~~
 \label{30}
 \label{ser}
 \ea
 If the tidally-perturbed and tide-raising secondaries are the same body, then
 $\,M_{sec}=M_{sec}^*\,$, and all the elements coincide with their counterparts
 with an asterisk. Hence the differences
 \ba
 \nonumber
 v^*_{{\it{l}}mpq}\,-\;v_{{\it{l}}mhj}\,=~~~~~~~~~~~~~~~~
 ~~~~~~~~~~~~~~~~~
 \ea
 \ba
 ({\it{l}}\,-\,2\,p\,+\,q)\;{\cal{M}}^*\,-\;({\it{l}}\,-\,2\,h\,+\,j)\,{\cal{M}}\;+\;
 m\,({\Omega}^*\, - {\Omega} )\;+\;{\it{l}}\;( {\omega}^*\,-\,{\omega} )\,-\,2\,p\,{\omega}^*
 \,+\,2\,h\,{\omega}~~~~~~
 \label{ho}
 \ea
 get simplified to
 \ba
 v^*_{{\it{l}}mpq}\,-\;v_{{\it{l}}mhj}\,=~(2\,h\,-\,2\,p\,+\,q\,-\,j)\;{\cal{M}}^*\,+
 \,(2\,h\,-\,2\,p)\,{\omega}^*~~~,~~~
 \label{}
 \ea
 an expression containing both short-period contributions proportional to the mean
 anomaly, and long-period contributions proportional to the argument of the pericentre.\\

 \subsubsection{The secular, the purely-short-period, and the mixed-period parts of the torque}

 Now we see that the terms entering series (\ref{ser}) can be split into three groups:\\
 ~\\
 (1)~~ The terms, in which indices $\,(p\,,\,q)\,$ coincide with $\,(h\,,\,j)\,$,
 constitute a secular part of the tidal torque, because in such
 terms $\,v_{{\it{l}}mhj}\,$ cancel with $\,v_{{\it l}mpq}^*\;$. This
 ${\cal{M}}$- and $\,\omega$-independent part is furnished by
 \ba
  \overline{\cal T}&=&
 \sum_{{\it{l}}=2}^{\infty}2~G~M_{sec}^{\textstyle{^{2}}}~
 \frac{R^{\textstyle{^{2{\it{l}}\,+\,1}}}}{
 a^{\textstyle{^{2\,{\it{l}}\,+\,2}}}}
 \sum_{m=0}^{\it l}
 \frac{({\it{l}}\,-\,m)!}{({\it{l}}\,+\,m)!}\;m
 \;\sum_{p=0}^{\it l}F^{\textstyle{^{2}}}_{{\it{l}}mp}(\inc)\sum^{\it \infty}_{q=-\infty}
 G^{\textstyle{^{2}}}_{{\it{l}}pq}(e)\;k_{\it l}\;\sin\epsilon_{{\it{l}}mpq}
 \;\;\;.~~~~~~~~~~~~~
 \label{31}
 \label{T31}
 \label{gedo}
 \ea
 ~\\
 ~\\
 (2)~~ The terms with $\,p\,=\,h\,$ and $\,q\,\neq\,j\,$ constitute a purely short-period part of the torque:
 \ba
 \nonumber
 \widetilde{\cal T}
 =-\sum_{{\it
 l}=2}^{\infty}\,2\,G\,M^{\textstyle{^2}}_{sec}\,\frac{R^{\textstyle{^{2{\it
 l}+1}}}}{a^{\textstyle{^{2{\it
 l}+2}}}} \;\sum_{m=0}^{\it l}\,\frac{({\it l} - m)!}{({\it l} + m)!}\;m\,
 \sum_{p=0}^{\it l}F^{\textstyle{^{2}}}_{{\it{l}}mp}(\inc)
 \sum_{q=-\infty~}^{\infty}
 {\sum_{\stackrel{\textstyle{^{~j=-\infty}}}{\textstyle{^{j\;\neq\;q}}}}^{\infty}}
 G_{{\it l}pq}(e)\;G_{{\it l}pj}(e)\;k_{\it l}\;\sin\left[\,\left(q  \right.\right.\\
 ~\nonumber\\
  \left.\left. -\,j\right)\;{\cal{M}}\;-\;\epsilon_{{\it l}mpq}\,\right]\,
 ~~_{\textstyle{_{\textstyle .}}}~~~~~~~~~~~~~~~~~~~~~~~~~~~~~~~~~~
 \label{tilde}
 \ea
 ~\\
 (3)~~ The remaining terms, ones with $\,p\,\neq\,h\,$, make a mixed-period part
 comprised of both short- and long-period terms:
 \ba
 \nonumber
 {\cal T}^{\textstyle{^{mixed}}}=~~~~~~~~~~~~~~~~~~~~~~~~~~~~~~~~~~~~~~~~~~~~~~~~~~~~~
 ~~~~~~~~~~~~~~~~~~~~~~~~~~~~~~~~~~~~~~~~~~~~~~~~~~~~~
 \ea
 \ba
 \nonumber
 -\;\sum_{{\it
 l}=2}^{\infty}2\,G\,M^{\textstyle{^2}}_{sec}\,\frac{R^{\textstyle{^{2{\it
 l}+1}}}}{a^{\textstyle{^{2{\it
 l}+2}}}} \sum_{m=0}^{\it l}\frac{({\it l} - m)!}{({\it l} + m)!}~ m
 \sum_{p=0}^{\it l} F_{{\it l}mp}(\inc)
 \sum_{\stackrel{\textstyle{^{h=0}}}{\textstyle{^{h\;\neq\;p}}}}^{\it l}
 F_{{\it l}mh}(\inc) \sum_{q=-\infty\,}^{\infty}
   \sum_{\stackrel{\textstyle{^{\,j=-\infty}}}{
  }}^{\infty} G_{{\it l}hq}(e)\,G_{{\it l}pj}(e)
 \,k_{\textstyle{_{\it l}}}\;\sin\left[\,\left(2\,h  \right.\right.\\
 \nonumber\\
 \left. \left.-\,2\,p\,+\,q\,-\,j\right)\;{\cal{M}}^*\,+
 \,(2\,h\,-\,2\,p)\,{\omega}^*\;-\;\epsilon_{{\it l}mpq}\,\right]
 ~~_{\textstyle{_{\textstyle .}}}~~~~~~~~~~~~~~~~~~~~~~~~~~~~~~~
 \label{mixed}
 \ea

 \subsubsection{The $\,l=2\,$ and $\,l=3\,$ terms in the $\,O(i^2)\,$ approximation}

 For $\,l=2\,$, index $\,m\,$ will take the values $\,0,1,2\,$ only.
 Although the $\,m=0\,$ terms enter the potential, they add nothing to the torque, because differentiation of (\ref{5})
 with respect to $\;-\theta\;$ furnishes the $\,m\,$ multiplier in (\ref{ser}). To examine the remaining terms, we should
 consider the inclination functions with subscripts $\,({\emph{l}}mp)\,=\,(220)\,,\,(210)\,,\,(211)\;$ only:
 \ba
 F_{220}(\inc)\,=\,3\,+\,O(\inc^2)~~~,~~~~~F_{210}(\inc)\,=\,\frac{3}{2}\;\sin\inc\,+\,
 O(\inc^2)\;\;\;,~~~~~F_{211}(\inc)\,=\;-\;\frac{3}{2}\;\sin\inc\,+\,O(\inc^2)~~~,~~~~~
 \label{}
 \ea
 all the other $\,F_{2mp}(\inc)\,$ being of order $\,O(\inc^2)\,$ or higher. Thence for $\,p=h\,$ (i.e., both in the secular and
 purely-short-period parts) it is sufficient, in the $\,O(\inc^2)\,$ approximation, to keep only the terms with
 $\,F_{220}^2(\inc)\,$, ignoring those with $\,F_{210}^2(\inc)\,$ and $\,F_{211}^2 (\inc)\,$. We see that in the $~O(\inc^2)~$ approximation
 \begin{itemize}
 \item{} among the $\,l=2\,$ terms, {{both in the secular and purely short-period parts}},\\ only the terms with $\;({\it{l}}mp)\,=\,(220)\;$ are relevant.
 \end{itemize}
 In the case of $\,p\neq h\,$, i.e., in the mixed-period part, the terms of the leading order in inclination are: $~\,F_{{\it{l}}mp}
 (\inc)\,F_{{\it{l}}mh}(\inc)~=~\,F_{210}(\inc)\,F_{211}(\inc)~$ and $~\,F_{{\it{l}}mp}(\inc)\,
 F_{{\it{l}}mh}(\inc)~=~\,F_{211}(\inc)\,F_{210}(\inc)~$, which happen to be equal to one another, and to be of order
 $~O(\inc^2)~$. This way, in the $~O(\inc^2)~$ approximation,
 \begin{itemize}
 \item{} the mixed-period part of the $\,l=2\,$ component may be omitted.
 \end{itemize}

 The inclination functions $\,F_{lmp}\,=\,F_{310}\,,\,F_{312}\,,\,F_{313}\,,\,F_{320}\,,\,F_{321}\,,\,F_{322}\,,\,F_{323}\,,\,F_{331}\,,
 \,F_{332}\,,\,F_{333}\,$ are of order $\,O(\inc)\,$ or higher. The terms containing the squares or cross-products of the these functions
 may thus be dropped. Specifically, the smallness of the cross-terms tells us that
 \begin{itemize}
 \item{} the mixed-period part of the $\,l=3\,$ component may be omitted.
 \end{itemize}

 What remains is the terms containing the squares of functions
 \ba
 F_{311}(\inc)\,=~-~\frac{3}{2}\,+\,O(\inc^2)~~~~~~\mbox{and}~~~~~~F_{330}(\inc)\,=\,15\,+\,O(\inc^2)\;\;\;~~~.~~~~~
 \label{bbv}
 \ea
 In other words,
 \begin{itemize}
 \item{} among the $\,l=3\,$ terms, both in the secular and purely short-period parts,\\ only the terms with $\;({\it{l}}mp)\,=\,(311)\;$
 and $\;({\it{l}}mp)\,=\,(330)\;$ are important.
 \end{itemize}
 ~\\
 All in all, for $\,l=2\,$ and $\,l=3\,$ the mixed-period parts of the torque may be neglected, in the $~O(\inc^2)~$ approximation. The
 surviving terms of the secular and the purely short-period parts will be developed up to $\,e^6\,$, inclusively.

 \subsection{Approximation for the secular and short-period parts of the tidal torque}

 As we just saw, both the secular and short-period parts of the torque may be approximated with the following degree-2 and degree-3
 components:
 \ba
 \nonumber
 \overline{\cal T}&=&\overline{\cal T}_{\textstyle{_{\textstyle_{\textstyle{_{l=2}}}}}}\,+~
 \overline{\cal T}_{\textstyle{_{\textstyle_{\textstyle{_{l=3}}}}}}\,+~O\left(\,\epsilon\,(R/a)^{9}\,\right)~\\
 \nonumber\\
 \nonumber\\
 &=&\overline{\cal T}_{\textstyle{_{\textstyle{_{(lmp)=(220)}}}}}\,+\left[\,\overline{\cal T}_{\textstyle{_{
 \textstyle{_{(lmp)=(311)}}}}}\,
 +~\overline{\cal T}_{\textstyle{_{\textstyle{_{(lmp)=(330)}}}}}  \right]~
  +~O(\epsilon\,i^2)    ~+~O\left(\,\epsilon\,(R/a)^9\,\right)~~~,\quad\quad
 ~~~~
 \label{}
 \ea
and
 \ba
 \nonumber
 \widetilde{\cal T}&=&\widetilde{\cal T}_{\textstyle{_{\textstyle{_{l=2}}}}}\,+~\widetilde{\cal T}_{\textstyle{_{\textstyle{_{l=3}}}}}\,
 +~O\left(\,\epsilon\,(R/a)^9\,\right)~\\
 \nonumber\\
 \nonumber\\
 &=&\widetilde{\cal T}_{\textstyle{_{\textstyle{_{(lmp)=(220)}}}}}\,+\left[\,\widetilde{\cal T}_{\textstyle{_{
 \textstyle{_{(lmp)=(311)}}}}}\,
 +~\widetilde{\cal T}_{\textstyle{_{\textstyle{_{(lmp)=(330)}}}}}  \right]~
  +~O(\epsilon\,i^2)    ~+~O\left(\,\epsilon\,(R/a)^9\,\right)~~~,\quad\quad
 \label{}
 \ea
 were the $\,l=2\,$ and $\,l=3\,$ inputs are of the order $\,(R/a)^{5}\,$ and $\,(R/a)^{7}\,$, accordingly; while the $\,l=4,\,5,\,.\,.
 \,.\,$ inputs constitute $\,O\left(\,\epsilon\,(R/a)^{9}\,\right)\,$.

 Expressions for $~\overline{\cal T}_{\textstyle{_{\textstyle{_{(lmp)=(220)}}}}}~$,
 $~\overline{\cal T}_{\textstyle{_{\textstyle{_{(lmp)=(311)}}}}}~$, and $~\overline{\cal T}_{\textstyle{_{
 \textstyle{_{(lmp)=(310)}}}}}~$ are furnished by formulae (\ref{T2}), (\ref{T31}), and (\ref{T32}) in Appendix \ref{secular}. As an
 example, here we provide one of these components:
  \ba
 \nonumber
 \overline{\cal T}_{\textstyle{_{\textstyle_{\textstyle{_{(lmp)=(220)}}}}}}
 &=&\frac{3}{2}~G\;M_{sec}^2\,R^5\,a^{-6}\,\left[~\frac{1}{2304}~e^6~k_2~\sin|\epsilon_{\textstyle{_{\textstyle{_{220~-3}}}}}|~~
 \mbox{sgn}\,\left(\,-\,n\,-\,2\,\dot{\theta}\,\right) \right.\\
 \nonumber\\
 \nonumber\\
 \nonumber
 &+&\left(~ \frac{1}{4}~e^2-\,\frac{1}{16}~e^4~+\,\frac{13}{768}~e^6 ~\right)~k_2~\sin|\epsilon_{\textstyle{_{\textstyle{_{220~-1}}}}}|
 ~~\mbox{sgn}\,\left(\,n\,-\,2\,\dot{\theta}\,\right)
 ~\\
 \nonumber\\
 \nonumber\\
 \nonumber
 &+&\left(1\,-\,5\,e^2\,+~\frac{63}{8}\;e^4\,-~\frac{155}{36}~e^6\right)~k_2~\sin|\epsilon_{\textstyle{_{\textstyle{_{2200}}}}}|~~
 ~~\mbox{sgn}\,\left(\,n\,-\,\dot{\theta}\,\right)\\
 \nonumber\\
 \nonumber\\
 \nonumber
 &+&\left(\frac{49}{4}~e^2-\;\frac{861}{16}\;e^4+~\frac{21975}{256}~e^6\right)~k_2~\sin|\epsilon_{\textstyle{_{\textstyle{_{2201}}}}}|
 ~~\mbox{sgn}\,\left(\,3\,n\,-\,2\,\dot{\theta}\,\right)\\
 \nonumber\\
 \nonumber\\
 \nonumber
 &+& \left(~ \frac{289}{4}\,e^4\,-\,\frac{1955}{6}~e^6 ~\right)~k_2~\sin|\epsilon_{\textstyle{_{\textstyle{_{2202}}}}}|
 ~\mbox{sgn}\,\left(\,2\,n\,-\,\dot{\theta}\,\right)\\
 \nonumber\\
 \nonumber\\
 &+&\left.\frac{714025}{2304}~e^6 ~k_2~\sin|\epsilon_{\textstyle{_{\textstyle{_{2203}}}}}|~~\mbox{sgn}\,\left(\,
5\,n\,-\,3\,\dot{\theta}\,\right)
 ~\,\right]+\,O(e^8\,\epsilon)\,+\,O(\inc^2\,\epsilon)~~~.\quad\quad\quad\quad
 \label{TGD}
 \ea
 Here each term changes its sign on crossing the appropriate resonance. The change of the sign
 takes place smoothly, as the value of the term goes through zero -- this can be seen from formula (\ref{lmpq}) and from the fact that
 the tidal mode $\,\omega_{\textstyle{_{lmpq}}}\,$ vanishes in the $\,lmpq\,$ resonance.

 Expressions for $~\widetilde{\cal T}_{\textstyle{_{\textstyle{_{(lmp)=(220)}}}}}~$, $~\widetilde{\cal T}_{\textstyle{_{\textstyle{_{
 (lmp)=(311)}}}}}~$, and $~\widetilde{\cal T}_{\textstyle{_{\textstyle{_{(lmp)=(330)}}}}}~$ are given by formulae (\ref{Til2} -
 \ref{Til30}) in Appendix \ref{short}. Although the average of the short-period part of the torque vanishes, this part does contribute to
 dissipation. Oscillating torques contribute also to variations of the surface of the tidally-distorted primary, the latter fact being of
 importance in laser-ranging experiments.

 Whether the short-period torque may or may not influence also the process of entrapment is worth exploring numerically. The reason why
 this issue is raised is that the frequencies $\,n(q-j)\,$ of the components of the short-period torque are integers of $\,n\,$ and thus
 are commensurate with the spin rate $\,\dot{\theta}\,$ near an $A/B$ resonance, $\,A\,$ and $\,B\,$ being integer. It may be
 especially interesting to check the role of this torque when $\,q-j=1\,$ and $\,A/B\,=\,N\,$ is integer.

 The hypothetical role of the short-period torque in the entrapment and libration dynamics has never been discussed so far, as
 the previous studies employed expressions for the tidal torque, which were obtained through averaging over the period of the
 secondary's orbital motion.

 \subsection{Librations}

Consider a tidally-perturbed body caught in a $\,A:B\,$ resonance with its tide-raising companion, $A$ and $B$ being integer. Then the
 spin rate of the body is
 \ba
 \dot{\theta}\;=\;\frac{A}{B}\;n\;+\;\dot{\psi}~~~,
 \label{}
 \ea
 where the physical-libration angle is
 \ba
 \psi\,=~-~\psi_0~\sin \left(\omega_{{_{PL}}}\,t\right)~~~,
 \label{}
 \ea
 $\,\omega_{{_{PL}}}\,$ being the physical-libration frequency. The oscillating tidal torque exerted on the body is comprised of
 the modes
 \ba
 \omega_{\textstyle{_{lmpq}}}\;=\;\left(l~-~2~p~+~q\right)~n~-~m~\dot{\theta}~=~
 \left(l~-~2~p~+~q~-~\frac{A}{B}~m\right)\,n~-~m~\dot{\psi}~~.~~~
 \label{du1}
 \ea


 In those $\,lmpq\,$ terms for which the combination $\,l-2p+q-\frac{\textstyle A}{\textstyle B}\,m\,$ is not zero, the
 small quantity $\,-m\dot{\psi}\,$ may be neglected.\footnote{~The physical-libration input $~-\,m\,\dot{\alpha}\,$ may be neglected
 in the expression for $\,\omega_{\textstyle{_{lmpq}}}\,$ even when the magnitude of the physical libration is comparable to that of the
 geometric libration (as in the case of Phobos).} The remaining terms will pertain to the geometric libration. The phase lags
 will be given by the standard formula $\,\epsilon_{\textstyle{_{lmpq}}}\,=\,\omega_{\textstyle{_{lmpq}}}\,\Delta t_{\textstyle{_{
 lmpq}}}\,$.

 In those $\,lmpq\,$ terms for which the combination $\,l-2p+q-\frac{\textstyle A}{\textstyle B}\,m\,$ vanishes, the physical-libration
 input $~-m\,\dot{\psi}\,$ is the only one left. Accordingly, the multiplier $~\sin\left[\,\left(v_{\textstyle{_{lmpq}}}^*\,
 -\,m\,\theta^*\,\right)\,-\right.$ $\left.\,\left(\,v_{\textstyle{_{lmpq}}}\,-\,m\,\theta\,\right)\,\right]~$ in the $\,lmpq\,$ term of the tidal torque will
 reduce to $~\sin\left[\,-\,m\,\left(\,\psi^*\,-\,\psi\,\right)\,\right]\,\approx~$ $-m\,\dot{\psi}\,\Delta t\,=\,m\,\psi_0\,
 \omega_{{_{PL}}}\,\Delta t\,\cos \omega_{{_{PL}}} t~$. Here the time lag $\,\Delta t\,$ is the one corresponding to the
 physical-libration frequency $\,\omega_{{_{PL}}}~$ which may be {\it{very different}}$\,$ from the usual tidal frequencies for
 nonsynchronous rotation -- see Williams \& Efroimsky (2012) for a comprehensive discussion.

 %
 %

 \section{Marking the minefield}

 The afore-presented expressions for the secular and purely short-period parts of the tidal torque look cumbersome when compared to the
 compact and elegant formulae employed in the literature hitherto. It will therefore be important to explain why those simplifications
 are impractical.

 \subsection{Perils of the conventional simplification}

 Insofar as the quality factor is large and the lag is small (i.e., insofar as $\,\sin\epsilon\,$ can be approximated with $\,\epsilon\,$),
 our expression (\ref{geoid}) assumes a simpler form:
 \ba
 ^{\textstyle{^{\tiny{(Q\,>\,10)\,}}}}\overline{\cal T}_{\textstyle{_{\textstyle_{\textstyle{_{l=2}}}}}}
 &=&\frac{3}{2}\;G\;M_{sec}^2\,R^5\,a^{-6}\,k_2\;\sum_{q=-3}^{3}\,
 G^{\textstyle{^{\,2}}}_{\textstyle{_{\textstyle{_{20\mbox{\it{q}}}}}}}(e)\,
 \epsilon_{\textstyle{_{\textstyle{_{220\mbox{\it{q}}}}}}}~+\,O(e^6\,\epsilon)\,+\,O(\inc^2\,\epsilon)\,
 +\,O(\epsilon^{3})~~~,~~~~~~~~~~~~
 \label{16b}
 \ea
 where the error $\,O(\epsilon^{3})\,$ originates from $\;\sin\epsilon = \epsilon + O(\epsilon^3)\;$.

 The simplification conventionally used in the literature ignores the frequency-dependence of the Love number and attributes the overall
 frequency-dependence to the lag. It also ignores the difference between the tidal lag $\,\epsilon\,$ and the lag in the material, $\,
 \delta\,$. This way, the conventional simplification makes $\,\epsilon\,$ obey the scaling law (\ref{generic_2}). At this point, most authors also
 set $\,\alpha = - 1\,$. Here we shall explore this approach, though shall keep $\,\alpha\,$ arbitrary.
 From the formula \footnote{~Let $~\frac{\textstyle 1}{\textstyle \sin\epsilon} = \left(\,{\cal E}\,\chi\right)^{\alpha}~$, where
 $\,{\cal{E}}\,$ is an empirical parameter of the dimensions of time, while $\,\epsilon\,$ is small enough, so $\,\sin
 \epsilon\approx\epsilon\,$.
 In combination with $\,\epsilon_{
 \textstyle{_{\textstyle{_{lmpq}}}}}\equiv\,\omega_{\textstyle{_{\textstyle{_{lmpq}}}}}\,\Delta t_{\textstyle{_{\textstyle{_{lmpq}}}}
 }\,$ and $\,\chi_{\textstyle{_{\textstyle{_{lmpq}}}}}\,=\,|\omega_{\textstyle{_{\textstyle{_{lmpq}}}}}|\,$, these formulae entail
 (\ref{631}).}
 \ba
 \Delta t_{\textstyle{_{\textstyle{_{lmpq}}}}}\;=\;{\cal{E}}^{\textstyle{^{-\,\alpha}}}\;
 \chi_{\textstyle{_{\textstyle{_{lmpq}}}}}^{\textstyle{^{-\,(\alpha+1)}}}
 \label{631}
 \ea
 derived by Efroimsky \& Lainey (2007) in the said approach, we see that the time
 lags are related to the principal-frequency lag $\;\Delta t_{\textstyle{_{\textstyle{_{2200}}}}}\;$ via:
 \ba
 \Delta t_{\textstyle{_{\textstyle{_{lmpq}}}}}\,=\,\Delta t_{\textstyle{_{\textstyle{_{2200}}}}}\;
 \left(\,\frac{\chi_{\textstyle{_{\textstyle{_{2200}}}}}}{\chi_{\textstyle{_{\textstyle{_{lmpq}}}}}}
 \,\right)^{\textstyle{^{\alpha+1}}}
 \;\;\;.
 \label{vv}
 \ea
 When the despinning is still going on and $\,\dot{\theta}\gg n\,$, the corresponding phase lags are:
 \ba
 \epsilon_{\textstyle{_{\textstyle{_{lmpq}}}}}\;\equiv\;
 \Delta t_{\textstyle{_{\textstyle{_{lmpq}}}}}\;\omega_{\textstyle{_{\textstyle{_{lmpq}}}}}\,=\,-~
 \Delta t_{\textstyle{_{\textstyle{_{2200}}}}}\;\chi_{\textstyle{_{\textstyle{_{lmpq}}}}}\;
 \left(\,\frac{\chi_{\textstyle{_{\textstyle{_{2200}}}}}}{\chi_{\textstyle{_{\textstyle{_{lmpq}}}}}}
 \,\right)^{\textstyle{^{\alpha+1}}}~=~
 -~\epsilon_{\textstyle{_{\textstyle{_{2200}}}}}\;
 \left(\,\frac{\chi_{\textstyle{_{\textstyle{_{2200}}}}}}{\chi_{\textstyle{_{\textstyle{_{lmpq}}}}}}
 \,\right)^{\textstyle{^{\alpha}}}
 ~~,\quad\quad
 \label{vvvvv}
 \ea
 which helps us to cast the secular part of the torque into the following convenient form:
 \footnote{~For $\,\dot{\theta}\gg 2n\,$, all the modes $\,\omega_{\textstyle{_{\textstyle{_{220q}}}}}\,$ are negative, so $\,\omega_{
 \textstyle{_{\textstyle{_{220q}}}}}\,=\,-\,\chi_{\textstyle{_{\textstyle{_{220q}}}}}\,$. Then, keeping in mind that $\,n/\dot{\theta}\ll 1\,$, we
 process (\ref{vvvvv}), for $q=1,$ like
 \ba
 \nonumber
 -\,\Delta t_{\textstyle{_{\textstyle{_{2200}}}}}\,\chi_{\textstyle{_{\textstyle{_{2200}}}}}~\left(\,\frac{\chi_{\textstyle{
 _{\textstyle{_{2200}}}}}}{\chi_{\textstyle{_{\textstyle{_{2201}}}}}}\,\right)^{\textstyle{^{\alpha}}}=\,-\,\Delta t
 _{\textstyle{_{\textstyle{_{2200}}}}}\,2\,|n-\dot{\theta}|\,\left(\frac{2|n-\dot{\theta}|}{|-2\dot{\theta}+3n|}\,
 \right)^{\textstyle{^{\alpha}}}=\,-\,\Delta t_{\textstyle{_{\textstyle{_{2200}}}}}\,2\,(\dot{\theta}-n)\,\left[1\,+\,\frac{
 \alpha}{2}\,\frac{n}{\stackrel{\bf\centerdot}{\theta\,}}\,+\,O(\,(n/\dot{\theta})^2\,)  \right]~~~,
 \ea
 and similarly for other $q\neq 0$, and then plug the results into (\ref{16b}). This renders us (\ref{653}).
 }
 \begin{subequations}
 \label{653}
 \ba
 \nonumber
 ^{\textstyle{^{\tiny{(Q\,>\,10)\,}}}}\overline{\cal T}_{\textstyle{_{\textstyle_{\textstyle{_{l=2}}}}}}
 \;=\;{\cal{Z}}
 \left[\;\,-\;\,\dot{\theta}\,\left(
 1\,+\,\frac{15}{2}\,e^2\,+\,\frac{105}{4}\,e^4\,+\,O(e^6)
 \right)\right.
 ~~~~~~~~~~~~~~~~~~~~~~~~~~~~~~~~~~~~~~~~~~~~~~~~~~~~~~~~~
 \\
 \nonumber\\
 \nonumber\\
 \label{632}
 +\;\left.n\left(1+\left(\frac{15}{2}-6\alpha\right)e^2+
 \left(\frac{105}{4}-\frac{363}{8}\alpha\right)e^4+O(e^6)\right)\right]
 +O(\inc^2/Q)+O(Q^{-3})+O(\alpha e^2 Q^{-1} n/\dot{\theta})\;\;\;~~~~
 \\
 \nonumber\\
 \nonumber\\
 \approx\;
 {\cal Z}\;
 \left[\,-\,\dot{\theta}\,\left(
 1\,+\,\frac{15}{2}\,e^2\right)
  +\,n\,\left(1\,+\,\left(\frac{15}{2}\,-\,6\,\alpha\right)\,e^2\,\right)\,\right]\;\;\;,
  ~~~~~~~~~~~~~~~~~~~~~~~~~~~~~~~~~~~~~~~~~~
 \label{633}
 \ea
 \end{subequations}
 where the overall factor reads as:
 \ba
 {\cal Z}~=~\frac{3\,G\,M_{sec}^{\textstyle{^{\,2}}}\;\,k_{2}\;\Delta t_{\textstyle{_{2200}}}}{R}\,~
 \frac{R^{\textstyle{^6}}}{a^6}~
 =~\frac{3\,n^2\,M_{sec}^{\textstyle{^{\,2}}}\;\,k_{2}\;\Delta t_{\textstyle{_{2200}}}}{
 (M_{prim}\,+\,M_{sec})}~\,\frac{R^{\textstyle{^5}}}{a^3}
 \,=\,\frac{3\,n\,M_{sec}^{\textstyle{^{\,2}}}~\,k_{2}}{Q_{\textstyle{_{2200}}}~(M_{prim}\,+\,M_{sec})}~\,
 \frac{R^{\textstyle{^5}}}{a^3}~\,\frac{n~~}{\chi_{\textstyle{_{2200}}}}~~,\quad\quad
 \label{634}
 \ea
 $M_{prim}\,$ and $\,M_{sec}\,$ being the primary's and secondary's masses.\footnote{~To arrive at the right-hand side of
 (\ref{634}), we recalled that $~\chi_{\textstyle{_{lmpq}}}\,\Delta t_{\textstyle{_{lmpq}}}\,=\,|\epsilon_{\textstyle{_{
 lmpq}}}|\,$ and that $~Q_{\textstyle{_{lmpq}}}^{-1}\,=\,|\epsilon_{\textstyle{_{lmpq}}}|\,+\,O(\epsilon^3)\,=\,|\epsilon
 _{\textstyle{_{lmpq}}}|\,+\,O(Q^{-3})\,$, according to formula (\ref{DI6}).} Dividing (\ref{634}) by the
 primary's principal moment of inertia $\,\xi\,M_{primary}\,R^2\,$, we obtain the contribution that this
 component of the torque brings into the angular deceleration rate $\;\ddot{\theta}\;$:
 \begin{subequations}
 \label{deceleration}
 \ba
 \nonumber
 \ddot{\theta}~=~{\cal{K}}~
 \left\{\,-\,\stackrel{\;\centerdot}{\theta}\,\left[~
 1\,+\,\frac{15}{2}\,e^2\,+\,\frac{105}{4}\,e^4\,+\,O(e^6)~
 \right]\right. ~+
 ~~~~~~~~~~~~~~~~~~~~~~~~~~~~~~~~~~~~~~~~~~~~~~~~~~~~~~~~~~~~~~
 \\
 \nonumber\\
 \nonumber\\
 \label{deceleration_1}
 \left.n\left[~1+\left(\frac{15}{2}-6\alpha\right)e^2+
 \left(\frac{105}{4}-\frac{363}{8}\alpha\right)e^4+O(e^6)~\right]\,\right\}
 +O(\inc^2/Q)+O(Q^{-3})+O(\alpha e^2 Q^{-1} n/\dot{\theta})\;~~~~~~~~~
 \\
 \nonumber\\
 \nonumber\\
 \approx~
 {\cal K}\;\left[\,-\,\stackrel{\;\centerdot}{\theta}\,\left(
 1\,+\,\frac{15}{2}\,e^2\right)
  +\,n\,\left(1\,+\,\left(\frac{15}{2}\,-\,6\,\alpha\right)\,e^2\,\right)\,\right]~~~,
  \quad\quad\quad\quad\quad\quad\quad\quad\quad\quad\quad\quad\quad\quad\quad\quad
 \label{deceleration_2}
 \ea
 \end{subequations}
 the factor $\,{\cal{K}}\,$ being given by
 \ba
 {\cal K}\,\equiv\,\frac{\cal Z}{\xi\,M_{prim}\,R^2}\,=\,\frac{3\,n^2\,M_{sec}^{\textstyle{^{\,2}}}\;\,k_{2}\;{\Delta t_{\textstyle{_{2200}}}}}{\xi\;
 M_{prim}\;(M_{prim}\,+\,M_{sec})}\;\frac{R^{\textstyle{^3}}}{a^3}
 \,=\,\frac{3\,n\,M_{sec}^{\textstyle{^{\,2}}}\;\,k_{2}}{\xi\;
 Q_{\textstyle{_{\textstyle{_{2200}}}}}\;M_{prim}\;(M_{prim}\,+\,M_{sec})}
 \;\frac{R^{\textstyle{^3}}}{a^3}\;\,
 \frac{n~~}{\chi_{\textstyle{_{\textstyle{_{2200}}}}}}\;\;\;,~\quad~
 \label{deceleration_3}
 \ea
 where $\,\xi\,$ is a multiplier emerging in the expression $\,\xi\,M_{primary}\,R^2\,$
 for the primary's principal moment of inertia ($\,\xi=2/5\,$ for a homogeneous sphere).

 In the special case of $\,\alpha\,=\,-\,1\;$, the above expressions enjoy agreement with the appropriate result stemming from the
 corrected MacDonald model -- except that our (\ref{653}) and (\ref{deceleration}) contain $\,\Delta t_{\textstyle{_{2200}}}\,$,
 $\,Q_{\textstyle{_{2200}}}\,$, and $\,\chi_{\textstyle{_{2200}}}\,$ instead of $\,\Delta t\,$, $\,Q\,$, and $\,\chi\,$ standing in
 formulae (44 - 47) from Williams \& Efroimsky (2012). Formula (\ref{deceleration_2}) tells us that the secular part of the tidal
 torque vanishes for
 \ba
 \stackrel{\;\centerdot}{\theta}\,-\;n\;=\;-\;6\;n\;e^2\;\alpha\;\;\;,
 \label{godot}
 \ea
 which coincides, for $\;\alpha= -1\;$, with the result obtained in Rodr\'{\i}guez et al. (2008, eqn. 2.4) and Williams \& Efroimsky
 (2012, eqn. 49). This coincidence however should not be taken at its face value, because it is {\emph{occasional}} or, possibly better
 to say, {\it{exceptional}}.

 Formulae (\ref{653} - \ref{634}) were obtained by insertion of the expressions for the eccentricity functions and the phase lags into
 (\ref{16b}), and {\underline{by assuming that $\,n\ll |\dot{\theta}|\,$}}. The latter caveat is a crucial element, not to be overlooked
 by the users of formulae (\ref{653} - \ref{634}) and of their corollary (\ref{deceleration} - \ref{deceleration_3}) for the tidal
 deceleration rate.

  {\underline{The case of $\,\alpha=-1\,$ is special, in that it permits derivation of
 (\ref{653} - \ref{godot}) {\it{without assuming}}}}\\
 {\underline{{\it{that}} $\,n\ll |\stackrel{\;\bf\centerdot}{\theta}|~$.}}
 ~However for $\,\alpha>-1\,$ the condition $\,n\ll |\stackrel{\;\bf\centerdot}{\theta}|\,$ remains
 mandatory, so formulae (\ref{653} - \ref{deceleration_3}) become {\it{inapplicable}} when
 $\,\stackrel{\;\centerdot}{\theta}\,$ reduces to values of about several $\,n\,$.

 Although formulae (\ref{632}) and (\ref{deceleration_1}) contain an absolute error $\,O(\alpha e^2 Q^{-1}n/\dot{\theta})\,$, this does
 {\underline{not}} mean that for $\,\stackrel{\;\centerdot}{\theta}\,$ comparable to $\,n\,$ the absolute error becomes $\,O(\alpha e^2
 Q^{-1} )\,$ and the relative one becomes $\,O(\alpha e^2)\,$. In reality, for $\,\stackrel{\;\centerdot}{\theta}\,$ comparable to $\,n\,$,
 {\it{the entire approximation falls apart}}, because formulae (\ref{vv} - \ref{vvvvv}) were derived from expression (\ref{631}), which
 is valid for $\,Q\gg1\,$ only (unless $\,\alpha=\,-\,1\,$). So these formulae become inapplicable in the vicinity of a commensurability.
 By ignoring this limitation, one can easily encounter unphysical paradoxes.\footnote{~For example, in the case of $\,\alpha>-1\,$,
 formulae (\ref{631} - \ref{vvvvv}) render infinite values for $\,\Delta t_{\textstyle{_{lmpq}}}\,$ and $\,\epsilon_{\textstyle{_{lmpq}}}
 \,$ on crossing the commensurability, i.e., when $\,\omega_{\textstyle{_{lmpq}}}\,$ goes through zero.}

 Thence, in all situations, except for the unrealistic rheology $\,\alpha=-1\,$, limitations of the approximation (\ref{653} -
 \ref{deceleration_3}) should be kept in mind. This approximation remains acceptable for $\,n\ll |\dot{\theta}|\,$, but becomes
 misleading on approach to the physically-interesting resonances.

 \subsection{An oversight in a popular formula}\label{oversight}

 The form in which our approximation (\ref{653} - \ref{deceleration_3}) is cast may appear awkward. The formula for the
 despinning rate $\,\ddot{\theta\,}\,$ is written as a function of $\,\dot{\theta}\,$ and $\,n\,$, multiplied by the overall factor
 $\,{\cal{K}}\,$. This form would be reasonable, were $\,{\cal{K}}\,$ a constant. That this is not the case  can be seen from the
 presence of the multiplier $\,\frac{\textstyle n}{\textstyle \chi_{\textstyle{_{2200}}}}\,=\,\frac{\textstyle n}{\textstyle 2\,|
 \stackrel{\bf\centerdot}{\theta}\,-\,n|}\,$ on the right-hand side of (\ref{deceleration_3}).

 Still, when written in this form, our result is easy to juxtapose with an appropriate formula from Correia \& Laskar (2004, 2009).
 There, the expression for the despinning rate looks similar to ours, up to an important detail: the overall factor is a constant,
 because it lacks the said multiplier $\frac{\textstyle n}{\textstyle \chi_{\textstyle{_{2200}}}}$. The multiplier was lost in those
 two papers, because the quality factor was introduced there as $\,1/( n\, \Delta t )\,$, see the line after formula (9) in
 Correia \& Laskar (2009). In reality, the quality factor $Q$ should, of course, be a function of the tidal frequency $\,\chi\,$,
 because $Q$ serves the purpose of describing the tidal damping at this frequency. Had the quality factor been taken as
 $\,1/( \chi\, \Delta t )\,$, it would render the corrected MacDonald model ($\,\alpha=\,-\,1\,$), and the missing multiplied would
 be there. Being unphysical,\footnote{~To be exact, the model is unphysical everywhere except in the closest vicinity of the
 resonance -- see formulae (\ref{aaaa} - \ref{cccc}).} the model is mathematically convenient, because it enables one to write
 down the secular part of the torque as one expression, avoiding the expansion into a series (Williams \& Efroimsky 2012). The model
 was pioneered by Singer (1968) and employed by Mignard (1979, 1980), Hut (1981) and other authors.

 Interestingly, in the special case of the 3:2 spin-orbit resonance, we have $\,\chi=n\,$. Still, the difference between $\,\chi\,$
 and $\,n\,$ in the vicinity of the resonance may alter the probability of entrapment of Mercury into this rotation mode.
 The difference between $\,\chi\,$ and $\,n\,$ becomes even more considerable near the other resonances of interest. So the
 probabilities of entrapment into those resonances must be recalculated.

 \section{Conclusions}

 The goal of this paper was to lay the ground for a reliable model of tidal entrapment into spin-orbital resonances. To this end, we
 approached the tidal theory from the first principles of solid-state mechanics. Starting from the expression for the material's
 compliance in the time domain, we derived the frequency-dependence of the Fourier components of the tidal torque. The other torque, one
 caused by the triaxiality of the rotator, is not a part of this study and will be addressed elsewhere.

 \begin{itemize}

 \item{} We base our work on the Andrade rheological model, providing arguments in favour of its applicability to the Earth's mantle,
 and therefore, very likely, to other terrestrial planets and moons. The model is also likely to apply to the icy moons (Castillo-Rogez
 et al. 2011). We have reformulated the model in terms of a characteristic anelastic timescale $\tau_{_A}$ (the Andrade time). The ratio
 of the Andrade time to the viscoelastic Maxwell time, $\zeta=\tau_{_A}/\tau_{_M}$, serves as a dimensionless free parameter of the
 rheological model.

 The parameters $\tau_{_A}$, $\tau_{_M}$, $\zeta$ cannot be regarded constant, though their values may be changing very slowly over vast
 bands of frequency. The shapes of these frequency-dependencies may depend on the dominating dissipation mechanisms and, thereby, on the
 magnitude of the load, as different damping mechanisms get activated under different loads.

 The main question here is whether, in the low-frequency limit, anelasticity becomes much weaker than viscosity. (That would imply an
 increase of $\,\tau_{_A}\,$ and $\,\zeta\,$ as the tidal frequency $\,\chi\,$ goes down.) The study of ices under weak loads, with
 friction caused mainly by lattice diffusion (Castillo-Rogez et al. 2011, Castillo-Rogez \& Choukroun 2010) has not shown such a decline
 of anelasticity. However, Karato \& Spetzler (1990) point out that it should be happening in the Earth's mantle, where the loads are
 much higher and damping is caused mainly by unpinning of dislocations. According to {\it{Ibid.}}, in the Earth, the decrease of the
 role of anelasticity happens abruptly as the frequency falls below the threshold $\,\chi_{\textstyle{_0}}\sim$\,1\,yr$^{-1}\,$. We then
 may expect a similar switch in the other terrestrial planets and the Moon, though there the threshold may be different as it is
 sensitive to the temperature of the mantle. The question, though, remains if this statement is valid also for the small bodies, in
 which the tidal stresses are weaker and dissipation is dominated by lattice diffusion.

 \item{} By direct calculation, we have derived the frequency dependencies of the factors $\,k_{\textstyle{_{l}}}\,\sin\epsilon_{
 \textstyle{_{l}}}\,$ emerging in the tidal theory. Naturally, the obtained dependencies of these factors upon the tidal frequency $\,
 \chi_{\textstyle{_{lmpq}}}\,$ (or, to put it more exactly, upon the tidal mode $\,\omega_{\textstyle{_{lmpq}}}\,$) mimic the
 frequency-dependence of the imaginary part of the complex compliance. They scale as $\,\sim\chi^{-\alpha}\,$ with $\,0<\alpha<1\,$, at
 higher frequencies; and look as $\,\sim\chi^{-1}\,$ at lower frequencies. However in the zero-frequency limit the factors
 $\,k_{\textstyle{_{l}}}\,\sin\epsilon_{\textstyle{_{l}}}\,$ demonstrate a behavior inherent in the tidal lagging and absent
 in the case of lagging in a sample: in a close vicinity of the zero frequency, these factors (and the appropriate
 components of the tidal torque) become linear in the frequency. This way, $\,k_{\textstyle{_{l}}}\,\sin\epsilon_{\textstyle{_{l}}}\,$
 first reaches a finite maximum, then decreases continuously to nil as the frequency approaches to zero, and then changes its sign. So
 the resonances are crossed continuously, with neither the tidal torque nor the tidal force diverging there. For example, the leading
 term of the torque vanishes at the synchronous orbit.

 This continuous traversing of resonances was pointed out in a heuristic manner by Efroimsky \& Williams (2009). Now we
 have derived this result directly from the expression for the compliance of the material of the rotating body. Our derivation, however,
 has a problem in it: the frequency, below which the factors $\,k_{\textstyle{_{l}}}\,\sin\epsilon_{
 \textstyle{_{l}}}\,$ and the appropriate components of the torque change their frequency-dependence to linear, is
 implausibly low (lower than $\,10^{-10}\,$Hz, if we take our formulae literally). The reason for this mishap is that in our formulae
 we kept using the known value of the Maxwell time $\,\tau_{_M}\,$ all the way down to the zero frequency. Possible changes of the
 viscosity and, accordingly, of the Maxwell time in the zero-frequency limit may broaden this region of linear dependence.

 \item{} We have offered an explanation of the unexpected frequency-dependence of dissipation in the Moon, discovered by LLR.  The
 main point of our explanation is that the LLR measures the {\it{tidal}} dissipation whose frequency-dependence is different from
 that of the {\it{seismic}} dissipation. Specifically, the ``wrong" sign of the exponent in the power dissipation law may indicate
 that the frequencies at which tidal friction was observed were below the frequency at which the lunar $~k_2\,\sin\epsilon_2~$ has
 its peak. Taken the relatively high frequencies of observation (corresponding to periods of order month to year), this explanation
 can be accepted only if the lunar mantle has a low mean viscosity. This may be the case, taken the presumably high concentration
 of the partial melt in the low mantle.

 \item{} We have developed a detailed formalism for the tidal torque, and have singled out its oscillating component.

 The studies of entrapment into spin-orbit resonances, performed in the past, took into account neither the afore-mentioned complicated
 frequency-dependence of the torque in the vicinity of a resonance, nor the oscillating part of the torque. We have written down a concise
 and practical expression for the oscillating part, and have raised the question whether it may play a role in the entrapment and libration
 dynamics.

 \end{itemize}


 \section{Acknowledgments}

  This paper stems largely from my prior research carried out in collaboration with James G. Williams and discussed on numerous occasions
 with Sylvio Ferraz Mello, to both of whom I am thankful profoundly.
 I am grateful to Benoît Noyelles, Julie Castillo-Rogez, Veronique Dehant, Shun-ichiro Karato, Val\'ery Lainey, Valeri Makarov,
 Francis Nimmo, Stan Peale, and Tim Van Hoolst for numerous enlightening exchanges and consultations on the theory of tides.

 I gladly acknowledge the help and inspiration which I obtained from reading the unpublished preprint by the late Vladimir Churkin
 (1998). In Appendix C, I present several key results from his preprint.

 I sincerely appreciate the support from my colleagues at the US Naval Observatory, especially from John Bangert.\\

  \pagebreak

\noindent
 {\underline{\textbf{\Large{Appendix.}}}}

 \appendix

 \section{Continuum mechanics.\\
 A celestial-mechanician's survival kit
  }\label{3AA3}

 Appendix \ref{3AA3} offers an extremely short introduction into continuum mechanics. The standard material, which normally occupies
 large chapters in books, is compressed into several pages.

 Subsection \ref{glossary} presents the necessary terms and definitions.
 Subsections \ref{A.1} explains the basic concepts employed in the theory of
 stationary deformation, while subsection \ref{her} explains extension of these methods to creep. These subsections also demonstrate the
 great benefits stemming from the isotropy and incompressibility assumptions. Subsection \ref{vis} introduces viscosity, while
 subsection \ref{viscoel} offers an example of how elasticity, viscosity get combined with hereditary reaction, into one expression.
 Subsection \ref{examples} renders several simple examples.

 \subsection{Glossary}\label{glossary}

 We start out with a brief guide elucidating the main terms employed in continuum mechanics.

 \begin{itemize}

 \item{} {\it Rheology} is the science of deformation and flow.

 \item{} {\it Elasticity:} ~~This is the most trivial reaction -- instantaneous, linear, and fully reversible after
     stressing is turned off.

 \item{} {\it Anelasticity:} ~~While still linear, this kind of deformation is not necessarily instantaneous, and can demonstrate
 ``memory", both under loading and when the load is removed. Importantly, the term {\it{anelasticity}} always implies reversibility:
 though with delay, the original shape is restored. Thus an anelastic material can assume two equilibrium states: one is the
 unstressed state, the other being the long-term relaxed state. Anelasticity is characterised by the difference in strain between
 the two states. It is also characterised by a relaxation time between these states, and by its inverse -- the frequency at which
 relaxation is most effective. The Hohenemser-Prager model, also called SAS (Standard Anelastic Solid), renders an example of
 anelastic behaviour.

 Anelasticity is an example of but not synonym to {\it{hereditary reaction}}. The latter includes also those kinds of delayed
 deformation, which are irreversible.

 \item{} {\it Inelasticity:} ~~This term embraces {\it{all}} kinds of irreversible deformation, i.e., deformation
     which stays, fully or partially, after the load is removed.

 \item{} {\it Unelasticity $\,$($\,$= Nonelasticity):} ~~These terms are very broad, in that they embrace any
     behaviour which is not elastic. In the literature, these terms are employed both for recoverable (anelastic) and
     unrecoverable (inelastic) deformations.

 \item{} {\it Plasticity:} ~~Some authors simply state that plastic deformation is deformation which is irreversible
     -- a very broad and therefore useless definition which makes plasticity sound like a synonym to inelasticity.

 More precisely, plastic is a stepwise behaviour: no strain is observed until the stress $\,\sigma\,$ reaches a threshold value
 $\,\sigma_{_Y}\,$ (called yield strength), while a steady flow begins as the stress reaches the said threshold. Plasticity can be
 either perfect (when deformation is going on without any increase in load) or with hardening (when increasingly higher stresses
 are needed to sustain the flow). It is always irreversible.

 In real life, plasticity shows itself in combination with elasticity or/and viscosity. Models describing these types of behaviour
 are called elastoplastic, viscoplastic, and viscoelastoplastic. They are all inelastic, in that they describe unrecoverable
 changes of shape.

 \item{} {\it Viscosity:}  ~~Another example of inelastic, i.e., irreversible behaviour. A viscous stress is
     proportional to the time derivative of the viscous strain.

 \item{} {\it Viscoelasticity:} ~~The term is customarily applied to all situations where both viscous and elastic
     (but not plastic) reactions are observed. One may then expect that the equations interrelating the
     viscoelastic stress to the strain would contain only the viscosity coefficients and the elastic moduli.
     However this is not necessarily true, as some other empirical constants may show up. For example, the Andrade
     model (\ref{I6412}) contains an elastic term, a viscous term, and an extra term
     responsible for a hereditary reaction (the ``memory"). Despite the presence of that extra term, the Andrade
     creep is still regarded viscoelastic. So it should be understood that viscoelasticity is, generally, more than
     just viscosity combined with elasticity. One might christen such deformations ``viscoelastohereditary", but
     such a term would sound awkward.\footnote{~Sometimes the term {\it{elastohereditary}} is used, but not
     {\it{viscoelastohereditary}} or {\it{elastoviscohereditary}}.}

     On many occasions, complex viscoelastic models can be illustrated with an infinite set of viscous and elastic
     elements. These serve to interpret the hereditary terms as manifestations of viscosity and elasticity only.
     While illustrative, these schemes with dashpots and springs have their limitations and should not be taken too
     literally. In some (not all) situations, the hereditary terms may be interpreted as time-dependent additions to
     the steady-state viscosity coefficient, the Andrade model being an example of such situation.

 \item{} {\it Viscoplasticity:} ~~These are all models wherein both viscosity and plasticity are present in some
     combination. In these situations, higher stresses have to be applied to increase the deformation rate. Just as
     in the case of viscoelasticity, viscoplastic models may, in principle, incorporate extra terms standing for
     hereditary reaction.

 \item{} {\it Elastoviscoplasticity $\,$($\,$= Viscoelastoplasticity):} ~~The same as above, though with elasticity
     present.

 \item{} {\it Hereditary reaction:} ~~While the term is self-explanatory, it would be good to limit its use to effects other
     than viscosity. In expression (\ref{permitted_1}) for the stress through strain, the distinction between the viscous
     and hereditary reactions is clear: while the viscous part of the stress is rendered instantaneously by the delta-function
     term of the kernel, the hereditary reaction is obtained through integration. In expression (\ref{I12}) for the strain
     through stress, though, the viscous part shows up, under the integral, in sum with the other delayed terms -- see, for
     example, the Andrade model (\ref{I6412}). This is one reason for which we shall use the term {\it{hereditary reaction}} in
     application to delayed behaviour different from the pure viscosity. Another reason is that viscous flow is always
     irreversible, while a hereditary reaction may be either irreversible (inelastic) or reversible (anelastic).

 \item{} {\it Creep:} ~~Widely accepted is the convention that this term signifies a slow-rate deformation under
     loads below the yield strength $\,\sigma_{_Y}\,$.

     Numerous authors, though, use the oxymoron {\it{plastic creep}}, thereby extending the applicability realm of
     the word {\it{creep}} to {\it{any}} slow deformation.

     Here we shall understand creep in the former sense, i.e., with no plasticity involved.

 \end{itemize}

 \noindent
 It would be important to distinguish between viscoelastic deformations, on the one hand, and viscoplastic (or, properly speaking,
 viscoelastoplastic) deformations on the other hand. Plasticity shows itself at larger stresses and is, typically, nonlinear. It comes
 into play when the linearity assertion fails. For most minerals, this happens when the strain approaches the threshold of $\,10^{-6}\,$.
 Although it is possible that this threshold is transcended in some satellites (for example, in the deeper layers of the Moon), we do not
 address plasticity in this paper.

 \subsection{Stationary linear deformation of isotropic incompressible media}\label{A.1}

 In the linear approximation, the tensor of elastic stress, $\,\stackrel{(e)}{\mathbb{S}}\,$, is proportional to the
 differences in displacement of the neighbouring elements of the medium. These differences are components of the {\it{tensor
 gradient}} $\,\nabla\otimes{\bf{u}}\,$, where ${\bf{u}}\,$ is the displacement vector.

 The tensor gradient can be decomposed, in an invariant way, into its symmetric and antisymmetric parts:
 \ba
 \nabla\otimes{\bf{u}}\;=\;\frac{\textstyle 1}{\textstyle 2}\,\left[\,\,\left(\nabla\otimes{\bf{u}}\right)\,+\,\left(
 \nabla\otimes{\bf{u}}\right)^{{^{T}}}\,\,\right]
 \,+~\frac{\textstyle 1}{\textstyle 2}\,\left[\,\,\left(\nabla\otimes{\bf{u}}\right)\,-\,\left(\nabla\otimes{\bf{u}}
 \right)^{{^{T}}}\,\,\right]~~~.
 \label{abb}
 \ea
 The decomposition being invariant, the two parts should contribute into the stress independently, at least in the
 linear approximation. However, as well known (Landau \& Lifshitz 1986), the antisymmetric part of (\ref{abb}) describes
 the displacement of the medium as a whole and thus brings nothing into the stress. This is why the linear dependence is
 normally written as
 \ba
 \stackrel{(e)}{\mathbb{S}}\,=\,{\mathbb{B}}\,\,{\mathbb{U}}~~~,
 \label{B}
 \ea
 where $\,{\mathbb{B}}\,$ is a four-dimensional tensor having $\,3^4\,=\,81\,$ components, while the strain tensor
 \ba
 {\mathbb{U}}\,\equiv\,\frac{\textstyle 1}{\textstyle 2}\,\left[\,\left(\nabla\otimes{\bf{u}}\right)\,+\,\left(\nabla
 \otimes{\bf{u}}\right)^{{^{T}}}\,\right]
 \label{}
 \ea
 is the symmetric part of the tensor gradient. Its components are related to the displacement vector $\,{\bf{u}}~$
 through $~u_{\alpha\beta}\,\equiv\,\frac{\textstyle 1}{\textstyle 2}\left(\frac{\textstyle{\partial u_{\alpha}}}{
 \textstyle{\partial x_\beta}} + \frac{\textstyle{\partial u_{\beta}}}{\textstyle{\partial x_\alpha}}\right)\,$.

 Although the matrix $\,{\mathbb{B}}\,$ is comprised of 81 empirical constants, in isotropic materials the description
 can be reduced to two constants only. To see this, recall that the expansion of the strain into a part with trace and
 a traceless part, $\,{\mathbb{U}}\,=\,\frac{\textstyle{1}}{\textstyle{3}}\,{\mathbb{I}}\;\mbox{Sp}\,{\mathbb{U}}\,+\,
 \left(\,{\mathbb{U}}\,-\,\frac{\textstyle{1}}{\textstyle{3}}\,{\mathbb{I}}\;\mbox{Sp}\,{\mathbb{U}}\,\right)\,$, is
 invariant. Here the trace of $\,{\mathbb{U}}\,$ is denoted with $\,\mbox{Sp}\,{\mathbb{U}}\,\equiv\,u_{\alpha\alpha}\,$,
 summation over repeated indices being implied. The notation $\,{\mathbb{I}}\,$ stands for the unity matrix consisting of
 elements $\,\delta_{\gamma\nu}\,$.

 In an isotropic medium, the elastic stress must be invariantly expandable into parts proportional to the afore-mentioned parts of the
 strain. The first part of the stress is proportional, with an empirical coefficient $\,3\,K\,$, to the trace part $\,\frac{\textstyle{1}
 }{\textstyle{3}}\,{\mathbb{I}}\;\mbox{Sp}\,{\mathbb{U}}\,$ of the strain. The second part of the stress will be proportional, with an
 empirical coefficient $\,2\,\mu\,$, to the traceless part $\,\left(\,{\mathbb{U}}\,-\,\frac{\textstyle{1}}{\textstyle{3}}\,{\mathbb{I}}~
 \mbox{Sp}\,{\mathbb{U}}\,\right)\,$ of the strain:
 \begin{subequations}
 \ba
 \stackrel{(e)}{\mathbb{S}}~=~K~{\mathbb{I}}~\mbox{Sp}\,{\mathbb{U}}\,+\,2\,\mu\left(\,{\mathbb{U}}\,-\,\frac{\textstyle{
 1}}{\textstyle{3}}\,{\mathbb{I}}\;\mbox{Sp}\,{\mathbb{U}}\,\right)~=~-~p~{\mathbb{I}}\,+\,2\,\mu\left(\,{\mathbb{U}}
 \,-\,\frac{\textstyle{1}}{\textstyle{3}}\,{\mathbb{I}}\;\mbox{Sp}\,{\mathbb{U}}\,\right)
 \label{SU_1}
 \ea
 or, in Cartesian coordinates:
 \ba
 \sigma_{\gamma\nu}~=~\,K~\delta_{\gamma\nu}\,u_{\alpha\alpha}\,+\,2\,\mu\,\left(\,u_{\gamma\nu}\,-\,\frac{\textstyle{1}}{
 \textstyle{3}}\,\delta_{\gamma\nu}\,u_{\alpha\alpha}\,\right)~=~-~p~\delta_{\gamma\nu}\,+\,2\,\mu\,\left(\,u_{\gamma\nu}
 \,-\,\frac{\textstyle{1}}{\textstyle{3}}\,\delta_{\gamma\nu}\,u_{\alpha\alpha}\,\right)~~~,\quad\quad
 \label{SU_2}
 \ea
 \label{SU}
 \end{subequations}
 where
 \ba
 p\,\equiv\,-\,\frac{1}{3}\;\mbox{Sp}\,{\mathbb{S}}\,=\,-\,K\,\mbox{Sp}\,{\mathbb{U}}
 \label{pressure}
 \ea
 is the hydrostatic pressure. Thus the elastic stress gets decomposed, in an invariant way, as:
 \ba
 \stackrel{(e)}{\mathbb{S}}~=~\stackrel{(e)}{\mathbb{S}}_{\textstyle{_{volumetric}}}\,+\;\stackrel{(e)}{\mathbb{S}}_{\textstyle{_{deviatoric}}}~~~,
 \label{total}
 \ea
 where the {\it{volumetric elastic stress}} is given by
 \ba
 \stackrel{(e)}{\mathbb{S}}_{\textstyle{_{volumetric}}}\;\,\equiv\;K~{\mathbb{I}}~\mbox{Sp}\,{\mathbb{U}}~=~{\mathbb{I}}~
 \frac{1}{3}~\mbox{Sp}\,{\mathbb{S}}\,=\,-\,p\,{\mathbb{I}}~~~,
 \label{isotropic}
 \ea
 while the {\it{deviatoric elastic stress}} is:
 \ba
 \stackrel{(e)}{\mathbb{S}}_{\textstyle{_{deviatoric}}}~\equiv~2\,\mu\left(\,{\mathbb{U}}\,-\,\frac{\textstyle{1}}{
 \textstyle{3}}\,{\mathbb{I}}\;\mbox{Sp}\,{\mathbb{U}}\,\right)~~~.
 \label{deviatoric}
 \ea

 Inverse to (\ref{SU_1} - \ref{SU_2}) are the following expressions for the strain tensor:
 \begin{subequations}
 \ba
 {\mathbb{U}}~=~\frac{1}{9\,K}~{\mathbb{I}}~\mbox{Sp}\,\stackrel{(e)}{\mathbb{S}}\,+\,\frac{1}{2\,\mu}\,\left(\,
 \stackrel{(e)}{\mathbb{S}}\,-\,\frac{\textstyle{1}}{\textstyle{3}}\,{\mathbb{I}}~\mbox{Sp}\,\stackrel{(e)}{
 \mathbb{S}}\,\right)
 \label{US_1}
 \ea
 and
 \ba
 u_{\gamma\nu}~=~\frac{1}{9\,K}~\delta_{\gamma\nu}~\stackrel{(e)}\sigma_{\alpha\alpha}~+~
 \frac{1}{2\,\mu}\,\left(\,\stackrel{(e)}\sigma_{\gamma\nu}\,-\,\frac{\textstyle{1}}{\textstyle{3}}\,\delta_{\gamma\nu}\,
 \stackrel{(e)}\sigma_{\alpha\alpha}\,\right)~~~,\quad
 \label{US_2}
 \ea
 \label{US}
 \end{subequations}
 where the term with trace, $\,\frac{\textstyle 1}{\textstyle 9\,K}\;{\mathbb{I}}\;\mbox{Sp}\,\stackrel{(e)}{\mathbb{S}}\,$,
 is called the {\it{volumetric strain}}, while the traceless term, $\,\frac{\textstyle{1}}{\textstyle{2\,\mu}}\,\left(\,
 \stackrel{(e)}{\mathbb{S}}\,-\,\frac{\textstyle{1}}{\textstyle{3}}\,{\mathbb{I}}\;\mbox{Sp}\,\stackrel{(e)}{\mathbb{S}}\,
 \right)\,$, is named the {\it{deviatoric strain}}. The quantity $\,K\,$ is called the {\it{bulk modulus}}, while $\,\mu\,$ is called the {\it{shear modulus}}.

 Expressions (\ref{SU}) trivially entail the following interrelation between traces:
 \ba
 \mbox{Sp}\,\stackrel{(e)}{\mathbb{S}}\,=\,3\,K\,\mbox{Sp}\,{\mathbb{U}}~\quad~\mbox{or, in terms of components:}\quad~\stackrel{(e)}\sigma_{\alpha\alpha}\,=\,
 3\,K\,u_{\alpha\alpha}~~~.
 \label{traces}
 \ea
 As demonstrated in many books (e.g., in Landau \& Lifshitz 1986), the trace of the strain is equal to the
 relative variation of the volume, experienced by the material subject to deformation: $~u_{\alpha\alpha}\,=\,\nabla\cdot{
 \bf{u}}\,=\,\frac{\textstyle{dV\,'-dV}}{\textstyle{dV}}~$, where $\,{\bf{u}}\,$ is the displacement vector. In the no-compressibility approximation, the trace of the strain and, according
 to (\ref{traces}), that of the stress become zero. Then, in the said approximation, the hydrostatic pressure
 (\ref{pressure}) and the volumetric elastic stress (\ref{isotropic}) become nil, and all we are left with is the deviatoric
 elastic stress (and, accordingly, the deviatoric part of the strain). Formulae (\ref{SU}) and (\ref{US}) get simplified to
 \ba
 \stackrel{(e)}{\mathbb{S}}~=~2\,\mu\,{\mathbb{U}}~~,\quad~~\mbox{which is the same as}~\quad\stackrel{(e)}\sigma_{\gamma\nu}~=~
 2\,\mu\,u_{\gamma\nu}~~~,\,\quad
 \label{SU_3}
 \ea
 and to
 \ba
 2\;{\mathbb{U}}~=~J\,\stackrel{(e)}{\mathbb{S}}~~,\quad~~\mbox{which is the same as}~\quad
 2\,u_{\gamma\nu}~=~J\,\stackrel{(e)}\sigma_{\gamma\nu}~~~,\quad
 \label{US_3}
 \ea
 the quantity $\,J\,\equiv\,1/\mu\,$ being called the {\it{compliance}} of the material.

 \subsection{Evolving linear deformation of isotropic incompressible isotropic media. Hereditary reaction}\label{her}

 Equations (\ref{B} - \ref{US_3}) were written for static deformation, so each of these equations can be assumed to connect the strain
 and the elastic stress taken at the same instant of time (for a static deformation their values stay constant anyway).

 Extension of this machinery is needed when one wants to describe evolving deformation of materials with ``memory". Thence the
 four-dimensional tensor $\,{\mathbb{B}}\,$ becomes a linear operator $\,\tilde{\mathbb{B}}\,$ acting on the strain tensor function as a
 whole. To render the value of the  stress at time $\,t\,$, the operator will ``consume" as arguments all the values of strain over the
 interval $\;t\,'\,\in\,\left(\right.-\infty\,,\,t\left.\right]\;$:
 \ba
 \stackrel{(h)}{\mathbb{S}}(t)\,=\,\tilde{\mathbb{B}}(t)\,\,{\mathbb{U}}~~~.
 \label{BB}
 \ea
 Thus $\,\tilde{\mathbb{B}}\,$ will be an integral operator, with integration going
 from $\,t\,'\,=\,-\,\infty\,$ through $\,t\,'\,=\,t\,$.

 In the static case, the linearity guaranteed elasticity, i.e., the ability of the body to regain its shape after the loading is
 turned off: no stress yields no strain. In a more general situation of materials with ``memory", this ability is no longer retained,
 as the material may demonstrate {\it{creep}}. This is why, in this section, the stress is called {\it{hereditary}} and is denoted with
 $\,\stackrel{(h)}{\mathbb{S}}\,$.

 Just as in the stationary case, we wish the properties of the medium to remain isotropic. As the decomposition of the strain into the
 trace and traceless parts remains invariant at each moment of time, these two parts will, separately, generate the trace and traceless
 parts of the hereditary stress in an isotropic medium. This means that, in such media, the four-dimensional tensor operator $\,\tilde{
 \mathbb{B}}\,$ gets reduced to two scalar linear operators $\,\tilde{K}\,$ and $\,\tilde{\mu}\,$:
 \ba
 \stackrel{(h)}{\mathbb{S}}~=~\stackrel{(h)}{\mathbb{S}}_{\textstyle{_{volumetric}}}\,+\;\stackrel{(h)}{\mathbb{S}}_{
 \textstyle{_{deviatoric}}}~=~\tilde{K}~{\mathbb{I}}~\mbox{Sp}\,{\mathbb{U}}\,+\,2\,\tilde{\mu}\left(\,{\mathbb{U}}\,-\,
 \frac{\textstyle{1}}{\textstyle{3}}\,{\mathbb{I}}\;\mbox{Sp}\,{\mathbb{U}}\,\right)~~~,
 \label{SU_11}
 \ea
 where both $\,\tilde{K}\,$ and $\,\tilde{\mu}\,$ are integral operators acting on the tensor function $\,u_{\gamma\nu}(t\,')\,$
 as a whole, i.e., with integration going from $\,t\,'\,=\,-\,\infty\,$ through $\,t\,'\,=\,t\,$.

 If we also assume that, under evolving load, the medium stays incompressible, the trace of the strain, $\,u_{\alpha\alpha}\,$, will
 stay zero. An operator generalisation of (\ref{traces}) now reads: $~\sigma_{\alpha\alpha}(t)\,=\,3\,\tilde{K}(t)\,u_{\alpha\alpha}~$.
 Under a reasonable assumption of $\,\sigma_{\alpha\alpha}\,$ being nil in the distant past, this integral operator renders $\,\sigma_{
 \alpha\alpha}\,=\,0\,$ at all times. This way, in a medium that is both isotropic and incompressible, we have:
 \ba
 {\mathbb{U}}~=~{\mathbb{U}}_{\textstyle{_{deviatoric}}}~\quad~\mbox{and, ~accordingly:}
 ~\quad~\stackrel{(h)}{\mathbb{S}}~=~\stackrel{(h)}{\mathbb{S}}_{\textstyle{_{deviatoric}}}\quad.\quad\quad
 \label{}
 \ea
 Then the time-dependent analogues to formulae (\ref{SU_3}) and (\ref{US_3}) will be:
 \ba
 \stackrel{(h)}{\mathbb{S}}(t)~=~2\,\tilde{\mu}(t)\,{\mathbb{U}}
 \label{SU_4}
 \ea
 and
 \ba
 2\,{\mathbb{U}}(t)~=~\hat{J}(t)\,\stackrel{(h)}{\mathbb{S}}~~~,
 \label{US_4}
 \ea
 where the compliance $\,\hat{J}\,$, too, has been promoted to operatorship and crowned with a caret.

 Formula (\ref{SU_4}) tells us that in a medium, which is both isotropic and incompressible, relation between the stress
 and strain tensors can be described with one scalar integral operator $\,\tilde{\mu}\,$ only, the complementary operator
 $\,\hat{J}\,$ being its inverse. (Here the adjective ``scalar" does not imply multiplication with a scalar. It means that
 the operator preserves its functional form under a change of coordinates.)

 Below we shall bring into the picture also the viscous component of the stress, a component related to the strain
 through a four-dimensional tensor whose $\,3^4=81\,$ components are differential operators. In that case too, the
 isotropy of the medium will enable us to reduce the 81-component tensor operator to two differential operators
 transforming as scalars. Besides, the incompressibility of the medium makes the viscous stress traceless. Thus it will
 turn out that, in an isotropic and incompressible medium, the viscous component of the stress can be described by only
 one scalar differential operator --  much like the elastic and hereditary parts of the stress. (Once again, ``scalar" means:
 indifferent to coordinate transformations.)

 Eventually, the elastic, hereditary, and viscous deformations will be united under the auspices of a general viscoelastic formalism. In
 an isotropic medium, this combined formalism will be reduced to two integro-differential operators only. In a medium which is both
 isotropic and incompressible, the formalism will be reduced to only one scalar integro-differential operator.

 \subsection{The viscous stress}\label{vis}

 While the elastic stress $\,\stackrel{{{(e)}}}{\mathbb{S}}\,$ is linear in the strain, the viscous stress
 $\,\stackrel{(v)}{\mathbb{S}}\,$ is linear in the first derivatives of the components of the velocity with respect to the coordinates:
 \ba
 \stackrel{(v)}{\mathbb{S}}\,= \,{\mathbb{A}}\,(\nabla\otimes {\bf{v}})\,
 \label{L36}
 \ea
 where $\,{\mathbb{A}}\,$ is the so-called viscosity tensor, $\,\nabla \otimes {\bf{v}}\,$ is the tensor gradient of the
 velocity. The velocity of a fluid parcel relative to its average position is connected to the displacement vector
 $\,{\bf u}\,$ through $\,{\bf{v}}\,=\,d{\bf u}/dt\,$.

 The tensor gradient of the velocity can be expanded, in an invariant way, into its antisymmetric and symmetric parts:
 \ba
 \label{ex}
 \nabla \otimes {\bf{v}}\;=\;\Omega\;+\;{\mathbb{E}}~~~,
 \ea
 where the antisymmetric part is furnished by the {\it{vorticity tensor}}
 \ba
 \Omega\;\equiv\;\frac{1}{2}~\left[\,(\nabla \otimes {\bf{v}})~-\;(\nabla \otimes {\bf{v}})^{^T}\,\right]~~~,
 \label{}
 \ea
 while the symmetric part is given by the {\it{rate-of-shear tensor}}
 \ba
 {\mathbb{E}}\;\equiv\;\frac{1}{2}~\left[\,(\nabla \otimes {\bf{v}})\,+\,(\nabla \otimes {\bf{v}})^{^T}\,\right]~~~.
 \label{}
 \ea
 The latter is obviously related to the strain tensor through
 \ba
 {\mathbb{E}}\;=\;\frac{\partial~}{\partial t}\;{\mathbb{U}}\;~~.
 \label{}
 \ea

 It can be demonstrated (e.g., Landau \& Lifshitz 1987) that the antisymmetric vorticity tensor describes the rotation
 of the medium as a whole\footnote{~This is why this tensor's components coincide with those of the angular velocity
 $\,\omegabold\,$ of the body. For example, $\;\Omega_{21}\,=\,\frac{\textstyle 1}{\textstyle 2}\,\left(\,\frac{
 \textstyle\partial v_2}{\textstyle\partial x_1}\,-\,\frac{\textstyle\partial v_1}{\textstyle\partial x_2}\,\right)\;$
 coincides with $\,\omega_3\,$.} and therefore contributes nothing to the stress.\footnote{Since expansion
 (\ref{ex}) of the tensor gradient into the vorticity and rate-of-shear tensors is invariant, then so is the conclusion
 about the irrelevance of the vorticity tensor for the stress picture.} For this reason, the viscous stress can be written as
 \ba
 \stackrel{(v)}{\mathbb{S}}\,=\,{\mathbb{A}}\,{\mathbb{E}}\,=\,{\mathbb{A}}\;\frac{\partial~}{\partial t}\;{\mathbb{U}}\;~~.
 \label{L36}
 \ea

 The matrix $\,{\mathbb{A}}\,$ is four-dimensional and contains $\,3^4=81\,$ components. Just as the matrix
 $\,{\mathbb{B}}\,$ emerging in expression (\ref{B}) for the elastic stress, the matrix $\,{\mathbb{A}}\,$
 can be reduced, in an isotropic medium, to only two empirical constants. To see this, mind that the rate-of-shear tensor
 can be decomposed, in an invariant manner, into two parts:
 \ba
 {\mathbb{E}}~=~\frac{1}{3}~{\mathbb{I}}~\nabla\cdot{\bf{v}}~+~\left(\,{\mathbb{E}}~-~\frac{1}{3}~{\mathbb{I}}~\nabla\cdot{\bf{v}}\,\right)~~~.
 \label{tensorial}
 \ea
 where the {\it{rate-of-expansion tensor}}
 \ba
 \frac{\textstyle 1}{\textstyle 3}~{\mathbb{I}}~\nabla\cdot{\bf{v}}
 \label{}
 \ea
 is diagonal and has a trace, while the combination
 \ba
 {\mathbb{E}}~-~\frac{1}{3}~{\mathbb{I}}~\nabla\cdot{\bf{v}}\;=\;\frac{1}{2}~\left[\,(\nabla \otimes {\bf{v}})\,+\,(\nabla \otimes {\bf{v}})^{^T}\,\right]~-~\frac{1}{3}~{\mathbb{I}}
 ~\nabla\cdot{\bf{v}}
 \label{}
 \ea
 is symmetric and traceless. These two parts contribute linearly proportional inputs into the stress. The first input
 is proportional, with an empirical coefficient $~3\,\zeta~$, to the rate-of-expansion term, while the second input
 into the stress is proportional, with an empirical coefficient $~2\,\eta~$, to the symmetric traceless combination:
 \ba
 \stackrel{(v)}{\mathbb{S}}\,=\,3\,\zeta\;\frac{\textstyle 1}{\textstyle 3}~{\mathbb{I}}~\nabla\cdot{\bf{v}}~+~2\,\eta\,\left(\,{\mathbb{E}}\;-\;\frac{\textstyle 1}{\textstyle 3}~{\mathbb{I}}~\nabla
 \cdot{\bf{v}}\,\right)~=~\zeta\,\frac{\textstyle\partial ~}{\textstyle\partial t}\,
 \left(\,{\mathbb{I}}~\mbox{Sp}\,{\mathbb{U}}\,\right)~+~
 2\,\eta\,\frac{\textstyle\partial ~}{\textstyle\partial t}\,\left(\,{\mathbb{U}}\;-\;\frac{\textstyle 1}{\textstyle 3}~
 {\mathbb{I}}~\mbox{Sp}\,{\mathbb{U}}\,\right)~~~.~\quad
 \label{lat}
 \ea
 Here we recalled that $\,\nabla\cdot{\bf{v}}\,=\,\frac{\textstyle\partial ~}{\textstyle\partial t}\,\nabla\cdot{\bf{u}}
 \,=\,\frac{\textstyle\partial ~}{\textstyle\partial t}\,u_{\alpha\alpha}\,=\,\frac{\textstyle\partial ~}{\textstyle\partial t}\,\mbox{Sp}\,{\mathbb{U}}\,$. Since $\,\mbox{Sp}\,{\mathbb{U}}\,$ is equal to the volume variation
 $~\frac{\textstyle{dV'-dV}}{\textstyle{dV}}~$ experienced by the material, we see that the first term
 in (\ref{lat}) is volumetric, the second being deviatoric.

 The quantity $\,\eta\,$ is called the {\it{first viscosity}} or the {\it{shear viscosity}}. The quantity $\,\zeta\,$
 is named the {\it{second viscosity}} or the {\it{bulk viscosity}}.

  \subsection{An example of approach to viscoelastic behaviour}\label{viscoel}

 In this subsection, we shall consider one possible approach to description of viscoelastic regimes. As we mentioned in
 subsection \ref{glossary}, the term {\it{viscoelasticity}} covers not only  combinations of elasticity and viscosity,
 but can also include other forms of delayed reaction. So the term {\it{viscoelastic}} is customarily used as a substitution
 for too long a term {\it{viscoelastohereditary}}.

 One possible approach would be to assume that the elastic, hereditary, and viscous stresses simply sum up, and that each of them is
 related the the same strain $\,{\mathbb{U}}\;$:
 an
 \begin{subequations}
 \ba
 \stackrel{(total)}{\mathbb{S}}\,=~
 \stackrel{{{(e)}}}{\mathbb{S}}\,+\,\stackrel{{{{(h)}}}}{\mathbb{S}}\,+\,\stackrel{{{{(v)}}}}{\mathbb{S}}~=~
 \left(\,{\mathbb{B}}~+~\tilde{\mathbb{B}}~+~{\mathbb{A}}\,\frac{\partial~}{\partial t}\,\right)\;{\mathbb{U}}~~~,
 \label{simple_1}
 \ea
 or simply
 \ba
 \stackrel{(total)}{\mathbb{S}}\;=\;\hat{\mathbb{B}}\,{\mathbb{U}}~\quad,\quad\quad\mbox{where}\quad \hat{\mathbb{B}}\,
 \equiv\,{\mathbb{B}}~+~\tilde{\mathbb{B}}~+~{\mathbb{A}}\,\frac{\partial~}{\partial t}~~~,
 \label{simple_2}
 \ea
 \label{simple}
 \end{subequations}
 where the three operators -- the integral operator $\,\tilde{\mathbb{B}}\,$, the differential operator $\,{\mathbb{A}}\,\frac{\textstyle\partial~}{
 \textstyle\partial t}\,$, and the operator of multiplication by matrix $\,{\mathbb{B}}\,$ -- comprise an integro-differential
 operator $\,\hat{\mathbb{B}}\,$.

 In an isotropic medium, each of the three matrices, $\,\tilde{\mathbb{B}}\,$, $\,{\mathbb{A}}\,\frac{\textstyle\partial~}{
 \textstyle\partial t}\,$, and $\,{\mathbb{B}}\,$, includes two terms only. This happens because in such a medium
 each of the three parts of the stress gets decomposed invariantly into its deviatoric and volumetric components:
 The elastic stress becomes:
 \ba
 \stackrel{(e)}{\mathbb{S}}\;=\;\stackrel{(e)}{\mathbb{S}}_{\textstyle{_{volumetric}}}\,+\,\stackrel{(e)}{\mathbb{S}}_{\textstyle{_{deviatoric}}}
 \,=\;3\,K\,
 \left(\frac{\textstyle 1}{\textstyle 3}~{\mathbb{I}}~\mbox{Sp}\,{\mathbb{U}}\,\right)~+~
 2\,\mu\,\left(\,{\mathbb{U}}\;-\;\frac{\textstyle 1}{\textstyle 3}~
 {\mathbb{I}}~\mbox{Sp}\,{\mathbb{U}}\,\right)\quad,\quad
 \label{LL203}
 \ea
 with $\,K\,$ and $\,\mu\,$ being the {\it{bulk elastic modulus}} and the {\it{shear elastic modulus}},
 correspondingly, $\;{\mathbb{I}}\,$ standing for the unity matrix, and $\,\mbox{Sp}\,$ denoting the trace of a matrix:
 $~\mbox{Sp}\,{\mathbb{U}}\,\equiv\,\sum_{i}U_{ii}\;$.

 The hereditary stress becomes:
 \ba
 \stackrel{(h)}{\mathbb{S}}~=~\stackrel{(h)}{\mathbb{S}}_{\textstyle{_{volumetric}}}\,+~\stackrel{(h)}
 {\mathbb{S}}_{\textstyle{_{deviatoric}}}~=\;3\,\tilde{K}~\left(\,\frac{\textstyle{1}}{\textstyle{3}}\,{\mathbb{I}}
 ~\mbox{Sp}\,{\mathbb{U}}\,\right)~+~2\,\tilde{\mu}\,\left(\,{\mathbb{U}}\,-\,\frac{\textstyle{1}}{\textstyle{3}}\,
 {\mathbb{I}}~\mbox{Sp}\,{\mathbb{U}}\,\right)~~~,~~
 \label{LL19}
 \ea
 where $\,\tilde{K}\,$ and $\,\tilde{\mu}\,$ are the {\it{bulk-modulus operator}} and the {\it{shear-modulus}} operator,
 accordingly.

 The viscous stress acquires the form:
 \ba
 \stackrel{(v)}{\mathbb{S}}\;=\;\stackrel{(v)}{\mathbb{S}}_{\textstyle{_{volumetric}}}\,+\,\stackrel{(v)}{\mathbb{S}}_{\textstyle{
 _{deviatoric}}}\,=\;3~\zeta\,\frac{\textstyle\partial ~}{\textstyle\partial t}\,
 \left(\frac{\textstyle 1}{\textstyle 3}~{\mathbb{I}}~\mbox{Sp}\,{\mathbb{U}}\,\right)~+~
 2\,\eta\,\frac{\textstyle\partial ~}{\textstyle\partial t}\,\left(\,{\mathbb{U}}\;-\;\frac{\textstyle 1}{\textstyle 3}~
 {\mathbb{I}}~\mbox{Sp}\,{\mathbb{U}}\,\right)\quad,\quad
 \label{LL20}
 \ea
 the quantities $\zeta$ and $\eta$ being termed as the {\it{bulk viscosity}} and the {\it{shear viscosity}},
 correspondingly

 The term $~\frac{\textstyle 1}{\textstyle 3}~{\mathbb{I}}~\mbox{Sp}\,{\mathbb{U}}~$ is called the {\it{volumetric}}
 part of the strain, while $~{\mathbb{U}}\,-\,\frac{\textstyle 1}{\textstyle 3}~{\mathbb{I}}~\mbox{Sp}\,{\mathbb{U}}~$
 is called the {\it{deviatoric}} part. Accordingly, in expressions (\ref{LL203} - \ref{LL20}) for the stresses, the pure-trace
 terms are called {\it{volumetric}}, the other term being named {\it{deviatoric}}.

 The total stress, too, can now be split into the total volumetric and the total deviatoric parts:
  \begin{subequations}
 \ba
 \nonumber
 \stackrel{(total)}{\mathbb{S}}&=&
 \overbrace{\left(\,\stackrel{(e)}{\mathbb{S}}_{\textstyle{_{volumetric}}}\,+\,\stackrel{(e)}{\mathbb{S}}_{\textstyle{_{deviatoric}}}\,\right)}^{\stackrel{\textstyle\stackrel{{{(e)}}}{\mathbb{S}}}{}}
 \,+\,
 \overbrace{\left(\,\stackrel{(v)}{\mathbb{S}}_{\textstyle{_{volumetric}}}\,+\,\stackrel{(v)}{\mathbb{S}}_{\textstyle{_{deviatoric}}}\,\right)}^{\stackrel{\textstyle\stackrel{{{(v)}}}{\mathbb{S}}}{}}
 \,+\,
 \overbrace{\left(\,\stackrel{(h)}{\mathbb{S}}_{\textstyle{_{volumetric}}}\,+\,\stackrel{(h)}{\mathbb{S}}_{\textstyle{_{deviatoric}}}\,\right)}^{\stackrel{\textstyle\stackrel{{{(h)}}}{\mathbb{S}}}{}}\\
 \nonumber\\
 \nonumber\\
 \nonumber
 &=&\overbrace{\left(\,\stackrel{(e)}{\mathbb{S}}_{\textstyle{_{volumetric}}}\,+\,\stackrel{(v)}{\mathbb{S}}_{\textstyle{_{volumetric}}}\,+\,\stackrel{(h)}{\mathbb{S}}_{\textstyle{_{volumetric}}}\,\right)}^{\stackrel{\textstyle{{{\mathbb{S}}_{\textstyle{_{volumetric}}}}}}{}}
 \,+\,
 \overbrace{\left(\,\stackrel{(e)}{\mathbb{S}}_{\textstyle{_{deviatoric}}}\,+\,\stackrel{(v)}{\mathbb{S}}_{\textstyle{_{deviatoric}}}\,+\,\stackrel{(h)}{\mathbb{S}}_{\textstyle{_{deviatoric}}}\,\right)}^{\stackrel{\textstyle{\mathbb{S}}_{\textstyle{_{deviatoric}}}}{}}\\
 \nonumber\\
 \nonumber\\
 &=&\left(3\,K\,+\,3\,\tilde{K}\,+\,3\,\zeta\,\frac{\partial\,}{\partial t}\right)\left(\frac{\textstyle{1}}{\textstyle{3}}\,
 {\mathbb{I}}~\mbox{Sp}\,{\mathbb{U}}\right)\,+\,\left(2\,\mu\,+\,2\,\tilde{\mu}\,+\,2\,\eta\,\frac{\partial\,}{\partial t}\right)
 \left(\,{\mathbb{U}}\,-\,\frac{\textstyle{1}}{\textstyle{3}}\,{\mathbb{I}}\;\mbox{Sp}\,{\mathbb{U}}\,\right)~\quad\quad\quad\quad
 \quad\quad\quad
 \label{simplissimo_1}\\
 \nonumber\\
 \nonumber\\
 &=&3~\hat{K}~\left(\,\frac{\textstyle{1}}{\textstyle{3}}\,{\mathbb{I}}~\mbox{Sp}\,{\mathbb{U}}\,\right)~+~2~\hat{\mu}~
 \left(\,{\mathbb{U}}\,-\,\frac{\textstyle{1}}{\textstyle{3}}\,{\mathbb{I}}\;\mbox{Sp}\,{\mathbb{U}}\,\right)~~~,
 \label{simplissimo_2}
 \ea
 \end{subequations}
 where
 \ba
 \hat{K}\;\equiv\;K\;+\;\tilde{K}\;+\;\zeta\;\frac{\partial~}{\partial t}~\quad\quad\mbox{and}\quad\quad
 \hat{\mu}\;\equiv\;\mu\;+\;\tilde{\mu}\;+\;\eta\;\frac{\partial~}{\partial t}~~~.\quad\quad\quad
 \label{}
 \ea
 As expected, a total linear deformation of an isotropic material can be described with two
 integro-differential operators, one acting on the volumetric strain, another on the deviatoric strain.

 If an isotropic medium is also incompressible, the relative change of the volume vanishes: $\,\mbox{Sp}\,{\mathbb{U}}=0\,$. Accordingly, the volumetric part of the strain becomes nil, and so do the
 volumetric parts of the elastic, hereditary, and viscous stresses. For such media, we end up with a simple relation which includes only deviators:
 \ba
 \stackrel{(total)}{\mathbb{S}}\,=\,{\mathbb{S}}_{\textstyle{_{deviatoric}}}\,=\,\stackrel{(e)}{\mathbb{S}}_{\textstyle{_{deviatoric}}}\,+\,
 \stackrel{(h)}{\mathbb{S}}_{\textstyle{_{deviatoric}}}\,+\,\stackrel{(v)}{\mathbb{S}}_{\textstyle{_{deviatoric}}}\,=\,
 2\,{\mu}~{\mathbb{U}}\,+\,2\,\tilde{\mu}~{\mathbb{U}}\,+\,2\,\eta\,\frac{\partial\,}{\partial t}~{\mathbb{U}}\quad\quad
 \label{LL21a}
 \ea
 or simply:
 \ba
 \stackrel{(total)}{\mathbb{S}}\,=\,{\mathbb{S}}_{\textstyle{_{deviatoric}}}\,=\,2\,\hat{\mu}~{\mathbb{U}}~~~,
 \label{}
 \ea
  where $\,{\mathbb{U}}\,$ contains only a deviatoric part, while
 \ba
 \hat{\mu}\,\equiv\,\mu\,+\,\tilde{\mu}\,+\,\eta\,\frac{\partial\,}{\partial t}
 \label{LLC}
 \ea
 is the total, integro-differential operator, which is mapping the preceding history and the present rate of change of the
 strain to the present value of the stress.

 It should be reiterated that the above approach is based on the assertion that the elastic, viscous, and hereditary
 stresses sum up, and that all three are related to the same total strain. A simple example of this approach,
 called the Kelvin-Voigt model, is rendered below in subsection \ref{Kelvin-Voigt}.

 A different approach would be to assume that the strain consists of three distinct parts -- elastic, hereditary, and viscous -- and that
 these components are related to the same overall stress. A simple example of this treatment, termed the Maxwell model, is set out in
 subsection \ref{Maxwe}. A more complex example of this approach is furnished by the Andrade model presented in subsection
 \ref{Andrade_section}. Another way of combining elasticity and viscosity (with no hereditary reaction involved) is
 implemented by the Hohenemser-Prager (SAS) model explained in subsection \ref{HOH} below.

 \subsection{Examples of viscoelastic behaviour with no hereditary reaction}\label{examples}

 \subsubsection{Elastic deformation}

 The truly simplest example of deformation is elastic:
 \ba
 \stackrel{(e)}{\mathbb{S}}\,=\,2\,\mu\,{\mathbb{U}}~\quad,\quad\quad{\mathbb{U}}~=
 ~J\,\stackrel{(e)}{\mathbb{S}}~~~,
 \label{this}
 \ea
 where $\,\mu\,$ and $\,J\,$ are the unrelaxed rigidity and compliance:
 \ba
 \mu\,=\,\mu(0)~~~,\quad~J\,=\,J(0)~~~,\quad~\mu\;=\;{1}/{J}~~~.
 \label{}
 \ea
 In the frequency domain, this relation assumes the same form as it would in the time domain:
 \ba
 \bar{\sigma}_{\gamma\nu}(\chi)\,=\,2\,\mu\,\bar{u}_{\gamma\nu}(\chi) ~\quad,\quad\quad 2\,\bar{u}_{\gamma\nu}(\chi)
 \,=\,J\,\bar{\sigma}_{\gamma\nu}(\chi)~~~.
 \label{}
 \ea

 \subsubsection{Viscous deformation}

 The next example is that of a purely viscous behaviour:
 \ba
 \stackrel{(v)}{\mathbb{S}}\,=\,2\,\eta\,\frac{\partial\,}{\partial t}~{\mathbb{U}}~~~.
 \label{LL39}
 \ea
 It is straightforward from (\ref{LL39}) and (\ref{phys}) that, in this regime, the Fourier components of the
 stress\footnote{~Although we no longer spell it out, the word {\it{stress}} everywhere means: {\it{deviatoric stress}},
 as we agreed to consider the medium incompressible.} are connected to those of the strain through
 \ba
 \bar{\sigma}_{\gamma\nu}(\chi)\,=\,2\,\bar{\mu}(\chi)\;\bar{u}_{\gamma\nu}(\chi) ~\quad,\quad
 \quad 2\,\bar{u}_{\gamma\nu}(\chi)\,=\,\bar{J}(\chi)\;\bar{\sigma}_{\gamma\nu}(\chi)~~~,
 \label{LL40}
 \ea
 where the complex rigidity and the complex compliance are given by
 \ba
 \bar{\mu}\,=\,\inc\,\eta\,\chi\quad,\quad\quad \bar{J}\,=\,-\,\frac{\inc}{\eta\,\chi}~~~.
 \label{}
 \ea

 \subsubsection{Viscoelastic deformation: a Kelvin-Voigt material}\label{Kelvin-Voigt}

 The Kelvin-Voigt model, also called the Voigt model, can be represented with a purely viscous damper and a
 purely elastic spring connected in parallel. Subject to the same elongation, these elements have their forces
 summed up. This illustrates the situation where the total, viscoelastic stress consists of a purely viscous
 and a purely elastic inputs called into being by the same strain:
 \ba
 {\mathbb{S}}\;=\;\stackrel{(ve)}{\mathbb{S}}\;=\;\stackrel{(v)}{\mathbb{S}}\,+\,\stackrel{(e)}{\mathbb{S}}
 ~,\,\quad\mbox{while}\,\quad{\mathbb{U}}\,=\;\stackrel{(v)}{\mathbb{U}}\,=\;\stackrel{(e)}{\mathbb{U}}~~~.
 \label{}
 \ea
 Then the total stress reads:
 \begin{subequations}
 \ba
 {\mathbb{S}}\,=\,\left(2\,{\mu}\,+\,2\,\eta\,\frac{\partial\,}{\partial t}\right)~{\mathbb{U}}~~~,
 \label{LL41_1}
 \ea
 which is often presented in the form of
 \ba
 {\mathbb{S}}\,=\,2\,\mu\,\left(\,{\mathbb{U}}\,+~\tau_{_V}\stackrel{\centerdot}{\mathbb{U}}\,\right)~~~,
 \label{LL41_2}
 \ea
 \label{LL41}
 \end{subequations}
 with the so-called {\emph{Voigt time}} defined as
 \ba
 \tau_{_V}\;\equiv\;{\eta}/{\mu}~~~.
 \label{Voigt}
 \ea
  Comparing (\ref{LL41}) with (\ref{permitted_1}), we understand that the kernel of the
  rigidity operator for the Kelvin-Voigt model can be written down as
  \ba
  \mu(t\,-\,t\,')\;=\;\mu\;+\;\eta\;\delta(t\,-\,t\,')~~~.
  \label{scra}
  \ea

 Suppose the strain is varying in time as
 \ba
 {u}_{\gamma\nu}(t)\;=\;\frac{\sigma_0}{2\,\mu}\;\left[\,1\;-\;\exp\left(\,-\,\frac{t\,-\,t_0}{\tau_{{_V}}}\,\right)\,\right]
 ~\Theta(t\,-\,t_0)~~~,
 \label{strain}
 \ea
 so that
 \ba
 \stackrel{\bf\centerdot}{u}_{\gamma\nu}(t)\;=\;\frac{\sigma_0}{2\,\mu\,}\;\frac{1}{\tau_{{_V}}}\;\exp\left(\,-\,\frac{t\,-\,t_0}{\tau_{
 {_V}}}\,\right)\,
 ~\Theta(t\,-\,t_0)~~~.
 \label{strainn}
 \ea

 Then insertion
 of (\ref{scra}) and (\ref{strainn}) into (\ref{permitted_2}) or, equivalently, insertion
 of (\ref{strain})
 into (\ref{LL41_1}) demonstrates that this strain results from a stress\footnote{~For example, plugging of (\ref{scra}) and
 (\ref{strainn}) into (\ref{permitted_2}) leads to:
 \ba
 \nonumber{}
 \sigma(t)&=&\int^{t\,'=\,t}_{t\,'=\,-\,\infty}\left[\,\mu\;+\;\eta\;\delta(t\,-\,t\,')\,\right]\;\frac{\sigma_0}{\mu}\;
 \frac{\textstyle 1}{\textstyle\tau_{{_V}}}\;\exp\left(\,-\,\frac{t\,'\,-\,t_0}{\tau_{{_V}}}\,\right)\,~\Theta(t\,'\,-\,t_0)\,dt\,'\\
 \nonumber\\
 \nonumber\\
 \nonumber
  &=&\left\{
  \begin{array}{c}
 \textstyle\sigma_0~\left[\;\int^{t\,'=\,t}_{t\,'=\,t_0}\exp
 \left(\,-\,\frac{\textstyle t\,'\,-\,t_0}{\textstyle \tau_{{_V}}}\,\right)\,
 \frac{\textstyle dt\,'}{\textstyle\tau_{{_V}}}\;+\;\exp
 \left(\,-\,\frac{\textstyle t\,-\,t_0}{\textstyle \tau_{{_V}}}\,\right)\right]
 ~~\quad\mbox{for}\quad t\,\geq\,t_0~~~~\\
 ~\\
 0\quad \quad\quad\quad\quad\quad\quad\quad \quad\quad\quad\quad\quad\quad\quad \quad\quad\quad\quad\quad\quad
 \quad \quad\quad\quad
 ~\,\mbox{for}\quad t\,<\,t_0 ~~~,
  \end{array}
  \right.
  \ea
  which is simply $\,\;\sigma_0\,\Theta(t\,-\,t_0)\;$.
 }
 \ba
 \sigma_{\gamma\nu}(t)\,=\,{\sigma_0}\;\Theta(t\,-\,t_0) ~~~.
 \label{stress}
 \ea
 It would however be a mistake to deduce from this that the compliance function is equal to $~\mu^{-1}\,\left[\,1\;-\;\exp
 \left(\,-\,\frac{\textstyle t\,-\,t\,'}{\textstyle\tau_{{_V}}}\,\right)\,\right]~$, even though such a misstatement
 is sometimes made in the literature. This expression furnishes the compliance function only in the special
 situation of a Heaviside-step stress (\ref{stress}), but not in the general case.

 As can be easily shown from (\ref{phys}), in the frequency domain model (\ref{LL41}) reads as (\ref{LL40}), except that
 the complex rigidity and the complex compliance are now given by
 \ba
 \bar{\mu}\,=\,{\mu}\,\left(\,1\,+\,\inc\,\chi\,\tau_{_V}\,\right)\,\quad,\quad\quad\bar{J}\,=\,
 \frac{J}{1\,+\,\inc\,\chi\,\tau_{_V}}~~~.
 \label{LL42}
 \ea
 Recall that, for brevity, here and everywhere we write $\,\mu\,$ and $\,J\,$ instead of $\,{\mu}(0)\,$ and $\,J(0)\,$.

 The Kelvin-Voigt material becomes elastic in the low-frequency limit, and viscous in the high-frequency limit.

 \subsubsection{Viscoelastic deformation: a Maxwell material}\label{Maxwe}\label{jjv}

 The Maxwell model can be represented with a viscous damper and an elastic spring connected in series. Experiencing the
 same force, these elements have their elongations summed up. This example illustrates the situation where the total, viscoelastic
 strain consists of a purely viscous and a purely elastic contributions generated by the same stress $\,{\mathbb{S}}\,$:
 \ba
 {\mathbb{U}}\,=\;\stackrel{(v)}{\mathbb{U}}\,+\,\stackrel{(e)}{\mathbb{U}}
  ~,\,\quad\mbox{where}\,\quad
 \stackrel{(e)}{\mathbb{S}}\,=\,2\,\mu\,\stackrel{(e)}{\mathbb{U}}~~~~\mbox{and}\quad~
 \stackrel{(v)}{\mathbb{S}}\,=\,2\,\eta\,\frac{\partial\,}{\partial t}~\stackrel{(v)}{\mathbb{U}}
  ~~~.
 \label{dddt}
 \ea
 Since in the Maxwell regime both contributions to the strain are generated by the same stress
 \ba
 {\mathbb{S}}\;=\;\stackrel{(ve)}{\mathbb{S}}\;=\;\stackrel{(v)}{\mathbb{S}}\,=\,\stackrel{(e)}{\mathbb{S}}~~~,
 \label{}
 \ea
 formula (\ref{dddt}) can be written down as
 \begin{subequations}
 \ba
 \stackrel{\centerdot}{\mathbb{U}}\,=\,\frac{1}{2\,\mu}\;\stackrel{\centerdot}{\mathbb{S}}\,+\,\frac{1}{2\,\eta}\;{\mathbb{S}}~~~
 \label{}
 \ea
 or, in a more conventional form:
 \ba
 \stackrel{\centerdot}{\mathbb{S}}\,+\;\frac{1\;}{\tau_{_M}}\,{\mathbb{S}}
 ~=~2\,\mu\,\stackrel{\centerdot}{\mathbb{U}}~~~,
 \label{}
 \ea
 \label{these}
 \end{subequations}
 with the so-called {\emph{Maxwell time}} introduced as
 \ba
 \tau_{_M}\;\equiv\;{\eta}/{\mu}~~~.
 \label{Maxwell}
 \ea
 Although formally the Maxwell time is given by an expression mimicking the definition of the Voigt time, the
 meaning of these times is different.

 Comparing (\ref{these}) with the general expression (\ref{I12_4}) for the compliance operator, we see that, for the
 Maxwell model, the compliance operator in the time domain assumes the form:
 \ba
 J(t\,-\,t\,')\,=\,\left[\,J\,+\,\left(t\;-\;t\,'\right)\;\frac{1}{\eta}\,\right]\;\Theta(t\,-\,t\,')~~~,
 \label{Max}
 \ea
 where $\,J\,\equiv\,1/\mu\,$. In the frequency domain, (\ref{these}) can be written down as (\ref{LL40}),
 with the complex rigidity and compliance given by
 \ba
 \bar{\mu}(\chi)\,=\,{\mu}\;\frac{\inc\,\chi\,\tau_M}{1\,+\,\inc\chi\tau_M}\,\quad,\quad\quad\bar{J}(\chi)\,=\,J\,\left(\,1\,-\,
 \frac{\inc}{\chi\,\tau_{_M}}\,\right)~=~J~-~\frac{\inc}{\chi\,\eta}~~~.
 \label{LL42}
 \ea
 Clearly, such a body becomes elastic in the high-frequency limit, and becomes viscous at low frequencies (the latter
 circumstance making the Maxwell model attractive to seismologists).

 \subsubsection{Viscoelastic deformation: the Hohenemser-Prager (SAS) model}\label{HOH}

 An attempt to combine the Kelvin-Voigt and Maxwell models leads to the Hohenemser-Prager model,
 also known as the Standard Anelastic Solid (SAS):
 \ba
 {\tau_{_M}}
 {\bf\dot{\mathbb{S}}}\,+\;\,{\mathbb{S}}
 ~=~2\,\mu\,\left(\,{\mathbb{U}}\,+\,\tau_{_V}\,
 {\bf\dot{\mathbb{U}}}  \,\right)~~~,
 \label{SAS}
 \ea
 In the limit of $\,\tau_{_M}\rightarrow 0\,$, this model approaches the one of Kelvin and Voigt
 (and $\,\tau_{_V}\,$ acquires the meaning of the Voigt time).

 A transition from the SAS to Maxwell model, however, can be achieved only through re-definition of
 parameters. One should set: $\;2\,\mu\rightarrow 0\;$ and $\;\tau_{_V}\rightarrow\infty\;$, along with
 $\,2\mu\tau_{_V}\rightarrow 2\eta\,$. Then (\ref{SAS}) will become (\ref{these}), with $\,\tau_{_M}\,$
 playing the role of the Maxwell time.

 In the frequency domain, (\ref{SAS}) can be put into the form of (\ref{LL40}), the complex rigidity and the complex
 compliance being expressed through the parameters as
 \ba
 \bar{\mu}\,=\,\mu\;\frac{\textstyle 1+i\tau_{_V}\chi}{\textstyle{1+i\tau_{_M}\chi}}\,\quad,\quad\quad
 \bar{J}\,=\,J\;\frac{\textstyle 1+i\tau_{_M}\chi}{\textstyle{1+i\tau_{_V}\chi}}~~~.
 \label{}
 \ea
 This entails: $~\tan\delta\,\equiv\,{\textstyle{{\cal{I}}{\it{m}}[
 \bar{\mu}]}}/{\textstyle{{\cal{R}}{\it{e}}[\bar{\mu}]}}\,=\,\frac{\textstyle{(\tau_{_V}-\tau_{_M})
 \,\chi}}{\textstyle{1\,+\,\tau_{_V}\tau_{_M}\chi^2}}\;$, whence it is easy to show that the tangent
 is related to its maximal value through
 \begin{equation}
 \nonumber
 \tan\delta\;=\;2\;\left[\tan\delta\right]_{max}\;\frac{\tau\;\chi}{1\,+\,\tau^{2}\,\chi^{2}}\,\quad,\quad\quad
 \mbox{where}\quad\tau\,\equiv\,\sqrt{\tau_{_M}\tau_{_V}}~~~.
 \end{equation}
 This is the so-called Debye peak, which is indeed observed in some materials.

 To prove that the SAS solid is indeed anelastic, one has to make sure
 that a Heaviside step stress $\,\Theta(t')\,$ entails a strain proportional to $\,1-\exp(-\Gamma t)\,$, and to
 demonstrate that a predeformed sample subject to stress $\,\Theta(-t')\,$ regains it shape as $\,\exp(-\Gamma t)\,$,
 where the relaxation constant $\,\Gamma\,$ is positive. In subsection \ref{chu} we shall do this for a SAS sphere.

 \section{Interconnection between the quality factor and the phase lag}\label{interconnection}

The power $\,P\,$ exerted by a tide-raising secondary on its primary can be written as
 \ba
 P\;=\;-\;\int \,\rho\;\Vbold\;\cdot\;\nabla W\;d^3x
 \label{A9}
 \ea
 $\rho\,,\;\Vbold\,$, and $\,W\,$ signifying the density, velocity, and tidal potential in the small volume $~d^3x~$ of the primary.
 The mass-conservation law $~\nabla\cdot(\rho\Vbold)\,+\frac{\textstyle \partial \rho}{\textstyle\partial t}\,=\,0\,~$ enables one to
 shape the dot-product into the form of
 \ba
 \rho\,\Vbold\cdot\nabla W\,=\,
 \nabla\cdot(\rho\,
 \Vbold\,W)\,-\,\rho\,W\,\nabla\cdot\Vbold\,-\,\Vbold\,W\,\nabla\rho\;\;~.\;\;\;\;
 \label{}
 \ea
 Under the realistic assumption of the primary's incompressibility, the term with $\,\nabla\cdot\Vbold\,$ may be omitted. To get rid
 of the term with $\,\nabla \rho\,$, one has to accept a much stronger approximation of the primary being homogeneous.
 Then the power will be rendered by
 \ba
 P\;=\;-\;\int\,\nabla\,\cdot\,(\rho\;\Vbold\;W)\,d^3x
 \;=\;-\;\int\,\rho\;W\;\Vbold\,\cdot\,{\vec{\bf{n}}}\;\,dS\;\;\;,
 \label{A9}
 \ea
 ${\vec{\bf{n}}}\,$ being the outward normal and $\,dS\,$ being an element of the surface area of the primary. This expression for the
 power (pioneered, probably, by Goldreich 1963) enables one to calculate the work through radial displacements only, in neglect of
 horizontal motion.

 Denoting the radial elevation with $\,\zeta\,$, we can write the power per unit mass, $~{\cal P}\equiv P/M~$, as:
 \ba
 {\cal P}\;=\;\left(-\,\frac{\partial W}{\partial r}\right)\;\Vbold\cdot{\vec{\bf{n}}}
 \;=\;\left(-\,\frac{\partial W}{\partial r}\right)\frac{d\zeta}{dt}\;\;\;.
 \label{}
 \ea

 A harmonic external potential
 \ba
 W~=~W_0~\cos(\,\omega_{\textstyle{_{lmpq}}} \,t\,)~~~,
 \label{AAAAA3}
 \ea
 applied at a point of the primary's surface, will elevate this point by
 \ba
 \zeta~=~h_2~\frac{W_o}{\mbox{g}}~\cos(\omega_{\textstyle{_{lmpq}}} \,t~-~\epsilon_{\textstyle{_{lmpq}}} )~=~h_2\;\frac{W_o}{\mbox{g}}~
 \cos (\omega_{\textstyle{_{lmpq}}} \,t\;-\;\omega_{\textstyle{_{lmpq}}}\,\Delta t_{\textstyle{_{lmpq}}} )~~~,
 \label{AAAAA4}
 \ea
 with $\,\mbox{g}\,$ being the surface gravity acceleration, and $\,h_2\,$ denoting the Love number.

 In formula (\ref{AAAAA4}), $\,\omega_{\textstyle{_{lmpq}}}\,$ is one of the modes (\ref{9}) showing up in the Darwin-Kaula expansion
 (\ref{5}). The quantity $\,\epsilon_{\textstyle{_{lmpq}}}\,=\,\omega_{\textstyle{_{lmpq}}}\,\Delta t_{\textstyle{_{lmpq}}}\,$ is
 the corresponding phase lag, while $\,\Delta t_{\textstyle{_{lmpq}}}\,$ is the positively defined time lag at this mode. Although the
 tidal modes $\,\omega_{\textstyle{_{lmpq}}}\,$ can assume any sign, both the potential $\,W\,$ and elevation $\,\zeta\,$ can be
 expressed via the positively defined forcing frequency $~\chi_{\textstyle{_{lmpq}}}\,=\,|\,\omega_{\textstyle{_{lmpq}}}\,|~$ and the
 absolute value of the phase lag:
 \ba
 W~=~W_0~\cos(\chi\,t)~~~,
 \label{AAA3}
 \ea
 \ba
 \zeta~=~h_2~\frac{W_o}{\mbox{g}}~\cos(\chi\,t~-~|\,\epsilon\,|)~~~,
 \label{AAA4}
 \ea
 subscripts $\,lmpq\,$ being dropped here and hereafter for brevity.

 The vertical velocity of the considered element of the primary's surface will be
 \ba
 \frac{d\zeta}{dt}\;=\;-\;h_2\;\chi\;\frac{W_o}{\mbox{g}}\;\sin (\chi t
 \;-\;|\epsilon|)\;=\;-\;h_2\;\chi\;\frac{W_o}{\mbox{g}}\;\left(\sin \chi
 t\;\cos |\epsilon|\;-\;\cos \chi t\; \sin |\epsilon|\right)\;\;.\;\;
 \label{A5}
 \label{472}
 \ea
 Introducing the notation $~A\,=\,h_2\,\frac{\textstyle W_0}{\textstyle\mbox{g}}\,\frac{\textstyle\partial W_0}{\textstyle\partial r}~$,
 we write the power per unit mass as
 \ba
 {\cal P}\;=\;\left(-\,\frac{\partial W}{\partial r}\right)\frac{d\zeta}{dt}\;=\;A~\chi~\cos(\chi\,t)~\sin (\chi t
 \;-\;|\epsilon|)\;\;\;,
 \label{}
 \ea
 and write the work $\,w\,$ per unit mass, performed over a time interval $\,(t_0\,,\;t)\,$, as:
 \ba
 w|^{
 \textstyle{^{~t}}}_{\textstyle{_{~t_0}}}=
 \int_{t_0}^t{\cal P}~dt=A\int^{\chi t}_{\chi t_0}\cos(\chi\,t)\,\sin (\chi t
 -|\epsilon|)d(\chi\,t)
 \nonumber
 =A\,\cos|\epsilon|\int_{\chi t_0}^{\chi t}\cos z\,\sin z\,dz-A\,\sin|\epsilon|\int_{\chi t_0}^{\chi t}\cos^2 z\,dz
 \ea
 \ba
 =~-~\frac{A}{4}~{\LARGE\left[ \right.}\,\cos(2\chi t-|\epsilon|)\,+\,2\;\chi\;t\;\sin|\epsilon|~{\LARGE\left.\right]}^{
 \textstyle{^{~t}}}_{\textstyle{_{~t_0}}}~~~.
 \label{dissipation}
 \ea
 Being cyclic, the first term in (\ref{dissipation}) renders the elastic energy stored in the body. The second term,
 being linear in time, furnishes the energy damped. This clear interpretation of the two terms was offered by Stan Peale
 [2011, personal communication].

 The work over a time period $\,T\,=\,2\pi/\chi\,$ is equal to the energy dissipated over the period:
 \ba
  w|^{\textstyle{^{~t=T}}}_{\textstyle{_{~t=0}}}=\Delta E_{\textstyle{_{cycle}}}\;=\;-\;A\;\pi\;\sin|\epsilon|~~~.
 \label{A6AA}
 \ea

 It can be shown that the peak {\it{work}} is obtained over the time span from $\,\pi\,$ to $\,|\epsilon|\,$ and assumes the value
 \ba
 E_{\textstyle{_{peak}}}^{\textstyle{^{(work)}}}\,=\;~\frac{A}{2}~\left[\,\cos|\epsilon|~-~\sin|\epsilon|~\left(\,\frac{\pi}{2}\,-\,
 |\epsilon|\,\right)\,\right]~~~,
 \label{}
 \ea
 whence the appropriate quality factor is given by:
 \ba
 Q^{-1}_{\textstyle{_{work}}}\,=~\frac{~-~\Delta E_{\textstyle{_{cycle}}}}{2\,\pi\,E_{\textstyle{_{peak}}}^{\textstyle{^{(work)}}}}~=~
 \frac{\tan|\epsilon|}{1\;-\;\left(\,\frac{\textstyle\pi}{\textstyle 2}\,-\,|\epsilon|\,\right)~\tan|\epsilon|}~~~.
 \label{}
 \ea

 To calculate the peak {\it{energy}} stored in the body, we would note that the first term in (\ref{dissipation}) is
 maximal when taken over the span from $\,\chi\,t\,=\,\pi/4\,+\,|\epsilon|/2\,$ through $\,\chi\,t\,=\,3\pi/4\,+\,|\epsilon|/2~$:
 \ba
 E_{\textstyle{_{peak}}}^{\textstyle{^{(energy)}}}\,=\;~\frac{A}{2}~~~,
 \label{}
 \ea
 and the corresponding quality factor is:
 \ba
 Q^{-1}_{\textstyle{_{energy}}}\,=~\frac{~-~\Delta E_{\textstyle{_{cycle}}}}{2\,\pi\,E_{\textstyle{_{peak}}}^{\textstyle{^{(energy)}}}}~=~
 \sin|\epsilon|~~~.
 \label{sin}
 \ea

 Goldreich (1963) suggested to employ the span $\,\chi\,t\,=\,(0\,,\;\pi/4)\,$. The absolute value of the resulting power, denoted
 in {\it{Ibid.}} as $\,E^*\,$, is equal to
 \ba
 E^*\,=\;~\frac{A}{2}~\cos|\epsilon|~~~
 \label{}
 \ea
 and is {\it{not}} the peak value of the energy stored nor of the work performed. Goldreich (1963) however employed it to define a
 quality factor, which we shall term $~Q_{\textstyle{_{Goldreich}}}~$. This factor is introduced via
 \ba
 Q^{-1}_{\textstyle{_{Goldreich}}}\,=~\frac{~-~\Delta E_{\textstyle{_{cycle}}}}{2\,\pi\,E^*}~=~\tan|\epsilon|~~~.
 \label{}
 \ea

 In our opinion, the quality factor $\,Q_{\textstyle{_{energy}}}\,$ defined through (\ref{sin}) is preferable, because the expansion of
 tides contains terms proportional to $\,k_l(\chi_{\textstyle{_{lmpq}}})~\sin\epsilon_l(\chi_{\textstyle{_{lmpq}}})~\,$. Since the
 long-established tradition suggests to substitute $\,\sin\epsilon\,$ with $\,1/Q\,$, it is advisable to define the $\,Q\,$ exactly as
 (\ref{sin}), and also to call it $\,Q_l\,$, to distinguish it from the seismic quality factor (Efroimsky 2012).

  \section{Tidal response of a homogeneous viscoelastic sphere (Churkin 1998)}\label{chur}

 This section presents some results from the unpublished preprint by Churkin (1998). We took the liberty of upgrading the
 notations\footnote{~Churkin (1998) employed the notation $\,k_{\it l}(\tau)\,$ for what we call $\,
 \stackrel{\bf\centerdot}{k}_{\textstyle{_l}}(\tau)\,$. Our notations are more convenient in that they amplify the close
 analogy between the Love functions and the compliance function.} and correcting some minor oversights.

 \subsection{A homogeneous Kelvin-Voigt spherical body}

 In combination with the Correspondence Principle, the formulae from subsection \ref{Kelvin-Voigt} furnish the following
 expression for the complex Love numbers of a Kelvin-Voigt body:
 \ba
 \bar{k}_{\textstyle{_l}}(\chi)\;=\;\frac{3}{2(l-1)}\;\,\frac{1}{\textstyle 1\;+\;A_{\textstyle{_l}}\,
 \left(\,1\;+\;\;\tau_{\textstyle{_V}}\;\inc\;\chi\,\right)\;}~~~,\quad
 \label{}
 \ea
 It then can be demonstrated, with aid of (\ref{L31}), that the time-derivative of the corresponding Love function is
 \ba
 \stackrel{\bf\centerdot}{k}_{\textstyle{_l}}(\tau)~=~
 \left\{
 \begin{array}{c}
 \frac{\textstyle 3}{\textstyle 2(l-1)}~\,\frac{\textstyle 1}{\textstyle A_{\textstyle{_{\textstyle{_l}}}}\;
 \tau_{\textstyle{_V}}}\;\exp(\textstyle \;-\,\tau\,\zeta_{\textstyle{_{\textstyle{_l}}}}\,)\;\Theta(\tau)\quad
 ~~\quad\mbox{for}\quad\tau_{\textstyle{_V}}\,>\,0~~~~\\
 ~\\
 \frac{\textstyle 3}{\textstyle 2(l-1)}~\,\frac{\textstyle 1}{\textstyle 1\;+\;A_{\textstyle{_{\textstyle{_l}}}}}
 \;\,\delta(\tau)\quad \quad\quad\quad\quad\quad\quad
 ~\,\mbox{for}\quad\tau_{\textstyle{_V}}\,=\,0 ~~~,
 \end{array}
 \right.
 \label{KV1}
 \ea
 while the Love function itself has the form of
 \ba
 \nonumber
 {k}_{\textstyle{_l}}(\tau)&=&\frac{3}{2(l-1)}~\,\frac{1}{A_{\textstyle{_{\textstyle{_l}}}}\;\zeta_{\textstyle{_{\textstyle{_l}
 }}}\;\tau_{\textstyle{_V}}}\;\left[\;1\;-\;\exp(\textstyle\;-\,\tau\,\zeta_{\textstyle{_{\textstyle{_l}}}}\,)\;\right]\;\Theta(\tau)\\
 \label{KV2}\\
 \nonumber
 &=&\frac{3}{2(l-1)}~\,\frac{1}{1\;+\;A_{\textstyle{_{\textstyle{_l}}}}}\;\left[\;1\;-\;\exp(\textstyle
 \;-\,\tau\,\zeta_{\textstyle{_{\textstyle{_l}}}}\,)\;\right]\;\Theta(\tau)\;\;\;,\quad\;
  \ea
 where
 \ba
 \zeta_{\textstyle{_{\textstyle{_l}}}}\;\equiv\;\frac{1\;+\;A_{\textstyle{_{\textstyle{_l}}}}}{A_{\textstyle{_{\textstyle{_l}}}}
 \;\tau_{\textstyle{_V}}}~~.
 \label{}
 \ea
 Formulae (\ref{KV1}) may look confusing, in that $\,\exp(\textstyle \;-\,\tau\,\zeta_{\textstyle{_{\textstyle{_l}}}}\,)\,$
 simply vanishes in the elastic limit, i.e., when $~\tau_{\textstyle{_V}}\rightarrow 0~$ and $~\zeta_{\textstyle{_l}}
 \rightarrow\infty~$. We however should not be misled by this mathematical artefact stemming from the nonanaliticity of
 the exponent function. Instead, we should keep in mind that a physical meaning is attributed not to the Love functions
 or their derivatives but to the results of the Love operator's action on realistic disturbances. For example, a Heaviside
 step potential
 \ba
 W_{\it{l}}(\eRbold\,,\;\erbold^{\;*}\,,\;t\,')\;=\;W\;\Theta(t\,')
 \label{}
 \ea
 applied to a homogeneous Kelvin-Voigt spherical body will furnish, through relation (\ref{chuk}), the following
 response of the potential:
 \ba
 \nonumber
 U_{\it l}(\erbold,\,t)&=&\left(\frac{R}{r}
 \right)^{{\it l}+1}\int_{t\,'=\,-\infty}^{\,t\,'=\,t} \stackrel{\centerdot}{k}_{\it l}(t-t\,')\;W\;
 \Theta(t\,')\,dt\,'
 \,=\;\left(\frac{R}{r}\right)^{{\it l}+1}\int_{t\,'=\,0}^{\,t\,'=\,t} \stackrel{\centerdot}{k}_{\it l}(t-t\,')\;W\;dt\,'\,
 \quad\quad\quad\quad
 ~\\
 \label{}\\
 \nonumber
 &=&W\;\left(\frac{R}{r}\right)^{{\it l}+1}\,\int_{\tau\,=\,0}^{\,\tau\,=\,t}\stackrel{\centerdot}{k}_{\it l}(\tau)
 \;d\tau
 \;=\;\frac{3}{2(l-1)}~\,\frac{\;1\;-\;\exp(\textstyle \;-\,t\,
 \zeta_{\textstyle{_{\textstyle{_l}}}}\,)\;}{1\;+\;A_{\textstyle{_l}}}\;\left(\frac{R}{r}\right)^{{\it l}+1}W
 ~~~.\quad\quad\quad\quad\quad
 \ea
 In the elastic limit, this becomes:
 \ba
 \tau_{\textstyle{_V}}\rightarrow 0\quad\Longrightarrow\quad\zeta_{\textstyle{_l}}\rightarrow\infty
 \quad\Longrightarrow\quad
  U_{\it l}(\,\erbold\,,\,\;t\,)\;\rightarrow\;\frac{3}{2(l-1)}~\,\frac{\;1\;}{1\;+\;A_{\textstyle{_l}}}\;\left(\frac{R}{r}
 \right)^{{\it l}+1}W~~~,\quad
 \label{as}
 \ea
 which reproduces the case described by the static Love number $\;k_{\textstyle{_l}}\,=\,\frac{\textstyle
 3}{\textstyle 2}~\,\frac{\textstyle\;1\;}{\textstyle 1\;+\;A_{\textstyle{_l}}}\;\,$.

 An alternative way of getting (\ref{as}) would be to employ formulae (\ref{KV2}) and (\ref{L30a}).

 \subsection{A homogeneous Maxwell spherical body}

 Using the formulae presented in the Appendix \ref{jjv}, and relying upon the Correspondence Principle, we write down
 the complex Love numbers for a Maxwell material as
 \ba
 \bar{k}_{\textstyle{_l}}(\chi)\;=\;\frac{3}{2(l-1)}\;\frac{1}{\;\textstyle 1\;+\;\frac{\textstyle A_{\textstyle{_l}}\;
 \tau_{\textstyle{_M}}\;\inc\;\chi}{\textstyle 1\;+\;\tau_{\textstyle{_M}}\;\inc\;\chi\;}\;}\;=\;\frac{3}{2(l-1)}\;\frac{
 1}{1\;+\;A_{\textstyle{_l}}}\;\left[\;1\;+\;\frac{A_{\textstyle{_l}}}{\textstyle 1\;+\;\left(\,1\;+\;A_{\textstyle{_l}}\,\right)
 \;\tau_{\textstyle{_M}}\;\inc\;\chi\;}\;\right]~~~,\quad
 \label{}
 \ea
 which corresponds, via (\ref{L31}), to
 \ba
 \stackrel{\bf\centerdot}{k}_{\textstyle{_l}}(\tau)~=~\frac{3}{2(l-1)}~\,\frac{\;\;\delta(\tau)\;+\;{A_{\textstyle{_l}}}\;
 \gamma_{\textstyle{_l}}\;
 \exp(\textstyle \;-\,\tau\,\gamma_{\textstyle{_{\textstyle{_l}}}}\,)
 \;\Theta(\tau)\;}{\textstyle 1~+~\textstyle A_{\textstyle{_l}
 }}
 \label{}
 \ea
 and
 \ba
 {k}_{\textstyle{_l}}(\tau)~=~\frac{3}{2(l-1)}~\,\frac{\;\;1\;+\;{A_{\textstyle{_l}}}\;
 \left[\,1\;
 -\;
 \exp(\textstyle \;-\,\tau\,
 \gamma_{\textstyle{_{\textstyle{_l}}}}\,)\;\right]\;}{\textstyle 1~+~\textstyle A_{\textstyle{_l}
 }}\;\,\Theta(\tau)~~~,
 \label{MBA}
 \ea
 where
 \ba
 \gamma_{\textstyle{_l}}\,\equiv\;\frac{1}{\textstyle \left(\,1\;+\;A_{\textstyle{_l}}\,\right)
 \;\tau_{\textstyle{_M}}\;}~~~.
 \label{}
 \ea

 A Heaviside step potential
 \ba
 W_{\it{l}}(\eRbold\,,\;\erbold^{\;*}\,,\;t\,')\;=\;W\;\Theta(t\,')
 \label{}
 \ea
 will, according to formula (\ref{chuk}), render the following response:
 \ba
 \nonumber
 U_{\it l}(\erbold,\,t)\;=\;\left(\frac{R}{r}
 \right)^{{\it l}+1}\int_{t\,'=\,-\infty}^{\,t\,'=\,t} \stackrel{\centerdot}{k}_{\it l}(t-t\,')\;W\;
 \Theta(t\,')\,dt\,'
 \,=\;\left(\frac{R}{r}\right)^{{\it l}+1}\int_{t\,'=\,0}^{\,t\,'=\,t} \stackrel{\centerdot}{k}_{\it l}(t-t\,')\;W\;dt\,'\,
 \quad\quad\quad\quad\quad
 ~\\
 \label{}\\
 \nonumber
 =W\;\left(\frac{R}{r}\right)^{{\it l}+1}\,\int_{\tau\,=\,0}^{\,\tau\,=\,t}\stackrel{\centerdot}{k}_{\it l}(\tau)
 \;d\tau
 \;=\;\frac{3}{2(l-1)}~\,\frac{\;1\;+\;
 A_{\textstyle{_l}}\;\left[1\;-\;\exp(\textstyle \;-\,t\,
 \gamma_{\textstyle{_{\textstyle{_l}}}}\,)\;\right]\;}{1\;+\;A_{\textstyle{_l}}}\;\left(\frac{R}{r}\right)^{{\it l}+1}W
 ~\Theta(t)~.\quad~\quad~
 \ea
 In the elastic limit, we obtain:
 \ba
 \tau_{\textstyle{_M}}\rightarrow\infty\quad\Longrightarrow\quad\gamma_{\textstyle{_l}}\rightarrow 0
 \quad\Longrightarrow\quad
  U_{\it l}(\,\erbold\,,\,\;t\,)\;\rightarrow\;\frac{3}{2(l-1)}~\,\frac{\;1\;}{1\;+\;A_{\textstyle{_l}}}\;\left(\frac{R}{r}
 \right)^{{\it l}+1}W~~~,\quad
 \label{}
 \ea
 which corresponds to the situation described by the static Love number $\;k_{\textstyle{_l}}\,=\,\frac{\textstyle
 3}{\textstyle 2(l-1)}~\,\frac{\textstyle\;1\;}{\textstyle 1\;+\;A_{\textstyle{_l}}}\;\,$.

 \subsection{A homogeneous Hohenemser-Prager (SAS) spherical body}\label{chu}

 The Correspondence Principle, along with the formulae from subsection \ref{HOH}, yields the following expression for the
 complex Love numbers of a Hohenemser-Prager (SAS) spherical body:
 \ba
 \bar{k}_{\textstyle{_l}}(\chi)\;=\;\frac{3}{2\,(l-1)}\;\,\frac{1}{\;1\;+\;A_{\textstyle{_l}}\;\frac{\textstyle 1\,+\,\inc\,\chi\,
 \tau_{\textstyle{_{\textstyle{_V}}}}}{\textstyle 1\,+\,\inc\,\chi\,\tau_{\textstyle{_{\textstyle{_M}}}}}\;}                  ~~~.
 \label{}
 \ea
 Combined with (\ref{L31}), this entails:
 \ba
 \stackrel{\bf\centerdot}{k}_{\textstyle{_l}}(\tau)\;=\;\frac{3}{2\,(l-1)}\;\,\frac{1}{\;1\;+\;A_{\textstyle{_l}}\;\,\frac{
 \textstyle\,\tau_{\textstyle{_{\textstyle{_V}}}}}{\,\textstyle \tau_{\textstyle{_{\textstyle{_M}}}}}\;}\;\left[\;\delta(\tau)
 \;+\;\frac{A_{\textstyle{_l}}}{\tau_{\textstyle{_{\textstyle{_M}}}}}\,\;\frac{\tau_{\textstyle{_{\textstyle{_V}}}}\,-\,\tau_{
 \textstyle{_{\textstyle{_M}}}}}{~\tau_{\textstyle{_{\textstyle{_M}}}}\,+\,A_{\textstyle{_l}}\,\tau_{\textstyle{_{\textstyle{_V}}}}}
 ~\,\exp\left(\,-\,\frac{\,1\;+\;A_{\textstyle{_l}}\,}{\,\tau_{\textstyle{_{\textstyle{_M}}}} \;+\;A_{\textstyle{_l}}\;\tau_{
 \textstyle{_{\textstyle{_V}}}}\,}\;\tau\,\right)\;\right]\quad\quad\quad
 \label{}
 \ea
 and
 \ba
 {k}_{\textstyle{_l}}(\tau)\;=\;\frac{3}{2\,(l-1)}\;\,\frac{\textstyle\;1
 ~-~\frac{\textstyle A_{\textstyle{_l}}}{\textstyle\,\tau_{\textstyle{_{\textstyle{_M}}}}}\,\;\frac{\textstyle\tau_{\textstyle{_{\textstyle{_V}}}}\,-\,\tau_{
 \textstyle{_{\textstyle{_M}}}}}{\textstyle ~1\,+\,A_{\textstyle{_l}}\,}
 \left[\;1-\;~\,\exp\left(\,-\,\frac{\textstyle\,1\;+\;A_{\textstyle{_l}}\,}{\textstyle\,\tau_{\textstyle{_{\textstyle{_M}}}} \;+\;A_{\textstyle{_l}}\;\tau_{
 \textstyle{_{\textstyle{_V}}}}\,}\;\tau\,\right)\;\right]\;}{\textstyle\;1\;+\;A_{\textstyle{_l}}\;\,\frac{
 \textstyle\,\tau_{\textstyle{_{\textstyle{_V}}}}}{\,\textstyle \tau_{\textstyle{_{\textstyle{_M}}}}}\;}\;~\Theta(\tau)\quad\quad
 \quad\label{}
 \ea
 A Heaviside step potential
 \ba
 W_{\it{l}}(\eRbold\,,\;\erbold^{\;*}\,,\;t\,')\;=\;W\;\Theta(t\,')
 \label{}
 \ea
 applied to a SAS spherical body will then result in the following variation of its potential:
 \ba
 U_{\it l}(\erbold,\,t)~=\quad\quad\quad\quad\quad\quad\quad\quad\quad\quad\quad\quad\quad\quad\quad\quad\quad\quad\quad\quad
 \quad\quad\quad\quad\quad\quad\quad\quad\quad\quad\quad\quad\quad\quad\quad\quad\quad\quad\quad\quad
  \nonumber
  \ea
  \ba
  \nonumber
 \left(\frac{R}{r}
 \right)^{{\it l}+1}\int_{t\,'=\,-\infty}^{\,t\,'=\,t} \stackrel{\centerdot}{k}_{\it l}(t-t\,')\,W\,
 \Theta(t\,')\,dt\,'
 =\left(\frac{R}{r}\right)^{{\it l}+1}\int_{t\,'=\,0}^{\,t\,'=\,t} \stackrel{\centerdot}{k}_{\it l}(t-t\,')\,W\,dt\,'
 =W\left(\frac{R}{r}\right)^{{\it l}+1}\,\int_{\tau\,=\,0}^{\,\tau\,=\,t}\stackrel{\centerdot}{k}_{\it l}(\tau)
 \,d\tau\quad\quad
 \ea
 \ba
 =\frac{3}{2(l-1)}\left(\frac{1}{{1+A_{\textstyle{_l}}\,\frac{\textstyle \tau_{{\textstyle{_{\textstyle{_V}}}
 }}}{\textstyle\tau_{{\textstyle{_{\textstyle{_M}}}}}}\,}
 }+\frac{A_{\textstyle{_l}}}{1+A_{\textstyle{_l}}}\,\frac{\tau_{{\textstyle{_{\textstyle{_V}}}
 }}\,-\,\tau_{{\textstyle{_{\textstyle{_M}}}
 }}}{\tau_{{\textstyle{_{\textstyle{_M}}}
 }}+A_{\textstyle{_l}}\,\tau_{{\textstyle{_{\textstyle{_V}}}
 }}}
 \right)\left[1-\exp\left(-\,\frac{\textstyle\,1\,+\,A_{\textstyle{_l}}\,}{\textstyle\,\tau_{\textstyle{_{\textstyle{_M}}}}+
 A_{\textstyle{_l}}
 \tau_{\textstyle{_{\textstyle{_V}}}}\,}\;t\right)\right]\left(\frac{R}{r}\right)^{{\it l}+1}W~\Theta(t)~.~~\quad
 \ea
 Within this model, the elastic limit is achieved by setting
 $\,\tau_{\textstyle{_{\textstyle{_M}}}}\,=\,\tau_{\textstyle{_{\textstyle{_V}}}}\,$, whence we obtain the case
 described by the static Love number $\;k_{\textstyle{_l}}\,=\,\frac{\textstyle
 3}{\textstyle 2}~\,\frac{\textstyle\;1\;}{\textstyle 1\;+\;A_{\textstyle{_l}}}\;\,$. Interestingly, the elastic regime is
 achieved even when these times are not zero. Their being equal to one another turns out to be sufficient.

 Repeating the above calculation for tidal disturbance $\,W\,\Theta(-t\,')\,$, we shall see that, after the tidal perturbation
 is removed, a tidally prestressed sphere regains its shape, the stress relaxing at a rate proportional to
 $~\exp\left(-\,~\frac{\textstyle\,1\,+\,A_{\textstyle{_l}}\,}{\textstyle\,\tau_{\textstyle{_{\textstyle{_M}}}}+A_{\textstyle{_l}}
 \tau_{\textstyle{_{\textstyle{_V}}}}\,}\,~t\right)~$.

 \section{The correspondence principle\\ (elastic-viscoelastic analogy)}\label{corpr}

  \subsection{The correspondence principle, for nonrotating bodies}\label{3.4}\label{4.3}

 While the static Love numbers depend on the static rigidity $\,\mu\,$ through (\ref{L4}), it is not immediately clear if
 a similar formula interconnects also $\,\bar{k}_{\it l}(\chi)\,$ with $\,\bar{\mu}(\chi)\,$. To understand why and when the relation
 should hold, recall that formulae (\ref{L4}) originate from the solution of a boundary-value problem for a system
 incorporating two  equations:
 \begin{subequations}
 \ba
 \sigma_{\textstyle{_{\beta\nu}}}&=&2\;\mu\;u_{\textstyle{_{\beta\nu}}}~~~,
 \label{L34}\\
 \nonumber\\
 0&=&\frac{\partial \sigma_{\textstyle{_{\beta\nu}}}}{\partial x_{\textstyle{_\nu}}}\;-\;
 \frac{\partial p}{\partial x_{\textstyle{_\beta}}}\;-\;\rho\;\frac{\partial (W\,+\,U)}{\partial
 x_{\textstyle{_\beta}}}~~~,
 \label{L35}
 \ea
 \label{stat}
 \end{subequations}
 the latter being simply the equation of equilibrium written for a {\emph{static}} viscoelastic medium, in neglect of
 compressibility and heat conductivity. The notations $\,\sigma_{\textstyle{_{\beta\nu}}}\,$ and $\,u_{\textstyle{_{\beta\nu}}}
 \,$ stand for the {\it{deviatoric}} stress and strain, $~p\,\equiv\,-\,\frac{\textstyle 1}{\textstyle 3}\,\mbox{Sp}\,
 {\mathbb{S}}~$ is the pressure (set to be nil in incompressible media), while $\,W\,$ and $\,U\,$ are the perturbing and perturbed potentials. By solving the
 system, one arrives at the static relation $\,U_{\it l}=k_{\it l}\,W_{\it l}\,$, with the customary static Love numbers
 $\,k_{\it l}\,$ expressed via $\,\rho\,$, $\,R\,$, and $\,\mu\,$ by (\ref{L4}).

 Now let us write equation like (\ref{L34} - \ref{L35}) for time-dependent deformation of a {\it{nonrotating}} body:
 \begin{subequations}
 \ba
 {\mathbb{S}}&=&2~\hat{\mu}~{\mathbb{U}}~~~,
 \label{}\\
 \nonumber\\
 \rho~{\bf\ddot{u}}&=&\nabla{\mathbb{S}}~-~\nabla p~-~\nabla(W\,+\,U)~~~
 \label{}
 \ea
 \end{subequations}
 or, in terms of components:
 \begin{subequations}
 \ba
 \sigma_{\textstyle{_{\beta\nu}}}&=&2~\hat{\mu}~u_{\textstyle{_{\beta\nu}}}~~~,
 \label{}\\
 \nonumber\\
 \rho\,\ddot{u}_{\textstyle{_\beta}}&=&\frac{\partial \sigma_{\textstyle{_{\beta\nu}}}}{\partial
 x_{\textstyle{_\nu}}}~-~\frac{\partial p}{\partial x_{\textstyle{_\beta}}}\;-\;\rho\;\frac{\partial (W\,+\,U)}{\partial
 x_{\textstyle{_\beta}}}~~~.
 \label{}
 \ea
 \end{subequations}
 In the frequency domain, this will look:
 \begin{subequations}
 \ba
 \bar{\sigma}_{\textstyle{_{\beta\nu}}}(\chi)&=&2~\bar{\mu}(\chi)~\bar{u}_{\textstyle{_{\beta\nu}}}(\chi)~~~,
 \label{}\\
 \nonumber\\
 \rho\,\chi^2\,\bar{u}_{\textstyle{_{\beta\nu}}}(\chi)&=&\frac{\partial \bar{\sigma}_{\textstyle{_{\beta\nu}}}(\chi)}{\partial
 x_{\textstyle{_\nu}}}~-~\frac{\partial \bar{p}(\chi)}{\partial x_{\textstyle{_\beta}}}\;-\;\rho\;\frac{\partial \left[\bar{W}(\chi)
 \,+\,\bar{U}(\chi)\right]}{\partial x_{\textstyle{_\beta}}}~~~,
 \label{gegg}
 \ea
 \label{suba}
 \end{subequations}
 where a bar denotes a spectral component for all functions except $\,\mu~$ -- recall that $\,\bar{\mu}\,$ is a spectral
 component not of the kernel $\,\mu(\tau)\,$ but of its time-derivative $\,{\bf{\dot{\mu}}}(\tau)\,$.

 Unless the frequencies are extremely high, we can neglect the body-fixed acceleration term $\,\chi^2\,\bar{u}_{\textstyle{_{
 \beta\nu}}}(\chi)\,$ in the second equation, in which case our system of equations for the spectral components will mimic
 (\ref{stat}). Thus we arrive at the so-called {\emph{correspondence principle}} (also known as the {\it{elastic-viscoelastic
 analogy}}), which maps a solution of a linear viscoelastic boundary-value problem to a solution of a corresponding elastic
 problem with the same initial and boundary conditions. As a result, the algebraic equations for the Fourier (or Laplace)
 components of the strain and stress in the viscoelastic case mimic the equations connecting the strain and stress in the
 appropriate elastic problem. So the viscoelastic operational moduli $\,\bar{\mu}(\chi)\,$ or $\,\bar{J}(\chi)\,$ obey the same
 algebraic relations as the elastic parameters $\,\mu\,$ or $\,J\,$.

 In the literature, there is no consensus on the authorship of this principle. For example, Haddad (1995) mistakenly attributes
 it to several authors who published in the 1950s and 1960s. In reality, the principle was pioneered almost a century earlier by
 Darwin (1879), for isotropic incompressible media. The principle was extended to more general types of media by Biot (1954, 1958),
 who also pointed out some limitations of this principle.

 \subsection{The correspondence principle, for rotating bodies}\label{4.4}

 Consider a body of mass $\,M_{prim}\,$, which is spinning at a rate $\,\omegabold\,$ and is also performing some orbital
 motion (for example, is orbiting, with its partner of mass $\,M_{sec}\,$, around their mutual centre of mass). Relative
 to some inertial coordinate system, the centre of mass of the body is located at $\,\xbold_{_{CM}}\,$, while a small parcel of
 its material is positioned at $\,\xbold\,$. Relative to the centre of mass of the body, the parcel is located at $\,\erbold\,=\,
 \xbold\,-\,\xbold_{_{CM}}\,$. The body being deformable, we can decompose $\,\erbold\,$ into its average value, $\,\erbold_0\,$,
 and an instantaneous displacement $\,\ubold~$:
 \ba
 \left.
 \begin{array}{c}
 \xbold\;=\;\xbold_{_{CM}}\,+\;\erbold~~\\
 ~\\
 \rbold\;=\;\rbold_{0}\,+\;\ubold~~~~
 \end{array}
 \right\}~\quad~\Longrightarrow\quad~\quad\xbold\;=\;\xbold_{_{CM}}\,+\;\erbold_0\,+\;\ubold~~.
 \label{oko}
 \ea
 Denote with $D/Dt$ the time-derivative in the inertial frame. The symbol $d/dt$ and its synonym, overdot,
 will be reserved for the time-derivative in the body frame, so $\,{d\rbold_0}/{dt}\,=\,0\,$. ~Then
 \ba
 \frac{D\rbold}{Dt}\;=\;\frac{d\rbold}{dt}~+~\omegabold\,\times\,\rbold
 \quad~\quad\mbox{and}\quad~\quad
 \frac{D^2\rbold}{Dt^2}\;=\;\frac{d^2\rbold}{dt^2}\;+\;2\;\omegabold\,\times\,\frac{d\rbold}{dt}\;+\;
 \omegabold\,\times\,\left(\omegabold\,\times\,\rbold\right)\;+\;\dotomegabold\,\times\,\rbold~~.\quad\quad
 \label{oj}
 \ea
 Together, the above formulae result in
 \ba
 \nonumber
 \frac{D^2\xbold}{Dt^2}\;=\;\frac{D^2\xbold_{_{CM}}}{Dt^2}\;+\;\frac{D^2\rbold}{Dt^2}
 &=&\frac{D^2\xbold_{_{CM}}}{Dt^2}\;+\;\frac{d^2\rbold}{dt^2}\;+\;2\;\omegabold\,\times\,\frac{d\rbold}{dt}\;+\;
 \omegabold\,\times\,\left(\omegabold\,\times\,\rbold\right)\;+\;\dotomegabold\,\times\,\rbold\\
 \nonumber\\
 &=&\frac{D^2\xbold_{_{CM}}}{Dt^2}\;+\;\frac{d^2\ubold}{dt^2}\;+\;2\;\omegabold\,\times\,\frac{d\ubold}{dt}\;+\;
 \omegabold\,\times\,\left(\omegabold\,\times\,\rbold\right)\;+\;\dotomegabold\,\times\,\rbold~~.\quad\quad\quad
 \label{ojo}
 \ea
 The equation of motion for a small parcel of the body's material will read as
 \ba
 \rho\;\frac{D^2\xbold}{Dt^2}\;=\;\nabla{\mathbb{S}}\;-\;\nabla p\;+\;\Fbold_{self}\;+\;\Fbold_{ext}~~~,
 \label{joj}
 \ea
 where $\,\Fbold_{ext}\,$ is the exterior gravity force {\it{per unit volume}}, while $\,\Fbold_{self}\,$ is the
 ``interior" gravity force {\it{per unit volume}}, i.e., the self-force wherewith the rest of the body is acting upon the
 selected parcel of medium. Insertion of (\ref{ojo}) in (\ref{joj}) furnishes:
 \ba
 \rho\,\left[\,\frac{D^2\xbold_{_{CM}}}{Dt^2}\,+\,\ddotubold\,+\,2\,\omegabold\,\times\,\dotubold\,+\,\omegabold\,\times\,
 \left(\omegabold\,\times\,\rbold\right)\,+\,\dotomegabold\,\times\,\rbold\,\right]\,=\,\nabla{\mathbb{S}}\,-\,\nabla p\,+\,
 \Fbold_{self}\,+\,\Fbold_{ext}~~.\quad\quad
 \label{}
 \ea
 At the same time, for the primary body as a whole, we can write:
 \ba
 M_{prim}\;\frac{D^2\xbold_{_{CM}}}{Dt^2}\;=\;\int_{V}\Fbold_{ext} \;d^3\rbold~~,
 \label{}
 \ea
 the integration being carried out over the volume $\,V\,$ of the primary. (Recall that $\,\Fbold_{ext}\,$ is a force per unit
 volume.) Combined together, the above two equations will result in
 \ba
 \rho\left[\,\ddotubold+\,2\,\omegabold\times\dotubold+\,\omegabold\times
 \left(\omegabold\times\rbold\right)+\dotomegabold\times\rbold\,\right]\,=\,\nabla{\mathbb{S}}\,-\,\nabla p\,+\,
 \Fbold_{self}\,+\,\Fbold_{ext}\,-\,\frac{\rho~}{\,M_{prim}\,}\int_{V}\Fbold_{ext} \;d^3\rbold~~.\quad~
 \label{}
 \ea
 For a spherically-symmetrical (not necessarily radially-homogeneous) body, the integral on the right-hand side clearly removes
 the Newtonian part of the force, leaving the harmonics intact:
 \ba
 \Fbold_{ext}\,-\,\frac{\rho~}{\,M_{prim}\,}\int_{V}\Fbold_{ext}~d^3\rbold~=~\rho~\sum_{l=2}^\infty\nabla W_l~~,
 \label{}
 \ea
 where the harmonics are given by
 \ba
 W_l(\erbold,\,\erbold^*)\;=\;-\;\frac{G\,M_{sec}}{r^*}\;\left(\frac{r}{r^*}\right)^l\,P_l(\cos\gamma)~~~,
 \label{}
 \ea
 $\erbold^{\,*}$ being the vector pointing from the centre of mass of the primary to that of the secondary, and $\,\gamma\,$
 being the angular separation between $\,\erbold\,$ and $\,\erbold^{\,*}\,$, subtended at the centre of mass of the primary.

 In reality, a tiny extra force $\,{\cal F}\,$, the tidal force per unit volume, is left over due to the body being
 slightly distorted:
 \ba
 \Fbold_{ext}\,-\,\frac{\rho~}{\,M_{prim}\,}\int_{V}\Fbold_{ext}~d^3\rbold~=~\rho~\sum_{l=2}^\infty\nabla W_l~+~{\cal F}~~.
 \label{}
 \ea
 Here $\,{\cal F}\,$ is the density multiplied by the average tidal acceleration experienced by the body as a whole.
 In neglect of $\,{\cal F}\,$, we arrive at
 \ba
 \rho\left[\,\ddotubold+\,2\,\omegabold\times\dotubold+\,\omegabold\times
 \left(\omegabold\times\rbold\right)+\dotomegabold\times\rbold\,\right]\,=\,\nabla{\mathbb{S}}\,-\,\nabla p\,-\,
 \rho~\sum_{l=2}^\infty\nabla(U_l\,+\;W_l)~~.\quad~
 \label{eqq}
 \ea
 Here, to each disturbing term of the exterior potential, $\,W_l\,$, corresponds a term $\,U_l\,$ of the self-potential, the
 self-force thus being expanded into $\,\Fbold_{self}\,=\,-\,\sum_{l=2}^\infty\nabla U_l\,$.

 Equation (\ref{eqq}) could as well have been derived in the body frame, where it would have assumed the same form.

 Denoting the tidal frequency with $\,\chi\,$, we see that the terms on the left-hand side have the order of
 $\,\rho\,\chi^2\,u~$, $~\rho\,\omega\,\chi\,u~$, $~\rho\,\omega^2\,r\,$, ~and $~\rho\,\dot{\chi}\,\omega\,r~$, correspondingly.
 In realistic situations, the first two terms, thus, can be neglected, and we end up with
 \ba
 0\,=\,\nabla{\mathbb{S}}\,-\,\nabla p\,-~\rho~\sum_{l=2}^\infty\nabla(U_l\,+\;W_l)
 ~-~\rho~\omegabold\times\left(\omegabold\times\rbold\right)
 ~-~\rho\,\dotomegabold\times\rbold~~,\quad~
 \label{equmotion}
 \ea
 the term $\;-\,\nabla p\,$ vanishing in an incompressible media.

 \subsection{The centripetal term and the zero-degree Love number}\label{rad}\label{4.5}

 The centripetal term in (\ref{equmotion}) can be split into a purely radial part and a part that can be incorporated into the $\,W_2\,$
 term of the tide-raising potential, as was suggested by Love (1909, 1911). Introducing the colatitude $\,\phi\,'\,$
 through
 $~
 \cos\phi\,'~=~\frac{\omegabold}{\,|\omegabold|\,}\cdot\frac{\erbold}{\,|\erbold|\,}
 ~$, we can write down the evident equality
 \ba
 \nonumber
 \omegabold\times\left(\omegabold\times\rbold\right)~=~\omegabold~(\omegabold\cdot\erbold)~-~\erbold~\omegabold^{\,2} =
 ~\nabla\left[\,\frac{1}{2}\,\left(\omegabold\cdot\erbold\right)^2\,-~\frac{1}{2}\,\omegabold^{\,2}\,\erbold^{\,2}\,\right]
 ~=~\nabla\left[~~\frac{1}{2}\,\omegabold^{\,2}\,\erbold^{\,2}\,\left(\,\cos^2\phi\,'~-~1\,\right)~~\right]~~~.\quad\quad\quad
 \quad\quad\quad
 \label{17}
 \ea
 ~The definition $~P_2(\cos\phi\,')\,=~\frac{\textstyle 1}{\textstyle 2}\left(3\,\cos^2\phi\,'\,-~1\right)~$ ~easily renders:
 $~\cos^2\phi\,'~=~\frac{\textstyle 2}{\textstyle 3}~P_2(\cos\phi\,')~+~\frac{\textstyle 1}{\textstyle 3}~$, whence:
 \ba
 \omegabold\times\left(\omegabold\times\rbold\right)~=~\nabla\left[~~\frac{1}{3}~\omegabold^{\,2}\,\erbold^{\,2}
 \,\left[\,P_2\left(\cos\phi\,'\right)~-~1~\right]~~\right]~~~.
 \ea
 We see that the centripetal force splits into a second-harmonic and purely-radial parts:
 \ba
 -~\rho~\omegabold\times\left(\omegabold\times\rbold\right)~=~-~\nabla\left[~\frac{\rho}{3}~\omegabold^{\,2}\,\erbold^{\,2}
 \,P_2\left(\cos\phi\,'\right)~\right]
 ~+~\nabla\;\left[~\frac{\rho}{3}~\omegabold^{\,2}\,\erbold^{\,2}~\right]~~~,
  \ea
 where we assume the body to be homogeneous. The second-harmonic part can be incorporated into
 the external potential. The response to this part will be proportional to the degree-2 Love number $\,k_2\,$.

 The purely radial part of the centripetal potential generates a radial deformation. This part of the potential is often
 ignored, the associated deformation being tacitly included into the equilibrium shape of the body. Compared to the main terms of the
 equation of motion, this radial term is of the order of $\,10^{-3}\,$ for the Earth, and is smaller for most other bodies. As the
 rotation variations of the Earth are of the order of $\,10^{-5}\,$, this term leads to a tiny change in the geopotential and to an
 associated displacement of the order of a micrometer. \footnote{~Tim Van Hoolst, private communication.}

 However, for other rotators the situation may be different. For example, in Phobos, whose libration magnitude is large (about 1 degree),
 the radial term may cause an equipotential-surface variation of about 10 cm. This magnitude is large enough to be observed by future
 missions and should be studied in more detail.\footnote{~Tim Van Hoolst, private communication.} The emergence of the purely radial
 deformation gives birth to the zero-degree Love number (Dahlen 1976, Matsuyama \& Bills 2010). Using Dahlen's results, Yoder (1982,
 eqns 21 - 22) demonstrated that the contribution of the radial part of the centripetal potential to the change in mean motion of Phobos
 is about 3\%, which is smaller than the uncertainty in our knowledge of Phobos' $\,k_2/Q\,$. It should be mentioned, however, that the
 calculations by Dahlen (1976) and Matsuyama \& Bills (2010) were performed for steady (or slowly changing) rotation, and not for
 libration. This means that Yoder's application of Dahlen's result to Phobos requires extra justification.

 What is important for us here is that the radial term does not interfere with the calculation of the Love number. Being independent of
 the longitude, this term generates no tidal torque either, provided the obliquity is neglected.

 \subsection{The toroidal term}\label{toro}\label{4.6}

 The inertial term $~~-~\rho~\dotomegabold\times\rbold\,$ in the equation of motion (\ref{equmotion}) can be cast into the form
 \ba
 -~\rho~\dotomegabold\times\rbold\;=\;\rho~\erbold\times\nabla(\dotomegabold\cdot\erbold)~~~,
 \label{toroidal}
 \ea
 whence we see that this term is of a toroidal type. Being almost nil for a despinning primary, this force becomes important for
 a librating object.

 In spherically-symmetric bodies, the toroidal force (\ref{toroidal}) generates toroidal deformation only. This deformation produces
 neither radial uplifts nor variations of the gravitational potential. Hence its presence does not influence the expressions for the Love
 numbers associated with vertical displacement ($\,h_{\textstyle{_l}}\,$) or the potential ($\,k_{\textstyle{_l}}\,$). As this deformation
 yields no change in the gravitational potential of the tidally-perturbed body, there is no tidal torque associated with this deformation.
 Being divergence-free, this deformation entails no contraction or expansion either, i.e., it is purely shear. Still, this deformation
 contributes to dissipation. Besides, since the toroidal forcing results in the toroidal deformation, it can, in principle, be associated
 with a ``toroidal" Love number.


 To estimate the dissipation caused by the toroidal rotational force, Yoder (1982) introduced an equivalent effective torque. He pointed
 out that this force becomes important when the magnitude of the physical libration is comparable to that of the optical libration.
 According to {\it{Ibid.}}, the toroidal force contributes to the change of the mean motion of Phobos about 1.6\%, which is less than
 the input from the purely radial part.

 \section{The Andrade and Maxwell models at different frequencies}

 \subsection{Response of a sample obeying the Andrade model}

 Within the Andrade model, the tangent of the phase lag demonstrates the so-called ``elbow dependence". At high frequencies, the
 tangent of the lag obeys a power law with an exponent equal to $\,-\alpha\,$, where $\,0\,<\,\alpha\,<\,1\,$. At low frequencies,
 the tangent of the lag once again obeys a power law, this time though with an exponent $~-\,(1-\alpha)\,$. This model fits well
 the behaviour of ices, metals, silicate rocks, and many other materials.

 However the applicability of the Andrade law may depend upon the intensity of the load and, accordingly, upon the damping mechanisms
 involved. Situations are known, when, at low frequencies, anelasticity becomes much less efficient than viscosity. In these cases,
 the Andrade model approaches, at low frequencies, the Maxwell model.

 \subsubsection{The high-frequency band}

 At high frequencies, expression (\ref{LL45b}) gets simplified. In the numerator, the term with $\,z^{-\alpha}\,$ dominates:
 $~z^{-\alpha}\,\gg\,z^{-1}\,\zeta\,$, ~which is equivalent to $\,z\,\gg\,\zeta^{\textstyle{^{\textstyle{\frac{1}{1-\alpha}}}}}\,$.
 In the denominator, the constant term dominates: $~1\,\gg\,z^{-\alpha}\,$, ~or simply: $\,z\,\gg\,1\,$. To know which of the
 two conditions, $\,z\,\gg\,\zeta^{\textstyle{^{\textstyle{\frac{1}{1-\alpha}}}}}\,$ or $\,z\,\gg\,1\,$, is stronger, we recall
 that at high frequencies anelasticity beats viscosity. So the $\,\alpha-$term in (\ref{I64}) is large enough. In other words, the
 Andrade timescale $\,\tau_{_A}\,$ should be smaller (or, at least, not much higher) than the viscoelastic time $\,\tau_{_M}\,$.
 Accordingly, at high frequencies, $\,\zeta\,$ is smaller (or, at least, not much higher) than unity. Hence, within the
 high-frequency band, either the condition $\,z\,\gg\,1\,$ is stronger than $\,z\,\gg\,\zeta^{\textstyle{^{\textstyle{\frac{1}{1-\alpha}
 }}}}\,$ or the two conditions are about equivalent. This, along with (\ref{sim}) and (\ref{ham}) enables us to write:
 \ba
 \tan \delta~\,\approx\,~(\chi\,\tau_{_A})^{\textstyle{^{-\alpha}}}\sin\left(\frac{\alpha\,\pi}{2}\right)~\Gamma(\alpha+1)~
 \quad~\quad~\mbox{for}\quad\chi\;\gg\;\tau^{-1}_{_A}\,=\,\tau_{_M}^{-1}\zeta^{-1}~~~.~~\,~
 \label{LL4433a}
 \ea
 The tangent being small, the expression for $\,\sin\delta\,$ looks identical:
 \ba
 \sin \delta\,~\approx\,~(\chi\,\tau_{_A})^{\textstyle{^{-\alpha}}}~\sin\left(\frac{\alpha\,\pi}{2}\right)~\Gamma(\alpha+1)~\quad~\quad~
 \mbox{for}~~~\chi\;\gg\;\tau^{-1}_{_A}\,=\,\tau_{_M}^{-1}\zeta^{-1}~~~.~~\,~
 \label{andrade_1}
 \label{LL43a}
 \ea

 \subsubsection{The intermediate region}

 In the intermediate region, the behaviour of the phase lag $\,\delta\,$ depends upon the frequency-dependence of $\,\zeta\,$. For
 example, if there happens to exist an interval of frequencies over which the conditions $\,1\,\gg\,z\,\gg\,\zeta^{\textstyle{^{
 \textstyle{\frac{1}{1-\alpha}}}}}\,$ are obeyed, then over this interval we shall have: $\,1\ll\,z^{-\alpha}\,$ and $\,z^{-\alpha}
 \gg\,\zeta z^{-1}\,$. Applying these inequalities to (\ref{LL45b}), we see that over such an interval of frequencies $\,\tan\delta\,$
 will behave as $\,z^{-2\alpha}\,\tan\left(\frac{\textstyle\alpha\,\pi}{\textstyle 2}\right)\,$.

 \subsubsection{The low-frequency band}

 At low frequencies, the term $\,z^{-1}\,\zeta\,$ becomes leading in the numerator of (\ref{LL45b}): $\,z^{-\alpha}\,\ll\,z^{-1}\,\zeta\,$,
 which requires $\,z\,\ll\,\zeta^{\textstyle{^{\textstyle{\frac{1}{1-\alpha}}}}}\,$. In the denominator, the term with $\,z^{-\alpha}\,$
 becomes the largest: $\,1\,\ll\,z^{-\alpha}\,$, whence $\,z\,\ll\,1\,$. Since at low frequencies the viscous term in (\ref{I64}) is
 larger than the anelastic term, we expect that for these frequencies $\,\zeta\,$ is larger (at least, not much smaller) than unity.
 Thence the condition $\,z\,\ll\,1\,$ becomes sufficient. Its fulfilment ensures the fulfilment of $\,z\,\ll\,\zeta^{\textstyle
 {^{\textstyle{\frac{1}{1-\alpha}}}}}\,$. Thus we state:
 \ba
 \tan \delta~\approx\,~(\chi\,\tau_{_A})^{\textstyle{^{-(1-\alpha)}}}\frac{\zeta}{~\cos\left(\frac{\textstyle\alpha\,\pi}{
 \textstyle 2}\right)~\Gamma(\alpha+1)}\quad\quad\quad\mbox{for}\quad\quad\chi~\ll~\tau^{-1}_{_A}\,=~\tau_{_M}^{-1}\,\zeta^{-1}~~~.~~\quad\quad
 \label{LL4433b}
 \ea
 The appropriate expression for $\,\sin\delta\,$ will be:
 \ba
 \sin \delta~\approx\,~1~-~O\left(~(\chi\,\tau_{_A})^{{{2(1-\alpha)}}}\,\zeta^{-2}\,\right)
  \quad\,\quad\quad\quad\,\mbox{for}\quad\quad\chi~\ll\;\tau_{_A}^{-1}\,=~\tau_{_M}^{-1}\,\zeta^{-1}~~~,~~
 \label{andrade_3}
 \label{LL43b}
 \ea
 \label{LL43}
 \label{andrade}

 It would be important to emphasise that the threshold $\,\tau^{-1}_{_A}\,=~\tau_{_M}^{-1}\,\zeta^{-1}\,$ standing in (\ref{LL4433a})
 and (\ref{LL43a}) is {\it{different}} from the threshold $\,\tau^{-1}_{_A}\,=~\tau_{_M}^{-1}\,\zeta^{-1}\,$ showing up in (\ref{LL4433b})
 and (\ref{LL43b}), even though these two thresholds are given by the same expression. The reason for this is that the timescales $\,
 \tau_{_A}\,$ and $\,\tau_{_M}\,$ are not fixed constants. While the Maxwell time is likely to be a very slow function of the frequency,
 the Andrade time may undergo a faster change over the transitional region: $\,\tau_{_A}\,$ must be larger than $\,\tau_{_M}\,$ at low
 frequencies (so anelasticity yields to viscosity), and must become shorter than or of the order of $\,\tau_{_M}\,$ at high frequencies (so anelasticity
 becomes stronger). This way, the threshold $\,\tau^{-1}_{_A}\,$ standing in (\ref{LL4433b} - \ref{LL43b}) is lower than the threshold
 $\,\tau^{-1}_{_A}\,$ standing in (\ref{LL4433a} - \ref{LL43a}). The gap between these thresholds is the region intermediate between the
 two pronounced power laws (\ref{LL4433a}) and (\ref{LL4433b}).

 \subsubsection{The low-frequency band: a special case, the Maxwell model}

 Suppose that, below some threshold $\,\chi_{\textstyle{_0}}\,$, anelasticity quickly becomes {\it{much less}} efficient than viscosity.
 This would imply a steep increase of $\,\zeta\,$ (equivalently, of $\,\tau_{_A}\,$) at low frequencies.
 Then, in (\ref{LL45b}), we shall have: $\,1\gg\,z^{-\alpha}\,$ and $\,z^{-\alpha}\ll\,\zeta z^{-1}\,$. This means that, for frequencies
 below $\,\chi_{\textstyle{_0}}\,$, the tangent will behave as
 \ba
 \tan \delta~\,\approx\,~z^{-1}\,\zeta~=~(\chi\,\tau_{_M})^{-1}~\quad~\quad~\mbox{for}\quad\quad
 \chi\,\ll\,\chi_{\textstyle{_0}}~~~.~~\,~
 \label{LL4433i}
 \ea
 the well-known viscous scaling law for the lag.

 The study of ices and minerals under weak loads (Castillo-Rogez et al. 2011, Castillo-Rogez 2011) has not shown such an abrupt
 vanishing of anelasticity. However, Karato \& Spetzler (1990) point out that this should be happening in the Earth's mantle, where
 the loads are much higher and anelasticity is caused by unpinning of dislocations.

 \subsection{The behaviour of $\,|k_{\textstyle{_l}}(\chi)|\;\sin\epsilon_{\textstyle{_l}}(\chi)=-\,{\cal{I}}{\it{m}}\left[
 \,\bar{k}_{\textstyle{_l}}(\chi)\,\right]\,$\\ within the Andrade and Maxwell models}\label{Andarde_Maxwell}

  As we explained in subsection 4.1, products $\,~k_{\textstyle{_l}}(\chi_{\textstyle{_{{\it{l}}mpq}}})\,\sin
 \epsilon_{\textstyle{_l}}(\chi_{\textstyle{_{{\it{l}}mpq}}})\,\;$ enter the $\;{\it l}mpq\;$ term of the Darwin-Kaula
 series for the tidal potential, force, and torque. Hence the importance to know the behaviour of
 these products as functions of the tidal frequency $~\chi_{\textstyle{_{{\it{l}}mpq}}}~$.


 \subsubsection{Prefatory algebra}

 It ensues from (\ref{k2bar_2}) that
 \ba
 \bar{k}_{\textstyle{_l}}(\chi)\;=\;
 \frac{3}{2\,(l-1)}\;\frac{\left(\;{\cal{R}}{\it{e}}\left[\bar{J}(\chi)\right]\;\right)^2
 \;+\;\left(\;{\cal{I}}{\it{m}}\left[\bar{J}(\chi)\right]\;\right)^2
 \;+\;A_{\textstyle{_l}}\;J\;{\cal{R}}{\it{e}}\left[\bar{J}(\chi)\right]
 \;+i\;A_{\textstyle{_l}}\;J\;{\cal{I}}{\it{m}}\left[\bar{J}(\chi)\right]
 }{
 \left(\;{\cal{R}}{\it{e}}\left[\bar{J}(\chi)\right]\;+\;A_{\textstyle{_l}}\;J\;\right)^2\;+\;
 \left(\;{\cal{I}}{\it{m}}\left[\bar{J}(\chi)\right]\;\right)^2 }~~~,~~~
 \label{DRR}
 \ea
 whence
 \ba
 |\bar{k}_{\textstyle{_l}}(\chi)|\;\sin\epsilon_{\textstyle{_l}}(\chi)\;=\;-\;{\cal{I}}{\it{m}}\left[\bar{k}_{\textstyle{_l}}(\chi)\right]\;=\;
 \frac{3}{2\,(l-1)}\;\frac{-\;A_{\textstyle{_l}}\;J\;{\cal{I}}{\it{m}}\left[\bar{J}(\chi)\right]}{\left(\;{\cal{R}}{\it{e}}
 \left[\bar{J}(\chi)\right]\;+\;A_{\textstyle{_l}}\;J\;\right)^2\;+\;\left(\;{\cal{I}}{\it{m}}\left[\bar{J}(\chi)
 \right]\;\right)^2} ~~~,~~~~~
 \label{A2c}
 \ea
 $J\,=\,J(0)\,\equiv\,1/\mu\,=\,1/\mu(0)\;$ being the unrelaxed compliance (the inverse of the unrelaxed shear modulus $\,\mu\,$). For an Andrade
 material, the compliance $\;\bar{J}\;$ in the frequency domain is rendered by (\ref{112}). Its imaginary and real parts are
 given by (\ref{A3} - \ref{A4}).
 It is then easier to rewrite (\ref{A2c}) as
  \ba
 \nonumber
 |\bar{k}_{\textstyle{_l}}(\chi)|\;\sin\epsilon_{\textstyle{_l}}(\chi)\;=\quad\quad\quad\quad\quad\quad\quad\quad\quad\quad\quad\quad
 \quad\quad\quad\quad\quad\quad\quad\quad\quad\quad\quad\quad\quad\quad\quad\quad\quad\quad\quad\quad\quad\\
 \nonumber\\
 \frac{3~A_{\textstyle{_l}}}{2\,(l-1)}~\frac{\zeta~z^{-1}~+~
 z^{-\alpha}~\sin\left(\frac{\textstyle\alpha\,\pi}{\textstyle 2}\right)~\Gamma(\alpha\,+\,1)}{\left[A_{\textstyle{_l}}+1+z^{-
 \alpha}\,\cos\left(\frac{\textstyle\alpha\,\pi}{\textstyle 2}\right)\,\Gamma(\alpha+1)\right]^2+\,\left[\zeta\,z^{-1}+\,
 z^{-\alpha}\,\sin\left(\frac{\textstyle\alpha\,\pi}{\textstyle 2}\right)\,\Gamma(1+\alpha)\right]^2}~~~,\quad\quad
 \label{A2cc}
 \ea
 where
 \ba
 z~\equiv~\chi~\tau_{_A}\,=~\chi~\tau_{_M}\,\zeta
 \label{}
 \ea
 and
 \ba
 \zeta~\equiv~\frac{\tau_{_A}}{\tau_{_M}}~~~.
 \label{}
 \ea
 For $\,\beta\,\rightarrow\,0\,$, i.e., for $\,\tau_{_A}\,\rightarrow\,\infty\,$, (\ref{A2cc}) coincides with the appropriate expression
 for a spherical Maxwell body.

 \subsubsection{The high-frequency band}

 Within the upper band, the term with $\,z^{-\alpha}\,$ dominates the numerator, while $\,A_{\textstyle{_l}}\,$ dominates the
 denominator. The domination of $\,z^{-\alpha}\,$ in the numerator requires that $\,z\gg\zeta^{\textstyle{^{\,{\textstyle\frac{1}{1-
 \alpha}}}}}\,$, which is the same as $\,\chi\gg\tau_{_M}^{-1}\,\zeta^{\textstyle{^{\textstyle\,\frac{\alpha}{1-\alpha}}}}\,$. The
 domination of $\,A_{\textstyle{_l}}\,$ in the denominator requires: $\,z\gg\,A^{\textstyle{^{\,-1/\alpha}}}\,$, which is the same as
 $\,\chi\gg\,\tau_{_M}^{-1}\,\zeta^{-1}\,A^{\textstyle{^{\,-\,1/\alpha}}}\,$. It also demands that $\,\zeta\,z^{-1}\ll A_{\textstyle{_{l}}}\,$,
 which is: $\,\chi\gg\tau^{-1}_{_M}\,A^{-1}_{\textstyle{_{l}}}\,$.

 For realistic values of $\,A_{\textstyle{_l}}\,$ (say, $10^3$) and $\,\alpha\,$ (say, 0.25), we have: $\,A^{\textstyle{^{\,-\,1/\alpha}}}
 \sim\,10^{-12}\,$. At high frequencies, anelasticity beats viscosity, so $\,\zeta\,$ is less than unity (or, at least, is not much larger
 than unity). On these grounds, the requirement
 $\,\chi\gg\tau_{_M}^{-1}\,\zeta^{\textstyle{^{\textstyle\,\frac{\alpha}{1-\alpha}}}}\,$ is the strongest here. Its fulfilment
 guarantees that of both $\,\chi\gg\,\tau_{_M}^{-1}\,\zeta^{-1}\,A^{\textstyle{^{\,-\,1/\alpha}}}\,$
 and $\,\chi\gg\tau_{_M}/A_{\textstyle{_{l}}}\,$.  Thus we have:
 \begin{subequations}
 \ba
 |\bar{k}_{\textstyle{_l}}(\chi)|\,\sin\epsilon_{\textstyle{_l}}(\chi)&\approx &\frac{3}{2\,(l-1)}\;\frac{1}{A_{\textstyle{_l}}}
 \,\sin\left(\frac{\alpha\,\pi}{2}\right)\,\Gamma(\alpha+1)~\,\zeta^{-\alpha}~\left(\,\tau_{_M}\,\chi\,\right)^{-\alpha}
 ~~\quad~\mbox{for}\quad\quad\chi\,\gg\,\tau_{_M}^{-1}\,\zeta^{\textstyle{^{
 \textstyle\,\frac{\alpha}{1-\alpha}}}}~~.\quad\quad\quad~\quad
 \label{a}
 \ea

 \subsubsection{The intermediate band}

 Within the intermediate band, the term $\,\zeta\,z^{-1}\,$ takes over in the numerator, while $\,A_{\textstyle{_l}}\,$ still dominates
 in the denominator. The domination of $\,\zeta\,z^{-\alpha}\,$ in the numerator implies that $\,z\ll\zeta^{\textstyle{^{\,{\textstyle
 \frac{1}{1-\alpha}}}}}\,$, which is equivalent to $\,\chi\gg\tau_{_M}^{-1}\,\zeta^{\textstyle{^{\textstyle\,\frac{\alpha}{1-\alpha}}}}\,
 $. The domination of $\,A_{\textstyle{_l}}\,$ in the denominator requires $\,\chi\gg\,\tau_{_M}^{-1}\,\zeta^{-1}\,A_{\textstyle{_l}}^{
 \textstyle{^{\,-\,1/\alpha}}}\,$ and $\,\chi\gg\tau_{_M}^{-1}A^{-1}_{\textstyle{_{l}}}\,$, as we just saw above.

 As we are considering the band where viscosity takes over anelasticity, we may expect that here $\,\zeta\,$ is about or, likely, larger
 than unity. Taken the large value of $\,A_{\textstyle{_l}}\,$, we see that the condition $\,\chi\gg\tau_{_M}^{-1}A^{-1}_{\textstyle{_{l}}}\,$ is
 stronger. Its fulfilment guarantees the fulfilment of $\,\chi\gg\,\tau_{_M}^{-1}\,\zeta^{-1}\,A_{\textstyle{_l}}^{
 \textstyle{^{\,-\,1/\alpha}}}\,$. This way, we obtain:
 \ba
 |\bar{k}_{\textstyle{_l}}(\chi)|\,\sin\epsilon_{\textstyle{_l}}(\chi)&\approx &\frac{3}{2\,(l-1)}~\frac{1}{A_{\textstyle{_l}}}~\left(
 \,\tau_{_M}\,\chi\,\right)^{-1}\quad\quad~\quad~\mbox{for}\quad\quad\tau_{_M}^{-1}\,\zeta^{\textstyle{^{\textstyle\,\frac{\alpha}{1-
 \alpha}}}}\,\gg\,\chi\,\gg\,\tau_{_M}^{-1}A_{\textstyle{_l}}^{-1}~~.\quad\quad\quad\quad\quad
 \label{b}
 \ea

 \subsubsection{The low-frequency band}

 For frequencies lower than $\,\tau_{_M}^{-1}A_{\textstyle{_l}}^{-1}\,$ the Andrade model renders the same frequency-dependency as that
 given (at frequencies below $\,\tau^{-1}_{_M}\,$) by the Maxwell model:
 \ba
 |\bar{k}_{\textstyle{_l}}(\chi)|\;\sin\epsilon_{\textstyle{_l}}(\chi)&\approx &\frac{3}{2\,(l-1)}~{A_{\textstyle{_l}}}~\,
 \tau_{_M}~\chi\quad\quad\quad~\quad~\quad\mbox{for}\quad\quad\tau_{_M}^{-1}\,A_{\textstyle{_l}}^{-1}\,\gg\,
 \chi~~~.\quad\,\quad\quad\quad\,\quad\quad\quad\,\quad\quad
 \label{c}
 \ea
 \end{subequations}

 \subsubsection{Interpretation}

 Formulae (\ref{a}) and (\ref{b})
 render a frequency-dependence mimicking that of $\,|\stackrel{\bf{-}}{J}(\chi)|\,\sin\delta(\chi)$ $=\,-\,{\cal{I}}\textit{m}\left[
 \bar{J}(\chi)\right]~$ in the high- and low-frequency bands. This can be seen from comparing (\ref{a}) and (\ref{b}) with (\ref{LL43a}).

 In contrast, (\ref{c}) reveals a peculiar feature inherent in the {\it{tidal}}~ lagging, and absent in the lagging in a sample.

 For terrestrial bodies, the condition $~\tau_{_M}^{-1}\,A_{\textstyle{_l}}^{-1}
 \,\gg\,\chi\;$ puts the values of $\,\chi\,$ below $\,10^{-10}\,${\it{Hz}}$\,$, $\,$ give or take several orders of
 magnitude. Hence $\,|\bar{k}_{\textstyle{_l}}(\chi)|\,\sin\epsilon_{\textstyle{_l}}(\chi)\,$ follows the linear scaling law
 (\ref{c}) only in an extremely close vicinity of the commensurability where the frequency $\,\chi\,$ vanishes. Nonetheless
 it is very important that $~|\bar{k}_2(\chi)|\,\sin\epsilon(\chi)~$ first reaches a finite maximum and then decreases
 continuously and vanishes, as the frequency goes to zero. This confirms that neither the tidal torque nor the tidal force
 becomes infinite in resonances.

  \section{The behaviour of $\;k_l(\chi)\equiv |\bar{k}_l(\chi)|\;$  in the\\ limit of vanishing tidal
  frequency $\;\chi\;\,$,\\
 within the Andrade and Maxwell models}

 From (\ref{k2bar_2}),
 we obtain:
  \ba
 |\bar{k}_l(\chi)|^2=\left(\frac{3}{2(l-1)}\right)^2\frac{|\,\stackrel{\bf{\_}}{J}(\chi)|\,}{
 |A_{\textstyle{_{l}}}\,J\,+\,\stackrel{\bf{\_}}{J}(\chi)|}
 \,=\left(\frac{3}{2(l-2)}\right)^2\,
 \frac{ \left(~{\cal{R}}{\it{e}}\left[\stackrel{\bf{\_}}{J}(\chi)\right]~\right)^2~+~
 \left(~{\cal{I}}{\it{m}}\left[\stackrel{\bf{\_}}{J}(\chi)\right]~\right)^2
 }{\left(\,{\cal{R}}{\it{e}}\left[\stackrel{\bf{\_}}{J}(\chi)\right]+\,A_{\textstyle{_l}}\,J\,\right)^2+
 \left(\,{\cal{I}}{\it{m}}\left[\stackrel{\bf{\_}}{J}(\chi)\right]\,\right)^2}
 ~~~.~~~
 \label{A8}
 \ea
 Bringing in expressions for the imaginary and real parts of the compliance, and introducing notations
 \ba
 E~\equiv~\eta^{-1}~~~,\quad B~\equiv~\beta~\sin\left(\,\frac{\alpha\,\pi}{2}\,\right)~\Gamma(1\,+\,\alpha)~~~,\quad
 D~\equiv~\beta~\cos\left(\,\frac{\alpha\,\pi}{2}\,\right)~\Gamma(1\,+\,\alpha)~~~,\quad\quad
 \label{}
 \ea
 we can write:
 \ba
 |k_l(\chi)|^2
 ~=~\left(\frac{3}{2(l-1)}\right)^2~\left[\,1\,-~\frac{\textstyle 2\,A_{\textstyle{_l}}\,J\,D\,}{E^2}~\chi^{2-\alpha}
 \,-\,\frac{\textstyle 2\,A_{\textstyle{_l}}\,J^2\,+\,A_{\textstyle{_l}}^2\,J^2\,}{E^2}~\chi^{2}\,+\,O(\chi^{3-2\alpha})\,\right]~~~
 \label{kkk3}
 \label{A9}
 \ea
 and
 \ba
 |k_l(\chi)|^{-2}
 ~=~\left(\frac{2(l-1)}{3}\right)^2~\left[\,1\,+~\frac{\textstyle 2\,A_{\textstyle{_l}}\,J\,D\,}{E^2}~\chi^{2-\alpha}
 \,+\,\frac{\textstyle 2\,A_{\textstyle{_l}}\,J^2\,+\,A_{\textstyle{_l}}^2\,J^2\,}{E^2}~\chi^{2}\,+\,O(\chi^{3-2\alpha})\,\right]~~~,
 \label{A10}
 \ea
 whence
 \ba
 |k_l(\chi)|
 ~=~\frac{3}{2(l-1)}~\left[\,1\,-\,A_{\textstyle{_l}}\,J~\frac{\textstyle D\,}{E^2}~\chi^{2-\alpha}
 \,-\,A_{\textstyle{_l}}\,J\,\frac{\textstyle J\,+\,A_{\textstyle{_l}}\,J/2\,}{E^2}~\chi^{2}\,+\,O(\chi^{3-2\alpha})\,\right]~~~
 \label{A11}
 \ea
 and
 \ba
 |k_l(\chi)|^{-1}
 ~=~\frac{2(l-1)}{3}~\left[\,1\,+\,A_{\textstyle{_l}}\,J~\frac{\textstyle D\,}{E^2}~\chi^{2-\alpha}
 \,+\,A_{\textstyle{_l}}\,J\,\frac{\textstyle J\,+\,A_{\textstyle{_l}}\,J/2\,}{E^2}~\chi^{2}\,+\,O(\chi^{3-2\alpha})\,\right]~~~,
 \label{A12}
 \ea
 the expansions being valid for $\,\chi\,J/E\,\ll\,1\,+\,A_{\textstyle{_l}}\,$, i.e., for $\,\chi\,\tau_{_M}\,\ll\,1\,+\,A_{\textstyle{_l}}\,$.

 Rewriting (\ref{k2bar_2}) as
 \ba
 \bar{k}_l(\chi)\,=\,\frac{3}{2(l-1)}\,\frac{
 \left( {\cal{R}}{\it{e}}\left[\stackrel{\bf{\_}}{J}(\chi)\right]~+~i~
 {\cal{I}}{\it{m}}\left[\stackrel{\bf{\_}}{J}(\chi)\right]\right)~
 \left( {\cal{R}}{\it{e}}\left[\stackrel{\bf{\_}}{J}(\chi)\right]~+~A_{\textstyle{_l}}~J~-~i~
 {\cal{I}}{\it{m}}\left[\stackrel{\bf{\_}}{J}(\chi)\right]\right)
 }{\left( {\cal{R}}{\it{e}}\left[\stackrel{\bf{\_}}{J}(\chi)\right]~+~A_{\textstyle{_l}}~J\right)^2~+~
 \left({\cal{I}}{\it{m}}\left[\stackrel{\bf{\_}}{J}(\chi)\right]\right)^2}~~~,\quad\quad\quad
 \label{A13}
 \ea
 we extract its real part:
 \ba
 \nonumber
 {\cal{R}}{\it{e}}\left[\bar{k}_l(\chi)\right]\,=\,\frac{3}{2(l-1)}\,\;\frac{
 \left( {\cal{R}}{\it{e}}\left[\stackrel{\bf{\_}}{J}(\chi)\right] \right)^2~+~
 \left({\cal{I}}{\it{m}}\left[\stackrel{\bf{\_}}{J}(\chi)\right]\right)^2~+~A_{\textstyle{_l}}~J~
 {\cal{R}}{\it{e}}\left[\stackrel{\bf{\_}}{J}(\chi)\right]
 }{\left( {\cal{R}}{\it{e}}\left[\stackrel{\bf{\_}}{J}(\chi)\right]~+~A_{\textstyle{_l}}~J\right)^2~+~
 \left({\cal{I}}{\it{m}}\left[\stackrel{\bf{\_}}{J}(\chi)\right]\right)^2}~~~~~~~~~~~~~~~~~~~
 \ea
 \ba
 =~\frac{3}{2(l-1)}~\left[~1~-~A_{\textstyle{_l}}~J~\frac{{\cal{R}}{\it{e}}\left[\stackrel{\bf{\_}}{J}(\chi)\right]
 ~+~A_{\textstyle{_l}}~J}{\left( {\cal{R}}{\it{e}}\left[\stackrel{\bf{\_}}{J}(\chi)\right]~+~A_{\textstyle{_l}}~J\right)^2~+~
 \left({\cal{I}}{\it{m}}\left[\stackrel{\bf{\_}}{J}(\chi)\right]\right)^2}~~\right]~~~.
 \label{A14}
 \ea
 Insertion of the expressions for $~{\cal{R}}{\it{e}}\left[\stackrel{\bf{\_}}{J}(\chi)\right]~$ and
 $~{\cal{I}}{\it{m}}\left[\stackrel{\bf{\_}}{J}(\chi)\right]~$ into the latter formula entails:
 \ba
 {\cal{R}}{\it{e}}\left[\bar{k}_l(\chi)\right]\,=~\frac{3}{2(l-1)}~\left[\,1\,-\,A_{\textstyle{_l}}\,J~
 \frac{\textstyle D\,}{E^2}~\chi^{2-\alpha}
 \,-\,A_{\textstyle{_l}}\,J~\frac{\textstyle J\,+\,A_{\textstyle{_l}}\,J}{E^2}~\chi^{2}\,+\,O(\chi^{3-2\alpha})\,\right]~~~.
 \label{A15}
 \ea

 Expressions (\ref{A12}) and (\ref{A15}) enable us to write down the cosine of the shape lag:
 \ba
 \cos\epsilon_l\;=\;\frac{{\cal R}{\it e} [k_l(\chi)]}{|k_l(\chi)|}\;=\;1\;-\;\frac{1}{2}\;\left(\,\frac{\textstyle A_{\textstyle{_l}}\;
 J}{\textstyle E}\,\right)^2\;\chi^2\;+\;O(\chi^{3-2\alpha})\;=\;1\;-\;\frac{1}{2}\;A_{\textstyle{_l}}^2\;(\tau_{_M}\chi)^2\;+\;
 O(\chi^{3-2\alpha}) \;\;.~~
 \label{A16}
 \ea

 Comparing this expression with (\ref{A11}), we see that, for the Andrade ($\alpha\neq 0$) model,
 the evolution of $\,k_l(\chi)\equiv|\bar{k}_2(\chi)|\,$ in the limit of small $\,\chi\,$,
 unfortunately, cannot be approximated with a convenient expression $\;k_l(\chi)\,\approx\,k_l(0)\,
 \cos\epsilon(\chi)\;$, which is valid for simpler models (like the one of Kelvin-Voigt or SAS).

 However, for the Maxwell model ($\beta=0$) expression (\ref{A11}) becomes:
 \ba
 |k_l(\chi)|=\frac{3}{2(l-1)}\left[1-\frac{\textstyle A_{\textstyle{_l}}J}{E^2}\left(J+A_{\textstyle{_l}}J/2
 \right)\chi^{2}+O(\chi^{3})\right]=\frac{3}{2(l-1)}\left[1-\textstyle A_{\textstyle{_l}}\left(1+A_{\textstyle{_l}}/2
 \right)(\tau_{_M}\chi)^{2}+O(\chi^{3})\right]
 ~~,~
 \label{A17}
 \ea
 which can be written as
 \ba
 |k_l(\chi)|~\approx~\frac{3}{2(l-1)}~\left[\,1\,-\;\frac{1}{2}\;\left(\,\frac{\textstyle A_{\textstyle{_l}}\,J}{E}\,\right)^2~\chi^{2}
 \,+\,O(\chi^{3})\,\right]~=~
 \frac{3}{2(l-1)}~\left[\,1\,-\;\frac{1}{2}\;A_{\textstyle{_l}}^2\;(\tau_{_M}\chi)^2\,+\,O(\chi^{3})\,\right]
 ~~,~~
 \label{A18}
 \ea
 insofar as $\;A_{\textstyle{_l}}\gg 1\;$. Comparing this with (\ref{A16}), we see that, for small terrestrial moons and planets (but
 not for superearths whose $\,A_l\,$ is small), the following convenient approximation is valid, provided the Maxwell model is employed:
 \ba
 k_l(\chi)\,\approx\,k_l(0)\,\cos\epsilon(\chi)\quad,\quad\mbox{for}\quad\,\chi\,\tau_{_M}\,\ll\,1\,+\,A_{\textstyle{_l}}\quad\quad
 \mbox{where}\quad A_{\textstyle{_l}}\gg 1~~~.
 \label{A19}
 \ea

 \section{The eccentricity functions}\label{functions}

 In our development, we take into account the expansions for $\,G^2_{\textstyle{_{lpq}}}(e)\,$ over the powers of eccentricity,
 keeping the terms up to $\,e^6\,$, inclusively. The table of the eccentricity functions presented in the book by Kaula (1966)
 is not sufficient for our purposes, because some of the $\,G_{\textstyle{_{lpq}}}(e)\,$ functions in that table are given with
 lower precision. For example, the $\,e^6\,$ term is missing in the approximation for $\,G_{\textstyle{_{200}}}(e)\,$. Besides,
 that table omits several functions which are of order $\,e^3\,$. So here we provide a more comprehensive table. The table is
 based on the information borrowed from Cayley (1861) who tabulated various expansions employed in astronomy. Among those, were
 series
 \ba
 \left(\,\frac{r}{a}\,\right)^{-(l+1)}
 \left[
 \begin{array}{cc}
 \left[\,\cos{}\right]^i~  \cos\\
 \left[\,\sin{}\right]^i~  \sin
 \end{array}
\right]j\nu~=~\sum_{i\,=\,-\,\infty}^{\infty}
 \left[
 \begin{array}{cc}
 \cos\\
 \sin
 \end{array}
 \right]i{\cal{M}}~~~,
 \label{}
 \ea
 $\nu\,$ and $\,{\cal{M}}\,$ signifying the true and mean anomalies, while $\,\left[\,\cos{}\right]^i\,$ and $\,\left[\,\sin{}\right]^i\,$
 denoting the coefficients tabulated by Cayley. These coefficients are polynomials in the eccentricity. Cayley's integer indices $~i\,,\,j~$
 are connected with Kaula's integers $~l\,,\,p\,,\,q~$ via
 \ba
 l\,-\,2\,p~=~j~\quad~,\quad\quad~l\,-\,2\,p\,+\,q~=~i~~~.
 \label{}
 \ea
 With the latter equalities kept in mind, the eccentricity functions, for $\,i\,\geq\,0\,$, are related to Cayley's coefficients by
 \ba
 G_{\textstyle{_{lpq}}}(e)~=~\left[\,\cos{}\right]^i\,+~\left[\,\sin{}\right]^i~\quad,\quad\mbox{for}\quad i~\geq~0~~~.
 \label{}
 \ea
 To obtain the eccentricity functions for $\,i\,<\,0\,$, one has to keep in mind that $\,\left[\,\cos{}\right]^{-i}\,=~
 \left[\,\cos{}\right]^{i}\,$, while $\,\left[\,\sin{}\right]^i\,=~-~\left[\,\sin{}\right]^i\,$. It is then possible to
 demonstrate that
 \ba
 G_{\textstyle{_{lpq}}}(e)~=~\left[\,\cos{}\right]^i\,-~\left[\,\sin{}\right]^i~\quad,\quad\mbox{for}\quad i~<~0~~~.
 \label{}
 \ea

 Then the following expressions, for $\,l=2\,$, ensue from Cayley's tables:
 \begin{subequations}
 \ba
 G_{\textstyle{_{20~-11}}}(e)&=&G_{\textstyle{_{20~-10}}}(e)~=~G_{\textstyle{_{20~-9}}}(e)~=~G_{\textstyle{_{20~-8}}}(e)~=~0~~~,
 \label{}\\ \nonumber\\
 G_{\textstyle{_{20~-7}}}(e)&=&\frac{15 625}{129 024}~e^7 ~~~,
 \label{}\\ \nonumber\\
 G_{\textstyle{_{20~-6}}}(e)&=&\frac{4}{45}~e^6 ~~~,
 \label{}\\ \nonumber\\
 G_{\textstyle{_{20~-5}}}(e)&=&   \frac{81}{1280}~e^5 ~+~   \frac{81}{2048}~e^7 ~~~,
 \label{}\\ \nonumber\\
 G_{\textstyle{_{20~-4}}}(e)&=&  \frac{1}{24}~e^4  \,+~  \frac{7}{240}~e^6 ~~~,
 \label{}\\ \nonumber\\
 G_{\textstyle{_{20~-3}}}(e)&=&  \frac{1}{48}~e^3 \,+~    \frac{11}{768}~e^5\,+~  \frac{313}{30720}~e^7 ~~~,
 \label{}\\ \nonumber\\
 G_{\textstyle{_{20~-2}}}(e)&=&  0 ~~~,
 \label{}\\ \nonumber\\
 G_{\textstyle{_{20~-1}}}(e)&=&  -~\frac{1}{2}~e ~+~    \frac{1}{16}~e^3 \,-~  \frac{5}{384}~e^5 \,-~   \frac{143}{18432}~e^7~~~,
 \label{}
 \ea
 \ba
 G_{\textstyle{_{200}}}(e)&=&1~-~  \frac{5}{2}~e^2 \,+~   \frac{13}{16}~e^4 \,-~ \frac{35}{288}~e^6  ~~~,
 \label{}\\ \nonumber\\
 G_{\textstyle{_{201}}}(e)&=&  \frac{7}{2}~e \,-~   \frac{123}{16}~e^3 \,+~ \frac{489}{128}~e^5 \,-~   \frac{1763}{2048}~e^7
 \label{}\\ \nonumber\\
 G_{\textstyle{_{202}}}(e)&=&  \frac{17}{2}~e^2 \,-~   \frac{115}{6}~e^4 \,+~ \frac{601}{48}~e^6 ~~~,
 \label{}
 \ea
 \ba
 G_{\textstyle{_{203}}}(e)&=&  \frac{845}{48}~e^3\,-~    \frac{32 525}{768}~e^5 \,+~  \frac{208 225}{6144}~e^7~~~,
 \label{}
 \ea
 \ba
 G_{\textstyle{_{204}}}(e)&=&  \frac{533}{16}~e^4  \,-~  \frac{13 827}{160}~e^6 ~~~,
  \label{}\\ \nonumber\\
 G_{\textstyle{_{205}}}(e)&=&  \frac{228 347}{3 840}~e^5 \,-~   \frac{3 071 075}{18 432}~e^7 ~~~,
 \label{}\\ \nonumber\\
 G_{\textstyle{_{206}}}(e)&=&  \frac{73 369}{720}~e^6 ~~~,
 \label{}\\ \nonumber\\
  G_{\textstyle{_{207}}}(e)&=&  \frac{12 144 273}{71 680}~e^7 ~~~,
 \label{}
 \ea
 \label{Gl=2}
 \end{subequations}
 the other values of $\,q\,$ generating polynomials $\,G_{\textstyle{_{20q}}}(e)\,$ whose leading terms are of order $\,e^8\,$ and higher.

 Since in our study we intend to employ the squares of these functions, with terms up to $\,e^6\,$ only,
 then we may completely omit
 the eccentricity functions with $\,|q|\,\geq\,4\,$. In our approximation, the squares of the eccentricity functions will look:
 \begin{subequations}
 \ba
 G^{\,2}_{\textstyle{_{\textstyle{_{20~-3}}}}}(e)&=&\frac{1}{2304}~e^6~+\,O(e^8)~~~,\\
 \nonumber\\
 G^{\,2}_{\textstyle{_{\textstyle{_{20~-2}}}}}(e)&=&0~~~,\\
 \nonumber\\
 G^{\,2}_{\textstyle{_{\textstyle{_{20~-1}}}}}(e)&=&\frac{1}{4}~e^2-\,\frac{1}{16}~e^4~+\,\frac{13}{768}~e^6+\,O(e^8)~~~,\\
 \nonumber\\
 G^{\,2}_{\textstyle{_{\textstyle{_{200}}}}}(e)~&=&1\,-\,5\,e^2\,+\;\frac{63}{8}\;e^4\;-\;\frac{155}{36}~e^6~
 +\;O(e^8)\;\;\;\,,\;\;\;\\
 \nonumber\\
 G^{\,2}_{\textstyle{_{\textstyle{_{20{{1}}}}}}}(e)~&=&\frac{49}{4}\;e^2\;-\;
 \frac{861}{16}\;e^4\;+~\frac{21975}{256}~e^6~+~O(e^8)~~~,\\
 \nonumber\\
 G^{\,2}_{\textstyle{_{\textstyle{_{20{{2}}}}}}}(e)~&=&\frac{289}{4}\,e^4\,-\,\frac{1955}{6}~e^6\,+~
 \,O(e^8)~~~,\\
 \nonumber\\
 G^{\,2}_{\textstyle{_{\textstyle{_{20{{3}}}}}}}(e)~&=&\frac{714025}{2304}\,e^6\,+~
 \,O(e^8)~~~,
 \ea
 \label{formulae1}
 \end{subequations}
 the squares of the others being of the order of $\,e^8\,$ or higher.

 Be mindful that, for $\,l=2\,$ we considered only the functions with $\,p=0\,$. This is dictated by the fact that the inclination functions
 $\,F_{\textstyle{_{lmp}}}\,=\,F_{\textstyle{_{22p}}}\,$ are of order $\,i\,$ (and, accordingly, their squares and cross-products are of
 order $\,i^2\,$) for all the values of $\,p\,$ except zero.

 For $\,l=3\,$, the situation changes. The inclination functions $\,F_{lmp}\,=\,F_{310}\,,\,F_{312}\,,\,F_{313}\,,\,F_{320}\,,$
 $F_{321}\,,
 \,F_{322}\,,\,F_{323}\,,\,F_{331}\,,
 \,F_{332}\,,\,F_{333}\,$ are of the order $\,O(\inc)\,$ or higher. The terms containing the squares or cross-products of the these functions
 may thus be omitted. (Specifically, by neglecting the cross-terms we get rid of the mixed-period part of the $\,l=3\,$ component.)
 What is left is the terms with $\,lmp\,=\,311\,$ and $\,lmp\,=\,330\,$. These terms contain the squares of functions
 \ba
 F_{311}(\inc)\,=~-~\frac{3}{2}\,+\,O(\inc^2)~~~~~~\mbox{and}~~~~~~F_{330}(\inc)\,=\,15\,+\,O(\inc^2)\;\;\;~~~,~~~~~
 \label{bbv}
 \ea
 accordingly. From here, we see that, for $\,l=3\,$, we shall need to employ the eccentricity functions $\,G_{\textstyle{_{lpq}}}(e)\,=\,
 G_{\textstyle{_{30q}}}(e)\,$ and $\,G_{\textstyle{_{lpq}}}(e)\,=\,G_{\textstyle{_{31q}}}(e)\,$.

 The following expressions, for $\,l=3\,$ and $\,p=0\,$, ensue from Cayley's tables:
 \begin{subequations}
 \ba
 G_{\textstyle{_{30~-11}}}(e)&=&G_{\textstyle{_{30~-10}}}(e)~=~G_{\textstyle{_{30~-9}}}(e)~=~G_{\textstyle{_{30~-8}}}(e)~=~0~~~,
 \label{}\\ \nonumber\\
 G_{\textstyle{_{30~-7}}}(e)&=&\frac{8}{315}~e^7 ~~~,
 \label{}\\ \nonumber\\
 G_{\textstyle{_{30~-6}}}(e)&=&\frac{81}{5120}~e^6 ~~~,
 \label{}\\ \nonumber\\
 G_{\textstyle{_{30~-5}}}(e)&=&\frac{1}{120}~e^5 ~+~   \frac{13}{1440}~e^7 ~~~,
 \label{}\\ \nonumber\\
 G_{\textstyle{_{30~-4}}}(e)&=&  \frac{1}{384}~e^4  \,+~  \frac{1}{384}~e^6 ~~~,
 \ea
 \ba
 G_{\textstyle{_{30~-3}}}(e)&=&  0 ~~~,
 \label{}\\ \nonumber\\
 G_{\textstyle{_{30~-2}}}(e)&=&  \frac{1}{8}~e^2 \,+~    \frac{1}{48}~e^4\,+~  \frac{55}{3072}~e^6 ~~~,
 \label{}\\ \nonumber\\
 G_{\textstyle{_{30~-1}}}(e)&=&  - ~e ~+~    \frac{5}{4}~e^3 \,-~  \frac{7}{48}~e^5 \,+~   \frac{23}{288}~e^7~~~,
 \label{}\\ \nonumber\\
 G_{\textstyle{_{300}}}(e)&=&1~-~6~e^2 \,+~   \frac{423}{64}~e^4 \,-~ \frac{125}{64}~e^6  ~~~,
 \label{}\\ \nonumber\\
 G_{\textstyle{_{301}}}(e)&=& 5~e \,-~ 22~e^3 \,+~ \frac{607}{24}~e^5 \,-~   \frac{98}{9}~e^7
 \label{}\\ \nonumber\\
 G_{\textstyle{_{302}}}(e)&=&  \frac{127}{8}~e^2 \,-~   \frac{3065}{48}~e^4 \,+~ \frac{243805}{3072}~e^6 ~~~,
 \label{}\\ \nonumber\\
 G_{\textstyle{_{303}}}(e)&=&  \frac{163}{4}~e^3\,-~    \frac{2577}{16}~e^5 \,+~  \frac{1089}{5}~e^7~~~,
 \label{}\\ \nonumber\\
 G_{\textstyle{_{304}}}(e)&=&  \frac{35413}{384}~e^4  \,-~  \frac{709471}{1920}~e^6 ~~~,
  \label{}\\ \nonumber\\
 G_{\textstyle{_{305}}}(e)&=&  \frac{23029}{120}~e^5 \,-~   \frac{35614}{45}~e^7 ~~~,
 \label{}
 \ea
 \ba
 G_{\textstyle{_{306}}}(e)&=&  \frac{385095}{1024}~e^6 ~~~,
 \label{}\\ \nonumber\\
  G_{\textstyle{_{307}}}(e)&=&  \frac{44377}{63}~e^7 ~~~,
 \label{}
 \ea
 \label{Gl=3}
 \end{subequations}
 the other values of $\,q\,$ generating polynomials $\,G_{\textstyle{_{30q}}}(e)\,$ and $\,G_{\textstyle{_{31q}}}(e)\,$, whose leading
 terms are of order $\,e^8\,$ and higher.

 The squares of some these functions, will read, up to $\,e^6\,$ terms inclusively, as:
 \begin{subequations}
 \ba
 G^{\,2}_{\textstyle{_{\textstyle{_{30~-3}}}}}(e)&=&0~~~,\\
 \nonumber\\
 G^{\,2}_{\textstyle{_{\textstyle{_{30~-2}}}}}(e)&=&\frac{1}{64}~e^4~+~\frac{1}{192}~e^6~+\,O(e^8)~~~,\\
 \nonumber\\
 G^{\,2}_{\textstyle{_{\textstyle{_{30~-1}}}}}(e)&=&e^2-\,\frac{5}{2}~e^4~+\,\frac{89}{48}~e^6+\,O(e^8)~~~,\\
 \nonumber\\
 G^{\,2}_{\textstyle{_{\textstyle{_{300}}}}}(e)~&=&1~-~12~e^2~+\;\frac{1575}{32}\;e^4\;-\;\frac{2663}{32}~e^6~
 +\;O(e^8)\;\;\;\,,\;\;\;\\
 \nonumber\\
 G^{\,2}_{\textstyle{_{\textstyle{_{30{{1}}}}}}}(e)~&=&25\;e^2\;-\;
 220\;e^4\;+~\frac{8843}{12}~e^6~+~O(e^8)~~~,\\
 \nonumber\\
 G^{\,2}_{\textstyle{_{\textstyle{_{30{{2}}}}}}}(e)~&=&\frac{16129}{64}\,e^4\,-\,\frac{389255}{192}~e^6\,+~
 \,O(e^8)~~~,\\
 \nonumber\\
 G^{\,2}_{\textstyle{_{\textstyle{_{30{{3}}}}}}}(e)~&=&\frac{26569}{16}~e^6\,+~
 \,O(e^8)~~~,
 \ea
 \label{formulae2}
 \end{subequations}
 the squares of the others being of the order of $\,e^8\,$ or higher.

 Finally, we write down the expressions for the eccentricity functions with $\,l=3\,$ and $\,p=1\,$:
 \begin{subequations}
 \ba
 G_{\textstyle{_{31~-9}}}(e)&=&G_{\textstyle{_{31~-8}}}(e)~=~0~~~,
 \label{}\\ \nonumber\\
 G_{\textstyle{_{31~-7}}}(e)&=&\frac{16337}{2240}~e^7 ~~~,
 \label{}\\ \nonumber\\
 G_{\textstyle{_{31~-6}}}(e)&=&\frac{48203}{9216}~e^6 ~~~,
 \label{}\\ \nonumber\\
 G_{\textstyle{_{31~-5}}}(e)&=&\frac{899}{240}~e^5 ~+~   \frac{2441}{480}~e^7 ~~~,
 \label{}\\ \nonumber\\
 G_{\textstyle{_{31~-4}}}(e)&=&  \frac{343}{128}~e^4  \,+~  \frac{2819}{640}~e^6 ~~~,
 \label{}\\ \nonumber\\
 G_{\textstyle{_{31~-3}}}(e)&=&  \frac{23}{12}~e^3~+~\frac{89}{24}~e^5~+~\frac{5663}{960}~e^7 ~~~,
 \label{}\\ \nonumber\\
 G_{\textstyle{_{31~-2}}}(e)&=&  \frac{11}{8}~e^2 \,+~    \frac{49}{16}~e^4\,+~  \frac{15665}{3072}~e^6 ~~~,
 \label{}
 \\ \nonumber\\
 G_{\textstyle{_{31~-1}}}(e)&=&  ~e ~+~    \frac{5}{2}~e^3 \,+~  \frac{35}{8}~e^5 \,+~   \frac{105}{16}~e^7~~~,
 \label{}
 \ea
 \ba
 G_{\textstyle{_{310}}}(e)&=&1~+~2~e^2 \,+~   \frac{239}{64}~e^4 \,+~ \frac{3323}{576}~e^6  ~~~,
 \label{}\\ \nonumber\\
 G_{\textstyle{_{311}}}(e)&=& 3~e \,+~\frac{11}{4}~e^3 \,+~ \frac{245}{48}~e^5 \,+~\frac{463}{64}~e^7
 \label{}
 \ea
 \ba
 G_{\textstyle{_{312}}}(e)&=&\frac{53}{8}~e^2 \,+~\frac{39}{16}~e^4 \,+~ \frac{7041}{1024}~e^6 ~~~,
 \label{}\\ \nonumber\\
 G_{\textstyle{_{313}}}(e)&=&  \frac{77}{6}~e^3\,-~    \frac{25}{48}~e^5 \,+~  \frac{4751}{480}~e^7~~~,
 \label{}\\ \nonumber\\
 G_{\textstyle{_{314}}}(e)&=&  \frac{2955}{128}~e^4  \,-~  \frac{3463}{384}~e^6 ~~~,
  \label{}\\ \nonumber\\
 G_{\textstyle{_{315}}}(e)&=&  \frac{3167}{80}~e^5 \,-~   \frac{8999}{320}~e^7 ~~~,
 \label{}\\ \nonumber\\
 G_{\textstyle{_{316}}}(e)&=&  \frac{3024637}{46080}~e^6 ~~~,
 \label{}\\ \nonumber\\
  G_{\textstyle{_{317}}}(e)&=&  \frac{178331}{1680}~e^7 ~~~,
 \label{}
 \ea
 \label{Gl}
 \end{subequations}
 and the squares:
\begin{subequations}
 \ba
 G^{\,2}_{\textstyle{_{\textstyle{_{31~-3}}}}}(e)&=&\frac{529}{144}~e^6~+\,O(e^8)~~~,\\
 \nonumber\\
 G^{\,2}_{\textstyle{_{\textstyle{_{31~-2}}}}}(e)&=&\frac{121}{64}~e^4~+~\frac{539}{64}~e^6~+\,O(e^8)~~~,\\
 \nonumber\\
 G^{\,2}_{\textstyle{_{\textstyle{_{31~-1}}}}}(e)&=&e^2+\,5~e^4~+\,15~e^6+\,O(e^8)~~~,\\
 \nonumber\\
 G^{\,2}_{\textstyle{_{\textstyle{_{310}}}}}(e)~&=&1~+\;4\;e^2\;+\;\frac{367}{32}~e^4\,+~\frac{7625}{288}~e^6~
 +\;O(e^8)\;\;\;\,,
 \ea
 \ba
 G^{\,2}_{\textstyle{_{\textstyle{_{31{{1}}}}}}}(e)~&=&9\;e^2\;+\;
 \frac{33}{2}\;e^4\;+~\frac{611}{16}~e^6~+~O(e^8)~~~,\\
 \nonumber\\
 G^{\,2}_{\textstyle{_{\textstyle{_{31{{2}}}}}}}(e)~&=&\frac{2809}{64}\,e^4\,+\,\frac{2067}{64}~e^6\,+~
 \,O(e^8)~~~,\\
 \nonumber\\
 G^{\,2}_{\textstyle{_{\textstyle{_{31{{3}}}}}}}(e)~&=&\frac{5929}{36}~e^6\,+~
 \,O(e^8)~~~,
 \ea
 \label{formulae3}
 \end{subequations}
 the squares of the other functions from this set being of the order $\,e^8\,$ or higher.

 \section{The $\,l=2\,$ and $\,l=3\,$ terms of the secular part of the torque}\label{secular}

 \subsection{The $\,l=2\,$ terms of the secular torque}

 Extracting the $\,l=2\,$ input from (\ref{gedo}), we recall that only the $\,(lmpq)=(220q)\,$ terms matter. Out of these terms, we need
 only the ones up to $\,e^6\,$. These are the terms with $\,|q|\,\leq\,3\,$. They are given by formulae (\ref{formulae1}) from Appendix
 \ref{functions}. Employing those formulae, we arrive at
 \begin{subequations}
 \ba
 \overline{\cal T}_{\textstyle{_{\textstyle_{\textstyle{_{l=2}}}}}}
 &=&\overline{\cal T}_{\textstyle{_{\textstyle_{\textstyle{_{(lmp)=(220)}}}}}}+\,O(\inc^2\,\epsilon)=\,\frac{3}{2}\,G\,M_{sec}^2\,R^5\,a^{-6}
 \sum_{q=-3}^{3}\,G^{\textstyle{^{\,2}}}_{\textstyle{_{\textstyle{_{20\mbox{\it{q}}}}}}}(e)~
 k_2\;\sin\epsilon_{\textstyle{_{\textstyle{_{220\mbox{\it{q}}}}}}}+\,O(e^8\,\epsilon)\,+\,O(\inc^2\,\epsilon)\quad\quad\quad\quad
 \label{geoid}\\
 \nonumber\\
 \nonumber\\
 \nonumber
 &=&\frac{3}{2}\;G\;M_{sec}^2\,R^5\,a^{-6}\,\left[~
 \frac{1}{2304}~e^6~k_2~\sin\epsilon_{\textstyle{_{\textstyle{_{220~-3}}}}}~+
 \left(~ \frac{1}{4}~e^2-\,\frac{1}{16}~e^4~+\,\frac{13}{768}~e^6 ~\right)~k_2~\sin\epsilon_{\textstyle{_{\textstyle{_{220~-1}}}}}\right.\\
 \nonumber\\
 \nonumber\\
 \nonumber
 &+&\left(1\,-\,5\,e^2\,+~\frac{63}{8}\;e^4\,-~\frac{155}{36}~e^6\right)\,k_2~\sin\epsilon_{\textstyle{_{\textstyle{_{2200}}}}}\,+
 \left(\frac{49}{4}~e^2-\;\frac{861}{16}\;e^4+~\frac{21975}{256}~e^6\right)\,k_2~\sin\epsilon_{\textstyle{_{\textstyle{_{2201}}}}}\\
 \nonumber\\
 \nonumber\\
   &+& \left.\left(~ \frac{289}{4}\,e^4\,-\,\frac{1955}{6}~e^6 ~\right)~k_2~\sin\epsilon_{\textstyle{_{\textstyle{_{2202}}}}}\,+
 \frac{714025}{2304}\,e^6 ~k_2~\sin\epsilon_{\textstyle{_{\textstyle{_{2203}}}}}
 \,\right]+\,O(e^8\,\epsilon)\,+\,O(\inc^2\,\epsilon)~~~,\quad\quad\quad\quad
 \label{161b}
 \ea
 \end{subequations}
 where the absolute error $\,O(e^8\,\epsilon)\,$ has emerged because of our neglect of terms with $\,|q|\,\geq\,4\,\,$, while the absolute
 error $\,O(i^2\,\epsilon)\,$ came into being after the truncation of terms with $\,p\,\geq\,1\,$.

 Recalling expression (\ref{1000000}), we can rewrite (\ref{161b}) in a form indicating explicitly at which resonance each term changes its
 sign. To this end, each $~k_l~\sin\epsilon_{\textstyle{_{
 \textstyle{_{lmpq}}}}}~$ will be rewritten as: $~k_l~\sin|\epsilon_{\textstyle{_{\textstyle{_{lmpq}}}}}|\,\mbox{sgn}\,\left[\,({\it l}-2p+q)\,n\,-
 \,m\,\dot{\theta}\,\right]~$. This will render:
 \ba
 \nonumber
 \overline{\cal T}_{\textstyle{_{\textstyle_{\textstyle{_{l=2}}}}}}
 &=&\frac{3}{2}~G\;M_{sec}^2\,R^5\,a^{-6}\,\left[~\frac{1}{2304}~e^6~k_2~\sin|\epsilon_{\textstyle{_{\textstyle{_{220~-3}}}}}|~~
 \mbox{sgn}\,\left(\,-\,n\,-\,2\,\dot{\theta}\,\right) \right.\\
 \nonumber\\
 \nonumber\\
 \nonumber
 &+&\left(~ \frac{1}{4}~e^2-\,\frac{1}{16}~e^4~+\,\frac{13}{768}~e^6 ~\right)~k_2~\sin|\epsilon_{\textstyle{_{\textstyle{_{220~-1}}}}}|
 ~~\mbox{sgn}\,\left(\,n\,-\,2\,\dot{\theta}\,\right)
 ~\\
 \nonumber\\
 \nonumber\\
 \nonumber
 &+&\left(1\,-\,5\,e^2\,+~\frac{63}{8}\;e^4\,-~\frac{155}{36}~e^6\right)~k_2~\sin|\epsilon_{\textstyle{_{\textstyle{_{2200}}}}}|~~
 ~~\mbox{sgn}\,\left(\,n\,-\,\dot{\theta}\,\right)\\
 \nonumber\\
 \nonumber\\
 \nonumber
 &+&\left(\frac{49}{4}~e^2-\;\frac{861}{16}\;e^4+~\frac{21975}{256}~e^6\right)~k_2~\sin|\epsilon_{\textstyle{_{\textstyle{_{2201}}}}}|
 ~~\mbox{sgn}\,\left(\,3\,n\,-\,2\,\dot{\theta}\,\right)\\
 \nonumber\\
 \nonumber\\
 \nonumber
 &+& \left(~ \frac{289}{4}\,e^4\,-\,\frac{1955}{6}~e^6 ~\right)~k_2~\sin|\epsilon_{\textstyle{_{\textstyle{_{2202}}}}}|
 ~\mbox{sgn}\,\left(\,2\,n\,-\,\dot{\theta}\,\right)\\
 \nonumber\\
 \nonumber\\
 &+&\left.\frac{714025}{2304}~e^6 ~k_2~\sin|\epsilon_{\textstyle{_{\textstyle{_{2203}}}}}|~~\mbox{sgn}\,\left(\,
5\,n\,-\,3\,\dot{\theta}\,\right)
 ~\,\right]+\,O(e^8\,\epsilon)\,+\,O(\inc^2\,\epsilon)~~~,\quad\quad\quad\quad
 \label{T2}
 \ea

 \subsection{The $\,l=3\,,~m=1\,$ terms of the secular torque}

 Getting the $\,l=3\,,\,m=1\,$ input from (\ref{gedo}) and leaving in it only the terms up to $\,e^6\,$, we obtain, with aid of
 formulae (\ref{formulae1}) from Appendix \ref{functions}, the following expression:
 \begin{subequations}
 \ba
 \overline{\cal T}_{\textstyle{_{\textstyle_{\textstyle{_{(lmp)=(311)}}}}}}&=&\frac{3}{8}\,G\,M_{sec}^2\,R^7\,a^{-8}\sum_{q=-3}^{3}\,
 G^{\textstyle{^{\,2}}}_{\textstyle{_{\textstyle{_{31\mbox{\it{q}}}}}}}(e)~
 k_3\;\sin\epsilon_{\textstyle{_{\textstyle{_{311\mbox{\it{q}}}}}}}~+\,O(e^8\,\epsilon)
 \label{}\\
 \nonumber\\
 \nonumber\\
 \nonumber
 &=&\frac{3}{8}\;G\;M_{sec}^2\,R^7\,a^{-8}\,\left[~\,
 \frac{529}{144}~e^6~k_3~\sin\epsilon_{\textstyle{_{\textstyle{_{311~-3}}}}}~+
 \left(~\frac{121}{64}~e^4~+~\frac{539}{64}~e^6~\right)~k_3~\sin\epsilon_{\textstyle{_{\textstyle{_{311~-2}}}}}\right.\\
 \nonumber\\
 \nonumber\\
 \nonumber
 &+&\left(e^2+\,5\,e^4+\,15\,e^6\right)\,k_3~\sin\epsilon_{\textstyle{_{\textstyle{_{311~-1}}}}}+\,
 \left(1\,+\,4\,e^2+\;\frac{367}{32}~e^4+~\frac{7625}{288}~e^6\right)\,k_3~\sin\epsilon_{\textstyle{_{\textstyle{_{3110}}}}}\\
 \nonumber\\
 \nonumber\\
 \nonumber
 &+&\left(9\;e^2\;+\;\frac{33}{2}\;e^4\;+~\frac{611}{16}~e^6\right)\,k_3~\sin\epsilon_{\textstyle{_{\textstyle{_{3111}}}}}~+~
 \left(\frac{2809}{64}\,e^4\,+\,\frac{2067}{64}~e^6\right)\,k_3~\sin\epsilon_{\textstyle{_{\textstyle{_{3112}}}}} \\
 \nonumber\\
 \nonumber\\
   &+& \left.
 \frac{5929}{36}\,e^6 \,k_3~\sin\epsilon_{\textstyle{_{\textstyle{_{3113}}}}}
 \,\right]+\,O(e^8\,\epsilon)~~~.\quad\quad\quad\quad\quad\quad\quad\quad\quad\quad\quad\quad\quad\quad\quad\quad\quad\quad\quad\quad
 \label{}
 \ea
 \end{subequations}
 With the signs depicted explicitly, this will look:
 \ba
 \nonumber
 \overline{\cal T}_{\textstyle{_{\textstyle_{\textstyle{_{(lmp)=(311)}}}}}}
 &=&\frac{3}{8}\;G\;M_{sec}^2\,R^7\,a^{-8}\,\left[~\,
 \frac{529}{144}~e^6~k_3~\sin|\epsilon_{\textstyle{_{\textstyle{_{311~-3}}}}}|~\,\mbox{sgn}\,\left(\,-\,2\,n\,-
 \,\dot{\theta}\,\right)\right.\\
 \nonumber\\
 \nonumber\\
 \nonumber
 &+&
 \left(~\frac{121}{64}~e^4~+~\frac{539}{64}~e^6~\right)~k_3~\sin|\epsilon_{\textstyle{_{\textstyle{_{311~-2}}}}}|~
 \,\mbox{sgn}\,\left(\,-\,n\,-\,\dot{\theta}\,\right)\\
 \nonumber\\
 \nonumber\\
 \nonumber
 &+&\left(e^2+\,5\,e^4+\,15\,e^6\right)\,k_3~\sin|\epsilon_{\textstyle{_{\textstyle{_{311~-1}}}}}|~
 \,\mbox{sgn}\,\left(\,-\,\dot{\theta}\,\right)\\
 \nonumber\\
 \nonumber\\
 \nonumber
 &+&
 \left(1\,+\,4\,e^2+\;\frac{367}{32}~e^4+~\frac{7625}{288}~e^6\right)\,k_3~\sin|\epsilon_{\textstyle{_{\textstyle{_{3110}}}}}|~
 \,\mbox{sgn}\,\left(\,n\,-
 \,\dot{\theta}\,\right)\\
 \nonumber\\
 \nonumber\\
 \nonumber
 &+&\left(9\;e^2\;+\;\frac{33}{2}\;e^4\;+~\frac{611}{16}~e^6\right)\,k_3~\sin|\epsilon_{\textstyle{_{\textstyle{_{3111}}}}}|~
 \,\mbox{sgn}\,\left(\,2\,n\,-
 \,\dot{\theta}\,\right)\\
 \nonumber\\
 \nonumber\\
 \nonumber
 &+&
 \left(\frac{2809}{64}\,e^4\,+\,\frac{2067}{64}~e^6\right)\,k_3~\sin|\epsilon_{\textstyle{_{\textstyle{_{3112}}}}}|~
 \,\mbox{sgn}\,\left(\,3\,n\,-
 \,\dot{\theta}\,\right) \\
 \nonumber\\
 \nonumber\\
   &+& \left.
 \frac{5929}{36}~e^6 \,k_3~\sin|\epsilon_{\textstyle{_{\textstyle{_{3113}}}}}|~\,\mbox{sgn}\,\left(\,4\,n\,-
 \,\dot{\theta}\,\right)
 \,\right]+\,O(e^8\,\epsilon)~~~.\quad\quad\quad\quad\quad\quad\quad\quad
 \label{T31}
 \ea

 \subsection{The $\,l=3\,,~m=3\,$ terms of the secular torque}

 The second relevant group of terms with $\,l=3\,$ will read:
 \begin{subequations}
 \ba
 \overline{\cal T}_{\textstyle{_{\textstyle_{\textstyle{_{(lmp)=(330)}}}}}}&=&\frac{15}{8}\;G\;M_{sec}^2\,R^7\,a^{-8}\,\sum_{q=-3}^{3}\,
 G^{\textstyle{^{\,2}}}_{\textstyle{_{\textstyle{_{30\mbox{\it{q}}}}}}}(e)~
 k_3\;\sin\epsilon_{\textstyle{_{\textstyle{_{330\mbox{\it{q}}}}}}}~+\,O(e^8\,\epsilon)
 \label{}\\
 \nonumber\\
 \nonumber\\
 \nonumber
 &=&\frac{15}{8}\;G\;M_{sec}^2\,R^7\,a^{-8}\,\left[~\,
 \left(~\frac{1}{64}~e^4~+~\frac{1}{192}~e^6~\right)~k_3~\sin\epsilon_{\textstyle{_{\textstyle{_{330~-2}}}}}\right.\\
 \nonumber\\
 \nonumber\\
 \nonumber
 &+&\left(e^2-\,\frac{5}{2}\,e^4+\,\frac{89}{48}\,e^6\right)\,k_3~\sin\epsilon_{\textstyle{_{\textstyle{_{330~-1}}}}}+\,
 \left(1\,-\,12\,e^2+\;\frac{1575}{32}~e^4-~\frac{2663}{32}~e^6\right)\,k_3~\sin\epsilon_{\textstyle{_{\textstyle{_{3300}}}}}\\
 \nonumber\\
 \nonumber\\
 \nonumber
 &+&\left(25~e^2\,-~220~e^4\,+~\frac{8843}{12}~e^6\right)\,k_3~\sin\epsilon_{\textstyle{_{\textstyle{_{3301}}}}}~+~
 \left(\frac{16129}{64}\,e^4\,-~\frac{389255}{192}~e^6\right)\,k_3~\sin\epsilon_{\textstyle{_{\textstyle{_{3302}}}}} \\
 \nonumber\\
 \nonumber\\
   &+& \left.
 \frac{26569}{16}\,e^6 \,k_3~\sin\epsilon_{\textstyle{_{\textstyle{_{3303}}}}}
 \,\right]+\,O(e^8\,\epsilon)\quad\quad\quad\quad\quad\quad\quad\quad\quad\quad\quad\quad\quad\quad\quad\quad\quad\quad\quad\quad
 \label{}
 \ea
 \end{subequations}
 or, with the signs shown explicitly:
 \ba
 \nonumber
 \overline{\cal T}_{\textstyle{_{\textstyle_{\textstyle{_{(lmp)=(330)}}}}}}
 &=&\frac{15}{8}\;G\;M_{sec}^2\,R^7\,a^{-8}\,\left[~\,
 \left(~\frac{1}{64}~e^4~+~\frac{1}{192}~e^6~\right)~k_3~\sin|\epsilon_{\textstyle{_{\textstyle{_{330~-2}}}}}|~
 \,\mbox{sgn}\,\left(\,-\,n\,-\,\dot{\theta}\,\right) \right.\\
 \nonumber\\
 \nonumber\\
 \nonumber
 &+&\left(e^2-\,\frac{5}{2}\,e^4+\,\frac{89}{48}\,e^6\right)\,k_3~\sin|\epsilon_{\textstyle{_{\textstyle{_{330~-1}}}}}|~
 \,\mbox{sgn}\,\left(\,-\,\dot{\theta}\,\right)\\
 \nonumber\\
 \nonumber\\
 \nonumber
 &+&
 \left(1\,-\,12\,e^2+\;\frac{1575}{32}~e^4-~\frac{2663}{32}~e^6\right)\,k_3~\sin|\epsilon_{\textstyle{_{\textstyle{_{3300}}}}}|~
 \,\mbox{sgn}\,\left(\,n\,-
 \,\dot{\theta}\,\right)\\
 \nonumber\\
 \nonumber\\
 \nonumber
 &+&\left(25~e^2\,-~220~e^4\,+~\frac{8843}{12}~e^6\right)\,k_3~\sin|\epsilon_{\textstyle{_{\textstyle{_{3301}}}}}|~
 \,\mbox{sgn}\,\left(\,2\,n\,-
 \,\dot{\theta}\,\right)\\
 \nonumber\\
 \nonumber\\
 \nonumber
 &+&
 \left(\frac{16129}{64}\,e^4\,-\,\frac{389255}{192}~e^6\right)\,k_3~\sin|\epsilon_{\textstyle{_{\textstyle{_{3302}}}}}|~
 \,\mbox{sgn}\,\left(\,3\,n\,-
 \,\dot{\theta}\,\right) \\
 \nonumber\\
 \nonumber\\
   &+& \left.
 \frac{26569}{16}~e^6 \,k_3~\sin|\epsilon_{\textstyle{_{\textstyle{_{3303}}}}}|~\,\mbox{sgn}\,\left(\,4\,n\,-
 \,\dot{\theta}\,\right)
 \,\right]+\,O(e^8\,\epsilon)~~~.\quad\quad\quad\quad\quad\quad\quad\quad
 \label{T33}
 \label{T32}
 \ea

 \section{The $\,l=2\,$ and $\,l=3\,$ terms of the short-period part of the torque}\label{short}

 The short-period part of the torque may be approximated with terms of degrees 2 and 3:
 \ba
 \nonumber
 \widetilde{\cal T}&=&\widetilde{\cal T}_{\textstyle{_{\textstyle{_{l=2}}}}}\,+~\widetilde{\cal T}_{\textstyle{_{\textstyle{_{l=3}}}}}\,
 +~O\left(\,\epsilon\,(R/a)^9\,\right)~\\
 \nonumber\\
 \nonumber\\
 &=&\widetilde{\cal T}_{\textstyle{_{\textstyle{_{(lmp)=(220)}}}}}\,+\left[\,\widetilde{\cal T}_{\textstyle{_{
 \textstyle{_{(lmp)=(311)}}}}}\,
 +~\widetilde{\cal T}_{\textstyle{_{\textstyle{_{(lmp)=(330)}}}}}  \right]~
  +~O(\epsilon\,i^2)    ~+~O\left(\,\epsilon\,(R/a)^9\,\right)~~~,\quad\quad
 \label{Ra}
 \ea
 where
 \begin{subequations}
 \ba
 \nonumber
 \widetilde{\cal T}_{\textstyle{_{\textstyle{_{(lmp)=(220)}}}}}=~3\,{G}\,M_{sec}^{\textstyle{^{2}}}\,R^{\textstyle{^{5}}}\,a^{-6}
 \sum_{q=-3~}^{3}{\sum_{\stackrel{\textstyle{^{~j=-3}}}{\textstyle{^{j~<~q}}}}^{3}}G_{20q}(e)~G_{20j}(e)~\left\{
 ~\cos\left[\,{\cal{M}}\,(q-j)\,\right]~\,k_{\textstyle{_2}}\,\sin\epsilon_{\textstyle{_{\textstyle{_{220q}}}}} \right.\quad\quad\quad\\
 \left.-~\sin\left[\,{\cal{M}}\,(q-j)\,\right]~\,k_{\textstyle{_2}}\,\cos\epsilon_{\textstyle{_{\textstyle{_{220q}}}}}~\right\}~+\,O(i^2\,
 \epsilon)\,+\,O(e^7\,\epsilon)~~,\quad\quad\quad\quad\quad
 \label{Til2_a}
 \ea
 \ba
 \left.\quad\quad~\quad\quad\right.=\,-\,3{G}\,M_{sec}^{\textstyle{^{2}}}\,R^{\textstyle{^{5}}}a^{-6}\sum_{q=-3~}^{3}{\sum_{\stackrel{\textstyle{^{~j=-3}}}{
 \textstyle{^{j~<~q}}}}^{3}}G_{20q}(e)\,G_{20j}(e)\,k_{\textstyle{_2}}\,\sin\left[{\cal{M}}\,(q-j)\right]+O(i^2\epsilon)+O(e\epsilon)~~,\quad~\quad\quad
 \label{Til2_b}
 \ea
 \label{Til2}
 \end{subequations}
 ~\\
 \begin{subequations}
 \ba
 \nonumber
 \widetilde{\cal T}_{\textstyle{_{\textstyle{_{(lmp)=(311)}}}}}=\,\frac{3}{4}\,{G}\,M_{sec}^{\textstyle{^{2}}}\,R^{\textstyle{^{7}}}\,a^{-8}
 \sum_{q=-3~}^{3}{\sum_{\stackrel{\textstyle{^{~j=-3}}}{\textstyle{^{j\;<\;q}}}}^{3}} G_{31q}(e)\;G_{31j}(e)
 ~\left\{\cos\left[\,{\cal{M}}\,(q-j)\,\right]~\,k_{\textstyle{_3}}\,\sin\epsilon_{\textstyle{_{\textstyle{_{311q}}}}}\right.\quad\quad\quad\\
 \left.-~\sin\left[\,{\cal{M}}\,(q-j)\,\right]~\,k_{\textstyle{_2}}\,\cos\epsilon_{\textstyle{_{\textstyle{_{311q}}}}}~\right\}+\,O(i^2\,\epsilon)\,+\,O(e^7\,\epsilon)~~,\quad\quad\quad\quad\quad
 \label{Til31_a}
 \ea
 \ba
 \left.\quad\quad~\quad\quad\right.=\,-\,\frac{3}{4}\,{G}\,M_{sec}^{\textstyle{^{2}}}\,R^{\textstyle{^{5}}}a^{-6}\sum_{q=-3~}^{3}{\sum_{\stackrel{\textstyle{^{~j=-3}}}{
 \textstyle{^{j~<~q}}}}^{3}}G_{31q}(e)\,G_{31j}(e)\,k_{\textstyle{_2}}\,\sin\left[{\cal{M}}\,(q-j)\right]+O(i^2\epsilon)+O(e\epsilon)~~,~~\quad~\quad
 \label{Til31_b}
 \ea
 \label{Til31}
 \end{subequations}
 ~\\
 \begin{subequations}
 \ba
 \nonumber
 \widetilde{\cal T}_{\textstyle{_{\textstyle{_{(lmp)=(330)}}}}}=\,\frac{15}{4}\,{G}\,M_{sec}^{\textstyle{^{2}}}\,R^{\textstyle{^{7}}}\,a^{-8}
 \sum_{q=-3~}^{3}{\sum_{\stackrel{\textstyle{^{~j=-3}}}{\textstyle{^{j\;<\;q}}}}^{3}}G_{30q}(e)\;G_{30j}(e)
 ~\cos\left[\,{\cal{M}}\,(q-j)\,\right]~\,k_{\textstyle{_3}}\,\sin\epsilon_{\textstyle{_{\textstyle{_{330q}}}}}\quad\quad\quad\\
  \left.-~\sin\left[\,{\cal{M}}\,(q-j)\,\right]~\,k_{\textstyle{_2}}\,\cos\epsilon_{\textstyle{_{\textstyle{_{330q}}}}}~\right\}+\,O(i^2\,\epsilon)\,+\,O(e^7\,\epsilon)~~,\quad\quad
 \label{Til30_a}
 \ea
 \ba
 \left.\quad\quad~\quad\quad\right.=\,-\,\frac{15}{4}\,{G}\,M_{sec}^{\textstyle{^{2}}}\,R^{\textstyle{^{5}}}a^{-6}\sum_{q=-3~}^{3}{\sum_{\stackrel{\textstyle{^{~j=-3}}}{
 \textstyle{^{j~<~q}}}}^{3}}G_{30q}(e)\,G_{30j}(e)\,k_{\textstyle{_2}}\,\sin\left[{\cal{M}}\,(q-j)\right]+O(i^2\epsilon)+O(e\epsilon)~~,~\quad~\quad
 \label{Til30_b}
 \ea
 \label{Til30}
 \end{subequations}
 the expressions for the eccentricity functions being provided in Appendix \ref{functions}. The overall numerical factors in (\ref{Til2} -
 \ref{Til30}) are twice the numerical factors in (\ref{tilde}), because in (\ref{Til2} - \ref{Til30}) we have $\,j<q\,$ and not $\,j\neq\,
 q\,$. The right-hand sides of (\ref{Til2} - \ref{Til30}) contain $\,O(e\epsilon)\,$ instead of $\,O(e^7\epsilon)\,$, because at the final
 step we approximated $~\cos\left[\,{\cal{M}}\,(q-j)\,\right]\,k_{\textstyle{_l}}\,\sin\epsilon_{\textstyle{_{\textstyle{_{lmpq}}}}}-\,\sin\left[\,{\cal{M}}\,(q-j)\,
 \right]\,k_{\textstyle{_l}}\,\cos\epsilon_{\textstyle{_{\textstyle{_{lmpq}}}}}~$ simply with $~-\sin\left[\,{\cal{M}}\,(q-j)\,\right]\,
 k_{\textstyle{_l}}~$. Doing so, we replaced the cosine with unity, because the entire Darwin-Kaula formalism is a linear approximation in
 the lags. We also neglected $\,k_{\textstyle{_l}}\,\sin\epsilon_{\textstyle{_{\textstyle{_{lmpq}}}}}\,$ and kept only the leading term
 with $\,k_{\textstyle{_l}}\,$. This neglect would be illegitimate in the secular part of the torque, but is probably acceptable in the
 purely short-period part, because the latter part has a zero average and therefore should be regarded as a small correction even in its
 leading order. The latter circumstance also will justify approximation of $\,k_{\textstyle{_l}}=\,k_{\textstyle{_l}}(\chi)\,$ with
 $\,k_{\textstyle{_l}}(0)\,$ in (\ref{Til2} - \ref{Til30}).


\begin{thebibliography}{}


 \bibitem{} Alterman, Z.; Jarosch, H.; and Pekeris, C. 1959. ``Oscillations of the Earth." {\it{Proceedings of the Royal
            Society of London, Series A}}, Vol. {\bf{252}}, pp. 80 - 95.

 \bibitem{} Andrade, E. N. da C. 1910. ``On the Viscous Flow in Metals, and Allied Phenomena."
            {\emph{Proceedings of the Royal Society of London. Series A.}} Vol. {\bf{84}},
            pp. 1 - 12

  \bibitem{} Benjamin, D.; Wahr, J. ; Ray, R. D.; Egbert, G. D.; and Desai, S. D. 2006. ``Constraints on
            mantle anelasticity from geodetic observations, and implications for the $\,J_2\,$ anomaly."
            {\emph{Geophysical Journal International}}, Vol. {\bf{165}}, pp. 3 - 16

 \bibitem{} Bills, B. G.; Neumann, G. A.; Smith, D.E.; and Zuber, M.T.
            2005. ``Improved estimate of tidal dissipation within Mars
            from MOLA observations of the shadow of Phobos."
            {\it{Journal of Geophysical Research}}, Vol. {\bf{110}}, pp.
            2376 - 2406. ~~doi:10.1029/2004JE002376, 2005

 \bibitem{} Biot, M. A. 1954. ``Theory of Stress-Strain Relaxation in Anisotropic Viscoelasticity
            and Relaxation Phenomena." {\it{Journal of Applied Physics}}, Vol. {\bf{25}}, pp. 1385
            - 1391\\
            http://www.pmi.ou.edu/Biot2005/papers/FILES/054.PDF

 \bibitem{} Biot, M. A. 1958. {\it{Linear thermodynamics and the mechanics of solids.}}
            Proceedings of the Third US National Congress of Applied Mechanics,
            held at Brown University,  pp. 1 - 18. Published by ASME, NY, June 1958\\
            http://www.pmi.ou.edu/Biot2005/papers/FILES/076.PDF

 \bibitem{} Birger, B. I. 2007. ``Attenuation of Seismic Waves and the Universal Rheological
            Model of the Earth's Mantle." {\emph{Izvestiya. Physics of the Solid Earth.}}
            Vol. {\bf{49}}, pp. 635 - 641

 \bibitem{} Castillo-Rogez, J. 2009. ``New Approach to Icy Satellite Tidal Response Modeling."  American Astronomical Society, DPS meeting 41, 61.07.

 \bibitem{} Castillo-Rogez, J. C.; Efroimsky, M., and Lainey, V. 2011. ``The tidal history of Iapetus. Dissipative spin dynamics in the
            light of a refined geophysical model". {\it{Journal of Geophysical Research -- Planets}}, Vol. {\bf{116}}, p. E09008\\
            doi:10.1029/2010JE003664

 \bibitem{} Castillo-Rogez, J. C., and Choukroun, M. 2010. ``Mars' Low Dissipation Factor at 11-h. Interpretation from an Anelasticity-Based
           Dissipation Model." American Astronomical Society, DPS Meeting No 42, Abstract 51.02. {\it{Bulletin of the American Astronomical
           Society}}, Vol. {\bf{42}}, p. 1069


 \bibitem{} Churkin, V. A. 1998. ``The Love numbers for the models of inelastic
            Earth." Preprint No 121. Institute of Applied Astronomy.
            St.Petersburg, Russia. /in Russian/

 \bibitem{} Correia, A. C. M., and Laskar, J. 2009. ``Mercury's capture into the $\,3/2\,$ spin-orbit resonance including
            the effect of core-mantle friction." {\emph{Icarus}}, Vol. {\bf{201}}, pp. 1 - 11

 \bibitem{} Correia, A. C. M., and Laskar, J. 2004. ``Mercury's capture into the
            $\,3/2\,$ spin-orbit resonance as a result of its chaotic dynamics."
            {\emph{Nature}}, Vol. {\bf{429}}, pp. 848 - 850


 \bibitem{} Cottrell, A. H., and Aytekin, V. 1947.  ``Andrade's creep law and the flow of zinc crystalls."
            {\it{Nature}}, Vol. {\bf{160}}, pp. 328 - 329

 \bibitem{} Dahlen, F. A. 1976. ``The passive influence of the oceans upon rotation of the Earth."
       {\it{Geophys. Journal of the Royal Astronomical Society.}} Vol. {\bf{46}}, pp. 363 - 406

 \bibitem{} Darwin, G. H. 1879. ``On the precession of a viscous spheroid
            and on the remote history of the Earth." {\it{Philosophical
            Transactions of the Royal Society of London}}, Vol.
            {\bf{170}}, pp. 447 - 530\\
            http://www.jstor.org/view/02610523/ap000081/00a00010/

 \bibitem{} Darwin, G. H. 1880. ``On the secular change in the elements
            of the orbit of a satellite revolving about a tidally
            distorted planet." {\it{Philosophical Transactions of the
            Royal Society of London}}, Vol. {\bf{171}}, pp. 713 - 891\\
            http://www.jstor.org/view/02610523/ap000082/00a00200

 \bibitem{} Defraigne, P., and Smits, I. 1999. ``Length of day variations due to zonal tides for an inelastic
            earth in non-hydrostatic equilibrium." {\it{Geophysical J. International}}, Vol. {\bf{139}}, pp. 563 - 572

 \bibitem{} Dehant V. 1987a. ``Tidal parameters for an inelastic Earth." {\it{Physics of the Earth and Planetary
            Interiors}}, Vol. {\bf{49}}, pp. 97 - 116

 \bibitem{} Dehant V. 1987b. ``Integration of the gravitational motion equations for an elliptical uniformly rotating Earth
            with an inelastic mantle." {\it{Physics of the Earth and Planetary Interiors}}, Vol. {\bf{49}}, pp. 242 - 258



 \bibitem{} Duval, P. 1976. ``Temporary or permanent creep laws of polycrystalline ice for
            different stress conditions." {\it{Annales de Geophysique}}, Vol. {\bf{32}},
            pp. 335 - 350

 \bibitem{} Eanes, R. J. 1995. {\emph{A study of temporal variations in Earth's gravitational
            field using LAGEOS-1 laser ranging observations}}. PhD thesis, University of Texas at Austin

 \bibitem{} Eanes, R. J., and Bettadpur, S. V. 1996. ``Temporal variability of Earth's gravitational
            field from laser ranging." In: Rapp, R. H., Cazenave, A. A., and Nerem, R. S. (Eds.)
            {\emph{Global gravity field and its variations. Proceedings of the International Association
            of Geodesy Symposium No 116 held in Boulder CO in July 1995.}} IAG Symposium Series. Springer 1997\\
            ISBN: 978-3-540-60882-0

 \bibitem{} Efroimsky, M., and V. Lainey. 2007. ``The Physics of Bodily Tides in Terrestrial Planets, and the Appropriate Scales
            of Dynamical Evolution." {\emph{Journal of Geophysical Research -- Planets}}, Vol. {\bf{112}}, p. E12003.
            ~~~doi:10.1029/2007JE002908

 \bibitem{} Efroimsky, M., and Williams, J. G. 2009. ``Tidal torques. A critical
            review of some techniques." {\emph{Celestial mechanics
            and Dynamical Astronomy,}} Vol. {\bf{104}}, pp. 257 - 289 \\
            arXiv:0803.3299

 \bibitem{} Efroimsky, M. 2012. ``Tidal dissipation compared to seismic dissipation: in small bodies, earths,
            and superearths." {\it{the Astrophysical Journal}}, Vol. {\bf{746}}, No 2, p. 150\\
            doi:10.1088/0004-637X/746/2/150\\
            arXiv:1105.3936

 \bibitem{} Ferraz-Mello, S.; Rodr\'{\i}guez, A.; and Hussmann, H. 2008.
            ``Tidal friction in close-in satellites and exoplanets: The Darwin theory
            re-visited." {\emph{Celestial mechanics and Dynamical Astronomy,}}
            Vol. {\bf{101}}, pp. 171 - 201

 \bibitem{}  Fontaine, F. R.; Ildefonse, B.; and Bagdassarov, N. 2005.
           ``Temperature dependence of shear wave attenuation in
             partially molten gabbronorite at seismic frequencies."
             {\it{Geophysical Journal International}}, Vol. {\bf{163}}, pp. 1025 {-} 1038


 \bibitem{} Goldreich, P. 1963. ``On the eccentricity of the satellite orbits in the Solar System." {\it{Monthly Notices of the Royal
 Astronomical Society of London}}, Vol. {\bf{126}}, pp. 257 - 268




 \bibitem{} Gooding, R.H., and Wagner, C.A. 2008. ``On the inclination functions and a rapid stable procedure for their evaluation
            together with derivatives." {\it{Celestial Mechanics and Dynamical Astronomy}}, Vol. {\bf{101}}, pp. 247 - 272

 \bibitem{} Gribb, T.T., and Cooper, R.F. 1998. ``Low-frequency shear attenuation in polycrystalline
            olivine: Grain boundary diffusion and the physical significance of the Andrade model for
            viscoelastic rheology." {{Journal of Geophysical Research -- Solid Earth}}, Vol. {\bf{103}},
            pp. 27267 - 27279

 \bibitem{} Haddad, Y. M. 1995. {\it{Viscoelasticity of Engineering Materials.}}
            Chapman and Hall, London UK, p. 279

 \bibitem{} Hut, P. 1981. ``Tidal evolution in close binary systems."
            {\emph{Astronomy and Astrophysics}},
            Vol. {\bf{99}}, pp. 126 - 140

 \bibitem{} Karato, S.-i. 2008. {\it{Deformation of Earth Materials.
            An Introduction to the Rheology of Solid Earth}}. Cambridge University Press, UK.

 \bibitem{} Karato, S.-i., and Spetzler, H. A. 1990. ``Defect Microdynamics
            in Minerals and Solid-State Mechanisms of Seismic Wave
            Attenuation and Velocity Dispersion in the Mantle."
            {\it{Reviews of Geophysics}}, Vol. {\bf{28}}, pp. 399 - 423

 \bibitem{} Kaula, W. M. 1961.  ``Analysis of gravitational and geometric
            aspects of geodetic utilisation of satellites." {\it{The
            Geophysical Journal}}, Vol. {\bf{5}}, pp. 104 - 133

 \bibitem{} Kaula, W. M. 1964.  ``Tidal Dissipation by Solid Friction and
            the Resulting Orbital Evolution." {\it{Reviews of Geophysics}},
            Vol. {\bf{2}}, pp. 661 - 684

 \bibitem{} Kaula, W. M. 1966. {\emph{Theory of Satellite Geodesy: Applications of
            Satellites to Geodesy.}} Blaisdell Publishing Co, Waltham MA.
            (Re-published in 2006 by Dover. ISBN: 0486414655.)



 \bibitem{} Landau, L., and Lifshitz, E. M. 1986. {\it{The Theory of Elasticity.}}
            Pergamon Press, Oxford 1986.

 \bibitem{} Landau, L., and Lifshitz, E. M. 1987. {\it{Fluid Mechanics.}}
            Pergamon Press, Oxford 1987.

 \bibitem{} Legros, H.; Greff, M.; and Tokieda, T. 2006 ``Physics inside the Earth. Deformation and Rotation."
            {\it{Lecture Notes in Physics}}, Vol. {\bf{682}}, pp. 23 - 66. Springer, Heidelberg

 \bibitem{} Love, A. E. H. 1909. ``The Yielding of the Earth to Disturbing Forces." {\it{Proceedings of the Royal Society of London.
            Series A,}} Vol. {\bf{82}}, pp. 73 - 88

 \bibitem{} Love, A. E. H. 1911. {\it{Some problems of geodynamics.}} Cambridge University Press, London.
            ~~~Reprinted by: Dover, NY 1967.

 \bibitem{} MacDonald, G. J. F. 1964. ``Tidal Friction." {\it{Reviews of
            Geophysics.}} Vol. {\bf{2}}, pp. 467 - 541

 \bibitem{} Matsuyama, I., and Bills, B. G. 2010. ``Global contraction of planetary bodies due to despinning.
            Application to Mercury and Iapetus." {\it{Icarus}}, Vol. {\bf{209}}, pp. 271 - 279

 \bibitem{} McCarthy, C.; Goldsby, D. L.; and Cooper, R. F. 2007. ``Transient and Steady-State Creep
            Responses of Ice-I/Magnesium Sulfate Hydrate Eutectic Aggregates."
            {\emph{38th Lunar and Planetary Science Conference XXXVIII}}, held on 12 - 16 March 2007
            in League City, TX. LPI Contribution No 1338, p. 2429



 \bibitem{} Mignard, F. 1979. ``The Evolution of the Lunar Orbit Revisited. I."
           {\it{The Moon and the Planets.}} Vol. {\bf{20}}, pp. 301 - 315.

 \bibitem{} Mignard, F. 1980. ``The Evolution of the Lunar Orbit Revisited. II."
           {\it{The Moon and the Planets.}} Vol. {\bf{23}}, pp. 185 - 201

 \bibitem{} Miguel, M.-C.; Vespignani, A.; Zaiser, M.; and Zapperi, S. 2002.
          ``Dislocation Jamming and Andrade Creep." {\emph{Physical Review Letters}},
           Vol. {\bf{89}}, pp. 165501 - 1655

 \bibitem{} Mitchell, B. J. 1995. ``Anelastic structure and evolution of the
            continental crust and upper mantle from seismic surface wave
            attenuation." {\emph{Reviews of Geophysics}}, Vol. {\bf{33}},
            No 4, pp. 441 - 462.

  \bibitem{} Munk, W. H., and MacDonald, G. J. F. 1960. {\it{The rotation of the earth; a geophysical discussion.}} Cambridge University Press, 323 pages.




\bibitem{} Nakamura, Y.; Latham, G.; Lammlein, D.; Ewing, M.; Duennebier, F.; and Dorman, J. 1974.
            ``Deep lunar interior inferred from recent seismic data."
            {\it{Geophysical Research Letters}}, Vol. {\bf{1}}, pp. 137 - 140

 \bibitem{} Nechada, H.; Helmstetterb, A.; El Guerjoumaa, R.; and Sornette, D. 2005.
          ``Andrade and critical time-to-failure laws in fiber-matrix composites.
            Experiments and model." {\emph{Journal of the Mechanics and Physics of solids}},
            Vol. {\bf{53}}, pp. 1099 - 1127.

 \bibitem{} Petit, G., and Luzum, B. (Eds.) 2010. ``IERS Conventions 2010. Technical Note No 36."
            Verlag des Bundesamts f{\"{u}}r Kartographie und Geod{\"{a}}sie.
            Frankfurt am Main 2010.\\
            http://www.iers.org/TN36/

 \bibitem{} Rambaux, N.; Castillo-Rogez, J. C.; Williams, J. G.; and Karatekin, \"{O}. 2010. ``The librational response of Enceladus."
            {\it{Geophysical Research Letters}}, Vol. {\bf{37}}, p. L04202\\
            doi:10.1029/2009GL041465

 \bibitem{} Remus, F.; Mathis, S.; and Zahn, J.-P. 2012a. ``The Equilibrium Tide in Stars and Giant Planets. I - The Coplanar Case."
            Submitted to: {\it{Astronomy \& Astrophysics}}

 \bibitem{} Remus, F.; Mathis, S.; Zahn, J.-P.; and Lainey, V. 2012b. ``Anelastic tidal dissipation in multi-layers planets."
            Submitted to: {\it{Astronomy \& Astrophysics}}

 \bibitem{} Remus, F.; Mathis, S.; Zahn, J.-P.; and Lainey, V. 2011. ``The elasto-viscous equilibrium tide in exoplanetary systems."
            EPSC-DPS Joint Meeting 2011, Abstract 1372.

 \bibitem{} Rodr{\'{\i}}guez; A., Ferraz Mello, S.; and Hussmann, H. 2008.
            ``Tidal friction in close-in planets."
            \\ In:
            Y.S.Sun, S.Ferraz-Mello and J.L.Zhou (Eds.) {\emph{Exoplanets:
            Detection, Formation and Dynamics. Proceedings of the IAU
            Symposium No 249,}} pp. 179 - 186\\
            doi:10.1017/S174392130801658X

 \bibitem{} Sabadini, R., and Vermeersen, B. 2004. {\it{Global Dynamics of the Earth: Applications
            of Normal Mode Relaxation Theory to Solid-Earth Geophysics.}} Kluwer, Dordrecht 2004

 \bibitem{} Shito, A.; Karato, S.-i.; and Park, J. 2004. ``Frequency
            dependence of $Q$ in Earth's upper mantle, inferred from
            continuous spectra of body wave." {\emph{Geophysical Research
            Letters}}, Vol. {\bf{31}}, No 12, p. L12603,
            doi:10.1029/2004GL019582

 \bibitem{} Singer, S. F. 1968. ``The Origin of the Moon and Geophysical Consequences." {\it{The
            Geophysical Journal of the Royal Astronomical Society}}, Vol. {\bf{15}}, pp. 205 - 226

 \bibitem{} Smith, M. 1974. ``The scalar equations of infinitesimal elastic-gravitational motion for a rotating, slightly
            elliptical Earth." {\it{The Geophysical Journal of the Royal Astronomical Society}}, Vol. {\bf{37}}, pp. 491 - 526

 \bibitem{} Stachnik, J. C.; Abers, G. A.; and Christensen, D. H. 2004. ``Seismic attenuation and mantle wedge temperatures in
            the Alaska subduction zone." {\emph{Journal of Geophysical Research -- Solid Earth}}, Vol. {\bf{109}}, No B10, p.
            B10304, doi:10.1029/2004JB003018

 \bibitem{} Tan, B. H.; Jackson, I.; and Fitz Gerald J. D. 1997.
          ``Shear wave dispersion and attenuation in fine-grained
            synthetic olivine aggregates: preliminary results."
            {\emph{Geophysical Research Letters}}, Vol. {\bf{24}},
            No 9, pp. 1055 - 1058, doi:10.1029/97GL00860

 \bibitem{} Taylor, P. A., and Margot, J.-L. 2010. ``Tidal evolution of close binary asteroid systems." {\it{Celestial Mechanics and
            Dynamical Astronomy}}, Vol. {\bf{108}}, pp. 315 - 338


  \bibitem{} Wahr, J.M. 1981a. ``A normal mode expansion for the forced response of a rotating Earth."
  {\it{The Geophysical Journal of the Royal Astronomical Society}}, Vol. {\bf{64}}, pp. 651 - 675

  \bibitem{} Wahr, J.M. 1981b. ``Body tides on an elliptical, rotating, elastic and oceanless Earth."
  {\it{The Geophysical Journal of the Royal Astronomical Society}}, Vol. {\bf{64}}, pp. 677 - 703

  \bibitem{} Wahr, J.M. 1981c. ``The forced nutations of an elliptical, rotating, elastic and oceanless Earth."
            {\it{The Geophysical Journal of the Royal Astronomical Society}}, Vol. {\bf{64}}, pp. 705 - 727

 \bibitem{} Weertman, J., and Weertman, J. R. 1975. ``High Temperature Creep of Rock and Mantle
             Viscosity." {\it{Annual Review of Earth and Planetary Sciences}}, Vol. {\bf{3}}, pp. 293 - 315

 \bibitem{} Weber, R. C.; Lin, Pei-Ying; Garnero, E.; Williams, Q.; and Lognonn\'{e}, P. 2011. ``Seismic Detection of the Lunar Core."
            {\it{Science}}, Vol. {\bf{331}}, Issue 6015, pp. 309 - 312

 \bibitem{} Williams, J. G., Boggs, D. H., Yoder, C. F., Ratcliff, J. T., and Dickey, J. O. 2001. ``Lunar rotational dissipation in
            solid-body and molten core." {\emph{The Journal of Geophysical Research -- Planets}}, Vol. {\bf{106}}, No E11, pp. 27933 - 27968.
            doi:10.1029/2000JE001396

 \bibitem{} Williams, J. G., Boggs, D. H., and Ratcliff, J. T. 2008. ``Lunar Tides, Fluid Core and Core/Mantle Boundary."
            The 39th Lunar and Planetary Science Conference, (Lunar and Planetary Science XXXIX), held on 10-14 March 2008 in
            League City, TX. LPI Contribution No. 1391., p. 1484\\
            http://www.lpi.usra.edu/meetings/lpsc2008/pdf/1484.pdf

 \bibitem{} Williams, J. G., and Boggs, D. H. 2009. ``Lunar Core and Mantle. What Does LLR See?"\\
 Proceedings of the 16th International Workshop on Laser Ranging, held on 12-17 October 2008 in Poznan, Poland. Edited by S. Schilliak.
 pp. 101 - 120\\
 http://cddis.gsfc.nasa.gov/lw16/docs/papers/sci$\_$1$\_$Williams$\_$p.pdf\\
 http://cddis.gsfc.nasa.gov/lw16/docs/papers/proceedings$\_$vol2.pdf

 \bibitem{} Williams, J. G., and Efroimsky, M. 2012. ``Bodily tides near the $\,1:1\,$ spin-orbit resonance. Correction to Goldreich's
            dynamical model." Submitted to {\it{Celestial Mechanics and Dynamical Astronomy}}.  ~~
            arXiv:........

 \bibitem{} Yoder, C. 1982. ``Tidal Rigidity of Phobos".
             {\emph{Icarus}}, Vol. {\bf{49}}, pp. 327 - 346

 \bibitem{} Zahn, J.-P. 1966. ``Les marées dans une étoile double serrée." {\it{Annales d'Astrophysique,}} Vol. {\bf{29}}, pp. 313 - 330

 \bibitem{} Zharkov, V.N., and Gudkova, T.V. 2009. ``The period and $Q$ of the Chandler wobble of
            Mars." {\it{Planetary and Space Science}}, Vol. {\bf 57}, pp. 288 - 295

 \bibitem{} Zschau, J. 1978. ``Phase shifts of tidal load deformations of the Earth's surface due to low-viscosity layers in the interior."\\
             In: ~M. Bonatz (Ed.), {\it{Proceedings of the 8th International Symposium on Earth Tides, held in Bonn on 19-24 September 1977}}


 \end{thebibliography}
 \end{document}